\documentclass[reprint,
superscriptaddress,
showpacs,
nofootinbib,
longbibliography,
prc,
aps,
floatfix,
]{revtex4-1}

\usepackage{amsmath,amssymb,amsfonts,amsthm}
\usepackage[english]{babel}
\usepackage{url}
\usepackage{bm}
\usepackage[latin1]{inputenc}
\usepackage{graphicx,rotating}
\usepackage[colorlinks=true,linkcolor=black, citecolor=blue]{hyperref}
\usepackage{booktabs}
\usepackage{multirow, array}
\usepackage{dcolumn}

\begin{document}
\title{First-forbidden transitions in the reactor anomaly}
\author{L. Hayen}
\email[Corresponding author: ]{leendert.hayen@kuleuven.be}
\affiliation{Instituut voor Kern- en Stralingsfysica, KU Leuven, Celestijnenlaan 200D, B-3001 Leuven, Belgium}

\author{J. Kostensalo}
\affiliation{Department of Physics, University of Jyv\"askyl\"a, P.O. Box 35, FI-40014 University of Jyv\"askyl\"a, Finland}

\author{N. Severijns}
\affiliation{Instituut voor Kern- en Stralingsfysica, KU Leuven, Celestijnenlaan 200D, B-3001 Leuven, Belgium}

\author{J. Suhonen}
\affiliation{Department of Physics, University of Jyv\"askyl\"a, P.O. Box 35, FI-40014 University of Jyv\"askyl\"a, Finland}

%
%
%
%

\date{\today}
\begin{abstract}
We describe here microscopic calculations performed on the dominant forbidden transitions in reactor antineutrino spectra above 4 MeV using the nuclear shell model. By taking into account Coulomb corrections in the most complete way, we calculate the shape factor with the highest fidelity and show strong deviations from allowed approximations and previously published results. Despite small differences in the ab initio electron cumulative spectra, large differences on the order of several percents are found in the antineutrino spectra. Based on the behaviour of the numerically calculated shape factors we propose a parametrization of forbidden spectra. Using Monte Carlo techniques we derive an estimated spectral correction and uncertainty due to forbidden transitions. We establish the dominance and importance of forbidden transitions in both the reactor anomaly and spectral shoulder analysis. Based on these results, we conclude that a correct treatment of forbidden transitions is indispensable in both the normalization anomaly and spectral shoulder.
\end{abstract}

\maketitle


\section{Introduction}
\label{sec:introduction}
The field associated with short baseline reactor neutrinos has seen tremendous activity in recent years. Faced both with long-standing issues (LNSD \cite{Athanassopoulos1998, Conrad2013} and GALLEX \& SAGE \cite{Kaether2010} collaborations) and more recently the reactor antineutrino anomaly (RAA) \cite{Aguilar-Arevalo2018a, Mention2011}, phenomenology proposes the existence of sterile neutrinos in an effort to solve these issues \cite{Abazajian2012, Jordan2019}. Besides the normalization anomaly, a spectral disagreement commonly referred to as the `5 MeV bump' remains after several years of intense work \cite{Hayes2015, Huber2016a, An2016, Abe2014, Seo2018}. Due to the magnitude of the problem in several regards, nuclear theory is pushing the boundaries in getting to grips with theoretical predictions and uncertainties \cite{Hayes2016}. 

A central element in the theoretical determination of the antineutrino flux is the theoretical shape of individual $\beta$ spectra. The original treatments by Huber and Mueller \emph{et al.} \cite{Huber2011, *Huber2012, Mueller2011} introduced strong approximations in their treatments of forbidden transitions. In the years following, the influence of forbidden transitions has, however, been discussed mostly in general terms \cite{Hayes2014}, with microscopic calculations performed only on 3 nuclei \cite{Fang2015}. While both of these studies showed a significant influence on the final result within the context of the RAA, its calculational difficulty presents a serious challenge for a more complete analysis. 

Following our earlier work \cite{Hayen2019}, we discuss here the result of a shell model calculation of the dominant forbidden transitions above 4 MeV.  This work represents both a more thorough explanation and discussion of our earlier work and an extension as more data was included and more sophisticated methods employed. We start off in Sec. \ref{sec:formalism} by revisiting the used formalism, and describe both the included corrections in this work and the breakdown of approximations made in the literature. We review the proper expressions for allowed shape factors and discuss several terms which are missing in previous descriptions and note their significance. In Sec. \ref{sec:data_selection_handling} we describe our selection and treatment of nuclear databases. We go on to describe the direct results of these calculations in Secs. \ref{sec:shape_factor_calc} and \ref{sec:results_numerical}, including an estimate of its uncertainties. We compare our findings to common approximations found in the literature and find strongly diverging results, which we interpret as the breakdown of approximations in the formalism of Sec. \ref{sec:formalism}. Further, in Sec. \ref{sec:generalization} we attempt an expansion of the numerical results in a statistical fashion and perform improved summation calculations. Finally, in Sec. \ref{sec:reactor_changes} we report on the consequences on both the reactor normalization anomaly and the spectral shoulder for the current generation of reactor antineutrino experiments.

\section{$\beta$ decay Formalism}
\label{sec:formalism}
The treatment of (forbidden) $\beta$ decays is a complex task, compounded by the large proton number of the fission fragments of interest. Its final description is an interplay between kinematic, nuclear and Coulomb terms with significant potential for cancellations. This leads to a wide variety of potential spectrum shapes and it serves one well to go back to the starting point of the $\beta$ decay description. Our discussion here will be relatively extensive since no such overview is currently present in the literature surrounding the RAA, even though a correct analysis hinges critically on a correct assessment of all intricacies and moving parts. In the case of forbidden decays this fact is amplified, as will become clear in later sections. All results are written in units natural for $\beta$ decay, i.e. $\hbar=c=m_e=1$, unless explicitly mentioned. 

Employing the usual Fermi contact interaction, the correct generalization of the tree-level transition matrix element in the presence of electromagnetic effects is given by \cite{Stech1964, Holstein1979b}
\begin{align}
    \mathcal{M}_{fi} &= \int \mathrm{d}^3 r\, \bar{\phi}_e(\vec{r}, \vec{p}_e)\gamma^\mu(1+\gamma^5)v(\vec{p}_{\bar{\nu}}) \nonumber \\
    & \times \int \frac{\mathrm{d}^3s}{(2\pi)^3}e^{i\vec{s}\cdot \vec{r}}\frac{1}{2}[\langle f(\vec{p}_f+\vec{p}_e-\vec{s})| V_\mu + A_\mu | i(\vec{p}_i) \rangle \nonumber \\
    & + \langle f(\vec{p}_f) | V_\mu + A_\mu | i(\vec{p}_i-\vec{p}_e+\vec{s}) \rangle ].
    \label{eq:generalized_matrix_element_SS}
\end{align}
where $\bar{\phi}_e$ is the solution to the Dirac equation in the static Coulomb potential of the final state and $V_\mu + A_\mu$ is the usual weak interaction current. Equation \ref{eq:generalized_matrix_element_SS} reveals two important, intertwined contributions: ($i$) Nuclear structure effects encoded in the weak interaction current in the inner integral; ($ii$) Coulomb influences represented by the outer integral through the use of the electron wave function in the static Coulomb potential of the final state.

Direct consequences of this form are a renormalization of the matrix element from extraction of the electron density at the origin\footnote{Rigorously, it corresponds to the extraction of the large components of of the $j=1/2$ ($s_{1/2}$ and $p_{1/2}$) wave functions at the origin. Small components, higher-$j$ components, radial dependence, etc. are commonly noted by `finite size corrections' which appear later in this work and are extensively discussed elsewhere \cite{Hayen2018}.}, $|\phi_e(0, \vec{p}_e)|^2$, resulting in the usual Fermi function. The electron continuum wave function varies significantly within the nuclear volume, however, so that its radial dependence couples directly to that of the nuclear weak interaction current. Besides the Fermi function then, the traditional nuclear structure terms can be modified significantly for higher $Z$ through the convolution with the electron continuum wave function. We will discuss the influence of both of these separately.

Combining Eq. (\ref{eq:generalized_matrix_element_SS}) with the available phase space, the $\beta$ spectrum shape is traditionally written as
\begin{align}
    \frac{dN}{dW} = &\frac{G_V^2V_{ud}^2}{2\pi^3}pW(W-W_0)^2 \nonumber \\
    &\times F(Z, W) C(Z, W) K(Z, W)
    \label{eq:spectrum_shape}
\end{align}
with $W = E_{kin}/m_ec^2 + 1$ the total electron energy in units of its rest mass, $W_0$ the spectral endpoint, $p = \sqrt{W^2-1}$ the electron momentum in units of $m_ec$, $Z$ the atomic number of the final state, $F(Z, W)$ the well-known Fermi function, $C(Z, W)$ the so-called shape factor and $K(Z, W)$ higher-order correction terms \cite{Hayen2018}. All nuclear structure information resides in the shape factor, $C$, which depends primarily on the degree of forbiddenness of the decay. The Fermi function and higher-order corrections are known to a sufficient level \cite{Hayes2016, Hayen2018}, making the shape factor, $C$, the primary target in this work.

One can generally write the shape factor as \cite{Behrens1971, Behrens1982}
\begin{align}
C(Z, W) = &\sum_{k_e,k_{\nu}, K}\lambda_{k_e}\left\{M_K^2(k_e, k_{\nu})+m_K^2(k_e,k_{\nu}) \right.\nonumber \\
&\left. - \frac{2\mu_{k_e}\gamma_{k_e}}{k_eW}M_K(k_e,k_{\nu})m_K(k_e,k_{\nu})\right\},
\label{eq:C_BB}
\end{align}
where
\begin{align}
\lambda_{k_e} &= \frac{\alpha^2_{-k_e}+\alpha^2_{+k_e}}{\alpha^2_{-1}+\alpha^2_{+1}}, \label{eq:lambda_k}\\
\mu_{k_e} &= \frac{\alpha^2_{-k_e}-\alpha^2_{+k_e}}{\alpha^2_{-k_e}+\alpha^2_{+k_e}}\frac{k_eW}{\gamma_{k_e}},
\label{eq:mu_k}
\end{align}
are Coulomb functions depending on the so-called Coulomb amplitudes $\alpha_\kappa$, which encode the value of the electron wave function at the origin. The integers $k_e, k_{\nu}$ are defined as $|\kappa_{e,\nu}|$ where $\kappa_{e, \nu}$ is related to the angular momenta in the usual way\footnote{Here $\kappa$ is the eigenvalue of the operator $K = \beta (\bm{\sigma} \bm{L} + 1)$, such that $k=|\kappa|=j+\frac{1}{2}, \kappa=-l-1$ if $l=j+\frac{1}{2}$, and $\kappa=l$ if $l=j-\frac{1}{2}$.}. Contributions from different $k_{e,\nu}$ come from the expansion of the lepton wave functions in spherical waves. The integer $K$ corresponds to the multipolarity of the relevant nuclear current, and must form a vector triangle with $j_e$ and $j_{\nu}$ as well as with the nuclear spins $J_i$ and $J_f$. We have then $|J_i-J_f| \leq K \leq J_i+J_f$ from the nuclear vector triangle. Finally, $M_K(k_e, k_{\nu})$ and $m_K(k_e, k_{\nu})$ contain the convolution of leptonic wave functions and nuclear structure information encoded as form factors. Appropriately, the capital letter contribution contains the dominant terms, so that typically one neglects the second term in Eq. (\ref{eq:C_BB}).

In conclusion then, the shape factor, $C$, as defined in Eqs. (\ref{eq:spectrum_shape}) and (\ref{eq:C_BB}) depends on three things: ($i$) the spin-change of the transitions and the corresponding appearance of kinematical factors ($W, p$) and form factors; ($ii$) finite size corrections proportional to $R^n$ resulting from the integration over the nuclear volume; ($iii$) Coulomb corrections proportional to $(\alpha Z)^n$ resulting from the expansion of the electron wave function. The final shape factor will be a combination of all three with various cross-terms.

\subsection{Nuclear structure}
In contrast to their nomenclature, so-called forbidden transitions correspond to $\beta$ decays for which the main Fermi and Gamow-Teller matrix elements are identically zero due to spin-parity requirements or internal nuclear structure. As a consequence, their decays are perpetuated by matrix elements that are typically strongly suppressed and are consequently heavily dependent on nuclear structure effects and prone to accidental cancellations. 

We briefly review a scheme to systematically classify their behaviour, the so-called \textit{elementary particle treatment}. This entails that initial and final nuclear states are treated as fundamental particles and all interaction dynamics is encoded through form factors which obey angular momentum conservation, $F(q^2)$, with $q$ the momentum transfer between initial and final nuclear states. It shines in the case of nuclear decays because of the (near-)spherical symmetry of the system at hand and the smallness of the momentum transfer with respect to the nuclear mass. The latter means that we are usually only concerned with the form factors near zero momentum transfer, $F(0) \equiv F$. The former implies that through conservation of angular momentum one can construct a multipole decomposition of both the nuclear and leptonic currents in terms of (vector) spherical harmonics for the timelike (spacelike) component. In the Behrens-B\"uhring formalism that we follow here \cite{Behrens1982}, this allows one to label the nuclear structure form factors using three numbers: $K$, $L$ and $s$, being the total and orbital angular momentum of the nuclear current and its timelike (0) or spacelike (1) nature, respectively. The form factors are denoted by ${}^{V/A}F_{KLs}$. The three quantum numbers form a vector triangle, and the parity requirement can be summarized as
\begin{equation}
\left.
\begin{array}{ll}
    \pi_i\pi_f = (-)^{L+s}  &  \text{vector contributions},\\
    \pi_i\pi_f = (-)^{L+s+1} & \text{axial vector contributions},
\end{array}
\right\}
\end{equation}
where $\pi$ is the parity of initial and final nuclear state. Conservation of angular momentum then limits the number of contributing form factors for a specific transition with spin-parity change $\Delta J^\pi$.

In this work we focus on first-forbidden $\beta$ transitions, i.e. $\Delta J = 0,1,2$ and $\pi_i\pi_f = -1$. To first order this limits the number of form factors to 6. In order to proceed with an actual calculation, each of these must be translated into nuclear matrix elements, ${}^{V/A}\mathcal{M}_{KLs}$. This is commonly done by introducing the impulse approximation, in which all nucleons inside a nucleus are treated as independent particles in a mean-field potential. This neglects multi-particle correlations and meson exchange effects, the effects of which are put in manually through effective interactions in the usual shell-model fashion \cite{Warburton1994, Baumann1998}. We briefly report on the expected shape factors for different $\Delta J$.

For a pure pseudoscalar transition ($0^+ \leftrightarrow 0^-$) only two form factors contribute. It is dominated by ${}^AF_{000}$ which translates into the traditional pseudoscalar matrix element ${}^A\mathcal{M}_{000} = - g_A \int \gamma_5$, and receives first-order corrections from ${}^AF_{011} \longrightarrow {}^A\mathcal{M}_{011} = - g_A \int i(\bm{\sigma}\cdot \bm{r})/R$. Here $R$ is the nuclear radius and is $\mathcal{O}(10^{-2})$ in our units. The shape factor can then be written as
\begin{equation}
    C_{0^-} \propto 1 + \frac{2R}{3W}b + \mathcal{O}(\alpha Z R, W_0R^2)
    \label{eq:C_pseudoscalar}
\end{equation}
after extraction of the main matrix element. Here $b = {}^A\mathcal{M}_{011}^{(0)}/{}^A\mathcal{M}_{000}^{(0)} \sim \mathcal{O}(-1)$ and $\alpha$ is the fine-structure constant.

Moving on to a pure pseudovector transition ($1^{+(-)} \leftrightarrow 0^{-(+)}$), three matrix elements contribute significantly and it is \textit{a priori} not possible to establish a hierarchy leading to an analogue of Eq. (\ref{eq:C_pseudoscalar}). Instead, we write
\begin{equation}
    C_{1^-} \propto 1 + aW + \mu_1 \gamma_1 \frac{b}{W} + cW^2
    \label{eq:C_pseudovector}
\end{equation}
inspired by the general form of Eq. (\ref{eq:C_BB}), where $a, b, c$ are free parameters.

In the case of unique forbidden decays, only one form factor contributes to first order and Eq. (\ref{eq:C_BB}) simplifies significantly, so that one is left with
\begin{equation}
    C_U \propto \sum_{k=1}^L\lambda_k \frac{p^{2(k-1)}q^{2(L-k)}}{(2k-1)![2(L-k)+1]!},
    \label{eq:C_unique_forbidden}
\end{equation}
after extraction of the prefactor, where $L$ is the maximum angular momentum change.


\subsection{Coulomb corrections}
The shape factor of Eq. (\ref{eq:C_BB}) is a result of the convolution of the leptonic and nucleonic wave functions written in Eq. (\ref{eq:generalized_matrix_element_SS}). The change due to the leptonic wave function, $\bar{\phi}_e$, resulting from the Coulomb interaction can be seen as ($i$) a renormalization at the origin, and ($ii$) a modified radial behaviour inside the nuclear volume. We discuss both in turn.

\subsubsection{Static Coulomb renormalization}
As noted at the start of this section, one traditionally extracts the large components of the $j=1/2$ electron wave function at the origin, denoted by $\alpha_\kappa$ in the Behrens-B\"uhring formalism. Here $\kappa$ takes the values $-1$ ($s_{1/2}$) or $+1$ ($p_{1/2}$) so that the Fermi function is defined as
\begin{equation}
    F_0(Z, W) = \frac{\alpha_{-1}^2+\alpha_{+1}^2}{2p^2}.
    \label{eq:Fermi_function_BB}
\end{equation}
Corrections from the small components or higher-$j$ components then introduce the $\mu_{k_e}$ (Eq. (\ref{eq:mu_k})) and $\lambda_{k_e}$ (Eq. (\ref{eq:lambda_k})) functions, respectively. In the region of interest, it is safe to set $\mu_{k_e}$ to unity \cite{Behrens1969} so that we focus our attention instead on $\lambda_{k_e}$. For a point-charge nucleus, it can be written as \cite{Behrens1982}
\begin{equation}
    \lambda_k = \frac{F_{k-1}}{F_0}\frac{k+\gamma_k}{k(1+\gamma_1)}
\end{equation}
where $k=|\kappa|$ and 
\begin{equation}
    \gamma_k = \sqrt{k^2-(\alpha Z)^2}
\end{equation}
is the generalized $\gamma$ parameter and
\begin{align}
    F_{k-1} = &[k(2k-1)!!]^2 4^k (2pR)^{2(\gamma_k-k)}\exp (\pi y)\nonumber \\
    &\times \left[|\Gamma(\gamma_k + iy)|/\Gamma(1+2\gamma_k) \right]^2
\end{align}
the generalized Fermi function, and
\begin{equation}
    y = \frac{\alpha Z W}{p}.
\end{equation}
While its influence is negligible in allowed decays except for extreme cases, it features quite prominently in forbidden transitions. The value of $\lambda_2$, for example, can exceed $10$ for very low momenta and does not converge to unity at large momenta like the Fermi function \cite{Behrens1969}. We explicitly discuss its influence in the following section in the context of unique decays.

\subsubsection{Coulombic convolution distortion}
Beyond the renormalization of the electron wave function at the origin, the radial behavior near the nucleus becomes modified due to the Coulomb potential. As the potential grows deeper with increasing $Z$, the electron density is greatly increased within the nuclear volume, so that the shape of the nuclear charge density also plays a role. One expands the electron wave function in terms of $(m_eR)^a, (WR)^b$ and $(\alpha Z)^c$, where the details of the Coulomb potential are encoded in functions\footnote{Here $m = a + b + c$ represents the total power of $mR$, $WR$, and $\alpha Z$, $n=b+c$ is the total power of $WR$ and $\alpha Z$, and $\rho$ is the power of $\alpha Z$. One has trivially that $I(k_e,m,n,0) = 1$.} $I(k_e, m, n, \rho)$. Following the result of Eq. (\ref{eq:generalized_matrix_element_SS}) this requires a generalization of the nuclear form factors and matrix elements according to the following notation
\begin{equation}
\mathcal{M}_{KLs}^{(n)} \longrightarrow \mathcal{M}_{KLs}^{(n)}(k_e, m, n, \rho)
\end{equation}
where now
\begin{align}
\mathcal{M}_{KLs}^{(n)}(k_e, m, n, \rho) = \int &dr ~ r^2 \phi_f(r) O_{KLs}^{(n)} \nonumber \\
&\times I(k_e, m, n, \rho; r) \phi_i(r),
\end{align}
where $O_{KLs}^{(n)}$ is the relevant operator and $\phi_{i.f}$ represent initial and final nuclear wave functions. The Coulomb shape functions, $I(k_e, m, n, \rho; r)$, are tabulated in \cite{Behrens1971} and depend on the charge distribution of the nucleus.
Terms with large values for $m,n,$ or $\rho$ are typically strongly suppressed, resulting in rather slight modifications of the main matrix elements. The modified matrix elements enter the shape factor of Eq. (\ref{eq:C_BB}), however, accompanied by factors of $\alpha Z$ and $W_0R$ resulting from the electron Coulomb-corrected wave function expansion. As such, the additional terms for nuclei in the fission fragment region are highly non-negligible. In case of cancellation effects, these Coulomb terms can even become the dominant contributions for the shape factor.

\subsection{Breakdown of usual approximations}
\label{sec:approximation_breakdown}
Some general remarks are essential at this point in order to both understand previous approximations and their breakdown discussed below.
\begin{itemize}
    \item Eq. (\ref{eq:C_pseudoscalar}) was derived assuming a \textit{pure} pseudoscalar transition. Many $\Delta J^\pi = 0^-$ transitions occur, however, between higher-spin partners meaning higher-order matrix elements can equally contribute. This can significantly change the energy dependence. Analogously, pseudovector contributions  can contain contributions from $\Delta J = 2$ matrix elements.
    \item Neglecting the electron mass and Coulomb interaction, Eq. (\ref{eq:C_unique_forbidden}) is symmetric when interchanging electron and antineutrino energies. This has been used as an argument to neglect forbidden transitions within the context of the RAA \cite{Huber2011, *Huber2012, Mueller2011}. This argument is invalid, however, for non-unique transitions (Eqs. (\ref{eq:C_pseudoscalar}) \& (\ref{eq:C_pseudovector})) which occur more frequently than anticipated as we shall see in Sec. \ref{sec:generalization}. Additionally, we will show explicitly that Coulomb corrections significantly distort the shape factor, breaking the purported symmetry, even for unique transitions.
\end{itemize}

We discuss the breakdown of the usual approximations both for non-unique transitions and unique transitions, which typically occur for different reasons.

\subsubsection{Non-unique forbidden transitions: The $\xi$ approximation}

In general the shape factor for non-unique decays is governed by 4 to 6 matrix elements for pseudovector and pseudoscalar transitions, respectively. It has long been known, however, that only certain linear combinations appear. Some of these contain the so-called Coulomb energy, $\alpha Z/R \equiv 2\xi$, its large magnitude making it useful as an expansion parameter\footnote{Remember that in our choice of units $R \sim \mathcal{O}(0.01)$.}. In the so-called $\xi$ approximation, one retains the shape factor only to order $\xi^2$ \cite{Morita1958, Schopper1966}. The particular benefit of this approximation is that it leaves the shape factor mostly energy-independent, as all kinematical terms contain lower powers of $\xi$. This leaves all quantities (such as the spectrum shape, the $\beta$-$\gamma$ correlation, etc.), equal to the results of allowed transitions up to order $1/\xi \sim 10\%$. Based on the general formulation of Eq. (\ref{eq:C_pseudovector}), it is a valid approximation when
\begin{equation}
    2 \xi = \frac{\alpha Z}{R} \gg W_0,
    \label{eq:xi_approximation}
\end{equation}
where $W_0$ is the endpoint of the transition. For the relevant fission fragments, however, this approximation is of questionable worth in the experimentally accessible regime. Using typical values for $Z$ encountered in a nuclear reactor one obtains rather $\alpha Z / 2R \sim W_0$ for endpoint energies of a few MeV. Substantial changes are expected to occur based on this breakdown alone. It is well-known, however, that even though Eq. (\ref{eq:xi_approximation}) might hold, the $\xi$ approximation can fail \cite{Kotani1959}. This is either due to cancellation effects, or through selection rules originating from the underlying nuclear structure and collective behaviour. We will demonstrate several examples of this occurrence in our discussion of the numerical results in Sec. \ref{sec:shape_factor_calc}.

\subsubsection{Unique transitions: Coulomb functions}
\begin{figure}[!ht]
    \centering
    \includegraphics[width=0.48\textwidth]{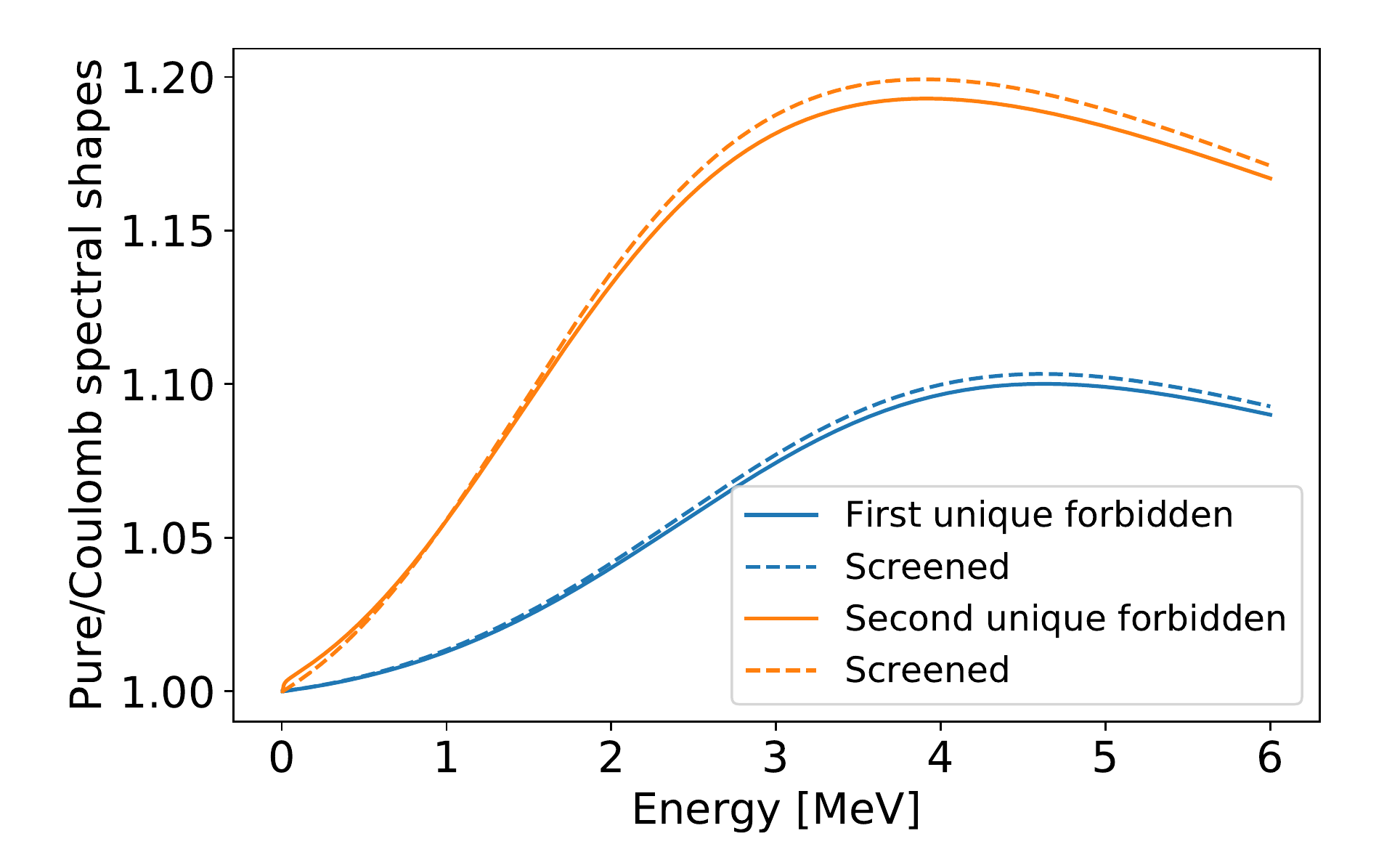}
    \caption{Change in the unique forbidden spectral shape when using the appropriate $\lambda_k$ Coulomb functions instead of approximating them as unity for a $\beta$ transition with $Z=50$ and $E_0=6$\,MeV. Full lines represent unscreened ratios, while dashed lines represent the screened ratios for the different $\lambda_k$.}
    \label{fig:spectral_change_coulomb_unique}
\end{figure}

Unique transitions have a particular simplicity as only one matrix element contributes to first order. Its shape factor for first forbidden transitions is simply
\begin{equation}
    C_{2^-} \propto p_\nu + \lambda_2p_e^2
    \label{eq:C_first_unique}
\end{equation}
where $p_\nu = W_0-W$ is the antineutrino momentum. As mentioned before, $\lambda_2$ (Eq. (\ref{eq:lambda_k})) is a Coulomb function of order unity. On the percent level precision, however, setting it to unity is unsatisfactory for the region of interest for the RAA. As an example, we consider the change in the spectrum shape due to the influence of these Coulomb functions on first and second unique forbidden decays. We consider a fictional transition in the region of interest, with $Z=50$ and endpoint energy $E_0 = 6$\,MeV. The relative change in spectral shapes can be seen in Fig. \ref{fig:spectral_change_coulomb_unique}, where we normalize the shape factor to unity at the start of the spectrum. Here we included, in addition, the results when introducing screening corrections to the Coulomb functions as described in Ref. \cite{Buhring1984}.

As can be observed, besides the clear deviation from allowed shapes in the parabolic expression of Eq. (\ref{eq:C_first_unique}), setting $\lambda_2$ to unity introduces additional discrepancies rising to 10-20\%. The increased numerical effort in including screening corrections is not expected to contribute substantially in cumulative $\beta$ spectra and will be omitted for the remainder of this work.

\subsection{On allowed shape factors}
\label{sec:allowed_shape_factors}
Several different expressions have been utilized for allowed shape factors throughout the literature within the context of the reactor antineutrino anomaly. As the results presented in this work depend not only on the shape factor of forbidden transitions but equally on its ratio to that of allowed transitions, we briefly review previous expressions and point out their deficiencies. 

In general, the shape factor is constructed in the rather opaque way of Eq. (\ref{eq:C_BB}). A particular advantage of allowed transitions, however, is their dominance of a single matrix element which simplifies its form dramatically. Compared to the main Fermi or Gamow-Teller matrix elements, corrections are usually on the order of only a few percent. This motivates one to write down the shape factor in its most crude form,
\begin{equation}
    C \approx 1.
    \label{eq:C_1}
\end{equation}
Within the context of the RAA, the original works by Mueller \textit{et al.} \cite{Mueller2011} and Huber \cite{Huber2011, *Huber2012} have gone beyond Eq. (\ref{eq:C_1}) to varying degree. While important differences appear for the `regular' finite size corrections (Eqs. (8) \& (9) in Ref. \cite{Huber2011, *Huber2012}; Eq. (8) in Ref. \cite{Mueller2011}), the correction due to induced currents is similar, and only takes into account a weak magnetism correction term:
\begin{equation}
    1+\delta_\text{wm} = 1+ \frac{4}{3M_n}\frac{b}{Ac}W,
    \label{eq:delta_wm_simple}
\end{equation}
where $M_n \approx 1830$ is the nucleon mass in our units, $A$ is the nuclear mass number and $b/c$ is the ratio of weak magnetism and Gamow-Teller form factors in the well-known Holstein formalism \cite{Holstein1974}. In impulse approximation the latter simplifies to
\begin{equation}
    \frac{b}{Ac} = \frac{1}{g_A}\left(g_M + g_V \frac{M_{L}}{M_{GT}} \right).
\end{equation}
Here $g_A = 1.27$ is the axial vector coupling constant, $g_M = 4.706$ is the weak magnetism coupling constant, $M_L = \langle f | \tau^\pm \vec{l} | i \rangle$ is the orbital angular momentum matrix element, and $M_{GT}$ the main Gamow-Teller matrix element. When proton and neutron Fermi surfaces are strongly separated, the ratio $M_{L}/M_{GT}$ is usually approximated as $-1/2$ \cite{Wang2017}, so that $b/Ac \approx 4.2/g_A$. In previous analyses \cite{Huber2011, Mueller2011}, a constant value was taken so that $dN/dE = 0.67\%$\,MeV$^{-1}$ extracted from an analysis of mirror decays for low masses. While large-scale calculations show significant variation around this value \cite{Wang2017}, we choose to use this value so that effects from forbidden decays can be cleanly separated.

The above expressions correspond to rather strong approximations. In fact, comparing to the full expressions (e.g. Eqs. (106a-d) in Ref. \cite{Hayen2018}), several differences appear which require some pause. Starting with the weak magnetism correction, we note that $\delta_\text{wm}$ should be written more completely
\begin{equation}
    \delta_\text{wm} = \frac{4}{3M_n}\frac{b}{Ac}\left(W-\frac{1}{2W} - \frac{W_0}{2} -\frac{3}{5}\frac{\alpha Z}{R}\right).
    \label{eq:delta_wm_complete}
\end{equation}
The last two terms are energy-independent but serve to renormalize the shape factor. The second term is energy-dependent and of opposite sign to the leading term. Its influence is most important in the low energy range, where in the context of the RAA it is unconstrained by the ILL data set. As with any low-energy effect, however, it shows up throughout the entire antineutrino spectrum and collectively changes the integrated antineutrino flux.

Further, in the case of allowed transitions the weak magnetism correction is not the only effect due to induced currents, as also the induced tensor term is non-zero for a general Gamow-Teller transition. One then requires an additional term which so far has never been taken into account
\begin{equation}
    \delta_\text{it} = \frac{1}{3M_n}\frac{d}{Ac}\left(W_0 + \frac{6}{5}\frac{\alpha Z}{R} - \frac{1}{W} \right).
\end{equation}
In general, $d/Ac$, is only identically zero for transitions within an isospin multiplet such as the mirror decays that were used for the weak magnetism correction by Huber \cite{Huber2011, *Huber2012}. For all remaining Gamow-Teller decays, however, $d/Ac$ is generally of comparable magnitude as $b/Ac$ and can easily exceed it by a factor $(-)5$ on a case-by-case basis.

Finally, allowed decays obtain corrections from another form factor with a similar structure as several finite size correction terms. In the Holstein formalism \cite{Holstein1974}, it is related to the induced pseudoscalar contribution, $h(q^2)$. Writing only the dominant term within the context of the RAA, the main terms are modified through the appearance of a $\Lambda^\prime$ contribution \cite{Hayen2018}
\begin{equation}
    \delta_\text{fs} \approx (\Lambda^\prime - 1)\left[\frac{21}{35}\alpha Z W R + \frac{4}{9}(W-W_0)WR^2\right],
    \label{eq:delta_fs_dominant}
\end{equation}
where
\begin{equation}
    \Lambda^\prime = \frac{\sqrt{2}}{3}\frac{\mathcal{M}_{121}}{\mathcal{M}_{101}}
\end{equation}
is of order unity and can vary substantially on a case-by-case basis. This $\Lambda^\prime$ contribution has so far never been taken into account. As such, the finite size corrections applied regularly in the RAA community (compare, e.g., Eq. \ref{eq:delta_fs_dominant} with Eq. (9) of \cite{Huber2011, *Huber2012}) can easily vanish or even change sign.

It should be clear by now that the shape factors used for allowed decays in the RAA analysis suffer from missing terms and an uncertain evaluation of the terms it does include. The effect on the anomaly itself and its uncertainty estimation will depend critically on a more careful evaluation, and is the topic of ongoing research.

In order to investigate the effect of the calculated forbidden shape factors presented here, we compare our results against, respectively, Eq. (\ref{eq:C_1}) and including only the weak magnetism correction as in Eq. (\ref{eq:delta_wm_simple}). This corresponds, approximately, to setting $\Lambda^\prime$ equal to zero and one, respectively, for $Z \sim 50$. The effects of Eqs. (\ref{eq:delta_wm_complete})-(\ref{eq:delta_fs_dominant}) are commented upon later and will be investigated in a future work.

\section{Data selection \& handling}
\label{sec:data_selection_handling}
The success of the summation approach hinges on the quality of the nuclear databases for fission yields and decay information \cite{Mueller2011, Fallot2012}. This is particularly true for our discussion here, as the impact of first-forbidden transitions depends critically on knowledge of nuclear level schemes with well-determined spin-parities and branching ratios. As such, we briefly discuss our selection and treatment of database information in the context of our later results in Secs. \ref{sec:flux_coverage}, \ref{sec:spectral_changes} and \ref{sec:generalization}.

\subsection{Database selection}
In terms of decay data, there are several evaluated public databases available. Of these, the Evaluated Nuclear Structure Data File (ENSDF) database is well-known but recent total absorption gamma spectroscopy (TAGS) \cite{Algora2010, Zakari-Issoufou2015, Rice2017} measurements have identified several discrepancies regarding branching ratios and level density. Previous measurements suffered from the so-called Pandemonium effect \cite{Hardy1977}, where due to the rapidly decreasing efficiency of Germanium detectors for high $\gamma$ energies de-excitations from highly excited states were missed, thereby overestimating $\beta$ branching ratios to low-lying states. This problem was apparent also in the context of the reactor anomaly in the significant overestimation of the flux in the high energy part \cite{Mueller2011, Fallot2012}. As such, we have opted here for the ENDF/B-VIII.0 (ENDF) decay data library \cite{Navratil2018}. In the latest version, some TAGS results were already incorporated. Additionally, consistency with reactor decay heat and a multitude of additional sources is checked \cite{VanderMarck2012}.

For the purpose of this work we are particularly interested in the spin-parities of nuclear levels. As such, we have made a combination of ENDF and ENSDF data in the following manner: Nuclear level energies and branching ratios are taken from ENDF and when a match is found with the ENSDF data we use the spin-parity information of the latter. This is because in ENDF, $\beta$ transitions are labeled explicitly only in the case of unique (forbidden) decays. In this way we benefit from pandemonium-corrected data but maintain nuclear level information.

For the fission yields we have used cumulative yields of the JEFF3.3 database \cite{Kellett2009}, which are to be preferred over those of ENDF \cite{FallotPC}. This is different compared to our previous work \cite{Hayen2019}, where the latter were used. For consistency, we report our results using both JEFF3.3 and ENDF fission yields together with the decay data of the latter as elaborated upon above. While differences arise for individual isotopes \cite{Hayes2015}, overall differences within the context of this work are minimal.

\subsection{Data treatment}
The nuclear databases are known to be incomplete for some regions of the nuclear chart. For some isotopes no (full) level schemes or branching ratios are known. If a particular isotope is populated in the fission process but contains incomplete or no data at all, we employ the so-called $Q_\beta$ approximation. The latter consists of filling the remaining $\beta$ branching by dividing it equally among a number of transitions. In the usual case, three branches are artificially created with endpoints at $\{Q, 2Q/3, Q/3 \}$ where $Q$ is the $Q$-value of the $\beta$ decay. In the case of the ENDF Decay Database, certain isotopes do not contain `discrete' information of transitions to specific final states but instead contain continuous spectrum data \cite{Herman2010, Chadwick2011,Kawano2008}. Within the context of reactor antineutrinos this poses a challenge for its inversion. We will treat this point more extensively in Sec. \ref{sec:ILL_reconstruction}.

When combining ENDF decay data with ENSDF level information, we assume the transition to be allowed if spin-parity determinations are incomplete or uncertain if the reported possibilities allow for it. Besides this, no information is replaced from the ENDF decay database.

\section{Shape factor calculation}
\label{sec:shape_factor_calc}
We proceed with the explicit calculation of a large sample of first-forbidden (non-)unique transitions using the nuclear shell model. Based on the discussion in Sec. \ref{sec:approximation_breakdown} we expect significant changes in the spectral shapes due to the breakdown of the usual approximations in the region of interest. Note that in the numerical results presented here, no approximations were made in the description of the shape factor, such as presented in Eqs. (\ref{eq:C_pseudoscalar})-(\ref{eq:C_unique_forbidden}).

\subsection{Selected transitions}
\label{sec:selected_transitions}

In the high energy region of the spectrum, i.e. larger than 4 MeV, the electron flux can be largely described using a limited number of $\beta$ branches. These have been compiled by Sonzogni et al. \cite{Sonzogni2015}, and in the following years several of these isotopes have been investigated using total absorption gamma spectroscopy (TAGS) \cite{Algora2010, Zakari-Issoufou2015, Rice2017}. This has for many isotopes resulted in a correction of branching ratios to high-lying states which had previously gone eluded due to the pandemonium effect \cite{Hardy1977}. Inspired by the compilation of Ref. \cite{Sonzogni2015}, we calculated 36 dominant forbidden transitions with the nuclear shell model, all of which are first forbidden. Note that we have included here more transitions than the 29 that were included in our previously published work \cite{Hayen2019}. A summary of their properties is shown in Table \ref{tab:summary_transitions_4MeV_235U}.  A large fraction of these are so-called pseudoscalar $\Delta J^\pi = 0^-$ transitions. Additionally, the initial and final states are either ground states or low-lying states, for which we can expect the nuclear shell model to perform adequately.

\begin{table}[!ht]
\caption{Dominant forbidden transitions above 4 MeV. Here $Q_{\beta}$ is the ground-state to ground-state $Q$-value, $E_{ex}$ the excitation energy of the daughter level, BR the branching ratio of the transition normalized to one decay and FY the cumulative fission yield of $^{235}$U from the ENDF database \cite{Chadwick2011}. Transitions with small fission yields shown here contribute substantially more for $^{238}$U and $^{241}$Pu.}
\label{tab:summary_transitions_4MeV_235U}
\begin{ruledtabular}
\begin{tabular}{l|cdcccc}
Nuclide & $Q_\beta$ & \multicolumn{1}{c}{$E_{ex}$} & \text{BR} & $J^\pi_i \to J^\pi_f$ & \text{FY} & $\Delta J$ \\
& \text{(MeV)} & \multicolumn{1}{c}{(MeV)} & (\%) & & (\%) & \\
\hline
\rule{0pt}{3ex}
$^{89}$Br & 8.3 & 0 & 16 & $3/2^- \to 3/2^+$ & 1.1 & 0\\
$^{90}$Rb & 6.6 & 0 & 33 & $0^- \to 0^+$ & 4.5 & 0\\
$^{91}$Kr & 6.8 & 0.11 & 18 & $5/2^+ \to 5/2^-$ & 3.5 & 0\\
$^{92}$Rb & 8.1 & 0 & 95.2 & $0^- \to 0^+$ & 4.8 & 0\\
$^{93}$Rb & 7.5 & 0 & 35 & $5/2^- \to 5/2^+$ & 3.5 & 0\\
$^{94}$Y & 4.9 & 0.92 & 39.6 & $2^- \to 2^+$ & 6.5 & 0\\
$^{95}$Rb$^\dagger$ & 9.3 & 0.68 & 5.9 & $5/2^- \to 5/2^+$ & 1.7 & 0\\
$^{95}$Sr & 6.1 & 0 & 56 & $1/2^+ \to 1/2^-$ & 5.3 & 0\\
$^{96}$Y & 7.1 & 0 & 95.5 & $0^- \to 0^+$ & 6.0 & 0\\
$^{97}$Y & 6.8 & 0 & 40 & $1/2^- \to 1/2^+$ & 4.9 & 0\\
$^{98}$Y & 9.0 & 0 & 18 & $0^- \to 0^+$ & 1.9 & 0\\
$^{133}$Sn & 8.0 & 0 & 85 & $7/2^- \to 7/2^+$ & 0.1 & 0\\
$^{135}$Te & 5.9 & 0 & 62 & $(7/2-) \to 7/2^+$ & 3.3 & 0\\
$^{135}$Sb & 8.1 & 0 & 47 & $(7/2+) \to (7/2^-)$ & 0.1 & 0\\
$^{136m}$I & 7.5 & 1.89 & 71 & $(6^-) \to 6^+$ & 1.3 & 0\\
$^{136m}$I & 7.5 & 2.26 & 13.4 & $(6^-) \to 6^+$ & 1.3 & 0\\
$^{137}$I & 6.0 & 0 & 45.2 & $7/2^+ \to 7/2^-$ & 3.1 & 0\\
$^{142}$Cs & 7.3 & 0 & 56 & $0^- \to 0^+$ & 2.7 & 0\\
$^{86}$Br & 7.3 & 0 & 15 & $(1^-) \to 0^+$ & 1.6 & 1 \\
$^{86}$Br & 7.3 & 1.6 & 13 & $(1^-) \to 2^+$ & 1.6 & 1\\
$^{87}$Se & 7.5 & 0 & 32 & $3/2^+ \to 5/2^-$ & 0.8 & 1\\
$^{89}$Br & 8.3 & 0.03 & 16 & $3/2^- \to 5/2^+$ & 1.1 & 1\\
$^{91}$Kr & 6.8 & 0 & 9 & $5/2^+ \to 3/2^-$ & 3.4 & 1\\
$^{95}$Rb$^\dagger$ & 9.3 & 0.56 & 6.0 & $5/2^- \to (7/2^+)$ & 1.7 & 1 \\
$^{95}$Rb & 9.3 & 0.68 & 5.9 & $5/2^- \to 3/2^+$ & 1.7 & 1\\
$^{134m}$Sb & 8.5 & 1.69 & 42 & $(7-) \to 6^+$ & 0.8 & 1\\
$^{134m}$Sb & 8.5 & 2.40 & 54 & $(7^-) \to (6^+)$ & 0.8 & 1\\
$^{136}$Te & 5.1 & 0 & 8.7 & $0^+ \to (1^-)$ & 3.7 & 1\\
$^{138}$I & 8.0 & 0 & 26 & $(1-) \to 0^+$ & 1.5 & 1\\
$^{140}$Xe & 4.0 & 0.08 & 8.7 & $0^+ \to 1^-$ & 4.9 & 1 \\
$^{140}$Cs & 6.2 & 0 & 36 & $1^- \to 0^+$ & 5.7 & 1 \\
$^{143}$Cs & 6.3 & 0 & 25 & $3/2^+ \to 5/2^-$ & 1.5 & 1 \\
$^{88}$Rb & 5.3 & 0 & 76.5 & $2^- \to 0^+$ & 3.6 & 2\\
$^{94}$Y & 4.9 & 0 & 41 & $2^- \to 0^+$ & 6.5 & 2\\
$^{95}$Rb & 9.3 & 0 & 0.1 & $5/2^- \to 1/2^+$ & 1.7 & 2\\
$^{139}$Xe & 5.1 & 0 & 15 & $3/2^- \to 7/2^+$ & 5.0 & 2
\end{tabular}
\end{ruledtabular}
\flushleft
{\scriptsize {}$^\dagger$ The spin-parity designation is uncertain, and shape factors were calculated for both options. Due to small branching ratios, the effect on the cumulative spectrum is negligible.}
\end{table}

\subsection{Flux coverage}
\label{sec:flux_coverage}
\begin{figure}[!ht]
    \includegraphics[width=0.48\textwidth]{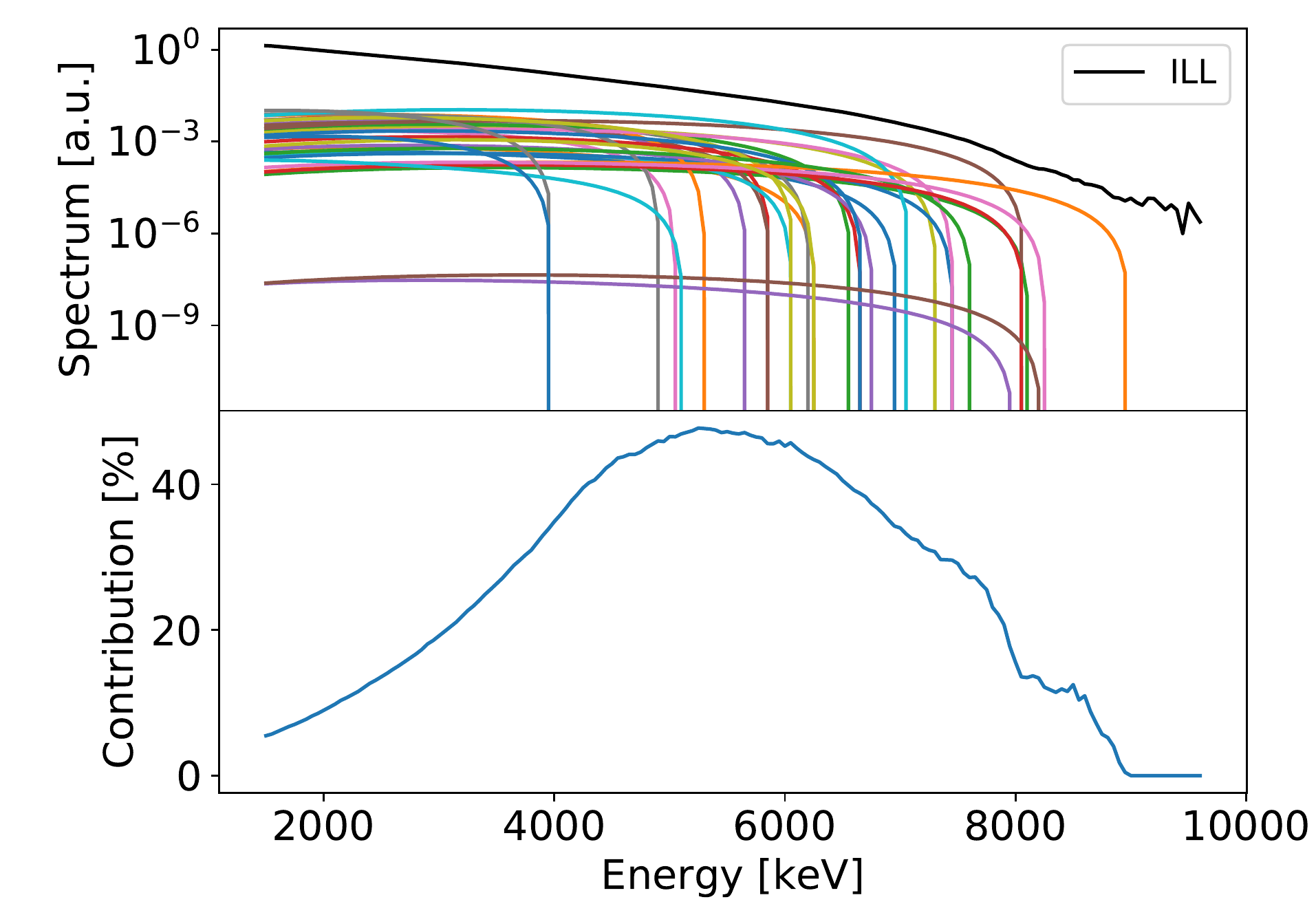}
        \caption{Contributions of individual $\beta$ transitions listed in Table \ref{tab:summary_transitions_4MeV_235U} and calculated as explained in the text and a comparison to the measured cumulative spectra measured at ILL for $^{235}$U. The upper panel shows the individual $\beta$ spectra, while the bottom panel shows the cumulative contribution of all calculated forbidden transitions relative to the ILL flux. The chosen transitions exceed 50\% of the flux around 6 MeV.}
    \label{fig:contribution_U235}
\end{figure}

The transitions of Table \ref{tab:summary_transitions_4MeV_235U} were selected for their large contribution to the total cumulative flux based on the compilation by Sonzogni \textit{et al.} \cite{Sonzogni2015}. In order to obtain a full spectrum shape for each transition, we combine the shape factor formalism of the previous section with the additional corrections to the $\beta$ spectrum shape \cite{Hayen2018, Hayen2019a} to form the full $\beta$ spectrum of Eq. (\ref{eq:spectrum_shape}). Summing the individual contributions of each of the transitions weighted by its fission yield and branching ratio discussed in Sec. \ref{sec:data_selection_handling}, we obtain a partial cumulative forbidden spectrum. Figure \ref{fig:contribution_U235} shows the contribution of the latter relative to the measured spectra at the ILL for $^{235}$U \cite{Haag2014}.

By including only 36 transitions, we reach 40\% of the total flux in the entire region between 4 to 7 MeV, while the maximum contribution exceeds 50\% around 6 MeV. Comparing with the results compiled by Sonzogni \textit{et al.} \cite{Sonzogni2015} we find that  inclusion of the dominant allowed $\beta$ spectra brings the total cumulative flux upwards of 80\% in this region. In conclusion, within the region of interest the chosen sample of transitions corresponds to a significant fraction of the total flux and our explicit calculation of their shape factor significantly influences the spectrum shape in this region.

\subsection{Nuclear shell model}
The shape factor for each of the transitions was calculated in the formalism by Behrens and B\"uhring using the nuclear shell model. No approximations were made concerning the formulation of the shape factors, so that the only dominant uncertainty comes from the shell model calculation of the nuclear matrix elements. These calculations were performed using the shell model code NUSHELLX@MSU \cite{Brown2014}. For nuclei with $A<100$ the effective interaction \texttt{glepn} \cite{Mach1990} was adopted in a full model space consisting of the proton orbitals $0f_{5/2}-1p-0g_{9/2}$ and the neutron orbitals $1d-2s$. The $^{86}$Br and $^{89}$Br cases were calculated using the interaction \texttt{jj45pna} \cite{Machleidt2001, Lalkovski2013}, in the full model space spanned by the proton orbitals $0f_{5/2}-1p-0g_{9/2}$ and the neutron orbitals $0g_{7/2}-2s-1d-0h_{11/2}$. For the nuclei with $A=$133--142 the Hamiltonian \texttt{jj56pnb} \cite{Brown2012} was used in the full model space spanned by the proton orbitals $0g_{7/2}-1d-2s-0h_{11/2}$  and neutron orbitals $0h_{9/2}-1f-2p-0i_{13/2}$ for $A<139$, while for the heavier nuclei the proton orbital $0h_{11/2}$ and the neutron orbital $0i_{13/2}$ were kept empty due to the enormous dimensions of a full model space calculation. 

The choice of a proper model space and Hamiltonian is crucial for meaningful shell model calculations. The region around $A\approx95$ is especially challenging, since taking full harmonic oscillator shells is currently not possible due to the enormous computational burden as well as a lack of a well tested Hamiltonian. Since a shell model Hamiltonian is fitted for a particular model space, it is always preferable to use a small enough model space to make the problem computationally reasonable without resorting to additional truncation of the model space. The model space chosen here for $A\approx95$ is small enough so that additional truncation of the model space is not necessary. In addition, this Hamiltonian is the natural choice for the reason that it was originally developed to describe one of the most important contributors to the cumulative beta spectrum here, namely the decay of $^{96}$Y \cite{Mach1990}. In principle all the decays with $A<100$ can be described using the interaction \texttt{glepn} but moving further away from $^{96}$Y the description of the nuclear structure starts to get more problematic. In the case of this study the lighter cases $^{86}$Br and $^{89}$Br turned out to be rather poorly described by this Hamiltonian, which is why the larger model space associated with the interaction \texttt{jj45pna} was used. It should be pointed out that agreement with the experimental half-life was also not reached with this interaction.

As is typical in the nuclear shell model a renormalization of fundamental coupling constants was used to account for meson exchange current and core polarization effects. For simple Gamow-Teller transitions the value of the axial charge coupling constant is changed to an effective value below $g_A = 1.27$. Also for the forbidden beta decays considered here, a quenching of the coupling constant $g_{\rm A}$ is necessary \cite{Suhonen2017a}. In the case of a pseudoscalar transition there is another nuance, as here the transition is dominated by the $^A\mathcal{M}_{000}^{(0)}$ nuclear matrix element, better known as the axial charge or $\gamma_5$ relativistic operator. Meson exchange current effects are known to be particularly strong for this operator, resulting in a well-known enhancement of this operator, which we denote here by $\epsilon_\text{MEC}$. For the pseudoscalar transitions, excluding the few problematic cases such as the bromide decays as well as the heavier cases where truncations are necessary, the experimental half-lives are reproduced with reasonable values of $g_{\rm A}$ and $\epsilon_{\text{MEC}}$. For example choices such as $g_{\rm A}=0.9$ and $\epsilon_{\text{MEC}}=1.4$ or $g_{\rm A}=0.75$ and $\epsilon_{\text{MEC}}=1.7$ give a good fit. For the pure $\Delta J =1$ transitions both $g_{\rm A}$ and $g_{\rm V}$ need to be quenched in order to reproduce the experimental half-lives. This is a well known issue which is usually attributed to core-polarization effects and is inline with previous research \cite{Suhonen2017a}. The excellent agreement using the experimental data with the usual assumptions is strong evidence that our calculations are indeed accurate for the majority of the decays, especially the most important ones.  Uncertainties due to $g_A$ quenching and meson exchange currents (MEC) in pseudoscalar transitions in the fission fragment region have been previously reported on \cite{Kostensalo2018, Kostensalo2017, Haaranen2017, Suhonen2017a, Hayen2019, Bodenstein-Dresler2018}, and will be discussed in the total uncertainty estimation of Sec. \ref{sec:uncertainty_estimation_shape_factors}.


\subsection{Numerical results}
\begin{figure}[!ht]
    \centering
    \includegraphics[width=0.49\textwidth]{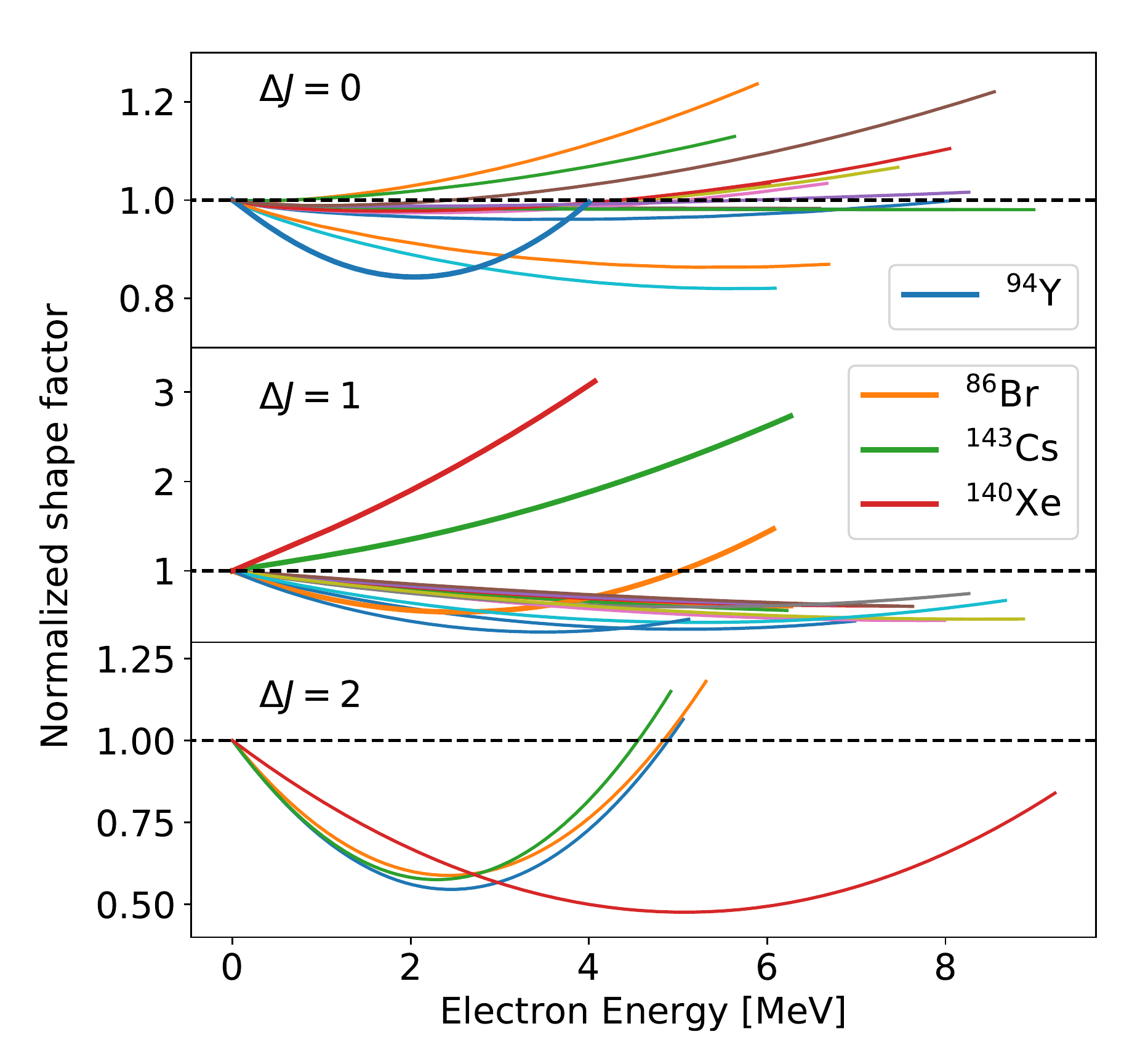}
    \caption{Calculated shape factors $C$ for the 36 first-forbidden transitions in Table \ref{tab:summary_transitions_4MeV_235U} versus electron kinetic energy, categorized according the spin-parity change of the transition. For allowed transitions $C\approx 1$, represented by the black dotted line. Each shape factor was normalized to its value at $E=0$. Results correspond to $g_A=0.9$ and $\epsilon_\text{MEC} = 1.4$, where applicable \cite{Kostensalo2018, Hayen2019}. Note the difference in scales on the $y$-axis. A few cases stand out and have been highlighted. These are discussed in the text.}
    \label{fig:shape_factors_collection}
\end{figure}

Taking the information of Table \ref{tab:summary_transitions_4MeV_235U} with the formalism of Sec. \ref{sec:formalism}, we calculate the numerical shape factors using a uniformly charged sphere for the charge density and nuclear wave functions from the nuclear shell model as described above. The results are shown in Fig. \ref{fig:shape_factors_collection}, categorized according to the spin change in the transition.

Almost all calculated shape factors deviate significantly from unity, including the pseudoscalar transitions. From Eq. (\ref{eq:C_pseudoscalar}) the behaviour of the latter should be trivial as $|bR| \sim 10^{-2}$, in an apparent contradiction. Many of these transitions connect initial and final states with spins larger than zero, meaning additional $\Delta J = 1, 2$ operators contribute. As such, in many cases the energy dependence is dominated by higher-order operators as is evident from the curves. This was already touched upon in Sec. \ref{sec:approximation_breakdown}. Additionally, because the $\xi$ approximation is not expected to hold for transitions with larger endpoints, this energy dependence is not suppressed. The pseudoscalar transition with the lowest endpoint energy, $^{94}$Y [$2^-$] to the first excited state of $^{94}$Zr [$2^+$], is of particular interest. Despite a reasonable argument in favor of the validity of the $\xi$ approximation (Eq. (\ref{eq:xi_approximation})), the calculated shape factor shows a strong parabolic behaviour reminiscent of a unique transition. Upon inspection of the level scheme of $^{94}$Zr, the first excited state can be interpreted as a consequence of collective behaviour of the nucleus in terms of a dipole vibration. Interpreted in the spherical shell model with explicit vibrational degrees of freedom, the nuclear wave function can be decomposed into a combination of Slater determinants and a vibrational wave function \cite{Davidson1968}. Neglecting higher-order corrections, the former is the same as that of the $0^+$ ground state. The $\beta$ decay operator acts only on the Slater determinants, so that the nature of the transition - and as a consequence the shape factor - resembles that of the ground-state ($^{94}$Y[$2^-$]) to ground-state ($^{94}$Zr[$0^+$]) unique $\beta$ decay. Residual interactions contaminate the vibrational wave functions, so that the change in vibrational states causes only a slowdown in the decay rate. This is an excellent example of the failure of the $\xi$ approximation due to the so-called selection rule effect \cite{Kotani1959}.

The pseudovector transitions show drastic deviations from unity for all studied transitions. For all transitions $\xi \sim W_0$, so that deviations are not wholly unexpected. Due to the nature of the fission process, almost all populated nuclei are heavily neutron-rich so that protons and neutrons reside in different major shells interpreted in the shell model. As a consequence, proton and neutron Fermi surfaces usually lie in regions of opposite parity so that many different possibilities arise for a parity-changing transition including $\Delta J = 2$. 

We discuss some cases that stand out from the pack. In the case of the $\beta$ transition of $^{86}$Br [$1^-$] to the first excited state of $^{86}$Kr [$2^+$] the $\Delta J = 2$ contribution is clearly seen to be dominant. While the excited state in $^{86}$Kr at 1.5\,MeV is possibly a good vibrational candidate, the higher-order band structure is not visible. The numerical results hint at a cancellation effect in the additional first-order matrix elements. Besides this, both $^{140}$Xe and $^{143}$Cs show strongly diverging shape factors compared to all others calculated. This will have important consequences in the parametrization described in Sec. \ref{sec:generalization}. It is not intuitively clear here why this occurs, as their results are particularly sensitive to cancellations. This can occur both due to nuclear structure considerations and contributions of various single-particle transitions of opposite sign, but also due to Coulomb effects. In the latter case, some matrix elements occur accompanied with factors of $\alpha Z$, so that changing the proton number has significant consequences. Because of this, even smaller matrix elements can end up dominating the shape factor due to cancellations between the main matrix elements. Regardless, all of these reasons are examples of an additional breakdown of the $\xi$ approximation.

Corrections to the unique shape factors are typically observed to be on the few percent level or lower when taking into account the appropriate Coulomb corrections factors as discussed in Sec. \ref{sec:approximation_breakdown}. Our numerical results confirm these findings in the studied transitions.

\subsection{Comparison with existing literature}
\begin{figure}[!ht]
    \centering
    \includegraphics[width=0.48\textwidth]{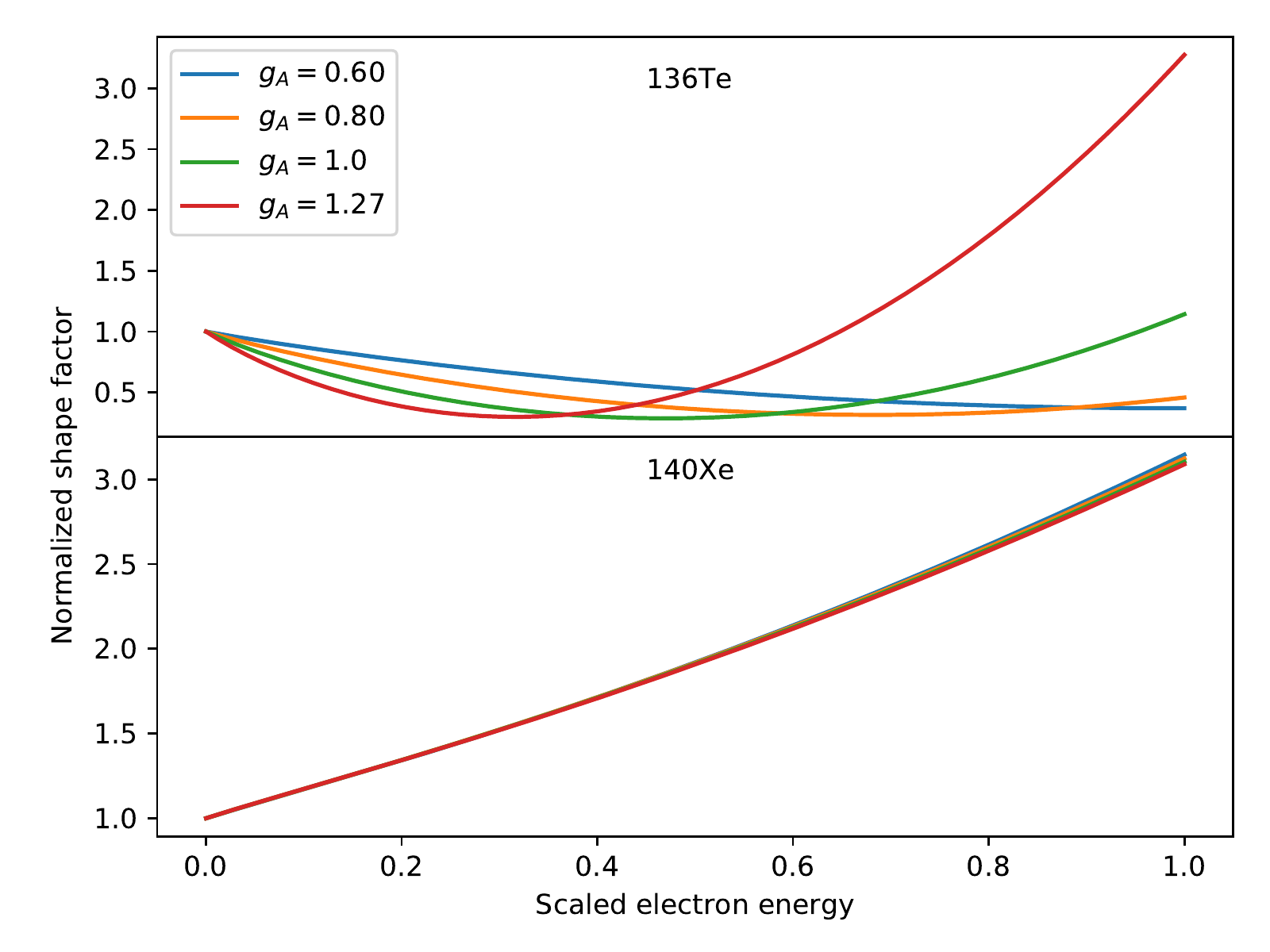}
    \caption{Numerical shape factors calculated with the nuclear shell model for the first-forbidden pseudovector transitions in $^{136}$Te (top) and $^{140}$Xe (bottom) for different values of the axial vector coupling constant. Comparison with the results by Fang and Brown \cite{Fang2015} are favourable for $^{140}$Xe, whereas for $^{136}$Te a slope with opposite sign is found for an equivalent value of $g_A$.}
    \label{fig:comp_Fang_Brown}
\end{figure}

As mentioned in the previous section, the chosen values for effective coupling constants $g_A$ and $\epsilon_\text{MEC}$ reproduce experimental half-lives nicely \cite{Suhonen2019}. In addition to these, there are data that we can compare our calculations to. While there has been a limited amount of study on the effect of forbidden transitions within the context of the reactor anomaly and shoulder \cite{Hayes2014, Hayes2015, Sonzogni2017, Sonzogni2015, Fang2015}, so far, there has only been a microscopic study on two nuclei \cite{Fang2015}: $^{136}$Te and $^{140}$Xe. As both of these are even-even nuclei, investigated decays occur from the $0^+$ ground state so that their first-forbidden transitions correspond to ``pure'' transitions. In Ref. \cite{Fang2015}, only $^{136}$Te was studied both in the shell model and the quasiparticle random phase approximation (QRPA), while $^{140}$Xe was computed only using the latter due to computational constraints \cite{FangPC}. Here we have calculated transitions from both nuclei in the nuclear shell model using the \texttt{jj56b} model space. Figure \ref{fig:comp_Fang_Brown} shows the calculated shape factors for different values of effective $g_A$ used.

The shape factor of $^{140}$Xe is almost insensitive to the choice of $g_A$ and agrees well with the results by Fang and Brown \cite{Fang2015}. The calculation for $^{136}$Te, on the other hand, shows a strong dependence on the effective value of $g_A$, in particular in connection with a quadratic component. 
However, when trying to replicate the shape factor of 136Te using the same Hamiltonian as reported in Ref. [17], we find a different slope. Interestingly, we can reproduce their results when manually changing the phase convention of either the single-particle matrix elements or one-body transition densities between Condon-Shotley (prevalent in shell model calculations) and Biedenharn-Rose (typical in QRPA codes) conventions. Additionally, we found that the shape factor is heavily dependent on the ratio $g_{\rm A}/g_{\rm V}$, the proper value of which is not well established for every Hamiltonian. In the original work, a value of $g_{A, eff} = 0.5\, g_A$ was used. Since also $g_V$ was quenched by the same amount, however, their ratio remains unchanged even though the half-life is naturally reproduced.

\section{Spectral changes}
\label{sec:spectral_changes}
Any spectral changes that occur from inclusion of our numerically calculated forbidden shape factors depend on the allowed shape factor that it is compared to. Following the discussion in Sec. \ref{sec:allowed_shape_factors}, we look at the difference in the spectral shapes of both electron and antineutrino spectra using both $C=1$ (Eq. (\ref{eq:C_1})) and the simplified weak magnetism correction of Eq. (\ref{eq:delta_wm_simple}).

\subsection{Results}

\begin{figure}[ht]
    \centering
    \includegraphics[width=0.48\textwidth]{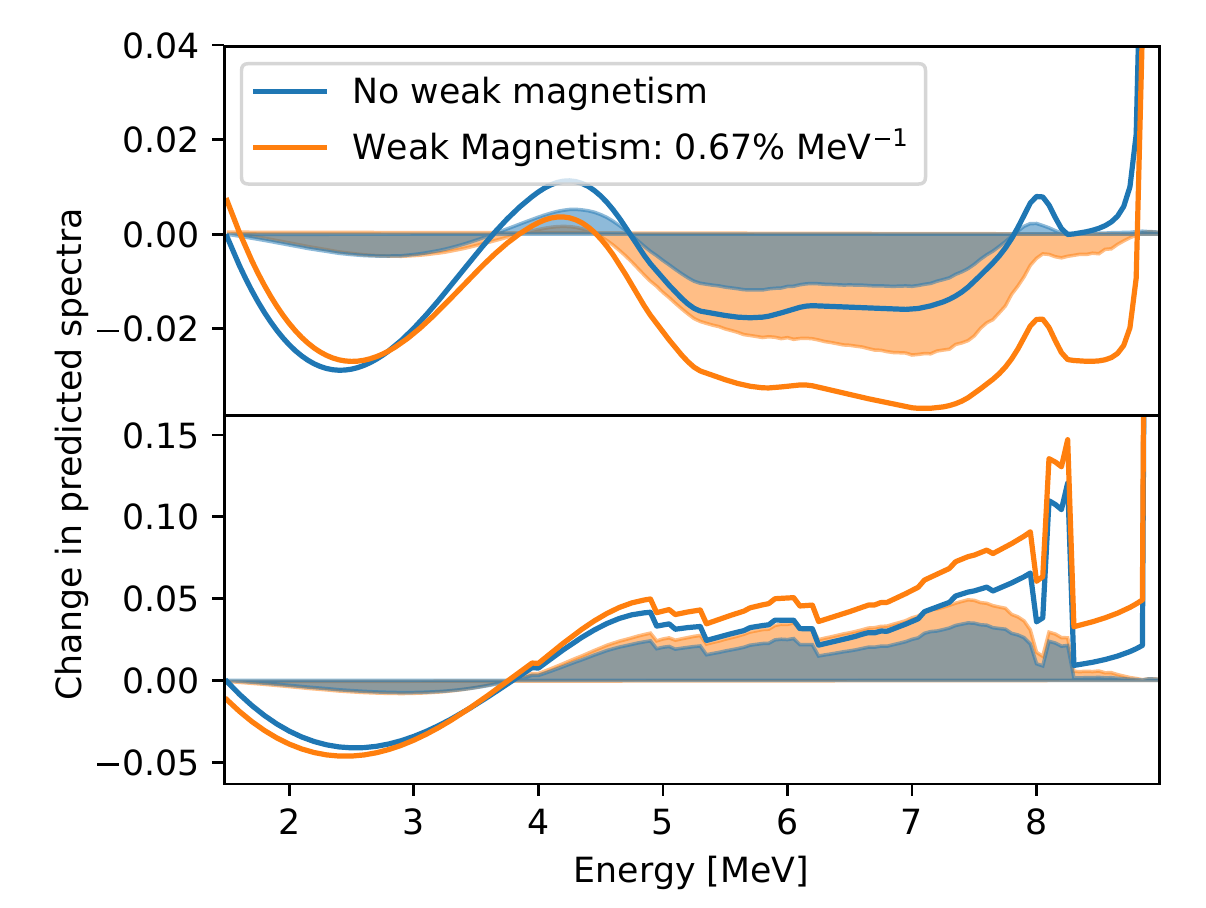}
    \caption{Top panel: Change in the predicted partial electron spectra of the considered transitions compared to the allowed approximation and with an optional weak magnetism correction. Bottom panel: Change in the predicted antineutrino spectrum compared to the allowed approximation. Shaded areas correspond to the results multiplied by the total spectral contribution compared to experimental flux results (Fig. \ref{fig:contribution_U235}). The energy axis refers to the kinetic energy of the electron (top) and antineutrino (bottom). All results are calculated using $g_A = 0.9$ and $\epsilon_\text{MEC} = 1.4$, for which good agreement was found with experimental lifetimes.}
    \label{fig:electron_antineutrino_prediction_change}
\end{figure}

We compare the effects of the forbidden shape factors taking into account the relative weights of the different transitions. Three different partial cumulative forbidden spectra are constructed using the forbidden shape factors of the previous section, the allowed approximation $C=1$, and the weak magnetism correction of Eq. (\ref{eq:delta_wm_simple}). Results for the ratio of forbidden to allowed calculations are shown for $^{235}$U in Fig. \ref{fig:electron_antineutrino_prediction_change} for both electron and antineutrino spectra. Shaded areas correspond to the partial spectral ratios weighted by the contribution of our included transitions to the total flux, as reported in Fig. \ref{fig:contribution_U235}, to estimate the effective change to the full spectrum.

Starting with relative changes in the electron spectrum, several quantitative features become immediately apparent. The first is the parabolic behaviour at energies below 4\,MeV, which originates from the unique forbidden transitions which dominate our transition selection (see also Fig. \ref{fig:shape_factors_collection}). Second is the lowering of the predicted electron flux in the higher energy window, for which the downward slope of the shape factors of the calculated pseudovector transitions are mainly responsible. Further, the strong increase at the highest energies is dominated by very few - or even a single - branch, for which strong deviations are expected near the end of the spectrum based on the results of Fig. \ref{fig:shape_factors_collection}. Finally, the tilt in the comparison between $C = 1$ and the weak magnetism correction comes from the positive linear slope in Eq. (\ref{eq:delta_wm_simple}). 

Quantitatively, clear changes are visible compared to the simple allowed approximation, and a general shift in predictions of roughly $-2\%$ is observed when comparing against the results obtained with a simple weak magnetism term. Due to the limited selection of the calculated transitions, contributions to the total flux swiftly recede to zero outside of the bump energy window, thereby quenching spectral changes.

The antineutrino spectra in the bottom panel of Fig. \ref{fig:electron_antineutrino_prediction_change} show several interesting features when compared to those of the electron. The fine structure in the spectrum is the consequence of the Fermi function, which lifts the $\beta$ spectrum shape above zero for near-vanishing electron energy. Besides this, the most interesting result resides in the magnitude of the induced discrepancies compared to that in the electron spectrum. For the antineutrino spectrum, a significant enhancement of the expected antineutrino flux is observed above 4\,MeV. Weighted results show enhancements of over 5\% around 6 and 7\,MeV, whereas changes are limited to 2\% in the equivalent electron window. The reason for this resides in the steep decrease of the total flux for increasing energy. A downward sloping shape factor such as those in Fig. \ref{fig:shape_factors_collection} pushes more of the flux to lower electron energies. The change to the total cumulative spectrum is minimal, however, due to the absolute magnitude of the spectrum being orders of magnitude larger at lower energies. The opposite goes for the antineutrino spectrum, resulting in stronger discrepancies. The downward trend below 4 MeV is mitigated due to the limited contribution of the considered forbidden spectra to the total flux.

\subsection{Uncertainty estimation}
\label{sec:uncertainty_estimation_shape_factors}
A trustworthy determination of the uncertainty of all sources included in the calculation is of paramount importance. On the other hand, estimation of theory uncertainties within nuclear structure calculations presents a tremendous challenge. Recently, some efforts have been made in the $sd$ shell, where a Bayesian analysis translated experimental and fit uncertainties into final uncertainties in nuclear matrix elements \cite{Yoshida2018}. Given the large model space and number of fit parameters, this procedure is not currently possible for our transitions of interest. As such, here we vary the available parameters used in tuning shell model results to obtain agreement with experimental lifetimes. In the most general case this corresponds to a modification of the axial coupling constant, $g_A$, whereas for the pseudoscalar transitions meson exchange currents strongly modify the so-called axial charge \cite{Kostensalo2018}. In order to take into account this effect, we additionally vary $g_A$ for pseudoscalar operators, which we note by $\epsilon_\text{MEC}$.

The results shown in Fig. \ref{fig:electron_antineutrino_prediction_change} were obtained for $g_A = 0.9$ and $\epsilon_\text{MEC} = 1.4$, for which good agreement was reached with experimental lifetimes for almost all transitions \cite{Kostensalo2018, Hayen2019}. In order to get a measure for the uncertainty on our results, we vary the coupling constants within a window as described below. This is done because for many isotopes the experimental half-life is the only quantity to which one can compare. When quenching both $g_V$ and $g_A$ (as was done by Fang and Brown \cite{Fang2015}), a degeneracy appears in their ratio as the experimental half-life can always be obtained after suitable quenching. For the axial vector coupling constant four different values were used, setting $g_A/g_V \in \{0.7, 0.9, 1.0, 1.27\}$. The meson exchange corrections to the axial charge were picked from the interval $\epsilon_\text{MEC} \in \{1.4, 1.7, 2.0\}$. 

There is, however, no unique way of choosing effective couplings for all transitions together. As a consequence, we choose the uncertainty to be the maximum of the deviation between fully correlated and random choices of $g_A$ and $\epsilon_\text{MEC}$ for all transitions. We do so only for the partial cumulative spectrum, as this is the only relevant theoretical input despite potential large differences in individual shape factors.

\begin{figure}[ht]
    \centering
    \includegraphics[width=0.48\textwidth]{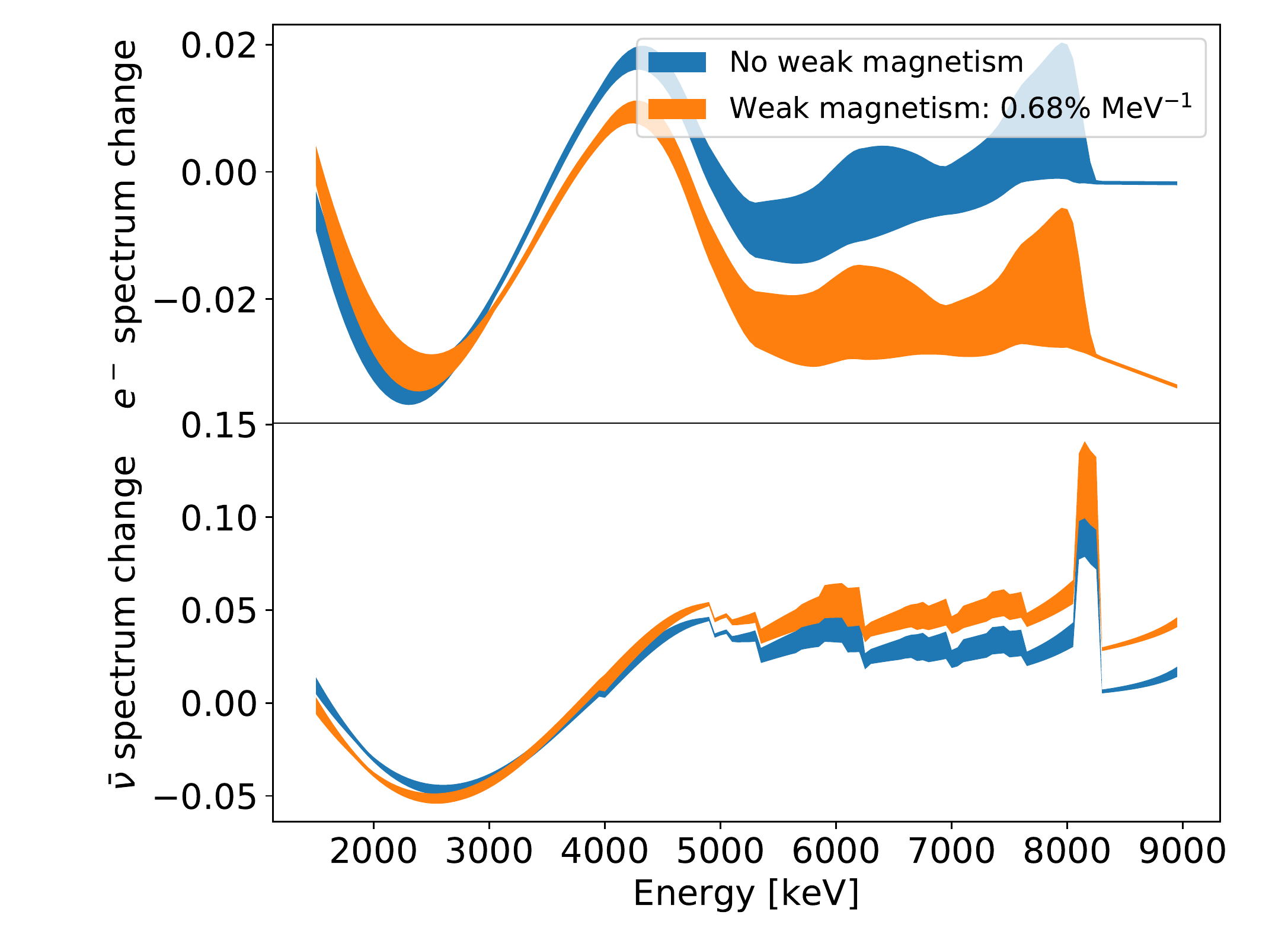}
    \caption{Top (bottom): Uncertainty in the relative change in the prediction of the electron (antineutrino) spectra when calculating the transitions using forbidden spectral shapes instead of simple allowed shapes using different values of $g_A$ and $\epsilon_\text{MEC}$ for the former (see text). The filled regions show the maximal deviations in results. A large part of this uncertainty comes from setting the axial vector coupling to the free nucleon value of $g_A = 1.27$.}
    \label{fig:electron_antineutrino_prediction_change_uncertainty}
\end{figure}

Figure \ref{fig:electron_antineutrino_prediction_change_uncertainty} shows the spread in the relative change of the partial cumulative electron and antineutrino spectra for both allowed approximations as before. In both cases the largest uncertainty appears in the higher end of the spectrum. The origin of this can mainly be traced back to the pseudovector transitions, where the slope of the shape factor is usually a combination of $\Delta J = 1$ and $\Delta J = 2$ operators with different $g_A$ dependence. These effects are limited to higher ends of the spectrum due to the selected transitions and their respective endpoints. The lower energy regions are mainly dominated by unique forbidden transitions, for which any deviations from Eq. (\ref{eq:C_unique_forbidden}) are already constrained to the per-cent level. The majority of the uncertainty comes from varying $g_A$. Effects from varying $\epsilon_\text{MEC}$ are only relevant for pseudoscalar transitions and are found to be sub-dominant. The reason for this can intuitively be understood, as it concerns changes to the $\Delta J = 0$ operators which have limited energy dependence (Eq. (\ref{eq:C_pseudoscalar})). A similar conclusion is reached for the antineutrino partial spectrum. Even so, the total uncertainty in the latter is about a factor 2 larger, putting the theory error at around 1\%, before multiplication with the forbidden flux contribution of Fig. \ref{fig:contribution_U235}.

The results of Fig. \ref{fig:electron_antineutrino_prediction_change_uncertainty} represent a bound rather than a confidence interval in the statistical sense. For the purpose of the discussion, however, we will treat the variation around the central value as a $1\sigma$ uncertainty. We shall see that this is not the dominant uncertainty when we generalize the approach of first-forbidden transitions for a more complete discussion in Sec. \ref{sec:generalization}. This is expanded upon in the Appendix.

\section{Improved forbidden transition treatment}
\label{sec:generalization}
Over the past several years, a lot of attention has gone towards an \textit{ab initio} treatment of the electron and antineutrino spectra, fueled by a strong experimental effort in TAGS measurements (e.g. \cite{Zakari-Issoufou2015}). Despite a significant number of uncertainties in nuclear databases, it provides an independent analysis path with a much more fine-grained control. Additionally, it is the only method available that can predict the electron and antineutrino spectra below $1.8\,$MeV with reasonable accuracy. Up to now, the treatment of non-unique forbidden transitions has proceeded by either approximating it as an allowed decay \cite{Huber2011, *Huber2012}, or as the shape of an $n-1$ unique forbidden decay for forbiddenness $n$ \cite{Mueller2011}. In the case of first forbidden decays, these are of course the same approximation. Based on the results of the previous section and the discussion of Sec. \ref{sec:approximation_breakdown}, the validity of these approximations appear unwarranted.

It is the question of this section to investigate the possibility of generalizing the information of the previous section and apply it to the remainder of (non-unique) forbidden decays present in the database. Before we embark on this journey, however, it is worthwhile to look at the structure of the electron and antineutrino flux. In doing so, we investigate the relative importance of forbidden transitions on the total flux. Following this, we attempt a parametrization of the results found in Sec. \ref{sec:shape_factor_calc}. Finally, we discuss how to use this information of the parametrization to obtain an uncertainty from the treatment of forbidden decays using Monte Carlo techniques.

\subsection{Forbidden flux coverage}
We investigate the composition of the cumulative electron spectrum. Table \ref{tab:breakdown_branches} shows the breakdown of the contributing $\beta$ branches following the fission of $^{235}$U.

\begin{table}[ht]
    \caption{Breakdown of the number of $\beta$ branches participating in the $^{235}$U electron flux. An arbitrary cut was made where the fission yield must be larger than $1 \cdot 10^{-6}$, bringing the total number to 8219. Exact numbers are not of importance, as several intermediate steps are required as described, e.g., in Sec. \ref{sec:data_selection_handling}.}
    \label{tab:breakdown_branches}
    \centering
    \begin{ruledtabular}
    \begin{tabular}{l|ccr}
        & Non-Unique & Unique & Total \\
        \hline
        Allowed & 3049 & 2648 & 5697 (69\%) \\
        1st forbidden & 1593 & 515 & 2108 (26\%) \\
        2nd forbidden & 235 & 97 & 332 (4\%) \\
        3rd forbidden & 52 & 12 & 64 (0.8\%) \\
        Other & 33 & 11 & 44 (0.5\%) 
    \end{tabular}
    \end{ruledtabular}
\end{table}

\begin{figure}[ht]
    \centering
    \includegraphics[width=0.48\textwidth]{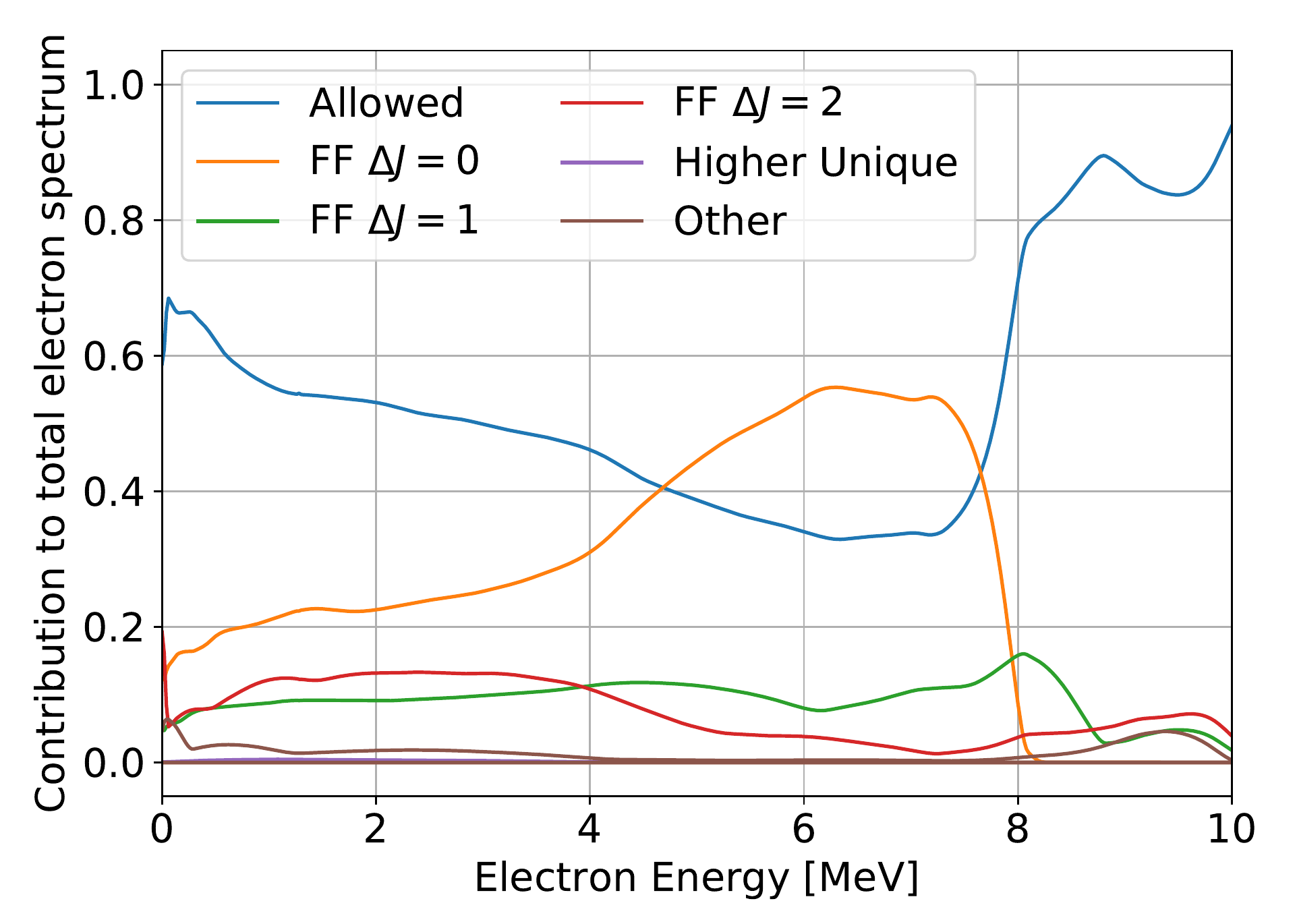}
    \caption{Overview of the spectral composition of the cumulative electron fluxes of $^{235}$U, calculated assuming allowed shape factors. It's clear that, despite weight in numbers, the contribution of allowed decays is greatly diminished in most of the region of interest. Above the inverse beta decay threshold at 1.8 MeV and below 8.5 MeV - precisely the experimental range of the ILL campaigns - the cumulative spectrum is dominated by forbidden decays.}
    \label{fig:spectral_constituents_electron}
\end{figure}

As is well-known by now, around $30\%$ of the transitions are forbidden. While several compilations have been made of the dominating branches or the number required to reach a certain flux \cite{Sonzogni2015}, a closer look at the underlying structure of the spectrum has been absent. In order to obtain a more realistic picture, the results of Table \ref{tab:breakdown_branches} must be adjusted to account for the branching ratio and fission yield of each transition. Figure \ref{fig:spectral_constituents_electron} shows the contributions of the various types of decays in the summed electron spectrum for $^{235}$U as a function of $\beta$ energy.

It is clear that the dominion of allowed spectra based on only their number is overestimated. Keeping in mind the inverse beta decay threshold at $1.8$ MeV and steep decrease in flux after 8 MeV, this conclusion becomes all the more relevant. In the 4-8 MeV region in particular, a clear dominance of forbidden spectra can be seen. This corresponds to the same region as the so-called bump or shoulder in the antineutrino spectra. 

Expected consequences for differences in cumulative spectrum shapes can be superficially deduced from the results of Sec. \ref{sec:shape_factor_calc}. While typically spectra for $\Delta J^\pi = 0^-$ closely correspond to equivalent allowed spectra, Fig. \ref{fig:shape_factors_collection} shows that significant variations can occur as higher-order operators often also contribute. Interesting to note is how the contribution of $\Delta J^\pi = 1^-$ transitions is reasonably constant around 15\% throughout the entire spectrum. As these decays in particular bring about a large change in predicted electron and antineutrino spectra (see Fig. \ref{fig:shape_factors_collection}), significant changes can be expected over the full range. Higher unique forbidden transitions for which shape factors can be very well calculated turn out to be negligible over the full range. Higher non-unique decays are equally insignificant over the full range.

There is an interesting structure in Fig. \ref{fig:spectral_constituents_electron}, which can at least superficially be understood from an intuitive nuclear physics point of view. The majority of neutron-rich fission fragments that are populated have $Q$ values around 4-8 MeV. Many of the transitions contributing in this window in Fig. \ref{fig:spectral_constituents_electron} correspond then to decays from initial ground states to final ground states or low-lying excited states. Due to the large proton-neutron asymmetry, these typically reside in adjacent major orbital shells. Most of these orbitals have opposite parity, so that ground state to ground state transitions are automatically forbidden. As a consequence, these are dominant in the flux in the 4-8 MeV window. Using the usual Woods-Saxon orbital properties as a reference, the structure within first-forbidden transitions can additionally be understood. As nuclei decay towards the line of stability, the $Q$ value decreases as the proton-neutron asymmetry lessens. Valence protons then populate the $\pi g_{9/2}$ orbital, whereas valence neutrons drop into $\nu d_{5/2}$ and $\nu g_{7/2}$ orbitals. One expects then a rise in unique first-forbidden ($\nu d_{5/2} \to \pi p_{1/2}$) and allowed ($\nu g_{7/2} \to \pi g_{9/2}$) decays, which is reflected in Fig. \ref{fig:spectral_constituents_electron}. Transitions to excited states complicate this picture significantly for higher excitation energies, and here we run into the limits of our simple picture. Similarly, the behaviour at high energies is dominated by very few branches from isotopes with very high $Q$ values. For many of the latter, spin-parities are unknown, meaning their $\beta$ branches are simply approximated to be allowed.

\subsection{Parametrization procedure}

From the results of Fig. \ref{fig:spectral_constituents_electron}, it is clear that the influence of forbidden transitions is non-negligible throughout the entire experimentally accessible spectrum. While the dominant contributions come from pseudoscalar transitions for which the shape factors resemble those of allowed decays, a significant contribution comes from higher forbidden decays with strikingly different shape factors. Additionally, from Fig. \ref{fig:shape_factors_collection} it can be gleaned that shape factors within the same $\Delta J$ category are reasonably similar, warranting a parametrization. It is with this observation in mind that we attempt to construct an effective correction to the spectra of both electron and antineutrino taking into account the underlying forbidden structure. Due to the larger sample of numerical shape factors presented here compared to our previous work, the parametrization procedure has evolved to better reflect the internal structure of the shape factor distribution. For completeness then, we outline both the procedure used in the previous work \cite{Hayen2019} and its current state.

\subsubsection{Parametrized forbidden shape factors}

The expected shape factor contribution from pseudoscalar operators is approximately equal to unity (see, e.g. Eq. (\ref{eq:C_pseudoscalar})), so that nearly all of the deviations observed in Fig. \ref{fig:shape_factors_collection} arise from higher-order operators contributing to the $J^\pi \to J^{-\pi}$ transition. Depending on the sign of these contributions one arrives at a positive or negative slope. As the number of terms contributing to the general shape factor is so large, combined with a near-statistical spread of the deviations from unity, we make no attempt at a \textit{smart} parametrization and simply fit each of the shape factors according to
\begin{equation}
C = 1 + aW + b/ W + c W^2
\label{eq:general_shape_factor_fit}
\end{equation}
inspired by the general form of the shape factor (Eq. (\ref{eq:C_BB})).

The behaviour of the $\Delta J^\pi = 1^-$ shape factors is more uniform as can be deduced from Fig. \ref{fig:shape_factors_collection}. As these operators now also carry a significant energy dependence, any energy-dependent change is not any more dominated by the influence of higher-order operators as it was for the pseudoscalar case. For nearly all transitions calculated, only two nuclear matrix elements contribute significantly: The time component of the first moment of the vector current, $^V\mathcal{M}_{110}$, and the space component of the first moment of the axial vector current, $^A\mathcal{M}_{111}$. While these are usually of similar magnitude, the possibility for cancellations stands in the way of a more insightful parametrization. The procedure is then analogous to that of the pseudoscalar transition, where we similarly fit all shape factors according to Eq. (\ref{eq:general_shape_factor_fit}).
Finally, the unique forbidden decays are well understood, with a shape factor that is approximately equal to that of Eq. (\ref{eq:C_unique_forbidden}). This result was obtained assuming the presence of only the dominant nuclear form factors, and deviations occur only at the percent-level. As this will not appreciatively influence the final uncertainty, we simply assume the approximate unique forbidden shape factor of Eq. (\ref{eq:C_unique_forbidden}).

Here we distinguish between the approaches followed for our previous work \cite{Hayen2019} and the current status. We discuss both in turn.

\begin{figure}[!ht]
    \centering
    \includegraphics[width=0.48\textwidth]{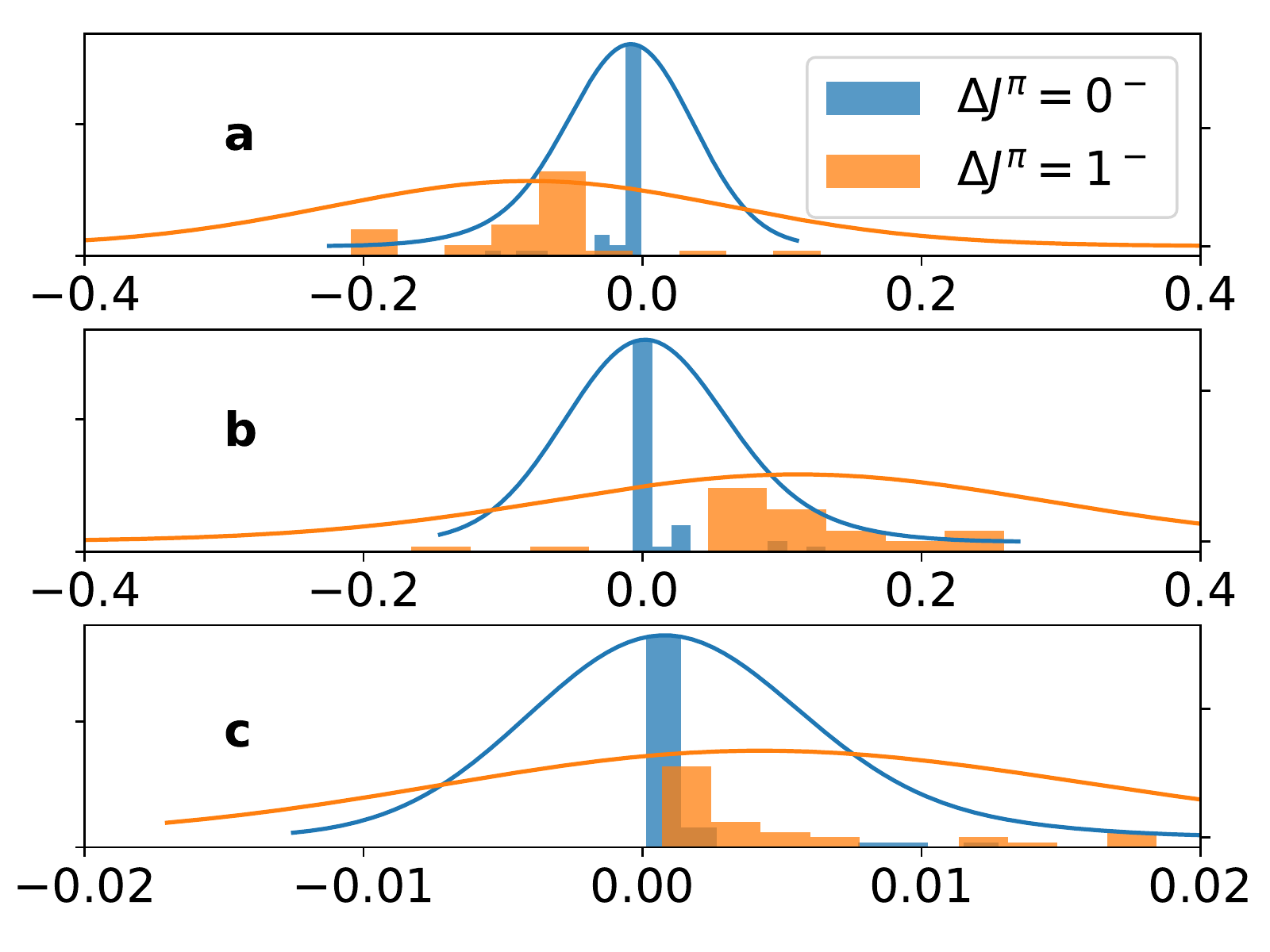}
    \caption{Distribution of fit parameters $a$, $b$ and $c$ (Eq. (\ref{eq:general_shape_factor_fit})) from the numerical results of Sec. \ref{sec:shape_factor_calc} for pseudoscalar and pseudovector transitions. Full lines represent an expectation of the underlying distribution using Gaussian kernel density estimation. This corresponds to the old approach used in Ref. \cite{Hayen2019}.}
    \label{fig:fit_parameter_dist_kernel_old}
\end{figure}

(\textit{Old}) After fitting all numerically calculated non-unique first forbidden shape factors using Eq. (\ref{eq:general_shape_factor_fit}), one obtains distributions of fit parameters for each $\Delta J$, including correlations between the fit parameters. Results are shown in Fig. \ref{fig:fit_parameter_dist_kernel_old} for the fit parameters with an average correlation matrix
\begin{align}
    \rho = \left(\begin{array}{ccc}
        1 & -0.62 & -0.98 \\
        -0.62 & 1 & 0.55 \\
        -0.98 & 0.55 & 1
    \end{array} \right).
    \label{eq:avg_corr}
\end{align}
Interestingly, the latter is almost identical for $\Delta J = 0$ or $1$ despite strong differences in the magnitude of the effect. Here all shape factors were included for the full range of $g_A$ and $\epsilon_\text{MEC}$. This way, both the uncertainty due to effective coupling constants and spread in calculated shape factors contributes to our effective knowledge of first-forbidden shape factors.

We apply one additional step to obtain a useful distribution to eventually sample from. By employing Gaussian kernel density estimation \cite{Scott1992}, one obtains a parameter probability density function. Doing so eliminates all knowledge one might have about the particular transition other than its degree of forbiddenness, so that this parametrization rather becomes a quantification of uncertainty due to non-unique first forbidden transitions in the electron and antineutrino spectra.

In performing this parametrization there is some freedom, hidden in the bandwidth estimate of the Gaussian kernel density estimation. While several rule-of-thumb bandwidth estimators exist in the literature, these are known to perform poorly for non-Gaussian or heavy-tail distributions. As such, we determine the bandwidth manually through comparison of the quantiles in the parametrized shape factors and the numerically calculated ones. By requiring all explicitly calculated shape factors to fall within $2\sigma$ of the procedural set, one arrives at a bandwidth of $h=2$. Using rule-of-thumb estimators such as `Silverman' or `Scott' \cite{Scott1992}, one finds much lower values for $h \approx 0.6$ and poor agreement with numerical results.

\begin{figure}[!ht]
    \centering
    \includegraphics[width=0.48\textwidth]{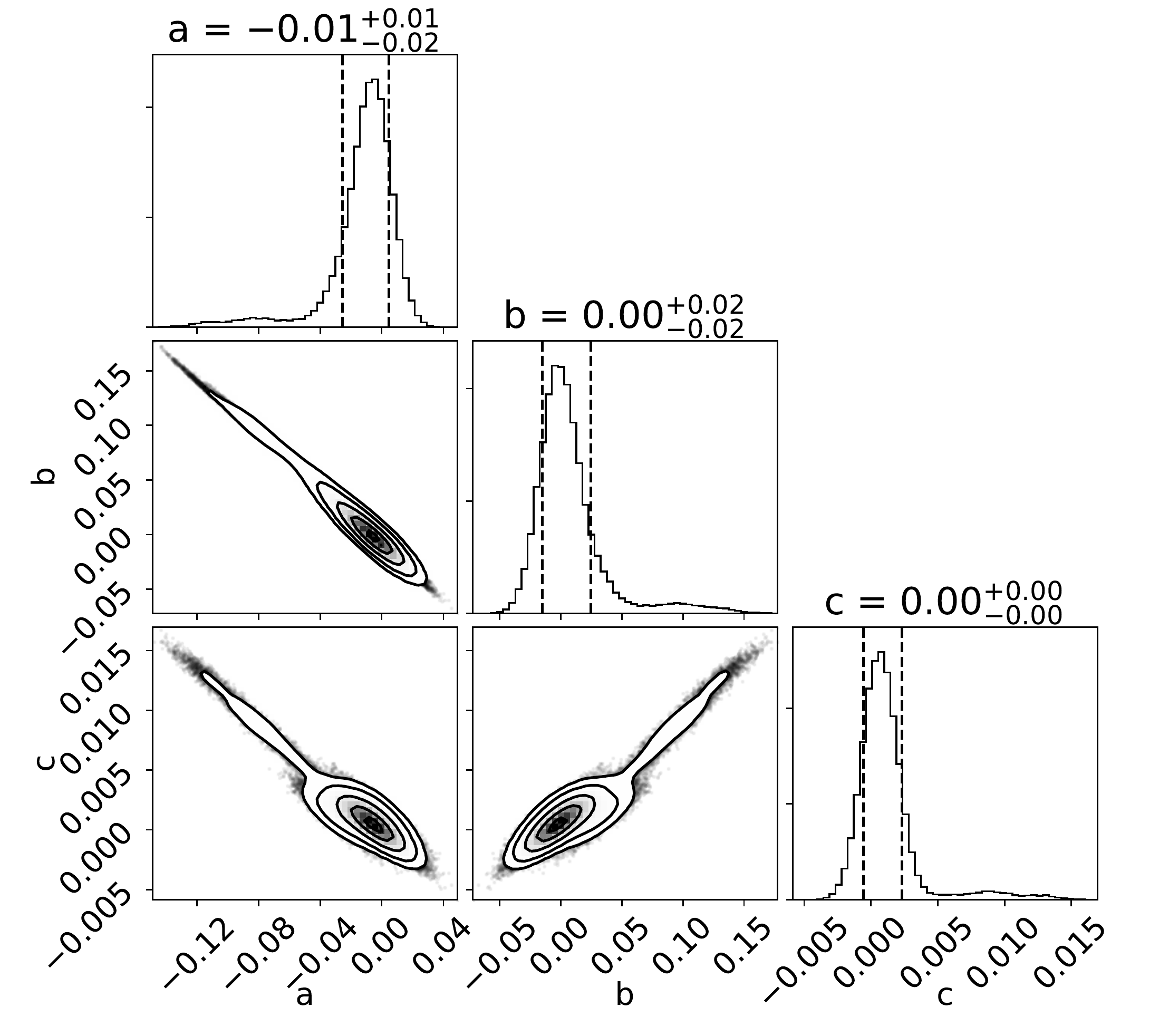}
    \includegraphics[width=0.48\textwidth]{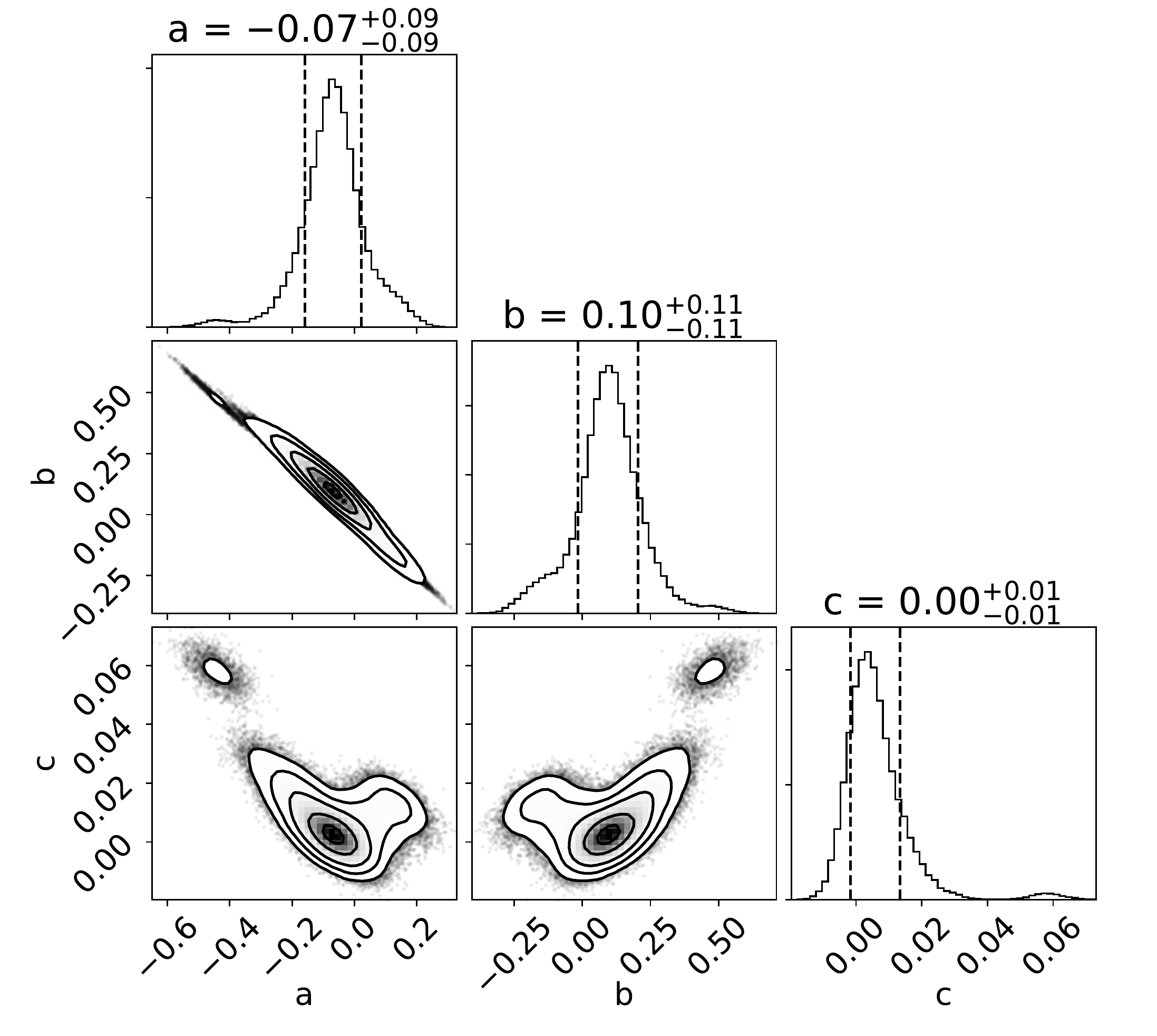}
    \caption{Distribution of fit parameters $a$, $b$ and $c$ (Eq. (\ref{eq:general_shape_factor_fit})) and their correlation projections from the numerical results of Sec. \ref{sec:shape_factor_calc} for pseudoscalar (top) and pseudovector (bottom) transitions. The appearance of heavy tails and multimodal distributions show the need for the improvement. Plots were made using Ref. \cite{Foreman-Mackey2016}.}
    \label{fig:fit_parameter_dist_kernel}
\end{figure}

(\textit{New}) Due to the inclusion of additional shape factors presented in Sec. \ref{sec:shape_factor_calc}, the old procedure discussed above is not optimal. One of the main reasons for this lies in the appearance of shape factors with large positive slopes (see Fig. \ref{fig:shape_factors_collection}) for $\Delta J=1$. Fit parameter distributions as shown in Fig. \ref{fig:fit_parameter_dist_kernel_old} become multimodal and substantial tails appear. As such, rather than approximating each parameter distribution individually as a single Gaussian related via an average correlation matrix (Eq. (\ref{eq:avg_corr})), we take into account all correlations without compromise. Figure \ref{fig:fit_parameter_dist_kernel} shows the results for both pseudoscalar and pseudovector transitions after application of Gaussian kernel density estimation using the `Scott' bandwidth estimator.

Both the appearance of heavy tails and multimodal distributions can clearly be seen, showing the necessity of the new approach. Using the results of Fig. \ref{fig:shape_factors_collection} some of the influences of the new shape factors on the parameter distribution can clearly be discerned. Using this information, one can now produce samples stochastically drawn from this three-dimensional probability distribution. This can be done either using Markov Chain Monte Carlo techniques or using the properties of Gaussian functions when applying Gaussian kernel density smoothing. Here, we have opted for the latter due to its computational simplicity.

\begin{figure}[!ht]
    \centering
    \includegraphics[width=0.48\textwidth]{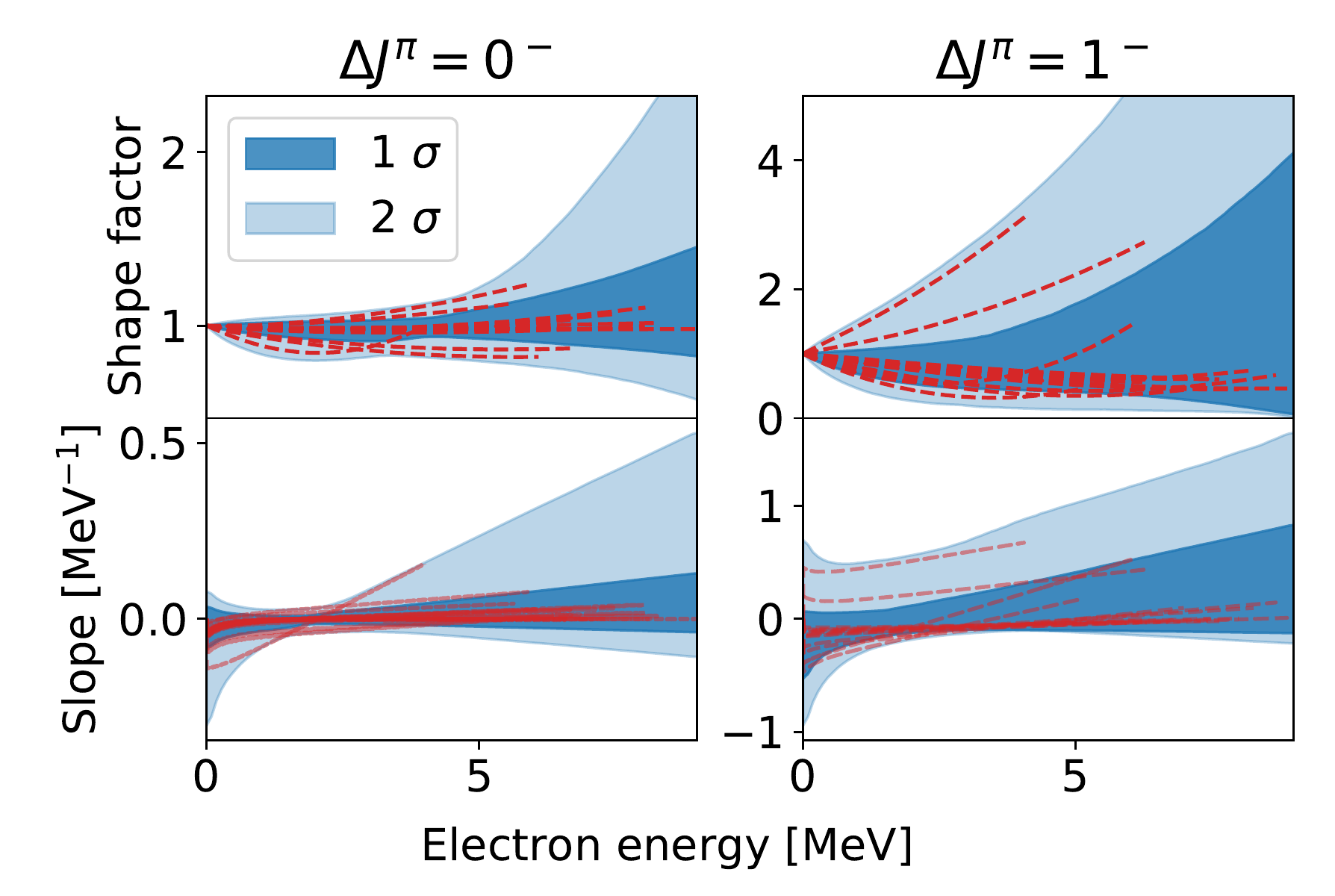}
    \caption{Assessment of the quality of parametrized shape factors through a comparison with the numerically calculated shapes of Sec. \ref{sec:shape_factor_calc}. The top rows shows the normalized shape factors, whereas the bottom row shows the slope. The left (right) column shows these for pseudoscalar (pseudovector). Intervals corresponding to $68\%$ and $95\%$ quantiles are shown as $1\sigma$ and $2\sigma$, respectively. Numerical shape factors are plotted using $g_A = 0.9$ and $\epsilon_\text{MEC} = 1.4$ where appropriate, as above.}
    \label{fig:parametrisation_comparison}
\end{figure}

In order to gauge how well the parametrization performs, we compare a generated ensemble against the numerical calculations of Sec. \ref{sec:shape_factor_calc}, shown in Fig. \ref{fig:parametrisation_comparison}. As we are mainly interested in the energy-dependence, we additionally plot the first derivative.

Excellent agreement is obtained for both $\Delta J = 0$ and $\Delta J = 1$, for the normalized shape factors as well as their first derivatives. The large range of possible shape factors for high endpoints, however, is related to the substantial variation in the quadratic component. In the pseudoscalar case this is, for example, because of transitions such as $^{94}$Y, where a strong quadratic component arises from a contribution of a $\Delta J = 2$ operator despite it being a pseudoscalar transition. In the case of pseudovector transitions, large ranges are obtained due to the appearance of both positive and negative slope shape factors. Using the old approach, it is not possible to achieve a good agreement as in Fig. \ref{fig:parametrisation_comparison} without drastically increasing the width of the Gaussian kernels, voiding the original intent.

In using a simple polynomial fit for all shape factors, however, the distinctions between the origin of the different kinematic terms are not made. This means, for example, that the quadratic component can be overestimated outside of the endpoint range for which the fit was made. Taking $^{94}$Y as an example once more, its parabolic behaviour arises mainly from the $\Delta J = 2$ operator, which will give rise to a parabolic shape no matter the endpoint. Due to the `blind' fit of Eq. (\ref{eq:general_shape_factor_fit}), however, such behaviour is not recognized and only the large quadratic component is recorded. Due to the limited number of data points we choose not to go further in this, and keep in mind that uncertainties could be too conservative in the high energy range.

As with any parametrization, its quality is only as good as the input data on which it is based. Following the discussion of Sec. \ref{sec:flux_coverage}, we argue that our selected transitions correspond to a representative sample of forbidden transitions within the region of interest. The parametrization proposed here thus corresponds to a generalization of our knowledge and our lack of it that we perceive to be realistic. 

\subsubsection{Monte Carlo procedure}
The jump to a generalized summation calculation taking into account all first forbidden transitions is now straightforward. The summation calculations proceed as normal, with the exception of forbidden transitions. Here, the spin-parity change is determined and the corresponding approximate shape factor is taken as described in the previous section, with the exception of the transitions described in Sec. \ref{sec:shape_factor_calc}. The nuclear-structure dependent change from the approximate shape factor is then assigned randomly according to the distribution of fit parameters as described above. Repeating the procedure many times results in a translation of the uncertainty of the shape factors into a spectral uncertainty. This uncertainty will become most apparent in regions where few branches contribute.

Based on the behaviour of the categorized shape factors of Fig. \ref{fig:shape_factors_collection}, we expect the deviations from pseudoscalar transitions to average out within an individual calculation in regions where many branches contribute. Sampling $\Delta J = 1^-$ shape factors as observed, we expect a decrease in the predicted electron flux at high energies and opposite for the antineutrino flux. Unique forbidden decays, finally, decrease both electron and antineutrino predictions in its central range, while providing only small increases at very low energies and their corresponding endpoints. This fact becomes increasingly strong for higher degrees of forbiddenness. 

Note that a single summation calculation like this samples the probability distributions roughly 1600 times (see Table \ref{tab:breakdown_branches}). In the results discussed below, 100 Monte Carlo calculations then correspond to a sampling of the parametrized probability distributions of 160000 times.

\section{Updated summation calculations}
We combine all information from the foregoing sections into a comprehensive spectral analysis. As it forms the only experimental data available, we commence the discussion with a comparison to the ILL data set \cite{Haag2014}. We move on to the spectral changes induced due to the enhanced treatment of forbidden transitions in the summation approach and continue to the uncertainty estimate. Finally, we discuss our results within the context of the reactor spectral bump and the flux anomaly. 

\subsection{ILL spectral reconstruction}
\label{sec:ILL_reconstruction}
Many authors have treated both summation and virtual branch methods in relation to the ILL data set \cite{Mueller2011, Mention2011, Huber2011, *Huber2012, Hayes2015, Dwyer2015}. While progress on the latter has been limited, improved summation calculations are made possible through an intensive research program employing Total Absorption Gamma Spectroscopy (e.g. \cite{Zakari-Issoufou2015}). Many cases troubled by Pandemonium \cite{Hardy1977} have been resolved, and very recently state-of-the-art calculations have achieved a correspondence with the ILL data set at the few percent level through intricate connections between a vast array of databases \cite{Estienne2019}. In this work we employ a simpler approach to clearly identify the impact of non-unique forbidden transitions on the antineutrino spectrum. As described in Sec. \ref{sec:data_selection_handling}, we choose to work here with a combination of the ENDF and ENSDF decay libraries, using the branching ratios of the former and spin-parities of the latter. This will be denoted as `ENDF+ENSDF'.

\begin{figure}[ht]
    \centering
    \includegraphics[width=0.48\textwidth]{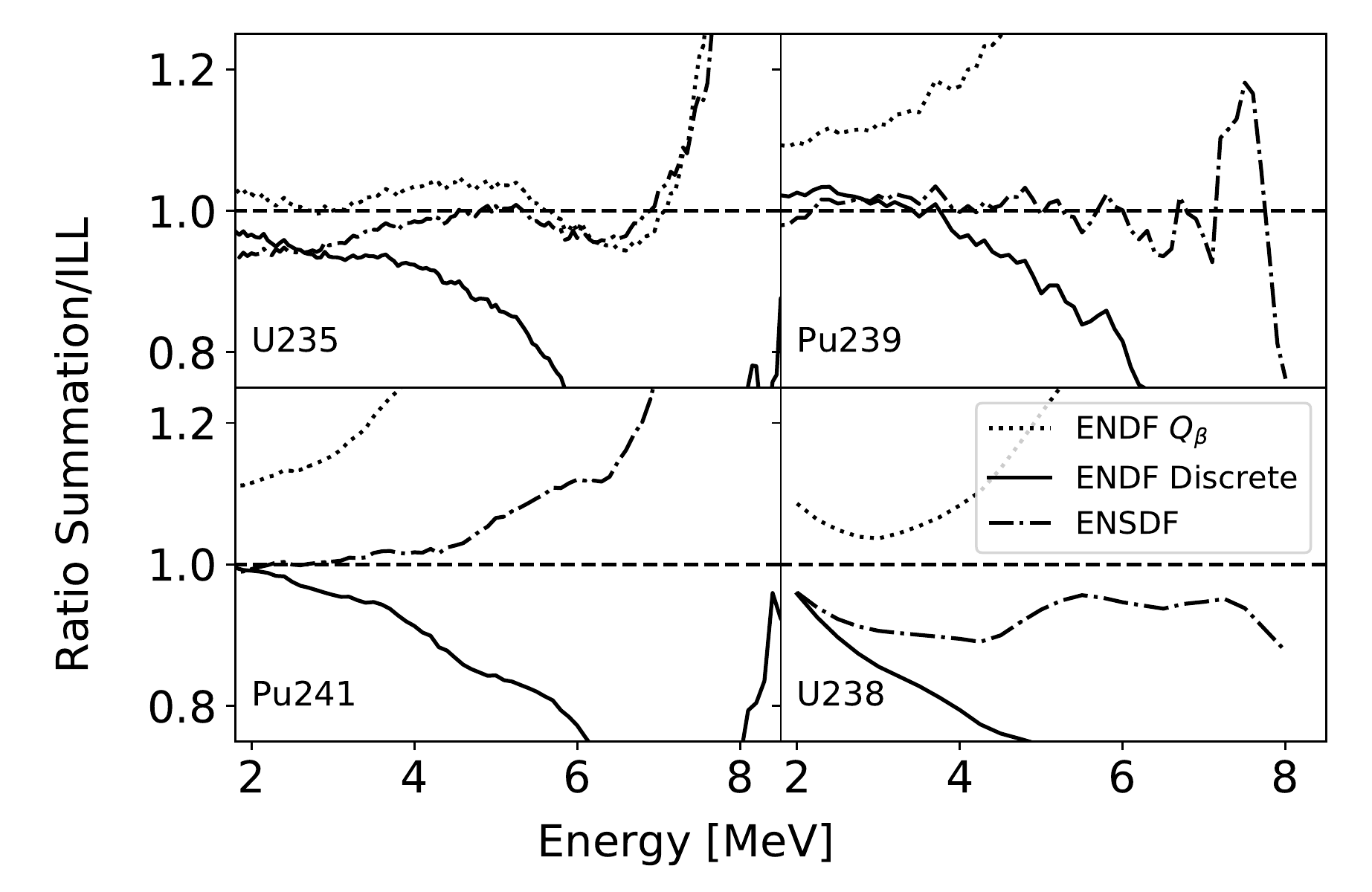}
     \caption{Comparison of different ways of combining the nuclear databases and the resulting agreement with the ILL experimental electron spectra. We use the JEFF3.3 database for fission yield, and the decay library of the ENDF database (see Sec. \ref{sec:data_selection_handling}). Here ENDF Discrete takes into account only transitions for which discrete level data was present.}
    \label{fig:Ratio_ILL_electron}
\end{figure}

Additionally, we return to the point of `continuous' data within the ENDF database for certain isotopes. This is often the case for isotopes far away from stability with relatively high $Q$ values \cite{Chadwick2011}. Its influence will then mainly be felt in the upper half ($> 5\,$MeV) of electron and antineutrino spectra. A proper spectral inversion relies on knowledge of the underlying structure, however, so that it is essentially a miniature form of the more general spectral inversion. Due to the degeneracy in input theoretical shapes \cite{Hayes2014} and the enormous amount of fitting parameters, we choose not to attempt individual conversions and use instead only the transitions for which discrete information is available (ENDF Discrete). Introducing instead the $Q_\beta$ approximation (Sec. \ref{sec:data_selection_handling}), poor agreement is obtained with the ILL data set with a significant overestimate of the total flux at energies higher than $5\,$MeV. A summary is presented in Fig. \ref{fig:Ratio_ILL_electron}.

As discussed previously, the omission of isotopes with continuous spectra (be they experimentally or theoretically obtained \cite{Chadwick2011, Kawano2008}) manifests itself as a deficit with respect to the ILL data sets mainly at higher energies. Additionally, beyond $7.5\,$MeV, $^{235}$U shows telltale signs of Pandemonium corruption. As this lies outside of the region of interest and is well under control through the inclusion of TAGS data, we instead move on with the `ENDF+ENSDF' discrete data set (ENDF Discrete in Fig. \ref{fig:Ratio_ILL_electron}).

\begin{figure}[!ht]
    \centering
    \includegraphics[width=0.48\textwidth]{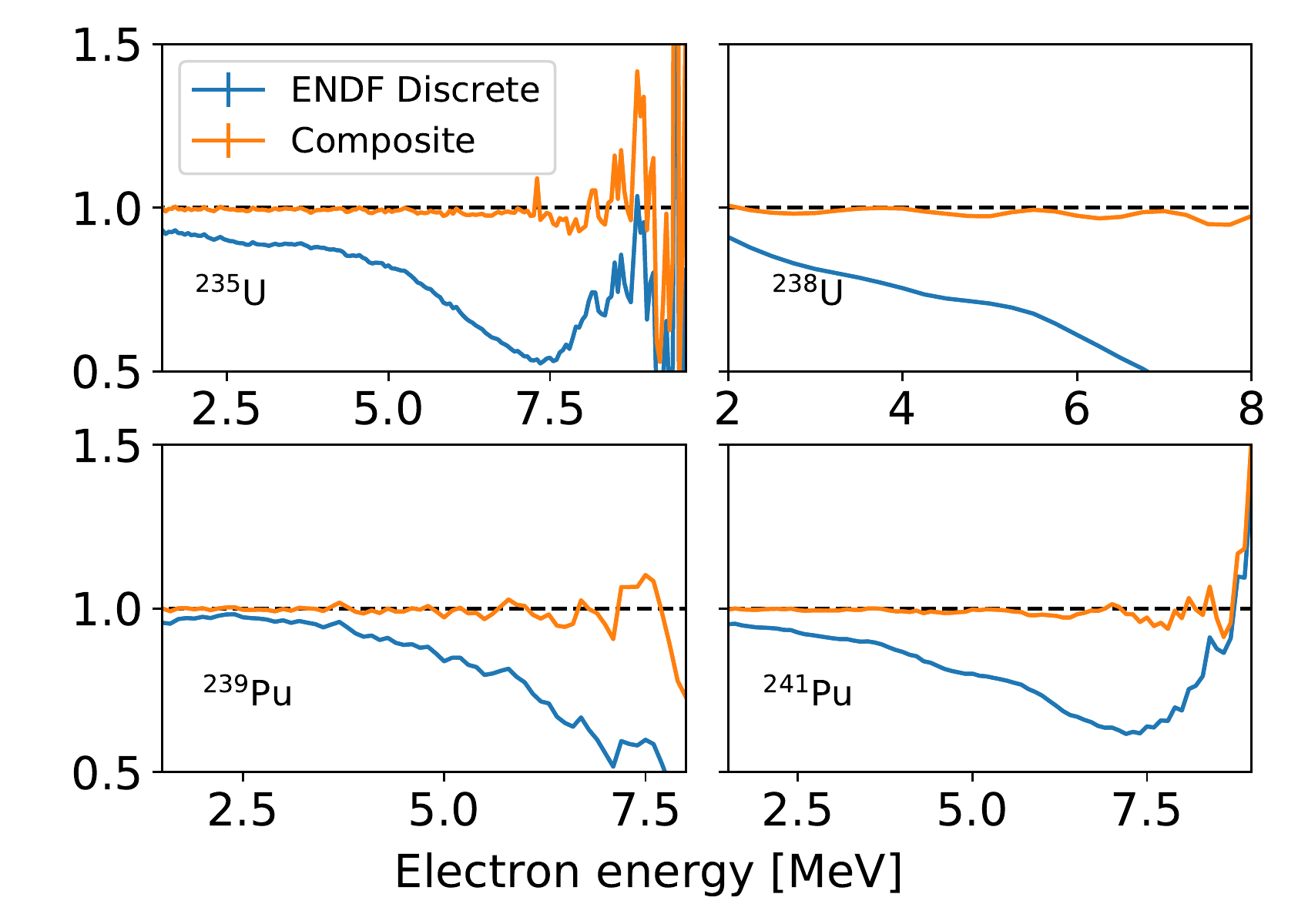}
    \caption{Comparison of the summation and composite approach when using only discrete spectral information from the ENDF database as a foundation. For all four isotopes, less than 8 virtual branches were used to obtain percent-level agreement in the composite approach.}
    \label{fig:composite_vs_summation_ENDF_discrete}
\end{figure}

In order to obtain correspondence with the ILL data sets, we extend the analysis using a composite approach. Here, we fit the residual electron flux between the ENDF+ENSDF data set and the experimental data using a limited number of virtual branches. These are then explicitly inverted for the antineutrino spectrum. A summary is shown in Fig. \ref{fig:composite_vs_summation_ENDF_discrete}. Good agreement is obtained for all fission actinides, with remaining residuals on the percent level. The fine structure that remains is a consequence of the summation part and can not be adequately compensated for using virtual branches. Since these are limited to the region beyond 7.5 MeV, these will be of no consequence to our final result.

\subsection{Spectral shape changes}
All the pieces are now in play to commence a final comparison of our results to those found in the literature. Throughout this work we have discussed various approximations made in earlier works (Sec. \ref{sec:approximation_breakdown} \& \ref{sec:allowed_shape_factors}) and effects of including additional corrections in our description of individual and cumulative spectra. Here, we will discuss the influence of these various effects on the cumulative electron and antineutrino spectra.

Throughout this section and the next, we will compare three different approaches to calculating the composite spectra: ($i$) treating first-forbidden transitions using the 36 calculated shape factors of Sec. \ref{sec:shape_factor_calc}, with and without parametrized results for the remaining forbidden branches; ($ii$) treating those forbidden transitions as allowed with $C = 1$ (Eq. (\ref{eq:C_1})); ($iii$) treating those transitions using the weak magnetism correction of 0.67\%\,MeV$^{-1}$ (Eq. (\ref{eq:delta_wm_simple})). We will report our result as the relative differences between these approaches, i.e. the difference between ($i$) and ($ii$), and ($i$) and ($iii$). Note that in each of these three approaches the summation contribution to the total electron flux will vary. As a consequence, the same applies for the virtual branch contribution in order to force correspondence with the ILL dataset. For the virtual branches, we allow the average $Z$ value to change within the uncertainties of the fit \cite{Huber2011, *Huber2012}, and randomize the endpoint energy of each virtual branch within the bin size. This results in a statistical uncertainty due to the conversion procedure, which contributes to the final spectral uncertainty reported in the Appendix.

\subsubsection{Numerical shape factors}
\label{sec:results_numerical}
The central result of this work is the direct effect of including the numerically calculated shape factors of Sec. \ref{sec:shape_factor_calc} into the summation and composite calculations. For the former, the results were already demonstrated in Fig. \ref{fig:electron_antineutrino_prediction_change}, where the shaded areas correspond to the total difference in both electron and antineutrino spectra. 

The remarkable finding is that the electron spectra can experience relatively minor changes of 2\% and lower, while the antineutrino spectrum can increase by up to 5\% in the same energy range. This is a coalescence of the steep decrease in magnitude of the cumulative electron flux and the composition profile of the flux as shown in Fig. \ref{fig:spectral_constituents_electron}. A downward sloping shape factor for a forbidden transition pushes electrons towards lower energies, but the relative change in the cumulative electron flux remains minor due to the flux being several orders of magnitude larger there. Any quadratic or energy-inverse component in the shape factor will enforce this result, as its effects are spread throughout the entire antineutrino spectrum due to the varying endpoints of the transitions. This is the central element common to all of our results.

\begin{figure}[!ht]
    \centering
    \includegraphics[width=0.48\textwidth]{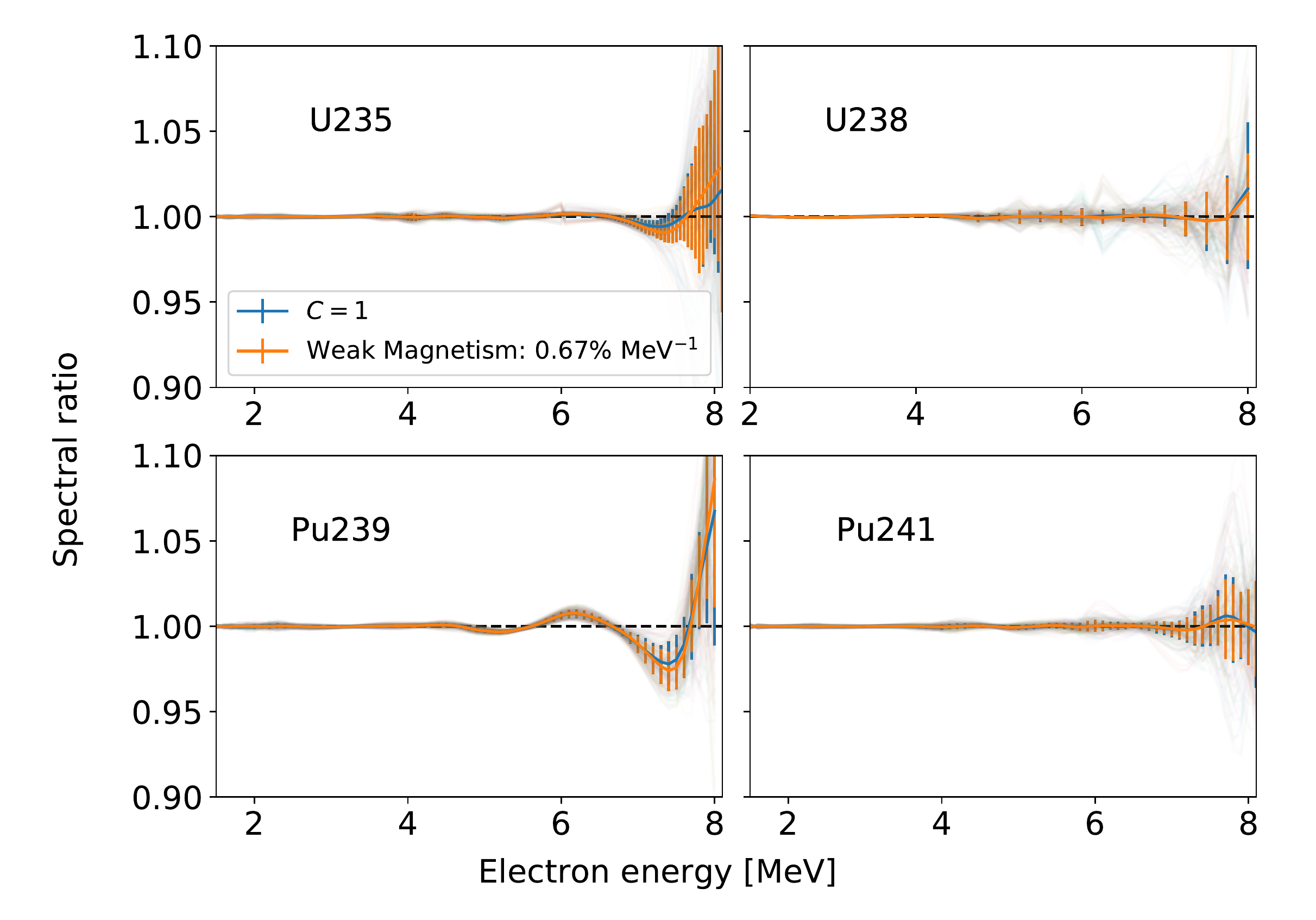}
    \caption{Relative change in the cumulative electron spectra in the composite approach when treating the transitions in Table \ref{tab:summary_transitions_4MeV_235U} as allowed and forbidden. The good agreement over the full range guarantees a good match to the ILL data set for all four actinides. Residuals from unity are taken into account as a bias uncertainty reported on in the Appendix.}
    \label{fig:composite_finnish_electron}
\end{figure}

Figure \ref{fig:composite_finnish_electron} shows the difference between cumulative electron spectra in the different approximations using the composite approach. Up to at least 7.5 MeV no statistically significant differences appear, meaning that our composite approach is able to fit successfully to the ILL spectra both when assuming the transitions of Table \ref{tab:summary_transitions_4MeV_235U} to be allowed and when using the forbidden shape factors. This procedure is successful within the percent level up to $\sim 7$ MeV. We take into account the remaining residuals in a so-called bias uncertainty, which is reported in the Appendix.

\begin{figure}[!ht]
    \centering
    \includegraphics[width=0.48\textwidth]{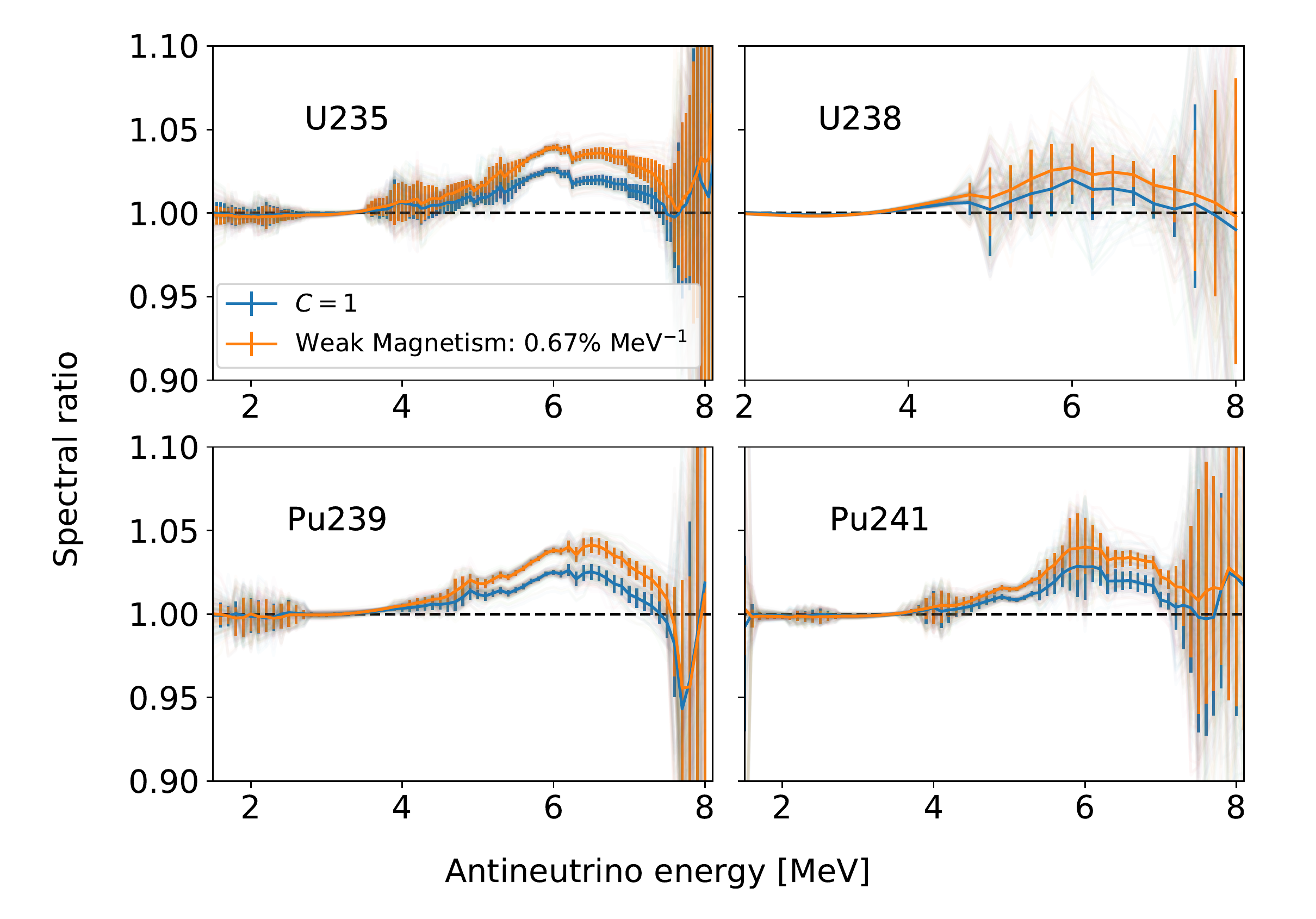}
    \caption{Relative change to the cumulative antineutrino spectra in the composite approach when treating the transitions in Table \ref{tab:summary_transitions_4MeV_235U} as allowed and forbidden. A bump appears between 4 and 7 MeV, with a magnitude of up to 4.5\%.}
    \label{fig:composite_finnish_neutrino}
\end{figure}

The good agreement of the electron spectra in the three different approaches is a necessary requirement for a clean interpretation of the results in the antineutrino spectrum, which are shown in Fig. \ref{fig:composite_finnish_neutrino}.

As anticipated (see Fig. \ref{fig:electron_antineutrino_prediction_change}), a similar pattern arises where significant changes occur in the antineutrino spectrum while the electron flux remains relatively unchanged. In performing the composite approach the latter is fixed, so that the decrease in the electron spectrum is compensated for through the virtual branches. This has then the effect that the change in the antineutrino spectrum change is even greater, and is approximately equal to the sum of the differences of antineutrino and electron flux in Fig. \ref{fig:electron_antineutrino_prediction_change}. As a consequence, a bump appears in the 4-7 MeV range with a magnitude of up to 4.5\% when comparing against the weak magnetism correction of Eq. (\ref{eq:delta_wm_simple}). When comparing against setting $C$ equal to unity the effect is less pronounced and the bump magnitude reaches only 2.5\%.

Due to their similar proton-to-neutron ratio, the fission fragment distributions are very similar for $^{235}$U and $^{239}$Pu, leading to near-identical results. For $^{241}$Pu, on the other hand, additional substructure is visible around 6 MeV. This appears to be an accidental combination of circumstances in the shape factor results and fission yield distributions.

\begin{figure}[!ht]
    \centering
    \includegraphics[width=0.48\textwidth]{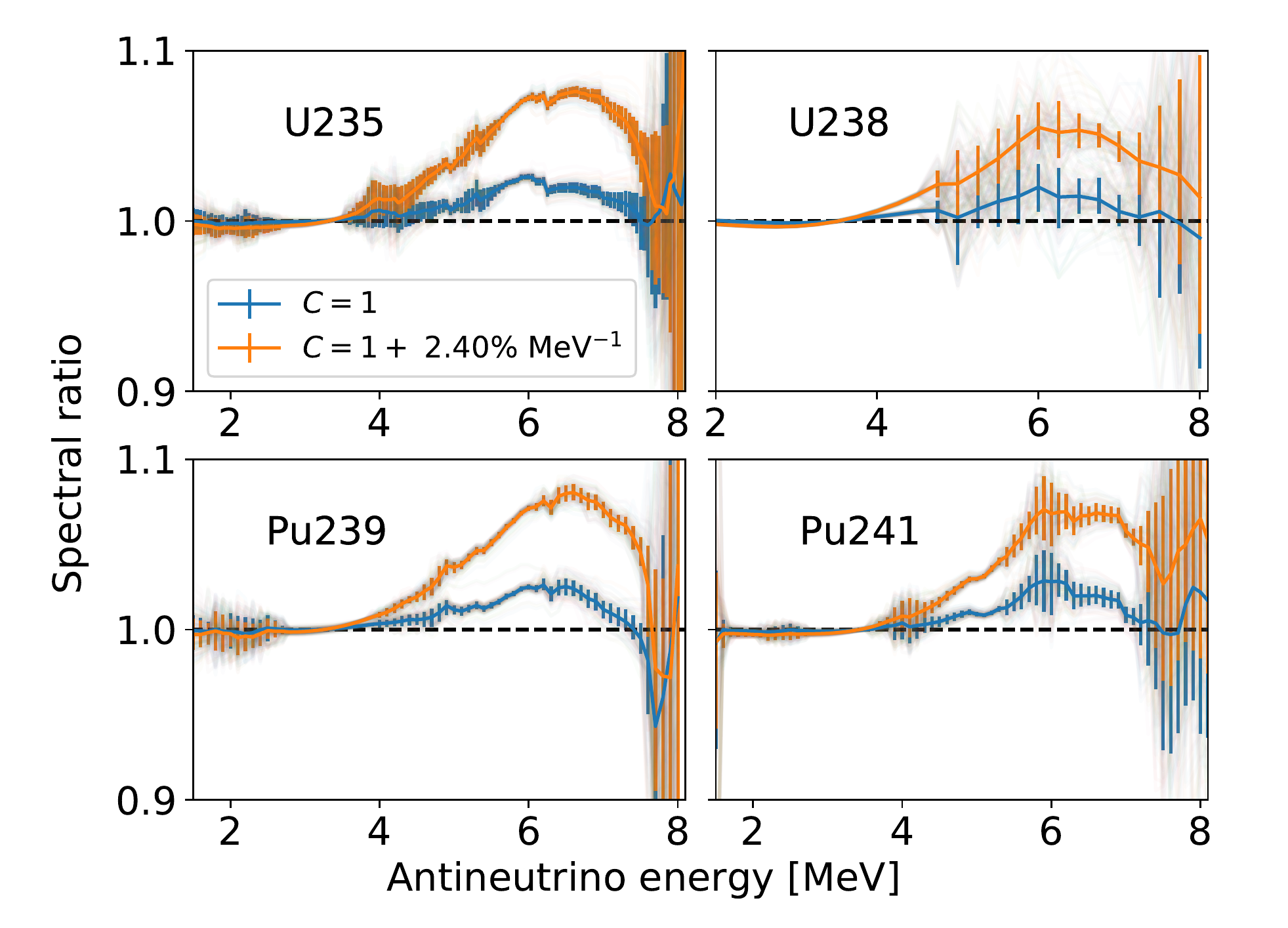}
    \caption{Relative change to the cumulative antineutrino spectra in the composite approach when comparing against a trivial allowed shape factor and one with a slope roughly four times the weak magnetism correction. The magnitude of the induced bump reaches over 8.5\%.}
    \label{fig:composite_finnish_neutrino_240}
\end{figure}

In the original work by Huber \cite{Huber2011, *Huber2012} it was noted that the chosen value for the weak magnetism correction had a strong influence on the final results of the reactor normalization anomaly. Further, it was estimated that an increase by a factor of four would be sufficient to eliminate the anomaly entirely. While a larger-scale study done specifically on weak magnetism found no such variations \cite{Wang2017}, from our discussion in Sec. \ref{sec:allowed_shape_factors} it is clear that such a slope difference can arise from a variety of other terms which were up to now forgotten. Additionally, we found above that the amplitude of the bump arising from a proper treatment of forbidden transitions depends on the magnitude of the slope in the allowed shape factor. Figure \ref{fig:composite_finnish_neutrino_240} shows the same results as above, but instead using an allowed shape factor with a slope that is four times larger than the weak magnetism correction used in the original analysis.

Unsurprisingly, the magnitude of the bump is larger than when compared to the normal weak magnetism correction and now reaches up to 8.5\% for all isotopes besides $^{238}$U. Interestingly, the magnitude of the correction now becomes very similar to that which is observed experimentally. It appears then that through a combination of proper treatment of forbidden transitions and a change in average slope of allowed transitions as discussed in Sec. \ref{sec:allowed_shape_factors}, both the normalization anomaly and the spectral shoulder can be solved at the same time. 

\subsubsection{Including parametrized forbidden shape factors}

\begin{figure}[!ht]
    \centering
    \includegraphics[width=0.48\textwidth]{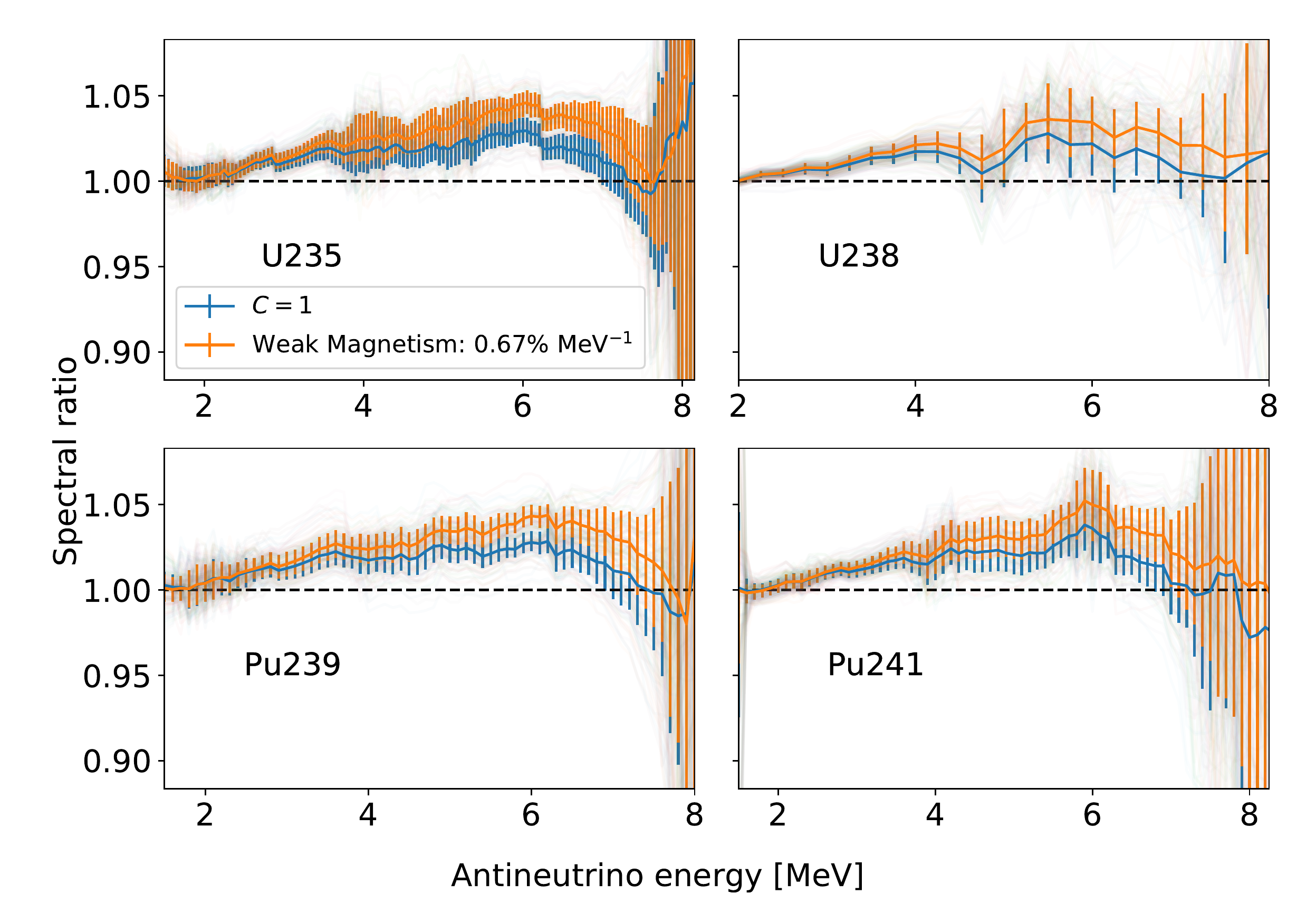}
    \caption{Relative change to the cumulative antineutrino spectra in the composite approach when treating the transitions in Table \ref{tab:summary_transitions_4MeV_235U} as allowed and forbidden and using the parametrization as described in Sec. \ref{sec:generalization} for the remaining forbidden branches. Statistics are based on 100 Monte Carlo cycles. Uncertainties are a combination of the virtual branch procedure (see Fig. \ref{fig:composite_finnish_neutrino}) and the parametrization.}
    \label{fig:composite_finnish_neutrino_parametrized}
\end{figure}

Following the discussion of Sec. \ref{sec:generalization}, we go one step further and use the parametrization derived there to look at the additional effects of including all other known forbidden transitions in a stochastic way. Since the explicitly calculated transitions already constitute a significant part of the total flux (see Fig. \ref{fig:contribution_U235}), and forbidden transitions take up about 60\% of the flux in the region of interest (see Fig. \ref{fig:spectral_constituents_electron}), the inclusion of the parametrized shape factors will mainly affect the spectral uncertainty rather than the magnitude. As before, the agreement with the ILL electron flux is required in our composite approach. While this succeeds, the uncertainty quickly grows to 10\% above 8 MeV due to the large range of parametrized shape factors at very high energies. For the purposes investigated here, however, this is sufficient. In Fig. \ref{fig:composite_finnish_neutrino_parametrized} we show the result of the calculation of 100 Monte Carlo samples. 

The most apparent change compared to Fig. \ref{fig:composite_finnish_neutrino} lies in the region between 2 and 4 MeV. According to the parametrization results, an increase in the theoretical flux is predicted over the whole range, which gives rise to a much wider shoulder. Whether this is a true verifiable feature or a limitation of our parametrization remains to be seen. In this region, the explicitly calculated transitions correspond to only 10-35\% of the total flux, even though according to Fig. \ref{fig:spectral_constituents_electron} about 50\% of the flux must originate from forbidden transitions. Unique transitions occur more prominently here, which could partially explain this increase. 

The uncertainties arising from the parametrization procedure are substantially larger than those from the conversion procedure of the spectrum residuals as discussed in the previous section. Depending on the isotope, $1\sigma$ uncertainties range from 1.5\% to 3\% around the 6 MeV range and quickly grow to more than 10\% at 8 MeV. These uncertainties must be added on top of those already present in the original procedure \cite{Mueller2011, Huber2011, *Huber2012}, as well as those originating from the uncertainty in the explicitly calculated shape factors of Sec. \ref{sec:shape_factor_calc}. This will be discussed in more detail in the Appendix.

\subsection{Integrated flux changes}

In the previous section we have summarized the changes to the total cumulative electron and antineutrino spectra. We will now use the same results to look at the change in the integrated cumulative and inverse $\beta$ decay (IBD) flux. For the cross section of the latter we use the expressions given by Ref. \cite{Hayes2016, Vogel1999, Strumia2003, Kurylov2003}. The strong energy dependence of the cross section forms a small counterweight against the steep decrease of the antineutrino flux. As a consequence, the change in theoretically predicted flux of Figs. \ref{fig:composite_finnish_neutrino}-\ref{fig:composite_finnish_neutrino_parametrized} will leave an imprint. In Table \ref{tab:flux_change_HM} we show the difference in antineutrino and IBD flux with respect to the Huber-Mueller predictions.

\begin{table}[!ht]
    \centering
    \caption{Difference in the integral and IBD flux compared to the Huber-Mueller results when using only the numerical shape factors as described in Secs. \ref{sec:shape_factor_calc} and \ref{sec:results_numerical}. Positive numbers indicate a larger calculated flux. Uncertainties quoted come only from the procedure explained here.}
    \label{tab:flux_change_HM}
    \begin{ruledtabular}
    \begin{tabular}{c|dddd}
         & \multicolumn{1}{c}{$^{235}$U} & \multicolumn{1}{c}{$^{238}$U} & \multicolumn{1}{c}{$^{239}$Pu} & \multicolumn{1}{c}{$^{241}$Pu} \\
         \hline
        $\phi$ & 0.2(2) & 0.4(5) & 0.2(2) & 0.3(2) \\
        $R_\text{IBD}$ & 0.8(5) & 2.3(10) & 0.7(5) & 0.7(6)
    \end{tabular}
    \end{ruledtabular}
\end{table}

As the antineutrino flux is dominated by its behaviour at low energies, relative changes to the Huber-Mueller model are minimal for all isotopes and correspond to a $1\sigma$ shift away from zero. The uncertainties are dominated by the changes to $g_A$ in the description of the shape factors as we consider them to be fully correlated across bins and isotopes, meaning deviations remain at the $1\sigma$ level even for different fuel compositions.

The IBD rate, on the other hand, picks up significant contributions from the expected increase in the bump region. The total effect is limited, however, to below one percent for the main contributors and constitutes a $\sim1.5\sigma$ effect. 

Table \ref{tab:flux_change} shows an overview of the change in total flux and IBD rate for the different possibilities of constructing a forbidden-corrected spectrum. All results are relative to including only the numerical shape factors of Sec. \ref{sec:shape_factor_calc}.

\begin{table*}[ht]
    \centering
    \caption{Integrated flux and IBD rate change due to the inclusion of forbidden transition shape factors for the different fission actinides. Here slope is the slope of the shape factor of allowed decays which are to be compared against (see Sec. \ref{sec:allowed_shape_factors}). Note that these are relative changes with respect to the improved treatment of forbidden transitions. For absolute changes with respect to the Huber-Mueller, one can take the difference with the results of Table \ref{tab:flux_change_HM}.}
    \label{tab:flux_change}
    \begin{ruledtabular}
    \begin{tabular}{ldddddddd}
         & \multicolumn{4}{c}{$\phi$} & \multicolumn{4}{c}{$R_\text{IBD}$} \\
         \cline{2-5}
         \cline{6-9}
         & \multicolumn{2}{c}{Numerical} & \multicolumn{2}{c}{Parametrized} &\multicolumn{2}{c}{Numerical} & \multicolumn{2}{c}{Parametrized} \\
         \cline{2-3}
         \cline{4-5}
         \cline{6-7}
         \cline{8-9}
        Slope & \multicolumn{1}{c}{0\%} & \multicolumn{1}{c}{0.67\%} & \multicolumn{1}{c}{0\%} & \multicolumn{1}{c}{0.67\%} & \multicolumn{1}{c}{0\%} & \multicolumn{1}{c}{0.67\%} & \multicolumn{1}{c}{0\%} & \multicolumn{1}{c}{0.67\%} \\
        \hline
        $^{235}$U & 0.0(2) & 0.2(2) & 0.8(7) & 1.5(7) & 0.6(4) & 1.0(6) & 1.7(12) & 2.4(12)\\
        $^{238}$U & 0.1(6) & 0.2(6) & 0.9(11) & 1.6(11) & 0.4(10) & 0.7(10) & 1.3(18) & 1.9(18)\\
        $^{239}$Pu & 0.1(3) & 0.2(3) & 1.3(8) & 1.6(8) & 0.6(5) & 0.9(6) & 1.7(13) & 2.4(13)\\
        $^{241}$Pu & 0.1(3) & 0.1(3) & 1.2(8) & 1.4(8) & 0.5(5) & 0.9(7) & 1.8(14) & 2.3(14)
    \end{tabular}
    \end{ruledtabular}
\end{table*}

As before, flux changes are minimal and within $1\sigma$ uncertainty when including only the numerical shape factors. Differences are larger for higher values of the slope  as the latter decreases the antineutrino spectrum yield. When including the parametrized shape factors both the central value and uncertainty increase significantly. Due to the increase in the expected flux starting at 3 MeV in the parametrized setup, flux changes exceed one percent and correspond to roughly a $\sim 2 \sigma$ effect. IBD rates, likewise, increase significantly as do the uncertainties. As with the effects of $g_A$ quenching before, the large uncertainty arises mainly from the bin-to-bin correlation for the parametrized shape factors.

\section{Reactor spectrum changes}
\label{sec:reactor_changes}
Up to now we have discussed changes in spectral features of individual fission isotopes, in particular for $^{235}$U. Depending on the type of reactor, the other three actinides contribute substantially to the total flux. In general, the flux from a nuclear reactor with thermal power $W_{\text{th}}$ can be given as
\begin{equation}
    S(E_{\nu}) = \frac{W_{\text{th}}}{\sum_iR_ie_i}\sum_iR_iS_i(E_{\nu})
\end{equation}
where $e_i$ is the energy released per fission by an actinide $i$, $R_i$ is the fractional contribution of each actinide and $S_i(E_{\nu})$ the corresponding antineutrino spectrum. Modern reactor experiments such as Daya Bay \cite{An2012}, RENO \cite{Ahn2012} and Double Chooz \cite{Abe2014}, all have different configurations of the fractional contributions $R_i$. All three experiments have, however, published results showing a bump in the 4-6 MeV region of the prompt positron energy ($E_{\text{prompt}} \approx E_{\nu} - 0.782$ MeV) spectrum relative to the Huber-Mueller theoretical predictions. It is currently unclear which isotope(s) contribute primarily to the spectral shoulder or normalization anomaly. Here we have used previously published values for the fuel composition of the three experiments, listed in Table \ref{tab:reactor_composition}.

\begin{table}[!ht]
    \centering
    \caption{Reactor fuel composition in the three modern reactor antineutrino experiments.}
    \label{tab:reactor_composition}
    \begin{ruledtabular}
    \begin{tabular}{c|dddd}
        Reactor & \multicolumn{1}{c}{$^{235}$U} & \multicolumn{1}{c}{$^{238}$U} & \multicolumn{1}{c}{$^{239}$Pu} & \multicolumn{1}{c}{$^{241}$Pu} \\
        \hline
        Daya Bay & 0.586 & 0.076 & 0.288 & 0.05 \\
        RENO & 0.62 & 0.12 & 0.21 & 0.05 \\
        Double Chooz & 0.496 & 0.087 & 0.351 & 0.066
    \end{tabular}
    \end{ruledtabular}
\end{table}

The main difference regarding total flux in these experiments is the contribution of $^{238}$U, as it provides the hardest antineutrino spectrum of the four fission actinides. An investigation of the fuel dependency of the results presented here will be a topic of further research. 

\subsection{Spectral shoulder}

\begin{figure}[!ht]
    \centering
    \includegraphics[width=0.48\textwidth]{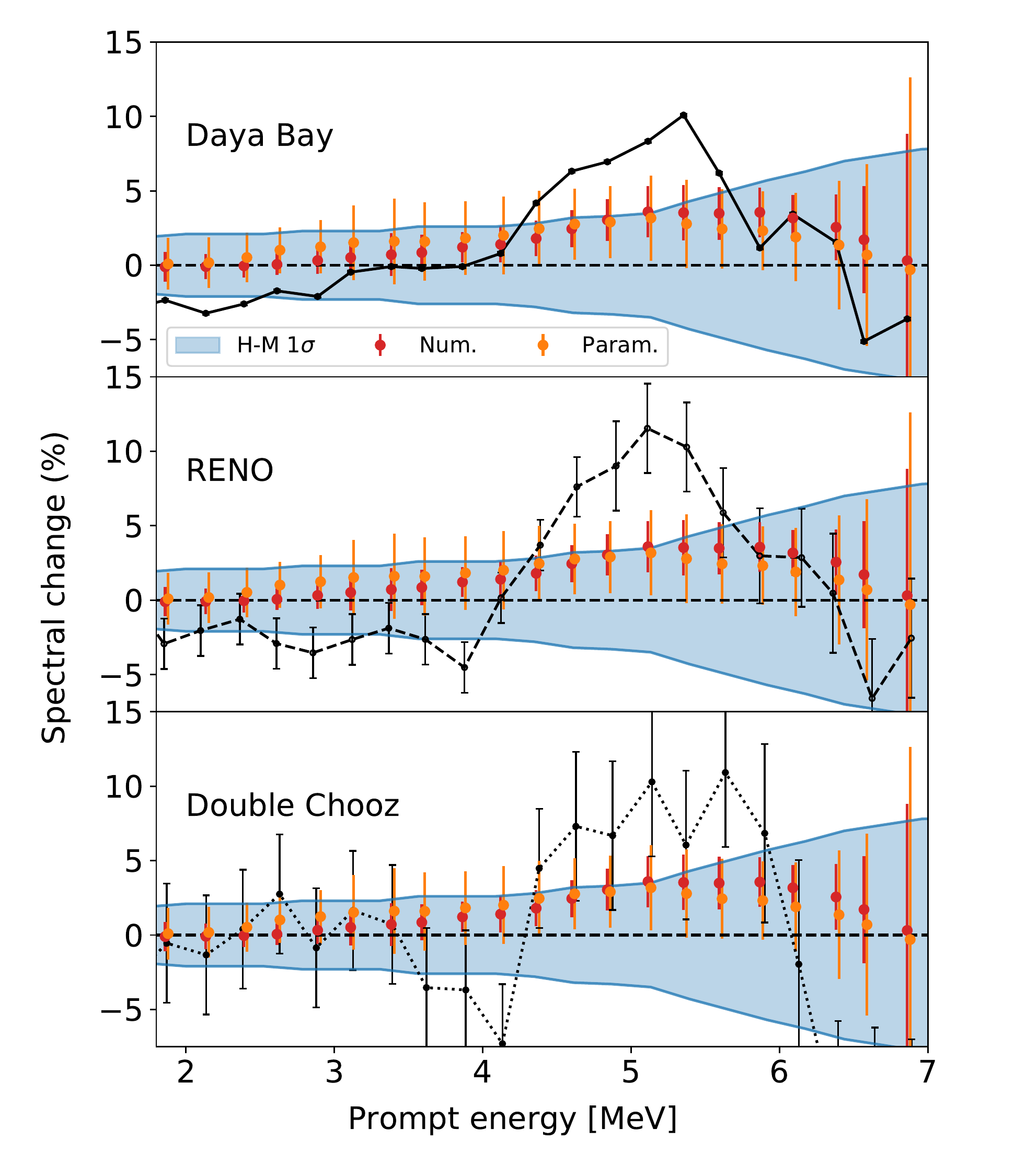}
    \caption{Comparison of the expected spectrum change due to forbidden transitions for the different reactors, together with the observed discrepancies of experimental data relative to the Huber-Mueller results. Here `Num.' stands for including only the shape factors of Sec. \ref{sec:shape_factor_calc} and `Param.' includes a parametrization of all other forbidden transitions as discussed in Sec. \ref{sec:generalization}. Both results are relative to an allowed shape with a weak magnetism correction of 0.67\%\,MeV$^{-1}$.}
    \label{fig:reactors_ratio_comp}
\end{figure}

In order to study the effect of the spectral shoulder, we focus only on the shape and leave the overall normalization a free parameter. Figure \ref{fig:reactors_ratio_comp} shows the spectral shape changes for the different reactors under the different approximations and treatments.

Both numerical only and parametrized version behave very similarly, despite the latter typically obtaining a larger deviation in absolute magnitude. This is due to the normalization requirement which pushes the results of the latter down. Both reach a magnitude of about 4\% relative to the Huber-Mueller predictions. This corresponds to slightly less than half of the total effect observed by all three modern experiments. Combined with the increased correlated uncertainty on every data point, the statistical significance of the spectral shoulder is strongly mitigated. Due to the similarity of the effect for the two main fission actinides, $^{235}$U and $^{239}$Pu, small changes in the fuel composition of Table \ref{tab:reactor_composition} do not appreciatively change the result due to forbidden transitions. 

This effect shown in Fig. \ref{fig:reactors_ratio_comp} is smaller than our previously reported findings \cite{Hayen2019}. Both our former and current results agree with each other within the uncertainties, however. The reason for this lies in part in the extended parametrization procedure discussed in Sec. \ref{sec:generalization}. Due to the inclusion of additional shape factors and the `blind' fit of Eq. (\ref{eq:general_shape_factor_fit}), the range for pseudoscalar and pseudovector shape factors is now substantially larger (see Fig. \ref{fig:parametrisation_comparison}), particularly for higher endpoints. As such, a significant number of parametrized shape factors occur with a positive slope, despite only 2 out of 36 explicitly calculated shape factors showing such a behaviour. As discussed above and in Sec. \ref{sec:generalization}, this is possibly a limitation of our current approach. Despite this, the spread in sampled shape factors and corresponding uncertainty in the cumulative antineutrino spectra remains of particular interest as it is an quantitative estimate of the true uncertainty of previous procedures due to the neglect of forbidden shape factors. Specifically, compared to the uncertainty estimates in Tables VII-X by Huber \cite{Huber2011, *Huber2012}, those arising from the parametrization (see Appendix) are larger or of similar size of the systematic effects presented there. We will elaborate upon this in the next section. 

\begin{figure}[!ht]
    \centering
    \includegraphics[width=0.48\textwidth]{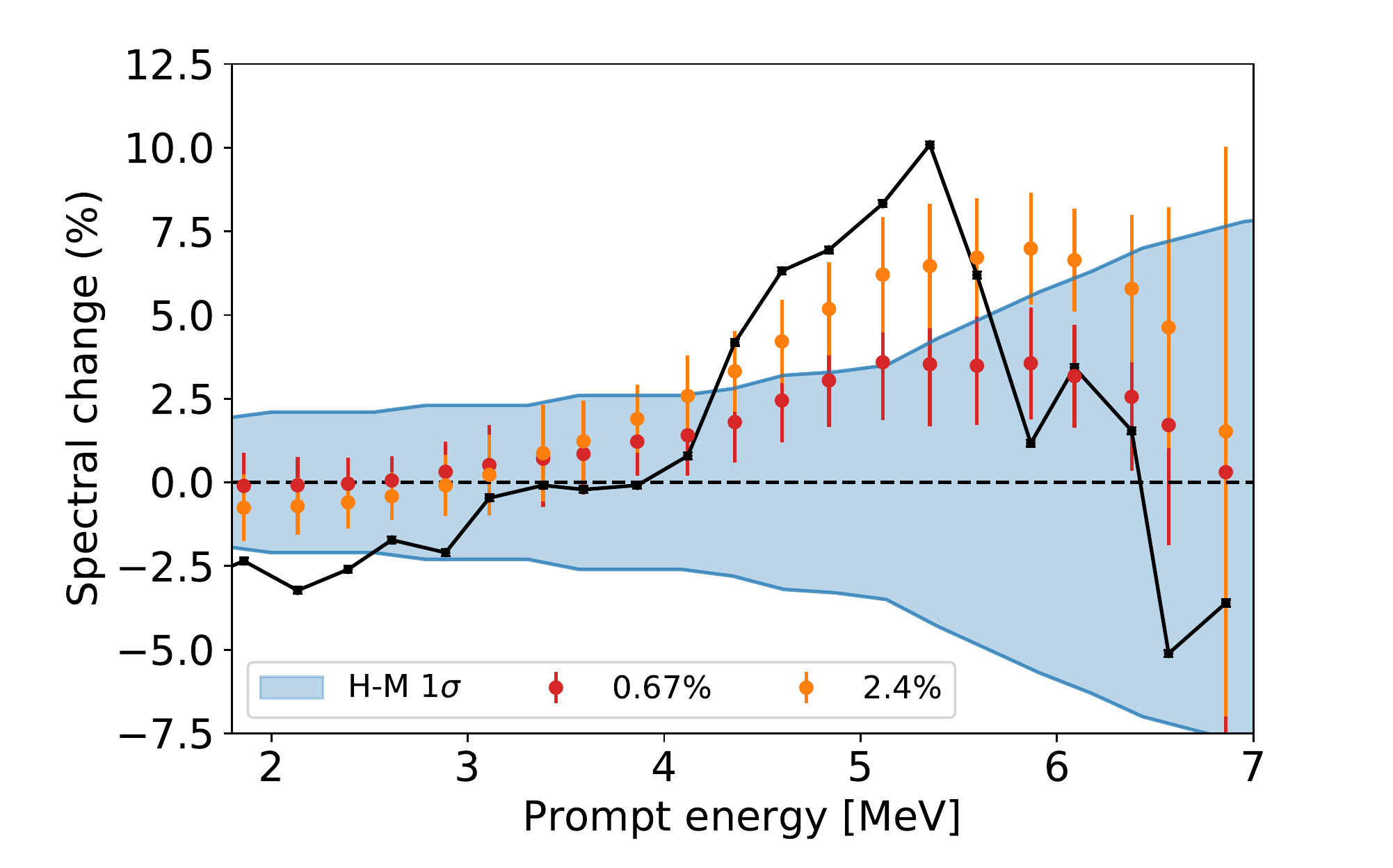}
    \caption{Comparison of the Daya Bay shape discrepancy with the change due to the forbidden shape factors of Sec. \ref{sec:shape_factor_calc} when compared to two different slopes of the allowed approximation: 0.67\% and 2.4\%, as discussed in Secs. \ref{sec:allowed_shape_factors} and \ref{sec:results_numerical}.}
    \label{fig:daya_bay_bump_067_24}
\end{figure}

We have discussed in Sec. \ref{sec:results_numerical} how an increase in the average slope of allowed transitions could combine with the results presented here to solve both the rate anomaly and the spectral shoulder at the same time. In Sec. \ref{sec:allowed_shape_factors} we have proposed how such an increase in the slope could arise. In Fig. \ref{fig:daya_bay_bump_067_24} we show the spectral shape change due to the forbidden shape factors of Sec. \ref{sec:shape_factor_calc} relative to an allowed shape factor with a slope of 2.4\%\,MeV$^{-1}$.

As mentioned above, the magnitude of the deviation is now comparable to what is observed in the modern reactor experiments. A more elaborate investigation of the precise behaviour of individual shape factors for the dominant allowed transitions will be able to shed more light on this. This is under current investigation.

\subsection{Rate anomaly}

Despite forbidden transitions introducing several percent deviations in the spectrum, its effect on the integrated flux and IBD rate is limited as shown in Tables \ref{tab:flux_change_HM} \& \ref{tab:flux_change}, due to the rapid decrease in antineutrino flux towards higher energies. As discussed above however, the additional uncertainty arising from this process is significant or even dominant relative to the other systematic effects taken into account in the original analyses \cite{Huber2011, *Huber2012, Mueller2011}. Additionally, due to the procedure used here several of the sources of uncertainty are fully correlated between bins \textit{and} isotopes. As a consequence, the uncertainty due to forbidden transitions can be significant even when compared to the statistical and normalization uncertainties due to the experimental electron data. Due to our simplified treatment of the databases in our composite approach (see Secs. \ref{sec:data_selection_handling} \& \ref{sec:ILL_reconstruction}), we will comment here only on relative changes due to the inclusion of forbidden transitions. Due to the similarity in fuel composition of the different reactors (Table \ref{tab:reactor_composition}), and the agreement in spectral changes among the main fission actinides presented here (Fig. \ref{fig:composite_finnish_neutrino}), our conclusions on the rate anomaly will be the same for all three reactors.

According to the Huber-Mueller model \cite{Huber2011, *Huber2012, Mueller2011, Mention2011} one found the following cross-section shifts for the four fission actinides: $3.7\%, 4.2\%, 4.7\%$ and $9.8\%$ for $^{235}$U, $^{239}$Pu, $^{241}$Pu and $^{238}$U, respectively. Following the reactor fuel composition, each contributes almost equally to generate a net 3-4\% total shift. If we limit ourselves to the results of Table \ref{tab:flux_change_HM} where we used only the explicitly calculated shape factors of Sec. \ref{sec:shape_factor_calc}, we see that each fission actinide has an increase in the theoretical prediction of the integrated flux of less than $10\%$ with an increased uncertainty of $\sim 5\%$. The IBD rate, on the other hand, is predicted to increase by $\sim 15$-$25\%$ compared to the original analysis with a doubled uncertainty. Relative to the rate uncertainties quoted by Huber \cite{Huber2011, *Huber2012}, however, this constitutes an increase of $\sim 10\%$ of uncertainty correlated between isotopes. When combining this with the spectral normalization uncertainty common to all isotopes originating from the ILL data set (see Tables VII-X in Ref. \cite{Huber2011, *Huber2012}), the model uncertainty on the rate anomaly is enlarged by only 4\% even though the disagreement shifts by +14\%.

Using instead the parametrized results together with the numerical shape factors of Sec. \ref{sec:shape_factor_calc}, both the shift and uncertainty increase dramatically as shown in Table \ref{tab:flux_change}. Due to the increase of the predicted flux between 2 and 4 MeV, where the IBD flux reaches a maximum, the total shift corresponds to 2.3(13)\%. Relative to the Huber predictions, this corresponds to an increase in the isotope shift of around 60\% even though the uncertainty increases only by 30\%. For the total rate anomaly this corresponds to a shift of almost $40\%$ even though the uncertainty increases only by $13\%$. 

Due to the limitations in the parametrization discussed above, one could argue using only the 36 explicitly calculated shape factors of Table \ref{tab:summary_transitions_4MeV_235U}, but consider the uncertainty due to the parametrization of non-unique forbidden shape factors to represent a good estimate of the uncertainty due to the neglect of all others present in the database. In this case, we combine the central value shifts of Table \ref{tab:flux_change_HM} with the uncertainties for the parametrized calculations of Table \ref{tab:flux_change}. This then corresponds to an increase of the IBD rate by 0.8(13)\% for every reactor. Once again combining this with the normalization uncertainties common to all isotopes, this leaves the significance of the rate anomaly unchanged as both central value and uncertainty shift by $\sim 14\%$.

%

\section{Conclusion}
We have, for the first time, calculated a significant number of dominant forbidden $\beta$ transitions above 4 MeV using the nuclear shell model. As anticipated, a large fraction of these show strongly deviating shape factors from the allowed approximation. Even for pseudoscalar transitions, i.e. $\Delta J = 0^-$, where shape factors are distributed around the allowed form, the limited number of contributing branches result in a net shift in the calculated electron and antineutrino spectra. This observation is strongly augmented by the results from transitions with $\Delta J = 1^-, 2^-$, as their shape factors are strongly divergent from the allowed approximation. A direct consequence of the inclusion of these results on the cumulative spectrum shapes is a net decrease of the predicted summation electron spectra by 1-2\% in the region between 5 and 7 MeV, depending on what weak magnetism correction is used to compare against. A similar increase on the order of 4-5\% is observed in the corresponding antineutrino spectra in the region between 4.5 and 7.5 MeV. We have investigated the uncertainty of these results by changing the renormalization of the axial vector coupling constant and axial charge, where appropriate, and found them to be on the order of 0.5\%. 

In addition to the precise calculations described here, we have shown that the contribution of allowed $\beta$ spectra dip (significantly) below 50\% over most of experimentally accessible range. This occurs through a combination of fission yields and branching ratios favouring forbidden spectra. This dominion of forbidden spectra is particularly apparent in the region of the spectral shoulder, which motivated us in an attempt to generalize the contribution due to forbidden spectra. In this spirit, we have attempted a parametrization of effective non-unique first forbidden shape factors according the spin change, $\Delta J$. Rather than assume an allowed shape, for each branch we now sample from a distribution of spectral shape based on the numerical results described above. While this parametrization has limitations due to the neglect of underlying nuclear structure considerations, the spectral uncertainty arising from it is comparable to or larger than the main systematic uncertainties in the Huber-Mueller model.

We have combined these results with the fuel compositions of the Daya Bay, RENO and Double Chooz experiments and interpreted our results in terms of the observed spectral shoulder and rate anomaly. For the former, we find a spectral distortion very similar in shape as to what is observed experimentally, albeit with a lower magnitude. Due to the similarity of the spectral changes of the different actinides and relatively minor changes in fuel composition, our results are the same for all three experiments. For the rate anomaly, we show that our results increase the expected theoretical flux to varying degree depending on whether or not the parametrization results are taken into account. Using only the explicitly calculated shape factors, we find an increase in the expected IBD rate of 0.8(5)\%. Using a proposed parametrization, this increase rises to 2.3(13)\%. When combining the central value from the explicitly calculated shape factors with the uncertainty of the parametrization, the statistical significance of the rate anomaly remains unaltered.

Finally, we have proposed that an increase in the average slope of allowed shape factors can yield a solution to both the rate anomaly and the spectral shoulder when combined with the results presented here. We have provided theoretical arguments and indications for why this could be the case. This is currently under investigation.

Based on the results presented above, it is clear that a proper understanding of forbidden shape factors is invaluable in the understanding of both the reactor anomaly and spectral bump.

\begin{acknowledgements}
One of the authors (L.H.) would like to thank Muriel Fallot and Alejandro Sonzogni for constructive discussion. This work has been partly funded by the Belgian Federal Science Policy Office, under Contract No. IUAP EP/12-c and the Fund for Scientific Research Flanders (FWO). This work has been partially supported by the Academy of Finland under the Academy project no. 318043. J. K. acknowledges the financial support from Jenny and Antti Wihuri Foundation.
\end{acknowledgements}

\onecolumngrid
\appendix

\section{Flux changes and uncertainties}
\label{app:flux_changes_uncertainty}

\begin{table*}[!ht]
    \centering
    \caption{Results for the $^{235}$U spectrum. The errors in column 4 are completely uncorrelated, while those of column 5, 6 and 7 are completely correlated. Column 4 is the statistical uncertainty in the conversion procedure resulting from the spread in virtual branch endpoints and average $Z$ values. The bias uncertainty is the difference from unity in the agreement between the electron cumulative spectra in the different approaches. Column 5 represents the uncertainty due to $g_A$ quenching as discussed in Sec. \ref{sec:uncertainty_estimation_shape_factors}. Columns 4-6 contribute to the total uncertainty for the numerical approach. The parametrized results receive additional uncertainty from the parametrization procedure, listed in column 7.}
    \label{tab:breakdown_changes_errors_U235}
    \begin{ruledtabular}
    \begin{tabular}{lddddccdc}
    \multicolumn{3}{c}{Value} & \multicolumn{6}{c}{$1\sigma$ errors}\\
    \cline{1-3}
    \cline{4-9}
    $E_\nu$ & \multicolumn{1}{c}{$\delta N$ Num} & \multicolumn{1}{c}{$\delta N$ Param} & \multicolumn{1}{c}{Conv.} & \multicolumn{1}{c}{Bias} & $g_A$ & Param. &  \multicolumn{1}{c}{Total Num.} & Total Param.\\
    (MeV) & \multicolumn{1}{c}{(\%)} & \multicolumn{1}{c}{(\%)} & \multicolumn{1}{c}{(\%)} & \multicolumn{1}{c}{(\%)} & (\%) & (\%) & \multicolumn{1}{c}{(\%)} & (\%)\\
    \hline
        2.0 & 0.1 & 0.4 & 0.3 & 0.0 & 0.1 & ${}^{+0.7}_{-0.5}$ & 0.1 & ${}^{+0.7}_{-0.5}$\\
        2.25 & -0.6 & 0.7 & 0.4 & 0.0 & 0.2 & ${}^{+1.0}_{-0.8}$ & 0.5 & ${}^{+1.1}_{-0.9}$\\
        2.5 & 0.0 & 0.7 & 0.2 & 0.0 & 0.2 &${}^{+0.3}_{-0.4}$ & 0.2 & ${}^{+0.4}_{-0.5}$\\
        2.75 & -0.1 & 1.2 & 0.0 & 0.0 & 0.3 &${}^{+0.5}_{-0.5}$ & 0.3 & 0.6\\
        3.0 & -0.1 & 1.1 & 0.0 & 0.0 & 0.3 & ${}^{+0.5}_{-0.4}$& 0.3 & ${}^{+0.6}_{-0.5}$\\
        3.25 & 0.0 & 1.6 & 0.0 & 0.0 & 0.3 & ${}^{+0.5}_{-0.4}$& 0.3 & ${}^{+0.6}_{-0.5}$\\
        3.5 & 0.1 & 2.1 & 0.0 & 0.0 & 0.2 &${}^{+0.4}_{-0.4}$ & 0.2 & 0.4\\
        3.75 & 0.5 & 2.0 & 0.6 & 0.0 & 0.2 &${}^{+0.5}_{-0.7}$ & 0.6 & ${}^{+0.8}_{-0.9}$\\
        4.0 & 0.7 & 2.9 & 0.8 & 0.0 & 0.5 &${}^{+1.3}_{-1.4}$ & 0.9 & ${}^{+1.6}_{-1.7}$\\
        4.25 & 0.8 & 2.3 & 0.9 & 0.0 & 0.4 &${}^{+1.3}_{-1.4}$ & 1.0 & ${}^{+1.6}_{-1.7}$\\
        4.5 & 0.9 & 2.5 & 0.3 & 0.0 & 0.3 &${}^{+1.3}_{-1.0}$ & 0.4 & ${}^{+1.3}_{-1.1}$\\
        4.75 & 1.5 & 2.7 & 0.6 & 0.0 & 0.2 &${}^{+0.9}_{-1.0}$ & 0.6 & ${}^{+1.1}_{-1.2}$\\
        5.0 & 1.3 & 2.6 & 0.6 & 0.0 & 0.1 &${}^{+1.0}_{-1.2}$ & 0.6 & ${}^{+1.2}_{-1.3}$\\
        5.25 & 2.1 & 3.6 & 0.4 & 0.0 & 0.3 &${}^{+1.0}_{-0.7}$ & 0.5 & ${}^{+1.1}_{-0.9}$\\
        5.5 & 2.9 & 3.5 & 0.3 & 0.0 &  0.5 &${}^{+0.8}_{-0.7}$ & 0.6 & ${}^{+1.0}_{-0.9}$\\
        5.75 & 3.4 & 3.9 & 0.2 & 0.0 & 0.7 &${}^{+0.8}_{-0.8}$ & 0.7 & 1.1\\
        6.0 & 3.9 & 4.2 & 0.2 & 0.0 & 1.0 &${}^{+0.9}_{-0.9}$ & 1.0 & 1.4\\
        6.25 & 3.2 & 3.2 & 0.3 & 0.2 & 0.4 & ${}^{+0.8}_{-0.7}$& 0.6 & ${}^{+1.0}_{-0.9}$\\
        6.5 & 3.6 & 3.4 & 0.4 & 0.2 & 0.5 & ${}^{+0.9}_{-0.8}$& 0.7 & 1.1\\
        6.75 & 3.5 & 3.0 & 0.4 & 0.2 & 0.7 &${}^{+1.0}_{-1.0}$ & 0.9 & 1.3\\
        7.0 & 2.9 & 2.5 & 0.6 & 0.1 & 0.6 &${}^{+1.4}_{-1.2}$ & 0.7 & ${}^{+1.6}_{-1.4}$\\
        7.25 & 2.5 & 2.2 & 0.9 & 0.6 & 0.7 &${}^{+2.3}_{-1.6}$ & 1.7 & ${}^{+2.8}_{-2.3}$\\
        7.50 & 0.6 & 0.6 & 2.0 & 1.3 & 0.7 &${}^{+3.7}_{-2.6}$ & 3.4 & ${}^{+5.0}_{-4.3}$\\
        7.75 & 0.7 & 1.0 & 6.3 & 3.6 & 0.4 &${}^{+4.1}_{-5.3}$ & 9.9 & ${}^{+10.7}_{-11.2}$\\
        8.0 & -1.0 & -4.0 & 19.0 & 5.5 & 0.6 &${}^{+11.0}_{-9.1}$ & 22.1 & ${}^{+24.7}_{-23.9}$
    \end{tabular}
    \end{ruledtabular}
\end{table*}

\begin{table*}[!ht]
    \centering
    \caption{Results for the $^{238}$U spectrum. The errors in column 4 are completely uncorrelated, while those of column 5, 6 and 7 are completely correlated. }
    \label{tab:breakdown_changes_errors_U238}
    \begin{ruledtabular}
    \begin{tabular}{lddddccdc}
    \multicolumn{3}{c}{Value} & \multicolumn{6}{c}{$1\sigma$ errors}\\
    \cline{1-3}
    \cline{4-9}
    $E_\nu$ & \multicolumn{1}{c}{$\delta N$ Num} & \multicolumn{1}{c}{$\delta N$ Param} & \multicolumn{1}{c}{Conv.} & \multicolumn{1}{c}{Bias} & $g_A$ & Param. &  \multicolumn{1}{c}{Total Num.} & Total Param.\\
    (MeV) & \multicolumn{1}{c}{(\%)} & \multicolumn{1}{c}{(\%)} & \multicolumn{1}{c}{(\%)} & \multicolumn{1}{c}{(\%)} & (\%) & (\%) & \multicolumn{1}{c}{(\%)} & (\%)\\
    \hline
        2.0 & -0.1 & 0.1 & 0.0 & 0.1 & 0.5 & ${}^{+0.1}_{-0.1}$ & 0.5 & ${}^{+0.5}_{-0.5}$\\
        2.25 & -0.1 & 0.4 & 0.0 & 0.0 & 0.2 & ${}^{+0.1}_{-0.1}$ & 0.2 & ${}^{+0.2}_{-0.2}$\\
        2.5 & -0.2 & 0.5 & 0.0 & 0.0 & 0.4 & ${}^{+0.1}_{-0.2}$ & 0.4 & ${}^{+0.4}_{-0.4}$\\
        2.75 & -0.2 & 0.8 & 0.0 & 0.0 & 0.4 & ${}^{+0.2}_{-0.2}$ & 0.4 & ${}^{+0.4}_{-0.4}$\\
        3.0 & -0.2 & 0.8 & 0.0 & 0.0 & 0.4 & ${}^{+0.3}_{-0.3}$& 0.4 & ${}^{+0.5}_{-0.5}$\\
        3.25 & 0.1 & 1.2 & 0.0 & 0.0 & 0.5 & ${}^{+0.4}_{-0.4}$& 0.5 & ${}^{+0.6}_{-0.6}$\\
        3.5 & 0.0 & 1.6 & 0.1 & 0.0 & 0.6 & ${}^{+0.4}_{-0.4}$ & 0.6 & ${}^{+0.7}_{-0.7}$\\
        3.75 & 0.2 & 1.7 & 0.1 & 0.0 & 0.6 & ${}^{+0.5}_{-0.4}$ & 0.6 & ${}^{+0.8}_{-0.7}$\\
        4.0 & 0.4 & 2.2 & 0.1 & 0.1 & 0.7 & ${}^{+0.6}_{-0.4}$ & 0.7 & ${}^{+0.9}_{-0.8}$\\
        4.25 & 0.6 & 2.3 & 0.1 & 0.1 & 0.7 & ${}^{+0.8}_{-0.6}$ & 0.7 & ${}^{+1.0}_{-0.9}$\\
        4.5 & 0.9 & 2.0 & 0.1 & 0.0 & 0.5 & ${}^{+1.0}_{-0.7}$ & 0.5 & ${}^{+1.1}_{-0.9}$\\
        4.75 & 1.5 & 1.6 & 0.4 & 0.2 & 0.4 & ${}^{+1.3}_{-0.6}$ & 0.6 & ${}^{+1.4}_{-0.8}$\\
        5.0 & 0.0 & 1.9 & 2.0 & 0.0 & 0.3 & ${}^{+2.0}_{-1.9}$ & 2.0 & ${}^{+2.8}_{-2.8}$\\
        5.25 & 1.3 & 3.2 & 1.2 & 0.1 & 0.2 & ${}^{+1.4}_{-1.3}$ & 1.2 & ${}^{+1.9}_{-1.8}$\\
        5.5 & 2.6 & 4.3 & 1.7 & 0.0 &  0.4 & ${}^{+1.7}_{-1.5}$ & 1.7 & ${}^{+2.4}_{-2.3}$\\
        5.75 & 2.8 & 3.6 & 1.3 & 0.2 & 0.6 & ${}^{+2.6}_{-2.8}$ & 1.4 & ${}^{+3.0}_{-3.2}$\\
        6.0 & 2.0 & 3.0 & 2.3 & 0.2 & 0.9 & ${}^{+2.2}_{-2.5}$ & 2.5 & ${}^{+3.3}_{-3.5}$\\
        6.25 & 1.5 & 3.0 & 2.4 & 0.2 & 0.5 & ${}^{+1.0}_{-1.0}$& 2.5 & ${}^{+2.7}_{-2.7}$\\
        6.5 & 2.3 & 3.4 & 1.6 & 0.0 & 0.6 & ${}^{+1.0}_{-1.3}$& 1.7 & ${}^{+2.0}_{-2.1}$\\
        6.75 & 2.1 & 2.9 & 1.1 & 0.2 & 0.8 & ${}^{+1.2}_{-1.5}$ & 1.4 & ${}^{+1.8}_{-2.0}$\\
        7.0 & 1.5 & 1.6 & 0.8 & 0.2 & 0.3 & ${}^{+2.0}_{-1.5}$ & 0.9 & ${}^{+2.2}_{-1.7}$\\
        7.25 & 2.3 & 1.6 & 2.2 & 0.1 & 0.3 & ${}^{+2.8}_{-2.9}$ & 2.2 & ${}^{+3.6}_{-3.7}$\\
        7.50 & -0.4 & -0.2 & 5.1 & 0.3 & 0.1 & ${}^{+5.2}_{-2.6}$ & 5.1 & ${}^{+7.3}_{-5.7}$\\
        7.75 & 1.3 & 2.3 & 5.3 & 0.3 & 0.3 & ${}^{+5.4}_{-6.4}$ & 5.3 & ${}^{+7.6}_{-8.3}$\\
        8.0 & -1.7 & 2.2 & 5.9 & 0.6 & 0.3 & ${}^{+6.7}_{-4.2}$ & 5.9 & ${}^{+9.0}_{-7.3}$
    \end{tabular}
    \end{ruledtabular}
\end{table*}

\begin{table*}[!ht]
    \centering
    \caption{Results for the $^{239}$Pu spectrum. The errors in column 4 are completely uncorrelated, while those of column 5, 6 and 7 are completely correlated. }
    \label{tab:breakdown_changes_errors_Pu239}
    \begin{ruledtabular}
    \begin{tabular}{lddddccdc}
    \multicolumn{3}{c}{Value} & \multicolumn{6}{c}{$1\sigma$ errors}\\
    \cline{1-3}
    \cline{4-9}
    $E_\nu$ & \multicolumn{1}{c}{$\delta N$ Num} & \multicolumn{1}{c}{$\delta N$ Param} & \multicolumn{1}{c}{Conv.} & \multicolumn{1}{c}{Bias} & $g_A$ & Param. &  \multicolumn{1}{c}{Total Num.} & Total Param.\\
    (MeV) & \multicolumn{1}{c}{(\%)} & \multicolumn{1}{c}{(\%)} & \multicolumn{1}{c}{(\%)} & \multicolumn{1}{c}{(\%)} & (\%) & (\%) & \multicolumn{1}{c}{(\%)} & (\%)\\
    \hline
        2.0 & -0.8 & 0.7 & 0.6 & 0.1 & 0.3 & ${}^{+0.8}_{-1.0}$ & 0.7 & ${}^{+1.0}_{-1.2}$\\
        2.25 & 0.0 & 0.3 & 0.8 & 0.0 & 0.4 & ${}^{+1.2}_{-0.8}$ & 0.9 & ${}^{+1.5}_{-1.2}$\\
        2.5 & -0.3 & 1.6 & 0.7 & 0.0 & 0.4 & ${}^{+0.9}_{-1.0}$ & 0.8 & ${}^{+1.2}_{-1.3}$\\
        2.75 & -0.1 & 1.3 & 0.1 & 0.0 & 0.3 & ${}^{+0.7}_{-0.6}$ & 0.3 & ${}^{+0.8}_{-0.7}$\\
        3.0 & -0.1 & 1.4 & 0.1 & 0.0 & 0.3 & ${}^{+0.8}_{-0.5}$ & 0.3 & ${}^{+0.9}_{-0.6}$\\
        3.25 & 0.0 & 1.9 & 0.1 & 0.0 & 0.2 & ${}^{+0.7}_{-0.6}$ & 0.2 & ${}^{+0.7}_{-0.6}$\\
        3.5 & 0.1 & 2.7 & 0.0 & 0.0 & 0.2 & ${}^{+0.7}_{-0.6}$ & 0.2 & ${}^{+0.7}_{-0.6}$\\
        3.75 & 0.4 & 2.9 & 0.0 & 0.1 & 0.2 & ${}^{+0.7}_{-0.6}$ & 0.2 & ${}^{+0.7}_{-0.6}$\\
        4.0 & 0.6 & 2.7 & 0.1 & 0.1 & 0.4 & ${}^{+1.2}_{-1.4}$ & 0.4 & ${}^{+1.3}_{-1.5}$\\
        4.25 & 0.9 & 3.2 & 0.2 & 0.0 & 0.3 & ${}^{+0.9}_{-0.9}$ & 0.4 & ${}^{+1.0}_{-1.0}$\\
        4.5 & 1.0 & 3.3 & 0.3 & 0.0 & 0.2 & ${}^{+0.7}_{-0.9}$ & 0.4 & ${}^{+0.8}_{-1.0}$\\
        4.75 & 1.5 & 3.9 & 0.5 & 0.1 & 0.2 & ${}^{+0.8}_{-1.0}$ & 0.5 & ${}^{+1.0}_{-1.1}$\\
        5.0 & 1.9 & 4.0 & 0.2 & 0.3 & 0.3 & ${}^{+0.7}_{-0.9}$ & 0.5 & ${}^{+0.8}_{-1.0}$\\
        5.25 & 2.2 & 4.4 & 0.1 & 0.3 & 0.5 & ${}^{+0.8}_{-0.8}$ & 0.6 & ${}^{+1.0}_{-1.0}$\\
        5.5 & 2.5 & 4.2 & 0.1 & 0.1 &  0.7 & ${}^{+0.7}_{-0.7}$ & 0.7 & ${}^{+1.0}_{-1.0}$\\
        5.75 & 3.2 & 4.5 & 0.1 & 0.2 & 0.8 & ${}^{+0.6}_{-0.6}$ & 0.8 & ${}^{+1.0}_{-1.0}$\\
        6.0 & 3.9 & 4.7 & 0.1 & 0.7 & 1.1 & ${}^{+0.7}_{-0.5}$ & 1.3 & ${}^{+1.5}_{-1.4}$\\
        6.25 & 4.2 & 3.7 & 0.3 & 0.7 & 1.1 & ${}^{+0.8}_{-0.7}$ & 1.3 & ${}^{+1.6}_{-1.5}$\\
        6.5 & 4.2 & 3.7 & 0.4 & 0.4 & 0.4 & ${}^{+0.7}_{-0.8}$ & 0.7 & ${}^{+1.0}_{-1.0}$\\
        6.75 & 3.9 & 3.4 & 0.7 & 0.1 & 0.5 & ${}^{+0.9}_{-0.8}$ & 0.9 & ${}^{+1.2}_{-1.2}$\\
        7.0 & 3.0 & 3.0 & 0.5 & 1.0 & 0.4 & ${}^{+1.2}_{-1.1}$ & 1.2 & ${}^{+1.7}_{-1.6}$\\
        7.25 & 2.4 & 2.5 & 0.3 & 2.0 & 0.5 & ${}^{+2.0}_{-1.6}$ & 2.1 & ${}^{+2.9}_{-2.6}$\\
        7.50 & 1.1 & 2.0 & 0.7 & 2.5 & 0.6 & ${}^{+3.6}_{-2.7}$ & 2.7 & ${}^{+4.5}_{-3.8}$\\
        7.75 & -1.1 & -1.4 & 7.3 & 0.0 & 0.4 & ${}^{+8.1}_{-7.1}$ & 7.3 & ${}^{+10.9}_{-10.2}$\\
        8.0 & -2.2 & 7.6 & 16.1 & 9.5 & 0.5 & ${}^{+20.1}_{-14.8}$ & 18.7 & ${}^{+27.5}_{-23.8}$
    \end{tabular}
    \end{ruledtabular}
\end{table*}

\begin{table*}[!ht]
    \centering
    \caption{Results for the $^{241}$Pu spectrum. The errors in column 4 are completely uncorrelated, while those of column 5, 6 and 7 are completely correlated. }
    \label{tab:breakdown_changes_errors_Pu241}
    \begin{ruledtabular}
    \begin{tabular}{lddddccdc}
    \multicolumn{3}{c}{Value} & \multicolumn{6}{c}{$1\sigma$ errors}\\
    \cline{1-3}
    \cline{4-9}
    $E_\nu$ & \multicolumn{1}{c}{$\delta N$ Num} & \multicolumn{1}{c}{$\delta N$ Param} & \multicolumn{1}{c}{Conv.} & \multicolumn{1}{c}{Bias} & $g_A$ & Param. &  \multicolumn{1}{c}{Total Num.} & Total Param.\\
    (MeV) & \multicolumn{1}{c}{(\%)} & \multicolumn{1}{c}{(\%)} & \multicolumn{1}{c}{(\%)} & \multicolumn{1}{c}{(\%)} & (\%) & (\%) & \multicolumn{1}{c}{(\%)} & (\%)\\
    \hline
        2.0 & -0.1 & 0.4 & 0.1 & 0.0 & 0.3 & ${}^{+0.2}_{-0.2}$ & 0.3 & ${}^{+0.4}_{-0.4}$\\
        2.25 & -0.2 & 0.6 & 0.4 & 0.0 & 0.3 & ${}^{+0.3}_{-0.3}$ & 0.5 & ${}^{+0.6}_{-0.6}$\\
        2.5 & -0.2 & 0.8 & 0.4 & 0.0 & 0.4 & ${}^{+0.5}_{-0.4}$ & 0.6 & ${}^{+0.8}_{-0.7}$\\
        2.75 & -0.1 & 1.1 & 0.8 & 0.0 & 0.4 & ${}^{+0.3}_{-0.3}$ & 0.9 & ${}^{+0.9}_{-0.9}$\\
        3.0 & -0.1 & 1.2 & 0.1 & 0.0 & 0.4 & ${}^{+0.2}_{-0.2}$& 0.4 & ${}^{+0.5}_{-0.5}$\\
        3.25 & -0.1 & 1.5 & 0.1 & 0.0 & 0.5 & ${}^{+0.2}_{-0.2}$& 0.5 & ${}^{+0.5}_{-0.5}$\\
        3.5 & 0.0 & 1.8 & 0.0 & 0.0 & 0.5 & ${}^{+0.3}_{-0.2}$ & 0.5 & ${}^{+0.6}_{-0.5}$\\
        3.75 & 0.2 & 1.8 & 0.1 & 0.1 & 0.5 & ${}^{+0.4}_{-0.6}$ & 0.5 & ${}^{+0.7}_{-0.8}$\\
        4.0 & 0.0 & 1.5 & 1.0 & 0.0 & 0.6 & ${}^{+0.6}_{-0.5}$ & 1.2 & ${}^{+1.3}_{-1.3}$\\
        4.25 & 0.6 & 2.0 & 0.4 & 0.1 & 0.5 & ${}^{+1.0}_{-1.0}$ & 0.7 & ${}^{+1.2}_{-1.2}$\\
        4.5 & 1.0 & 2.0 & 0.2 & 0.1 & 0.4 & ${}^{+1.4}_{-1.1}$ & 0.5 & ${}^{+1.5}_{-1.2}$\\
        4.75 & 1.4 & 2.3 & 0.1 & 0.0 & 0.3 & ${}^{+1.6}_{-1.3}$ & 0.3 & ${}^{+1.6}_{-1.3}$\\
        5.0 & 1.6 & 2.4 & 0.1 & 0.1 & 0.3 & ${}^{+1.3}_{-1.4}$ & 0.3 & ${}^{+1.3}_{-1.4}$\\
        5.25 & 1.8 & 2.7 & 0.2 & 0.1 & 0.5 & ${}^{+1.0}_{-1.1}$ & 0.5 & ${}^{+1.1}_{-1.2}$\\
        5.5 & 2.5 & 3.6 & 0.5 & 0.0 & 0.7 & ${}^{+0.7}_{-0.7}$ & 0.9 & ${}^{+1.1}_{-1.1}$\\
        5.75 & 3.6 & 4.9 & 0.3 & 0.0 & 0.9 & ${}^{+1.3}_{-1.1}$ & 0.9 & ${}^{+1.6}_{-1.5}$\\
        6.0 & 4.3 & 5.1 & 0.9 & 0.1 & 1.3 & ${}^{+2.0}_{-2.0}$ & 1.6 & ${}^{+2.6}_{-2.6}$\\
        6.25 & 3.9 & 3.7 & 0.9 & 0.1 & 1.4 & ${}^{+1.4}_{-1.4}$& 1.7 & ${}^{+2.2}_{-2.2}$\\
        6.5 & 3.6 & 3.9 & 0.5 & 0.1 & 0.6 & ${}^{+1.2}_{-1.1}$& 0.8 & ${}^{+1.4}_{-1.4}$\\
        6.75 & 3.5 & 3.5 & 0.5 & 0.0 & 0.9 & ${}^{+1.4}_{-1.1}$ & 1.0 & ${}^{+1.7}_{-1.5}$\\
        7.0 & 2.6 & 2.4 & 0.4 & 0.3 & 0.3 & ${}^{+2.0}_{-1.4}$ & 0.6 & ${}^{+2.1}_{-1.5}$\\
        7.25 & 1.8 & 1.1 & 0.5 & 0.3 & 0.4 & ${}^{+3.1}_{-3.3}$ & 0.7 & ${}^{+3.2}_{-3.4}$\\
        7.50 & 0.8 & -0.2 & 3.6 & 0.5 & 0.4 & ${}^{+4.2}_{-4.3}$ & 3.7 & ${}^{+5.6}_{-5.7}$\\
        7.75 & 2.2 & 5.0 & 5.1 & 0.5 & 0.2 & ${}^{+7.5}_{-7.9}$ & 5.1 & ${}^{+9.1}_{-9.4}$\\
        8.0 & 4.6 & 0.3 & 4.0 & 0.5 & 0.3 & ${}^{+10.5}_{-9.7}$ & 4.0 & ${}^{+11.3}_{-10.5}$
    \end{tabular}
    \end{ruledtabular}
\end{table*}

\twocolumngrid

\bibliography{library}

\begin{thebibliography}{72}%
\makeatletter
\providecommand \@ifxundefined [1]{%
 \@ifx{#1\undefined}
}%
\providecommand \@ifnum [1]{%
 \ifnum #1\expandafter \@firstoftwo
 \else \expandafter \@secondoftwo
 \fi
}%
\providecommand \@ifx [1]{%
 \ifx #1\expandafter \@firstoftwo
 \else \expandafter \@secondoftwo
 \fi
}%
\providecommand \natexlab [1]{#1}%
\providecommand \enquote  [1]{``#1''}%
\providecommand \bibnamefont  [1]{#1}%
\providecommand \bibfnamefont [1]{#1}%
\providecommand \citenamefont [1]{#1}%
\providecommand \href@noop [0]{\@secondoftwo}%
\providecommand \href [0]{\begingroup \@sanitize@url \@href}%
\providecommand \@href[1]{\@@startlink{#1}\@@href}%
\providecommand \@@href[1]{\endgroup#1\@@endlink}%
\providecommand \@sanitize@url [0]{\catcode `\\12\catcode `\$12\catcode
  `\&12\catcode `\#12\catcode `\^12\catcode `\_12\catcode `\%12\relax}%
\providecommand \@@startlink[1]{}%
\providecommand \@@endlink[0]{}%
\providecommand \url  [0]{\begingroup\@sanitize@url \@url }%
\providecommand \@url [1]{\endgroup\@href {#1}{\urlprefix }}%
\providecommand \urlprefix  [0]{URL }%
\providecommand \Eprint [0]{\href }%
\providecommand \doibase [0]{http://dx.doi.org/}%
\providecommand \selectlanguage [0]{\@gobble}%
\providecommand \bibinfo  [0]{\@secondoftwo}%
\providecommand \bibfield  [0]{\@secondoftwo}%
\providecommand \translation [1]{[#1]}%
\providecommand \BibitemOpen [0]{}%
\providecommand \bibitemStop [0]{}%
\providecommand \bibitemNoStop [0]{.\EOS\space}%
\providecommand \EOS [0]{\spacefactor3000\relax}%
\providecommand \BibitemShut  [1]{\csname bibitem#1\endcsname}%
\let\auto@bib@innerbib\@empty
\bibitem [{\citenamefont {Athanassopoulos}\ \emph {et~al.}(1998)\citenamefont
  {Athanassopoulos}, \citenamefont {Auerbach}, \citenamefont {Burman},
  \citenamefont {Caldwell}, \citenamefont {Church}, \citenamefont {Cohen},
  \citenamefont {Donahue}, \citenamefont {Fazely}, \citenamefont {Federspiel},
  \citenamefont {Garvey}, \citenamefont {Gunasingha}, \citenamefont {Imlay},
  \citenamefont {Johnston}, \citenamefont {Kim}, \citenamefont {Louis},
  \citenamefont {Majkic}, \citenamefont {McIlhany}, \citenamefont {Mills},
  \citenamefont {Reeder}, \citenamefont {Sandberg}, \citenamefont {Smith},
  \citenamefont {Stancu}, \citenamefont {Strossman}, \citenamefont {Tayloe},
  \citenamefont {VanDalen}, \citenamefont {Vernon}, \citenamefont {Wadia},
  \citenamefont {Waltz}, \citenamefont {White}, \citenamefont {Works},
  \citenamefont {Xiao},\ and\ \citenamefont {Yellin}}]{Athanassopoulos1998}%
  \BibitemOpen
  \bibfield  {author} {\bibinfo {author} {\bibfnamefont {C.}~\bibnamefont
  {Athanassopoulos}}, \bibinfo {author} {\bibfnamefont {L.~B.}\ \bibnamefont
  {Auerbach}}, \bibinfo {author} {\bibfnamefont {R.~L.}\ \bibnamefont
  {Burman}}, \bibinfo {author} {\bibfnamefont {D.~O.}\ \bibnamefont
  {Caldwell}}, \bibinfo {author} {\bibfnamefont {E.~D.}\ \bibnamefont
  {Church}}, \bibinfo {author} {\bibfnamefont {I.}~\bibnamefont {Cohen}},
  \bibinfo {author} {\bibfnamefont {J.~B.}\ \bibnamefont {Donahue}}, \bibinfo
  {author} {\bibfnamefont {A.}~\bibnamefont {Fazely}}, \bibinfo {author}
  {\bibfnamefont {F.~J.}\ \bibnamefont {Federspiel}}, \bibinfo {author}
  {\bibfnamefont {G.~T.}\ \bibnamefont {Garvey}}, \bibinfo {author}
  {\bibfnamefont {R.~M.}\ \bibnamefont {Gunasingha}}, \bibinfo {author}
  {\bibfnamefont {R.}~\bibnamefont {Imlay}}, \bibinfo {author} {\bibfnamefont
  {K.}~\bibnamefont {Johnston}}, \bibinfo {author} {\bibfnamefont {H.~J.}\
  \bibnamefont {Kim}}, \bibinfo {author} {\bibfnamefont {W.~C.}\ \bibnamefont
  {Louis}}, \bibinfo {author} {\bibfnamefont {R.}~\bibnamefont {Majkic}},
  \bibinfo {author} {\bibfnamefont {K.}~\bibnamefont {McIlhany}}, \bibinfo
  {author} {\bibfnamefont {G.~B.}\ \bibnamefont {Mills}}, \bibinfo {author}
  {\bibfnamefont {R.~A.}\ \bibnamefont {Reeder}}, \bibinfo {author}
  {\bibfnamefont {V.}~\bibnamefont {Sandberg}}, \bibinfo {author}
  {\bibfnamefont {D.}~\bibnamefont {Smith}}, \bibinfo {author} {\bibfnamefont
  {I.}~\bibnamefont {Stancu}}, \bibinfo {author} {\bibfnamefont
  {W.}~\bibnamefont {Strossman}}, \bibinfo {author} {\bibfnamefont
  {R.}~\bibnamefont {Tayloe}}, \bibinfo {author} {\bibfnamefont {G.~J.}\
  \bibnamefont {VanDalen}}, \bibinfo {author} {\bibfnamefont {W.}~\bibnamefont
  {Vernon}}, \bibinfo {author} {\bibfnamefont {N.}~\bibnamefont {Wadia}},
  \bibinfo {author} {\bibfnamefont {J.}~\bibnamefont {Waltz}}, \bibinfo
  {author} {\bibfnamefont {D.~H.}\ \bibnamefont {White}}, \bibinfo {author}
  {\bibfnamefont {D.}~\bibnamefont {Works}}, \bibinfo {author} {\bibfnamefont
  {Y.}~\bibnamefont {Xiao}}, \ and\ \bibinfo {author} {\bibfnamefont
  {S.}~\bibnamefont {Yellin}},\ }\href {\doibase 10.1103/PhysRevLett.81.1774}
  {\bibfield  {journal} {\bibinfo  {journal} {Physical Review Letters}\
  }\textbf {\bibinfo {volume} {81}},\ \bibinfo {pages} {1774} (\bibinfo {year}
  {1998})},\ \Eprint {http://arxiv.org/abs/9709006} {arXiv:9709006 [nucl-ex]}
  \BibitemShut {NoStop}%
\bibitem [{\citenamefont {Conrad}\ \emph {et~al.}(2013)\citenamefont {Conrad},
  \citenamefont {Louis},\ and\ \citenamefont {Shaevitz}}]{Conrad2013}%
  \BibitemOpen
  \bibfield  {author} {\bibinfo {author} {\bibfnamefont {J.~M.}\ \bibnamefont
  {Conrad}}, \bibinfo {author} {\bibfnamefont {W.~C.}\ \bibnamefont {Louis}}, \
  and\ \bibinfo {author} {\bibfnamefont {M.~H.}\ \bibnamefont {Shaevitz}},\
  }\href {\doibase 10.1146/annurev-nucl-102711-094957} {\bibfield  {journal}
  {\bibinfo  {journal} {Annual Review of Nuclear and Particle Science}\
  }\textbf {\bibinfo {volume} {63}},\ \bibinfo {pages} {45} (\bibinfo {year}
  {2013})},\ \Eprint {http://arxiv.org/abs/1306.6494} {arXiv:1306.6494}
  \BibitemShut {NoStop}%
\bibitem [{\citenamefont {Kaether}\ \emph {et~al.}(2010)\citenamefont
  {Kaether}, \citenamefont {Hampel}, \citenamefont {Heusser}, \citenamefont
  {Kiko},\ and\ \citenamefont {Kirsten}}]{Kaether2010}%
  \BibitemOpen
  \bibfield  {author} {\bibinfo {author} {\bibfnamefont {F.}~\bibnamefont
  {Kaether}}, \bibinfo {author} {\bibfnamefont {W.}~\bibnamefont {Hampel}},
  \bibinfo {author} {\bibfnamefont {G.}~\bibnamefont {Heusser}}, \bibinfo
  {author} {\bibfnamefont {J.}~\bibnamefont {Kiko}}, \ and\ \bibinfo {author}
  {\bibfnamefont {T.}~\bibnamefont {Kirsten}},\ }\href {\doibase
  10.1016/j.physletb.2010.01.030} {\bibfield  {journal} {\bibinfo  {journal}
  {Physics Letters, Section B: Nuclear, Elementary Particle and High-Energy
  Physics}\ }\textbf {\bibinfo {volume} {685}},\ \bibinfo {pages} {47}
  (\bibinfo {year} {2010})},\ \Eprint {http://arxiv.org/abs/1001.2731v1}
  {arXiv:1001.2731v1} \BibitemShut {NoStop}%
\bibitem [{\citenamefont {Aguilar-Arevalo}\ \emph {et~al.}(2018)\citenamefont
  {Aguilar-Arevalo}, \citenamefont {Brown}, \citenamefont {Bugel},
  \citenamefont {Cheng}, \citenamefont {Conrad}, \citenamefont {Cooper},
  \citenamefont {Dharmapalan}, \citenamefont {Diaz}, \citenamefont {Djurcic},
  \citenamefont {Finley}, \citenamefont {Ford}, \citenamefont {Garcia},
  \citenamefont {Garvey}, \citenamefont {Grange}, \citenamefont {Huang},
  \citenamefont {Huelsnitz}, \citenamefont {Ignarra}, \citenamefont {Johnson},
  \citenamefont {Karagiorgi}, \citenamefont {Katori}, \citenamefont
  {Kobilarcik}, \citenamefont {Louis}, \citenamefont {Mariani}, \citenamefont
  {Marsh}, \citenamefont {Mills}, \citenamefont {Mirabal}, \citenamefont
  {Monroe}, \citenamefont {Moore}, \citenamefont {Mousseau}, \citenamefont
  {Nienaber}, \citenamefont {Nowak}, \citenamefont {Osmanov}, \citenamefont
  {Pavlovic}, \citenamefont {Perevalov}, \citenamefont {Ray}, \citenamefont
  {Roe}, \citenamefont {Russell}, \citenamefont {Shaevitz}, \citenamefont
  {Spitz}, \citenamefont {Stancu}, \citenamefont {Tayloe}, \citenamefont
  {Thornton}, \citenamefont {Tzanov}, \citenamefont {{Van de Water}},
  \citenamefont {White}, \citenamefont {Wickremasinghe},\ and\ \citenamefont
  {Zimmerman}}]{Aguilar-Arevalo2018a}%
  \BibitemOpen
  \bibfield  {author} {\bibinfo {author} {\bibfnamefont {A.~A.}\ \bibnamefont
  {Aguilar-Arevalo}}, \bibinfo {author} {\bibfnamefont {B.~C.}\ \bibnamefont
  {Brown}}, \bibinfo {author} {\bibfnamefont {L.}~\bibnamefont {Bugel}},
  \bibinfo {author} {\bibfnamefont {G.}~\bibnamefont {Cheng}}, \bibinfo
  {author} {\bibfnamefont {J.~M.}\ \bibnamefont {Conrad}}, \bibinfo {author}
  {\bibfnamefont {R.~L.}\ \bibnamefont {Cooper}}, \bibinfo {author}
  {\bibfnamefont {R.}~\bibnamefont {Dharmapalan}}, \bibinfo {author}
  {\bibfnamefont {A.}~\bibnamefont {Diaz}}, \bibinfo {author} {\bibfnamefont
  {Z.}~\bibnamefont {Djurcic}}, \bibinfo {author} {\bibfnamefont {D.~A.}\
  \bibnamefont {Finley}}, \bibinfo {author} {\bibfnamefont {R.}~\bibnamefont
  {Ford}}, \bibinfo {author} {\bibfnamefont {F.~G.}\ \bibnamefont {Garcia}},
  \bibinfo {author} {\bibfnamefont {G.~T.}\ \bibnamefont {Garvey}}, \bibinfo
  {author} {\bibfnamefont {J.}~\bibnamefont {Grange}}, \bibinfo {author}
  {\bibfnamefont {E.-C.}\ \bibnamefont {Huang}}, \bibinfo {author}
  {\bibfnamefont {W.}~\bibnamefont {Huelsnitz}}, \bibinfo {author}
  {\bibfnamefont {C.}~\bibnamefont {Ignarra}}, \bibinfo {author} {\bibfnamefont
  {R.~A.}\ \bibnamefont {Johnson}}, \bibinfo {author} {\bibfnamefont
  {G.}~\bibnamefont {Karagiorgi}}, \bibinfo {author} {\bibfnamefont
  {T.}~\bibnamefont {Katori}}, \bibinfo {author} {\bibfnamefont
  {T.}~\bibnamefont {Kobilarcik}}, \bibinfo {author} {\bibfnamefont {W.~C.}\
  \bibnamefont {Louis}}, \bibinfo {author} {\bibfnamefont {C.}~\bibnamefont
  {Mariani}}, \bibinfo {author} {\bibfnamefont {W.}~\bibnamefont {Marsh}},
  \bibinfo {author} {\bibfnamefont {G.~B.}\ \bibnamefont {Mills}}, \bibinfo
  {author} {\bibfnamefont {J.}~\bibnamefont {Mirabal}}, \bibinfo {author}
  {\bibfnamefont {J.}~\bibnamefont {Monroe}}, \bibinfo {author} {\bibfnamefont
  {C.~D.}\ \bibnamefont {Moore}}, \bibinfo {author} {\bibfnamefont
  {J.}~\bibnamefont {Mousseau}}, \bibinfo {author} {\bibfnamefont
  {P.}~\bibnamefont {Nienaber}}, \bibinfo {author} {\bibfnamefont
  {J.}~\bibnamefont {Nowak}}, \bibinfo {author} {\bibfnamefont
  {B.}~\bibnamefont {Osmanov}}, \bibinfo {author} {\bibfnamefont
  {Z.}~\bibnamefont {Pavlovic}}, \bibinfo {author} {\bibfnamefont
  {D.}~\bibnamefont {Perevalov}}, \bibinfo {author} {\bibfnamefont
  {H.}~\bibnamefont {Ray}}, \bibinfo {author} {\bibfnamefont {B.~P.}\
  \bibnamefont {Roe}}, \bibinfo {author} {\bibfnamefont {A.~D.}\ \bibnamefont
  {Russell}}, \bibinfo {author} {\bibfnamefont {M.~H.}\ \bibnamefont
  {Shaevitz}}, \bibinfo {author} {\bibfnamefont {J.}~\bibnamefont {Spitz}},
  \bibinfo {author} {\bibfnamefont {I.}~\bibnamefont {Stancu}}, \bibinfo
  {author} {\bibfnamefont {R.}~\bibnamefont {Tayloe}}, \bibinfo {author}
  {\bibfnamefont {R.~T.}\ \bibnamefont {Thornton}}, \bibinfo {author}
  {\bibfnamefont {M.}~\bibnamefont {Tzanov}}, \bibinfo {author} {\bibfnamefont
  {R.~G.}\ \bibnamefont {{Van de Water}}}, \bibinfo {author} {\bibfnamefont
  {D.~H.}\ \bibnamefont {White}}, \bibinfo {author} {\bibfnamefont {D.~A.}\
  \bibnamefont {Wickremasinghe}}, \ and\ \bibinfo {author} {\bibfnamefont
  {E.~D.}\ \bibnamefont {Zimmerman}},\ }\href {\doibase
  10.1103/PhysRevLett.121.221801} {\bibfield  {journal} {\bibinfo  {journal}
  {Physical Review Letters}\ }\textbf {\bibinfo {volume} {121}},\ \bibinfo
  {pages} {221801} (\bibinfo {year} {2018})},\ \Eprint
  {http://arxiv.org/abs/1805.12028} {arXiv:1805.12028} \BibitemShut {NoStop}%
\bibitem [{\citenamefont {Mention}\ \emph {et~al.}(2011)\citenamefont
  {Mention}, \citenamefont {Fechner}, \citenamefont {Lasserre}, \citenamefont
  {Mueller}, \citenamefont {Lhuillier}, \citenamefont {Cribier},\ and\
  \citenamefont {Letourneau}}]{Mention2011}%
  \BibitemOpen
  \bibfield  {author} {\bibinfo {author} {\bibfnamefont {G.}~\bibnamefont
  {Mention}}, \bibinfo {author} {\bibfnamefont {M.}~\bibnamefont {Fechner}},
  \bibinfo {author} {\bibfnamefont {T.}~\bibnamefont {Lasserre}}, \bibinfo
  {author} {\bibfnamefont {T.~A.}\ \bibnamefont {Mueller}}, \bibinfo {author}
  {\bibfnamefont {D.}~\bibnamefont {Lhuillier}}, \bibinfo {author}
  {\bibfnamefont {M.}~\bibnamefont {Cribier}}, \ and\ \bibinfo {author}
  {\bibfnamefont {A.}~\bibnamefont {Letourneau}},\ }\href {\doibase
  10.1103/PhysRevD.83.073006} {\bibfield  {journal} {\bibinfo  {journal}
  {Physical Review D}\ }\textbf {\bibinfo {volume} {83}},\ \bibinfo {pages}
  {073006} (\bibinfo {year} {2011})},\ \Eprint {http://arxiv.org/abs/1101.2755}
  {arXiv:1101.2755} \BibitemShut {NoStop}%
\bibitem [{\citenamefont {Abazajian}\ \emph {et~al.}(2012)\citenamefont
  {Abazajian}, \citenamefont {Acero}, \citenamefont {Agarwalla}, \citenamefont
  {Aguilar-Arevalo}, \citenamefont {Albright}, \citenamefont {Antusch},
  \citenamefont {Arguelles}, \citenamefont {Balantekin}, \citenamefont
  {Barenboim}, \citenamefont {Barger}, \citenamefont {Bernardini},
  \citenamefont {Bezrukov}, \citenamefont {Bjaelde}, \citenamefont {Bogacz},
  \citenamefont {Bowden}, \citenamefont {Boyarsky}, \citenamefont {Bravar},
  \citenamefont {Berguno}, \citenamefont {Brice}, \citenamefont {Bross},
  \citenamefont {Caccianiga}, \citenamefont {Cavanna}, \citenamefont {Chun},
  \citenamefont {Cleveland}, \citenamefont {Collin}, \citenamefont {Coloma},
  \citenamefont {Conrad}, \citenamefont {Cribier}, \citenamefont {Cucoanes},
  \citenamefont {D'Olivo}, \citenamefont {Das}, \citenamefont {de~Gouvea},
  \citenamefont {Derbin}, \citenamefont {Dharmapalan}, \citenamefont {Diaz},
  \citenamefont {Ding}, \citenamefont {Djurcic}, \citenamefont {Donini},
  \citenamefont {Duchesneau}, \citenamefont {Ejiri}, \citenamefont {Elliott},
  \citenamefont {Ernst}, \citenamefont {Esmaili}, \citenamefont {Evans},
  \citenamefont {Fernandez-Martinez}, \citenamefont {Figueroa-Feliciano},
  \citenamefont {Fleming}, \citenamefont {Formaggio}, \citenamefont {Franco},
  \citenamefont {Gaffiot}, \citenamefont {Gandhi}, \citenamefont {Gao},
  \citenamefont {Garvey}, \citenamefont {Gavrin}, \citenamefont {Ghoshal},
  \citenamefont {Gibin}, \citenamefont {Giunti}, \citenamefont {Gninenko},
  \citenamefont {Gorbachev}, \citenamefont {Gorbunov}, \citenamefont
  {Guenette}, \citenamefont {Guglielmi}, \citenamefont {Halzen}, \citenamefont
  {Hamann}, \citenamefont {Hannestad}, \citenamefont {Haxton}, \citenamefont
  {Heeger}, \citenamefont {Henning}, \citenamefont {Hernandez}, \citenamefont
  {Huber}, \citenamefont {Huelsnitz}, \citenamefont {Ianni}, \citenamefont
  {Ibragimova}, \citenamefont {Karadzhov}, \citenamefont {Karagiorgi},
  \citenamefont {Keefer}, \citenamefont {Kim}, \citenamefont {Kopp},
  \citenamefont {Kornoukhov}, \citenamefont {Kusenko}, \citenamefont {Kyberd},
  \citenamefont {Langacker}, \citenamefont {Lasserre}, \citenamefont {Laveder},
  \citenamefont {Letourneau}, \citenamefont {Lhuillier}, \citenamefont {Li},
  \citenamefont {Lindner}, \citenamefont {Link}, \citenamefont {Littlejohn},
  \citenamefont {Lombardi}, \citenamefont {Long}, \citenamefont {Lopez-Pavon},
  \citenamefont {Louis}, \citenamefont {Ludhova}, \citenamefont {Lykken},
  \citenamefont {Machado}, \citenamefont {Maltoni}, \citenamefont {Mann},
  \citenamefont {Marfatia}, \citenamefont {Mariani}, \citenamefont {Matveev},
  \citenamefont {Mavromatos}, \citenamefont {Melchiorri}, \citenamefont
  {Meloni}, \citenamefont {Mena}, \citenamefont {Mention}, \citenamefont
  {Merle}, \citenamefont {Meroni}, \citenamefont {Mezzetto}, \citenamefont
  {Mills}, \citenamefont {Minic}, \citenamefont {Miramonti}, \citenamefont
  {Mohapatra}, \citenamefont {Mohapatra}, \citenamefont {Montanari},
  \citenamefont {Mori}, \citenamefont {Mueller}, \citenamefont {Mumm},
  \citenamefont {Muratova}, \citenamefont {Nelson}, \citenamefont {Nico},
  \citenamefont {Noah}, \citenamefont {Nowak}, \citenamefont {Smirnov},
  \citenamefont {Obolensky}, \citenamefont {Pakvasa}, \citenamefont {Palamara},
  \citenamefont {Pallavicini}, \citenamefont {Pascoli}, \citenamefont
  {Patrizii}, \citenamefont {Pavlovic}, \citenamefont {Peres}, \citenamefont
  {Pessard}, \citenamefont {Pietropaolo}, \citenamefont {Pitt}, \citenamefont
  {Popovic}, \citenamefont {Pradler}, \citenamefont {Ranucci}, \citenamefont
  {Ray}, \citenamefont {Razzaque}, \citenamefont {Rebel}, \citenamefont
  {Robertson}, \citenamefont {Rodejohann}, \citenamefont {Rountree},
  \citenamefont {Rubbia}, \citenamefont {Ruchayskiy}, \citenamefont {Sala},
  \citenamefont {Scholberg}, \citenamefont {Schwetz}, \citenamefont {Shaevitz},
  \citenamefont {Shaposhnikov}, \citenamefont {Shrock}, \citenamefont {Simone},
  \citenamefont {Skorokhvatov}, \citenamefont {Sorel}, \citenamefont {Sousa},
  \citenamefont {Spergel}, \citenamefont {Spitz}, \citenamefont {Stanco},
  \citenamefont {Stancu}, \citenamefont {Suzuki}, \citenamefont {Takeuchi},
  \citenamefont {Tamborra}, \citenamefont {Tang}, \citenamefont {Testera},
  \citenamefont {Tian}, \citenamefont {Tonazzo}, \citenamefont {Tunnell},
  \citenamefont {{Van de Water}}, \citenamefont {Verde}, \citenamefont
  {Veretenkin}, \citenamefont {Vignoli}, \citenamefont {Vivier}, \citenamefont
  {Vogelaar}, \citenamefont {Wascko}, \citenamefont {Wilkerson}, \citenamefont
  {Winter}, \citenamefont {Wong}, \citenamefont {Yanagida}, \citenamefont
  {Yasuda}, \citenamefont {Yeh}, \citenamefont {Yermia}, \citenamefont
  {Yokley}, \citenamefont {Zeller}, \citenamefont {Zhan},\ and\ \citenamefont
  {Zhang}}]{Abazajian2012}%
  \BibitemOpen
  \bibfield  {author} {\bibinfo {author} {\bibfnamefont {K.~N.}\ \bibnamefont
  {Abazajian}}, \bibinfo {author} {\bibfnamefont {M.~A.}\ \bibnamefont
  {Acero}}, \bibinfo {author} {\bibfnamefont {S.~K.}\ \bibnamefont
  {Agarwalla}}, \bibinfo {author} {\bibfnamefont {A.~A.}\ \bibnamefont
  {Aguilar-Arevalo}}, \bibinfo {author} {\bibfnamefont {C.~H.}\ \bibnamefont
  {Albright}}, \bibinfo {author} {\bibfnamefont {S.}~\bibnamefont {Antusch}},
  \bibinfo {author} {\bibfnamefont {C.~A.}\ \bibnamefont {Arguelles}}, \bibinfo
  {author} {\bibfnamefont {A.~B.}\ \bibnamefont {Balantekin}}, \bibinfo
  {author} {\bibfnamefont {G.}~\bibnamefont {Barenboim}}, \bibinfo {author}
  {\bibfnamefont {V.}~\bibnamefont {Barger}}, \bibinfo {author} {\bibfnamefont
  {P.}~\bibnamefont {Bernardini}}, \bibinfo {author} {\bibfnamefont
  {F.}~\bibnamefont {Bezrukov}}, \bibinfo {author} {\bibfnamefont {O.~E.}\
  \bibnamefont {Bjaelde}}, \bibinfo {author} {\bibfnamefont {S.~A.}\
  \bibnamefont {Bogacz}}, \bibinfo {author} {\bibfnamefont {N.~S.}\
  \bibnamefont {Bowden}}, \bibinfo {author} {\bibfnamefont {A.}~\bibnamefont
  {Boyarsky}}, \bibinfo {author} {\bibfnamefont {A.}~\bibnamefont {Bravar}},
  \bibinfo {author} {\bibfnamefont {D.~B.}\ \bibnamefont {Berguno}}, \bibinfo
  {author} {\bibfnamefont {S.~J.}\ \bibnamefont {Brice}}, \bibinfo {author}
  {\bibfnamefont {A.~D.}\ \bibnamefont {Bross}}, \bibinfo {author}
  {\bibfnamefont {B.}~\bibnamefont {Caccianiga}}, \bibinfo {author}
  {\bibfnamefont {F.}~\bibnamefont {Cavanna}}, \bibinfo {author} {\bibfnamefont
  {E.~J.}\ \bibnamefont {Chun}}, \bibinfo {author} {\bibfnamefont {B.~T.}\
  \bibnamefont {Cleveland}}, \bibinfo {author} {\bibfnamefont {A.~P.}\
  \bibnamefont {Collin}}, \bibinfo {author} {\bibfnamefont {P.}~\bibnamefont
  {Coloma}}, \bibinfo {author} {\bibfnamefont {J.~M.}\ \bibnamefont {Conrad}},
  \bibinfo {author} {\bibfnamefont {M.}~\bibnamefont {Cribier}}, \bibinfo
  {author} {\bibfnamefont {A.~S.}\ \bibnamefont {Cucoanes}}, \bibinfo {author}
  {\bibfnamefont {J.~C.}\ \bibnamefont {D'Olivo}}, \bibinfo {author}
  {\bibfnamefont {S.}~\bibnamefont {Das}}, \bibinfo {author} {\bibfnamefont
  {A.}~\bibnamefont {de~Gouvea}}, \bibinfo {author} {\bibfnamefont {A.~V.}\
  \bibnamefont {Derbin}}, \bibinfo {author} {\bibfnamefont {R.}~\bibnamefont
  {Dharmapalan}}, \bibinfo {author} {\bibfnamefont {J.~S.}\ \bibnamefont
  {Diaz}}, \bibinfo {author} {\bibfnamefont {X.~J.}\ \bibnamefont {Ding}},
  \bibinfo {author} {\bibfnamefont {Z.}~\bibnamefont {Djurcic}}, \bibinfo
  {author} {\bibfnamefont {A.}~\bibnamefont {Donini}}, \bibinfo {author}
  {\bibfnamefont {D.}~\bibnamefont {Duchesneau}}, \bibinfo {author}
  {\bibfnamefont {H.}~\bibnamefont {Ejiri}}, \bibinfo {author} {\bibfnamefont
  {S.~R.}\ \bibnamefont {Elliott}}, \bibinfo {author} {\bibfnamefont {D.~J.}\
  \bibnamefont {Ernst}}, \bibinfo {author} {\bibfnamefont {A.}~\bibnamefont
  {Esmaili}}, \bibinfo {author} {\bibfnamefont {J.~J.}\ \bibnamefont {Evans}},
  \bibinfo {author} {\bibfnamefont {E.}~\bibnamefont {Fernandez-Martinez}},
  \bibinfo {author} {\bibfnamefont {E.}~\bibnamefont {Figueroa-Feliciano}},
  \bibinfo {author} {\bibfnamefont {B.~T.}\ \bibnamefont {Fleming}}, \bibinfo
  {author} {\bibfnamefont {J.~A.}\ \bibnamefont {Formaggio}}, \bibinfo {author}
  {\bibfnamefont {D.}~\bibnamefont {Franco}}, \bibinfo {author} {\bibfnamefont
  {J.}~\bibnamefont {Gaffiot}}, \bibinfo {author} {\bibfnamefont
  {R.}~\bibnamefont {Gandhi}}, \bibinfo {author} {\bibfnamefont
  {Y.}~\bibnamefont {Gao}}, \bibinfo {author} {\bibfnamefont {G.~T.}\
  \bibnamefont {Garvey}}, \bibinfo {author} {\bibfnamefont {V.~N.}\
  \bibnamefont {Gavrin}}, \bibinfo {author} {\bibfnamefont {P.}~\bibnamefont
  {Ghoshal}}, \bibinfo {author} {\bibfnamefont {D.}~\bibnamefont {Gibin}},
  \bibinfo {author} {\bibfnamefont {C.}~\bibnamefont {Giunti}}, \bibinfo
  {author} {\bibfnamefont {S.~N.}\ \bibnamefont {Gninenko}}, \bibinfo {author}
  {\bibfnamefont {V.~V.}\ \bibnamefont {Gorbachev}}, \bibinfo {author}
  {\bibfnamefont {D.~S.}\ \bibnamefont {Gorbunov}}, \bibinfo {author}
  {\bibfnamefont {R.}~\bibnamefont {Guenette}}, \bibinfo {author}
  {\bibfnamefont {A.}~\bibnamefont {Guglielmi}}, \bibinfo {author}
  {\bibfnamefont {F.}~\bibnamefont {Halzen}}, \bibinfo {author} {\bibfnamefont
  {J.}~\bibnamefont {Hamann}}, \bibinfo {author} {\bibfnamefont
  {S.}~\bibnamefont {Hannestad}}, \bibinfo {author} {\bibfnamefont
  {W.}~\bibnamefont {Haxton}}, \bibinfo {author} {\bibfnamefont {K.~M.}\
  \bibnamefont {Heeger}}, \bibinfo {author} {\bibfnamefont {R.}~\bibnamefont
  {Henning}}, \bibinfo {author} {\bibfnamefont {P.}~\bibnamefont {Hernandez}},
  \bibinfo {author} {\bibfnamefont {P.}~\bibnamefont {Huber}}, \bibinfo
  {author} {\bibfnamefont {W.}~\bibnamefont {Huelsnitz}}, \bibinfo {author}
  {\bibfnamefont {A.}~\bibnamefont {Ianni}}, \bibinfo {author} {\bibfnamefont
  {T.~V.}\ \bibnamefont {Ibragimova}}, \bibinfo {author} {\bibfnamefont
  {Y.}~\bibnamefont {Karadzhov}}, \bibinfo {author} {\bibfnamefont
  {G.}~\bibnamefont {Karagiorgi}}, \bibinfo {author} {\bibfnamefont
  {G.}~\bibnamefont {Keefer}}, \bibinfo {author} {\bibfnamefont {Y.~D.}\
  \bibnamefont {Kim}}, \bibinfo {author} {\bibfnamefont {J.}~\bibnamefont
  {Kopp}}, \bibinfo {author} {\bibfnamefont {V.~N.}\ \bibnamefont
  {Kornoukhov}}, \bibinfo {author} {\bibfnamefont {A.}~\bibnamefont {Kusenko}},
  \bibinfo {author} {\bibfnamefont {P.}~\bibnamefont {Kyberd}}, \bibinfo
  {author} {\bibfnamefont {P.}~\bibnamefont {Langacker}}, \bibinfo {author}
  {\bibfnamefont {T.}~\bibnamefont {Lasserre}}, \bibinfo {author}
  {\bibfnamefont {M.}~\bibnamefont {Laveder}}, \bibinfo {author} {\bibfnamefont
  {A.}~\bibnamefont {Letourneau}}, \bibinfo {author} {\bibfnamefont
  {D.}~\bibnamefont {Lhuillier}}, \bibinfo {author} {\bibfnamefont {Y.~F.}\
  \bibnamefont {Li}}, \bibinfo {author} {\bibfnamefont {M.}~\bibnamefont
  {Lindner}}, \bibinfo {author} {\bibfnamefont {J.~M.}\ \bibnamefont {Link}},
  \bibinfo {author} {\bibfnamefont {B.~L.}\ \bibnamefont {Littlejohn}},
  \bibinfo {author} {\bibfnamefont {P.}~\bibnamefont {Lombardi}}, \bibinfo
  {author} {\bibfnamefont {K.}~\bibnamefont {Long}}, \bibinfo {author}
  {\bibfnamefont {J.}~\bibnamefont {Lopez-Pavon}}, \bibinfo {author}
  {\bibfnamefont {W.~C.}\ \bibnamefont {Louis}}, \bibinfo {author}
  {\bibfnamefont {L.}~\bibnamefont {Ludhova}}, \bibinfo {author} {\bibfnamefont
  {J.~D.}\ \bibnamefont {Lykken}}, \bibinfo {author} {\bibfnamefont {P.~A.~N.}\
  \bibnamefont {Machado}}, \bibinfo {author} {\bibfnamefont {M.}~\bibnamefont
  {Maltoni}}, \bibinfo {author} {\bibfnamefont {W.~A.}\ \bibnamefont {Mann}},
  \bibinfo {author} {\bibfnamefont {D.}~\bibnamefont {Marfatia}}, \bibinfo
  {author} {\bibfnamefont {C.}~\bibnamefont {Mariani}}, \bibinfo {author}
  {\bibfnamefont {V.~A.}\ \bibnamefont {Matveev}}, \bibinfo {author}
  {\bibfnamefont {N.~E.}\ \bibnamefont {Mavromatos}}, \bibinfo {author}
  {\bibfnamefont {A.}~\bibnamefont {Melchiorri}}, \bibinfo {author}
  {\bibfnamefont {D.}~\bibnamefont {Meloni}}, \bibinfo {author} {\bibfnamefont
  {O.}~\bibnamefont {Mena}}, \bibinfo {author} {\bibfnamefont {G.}~\bibnamefont
  {Mention}}, \bibinfo {author} {\bibfnamefont {A.}~\bibnamefont {Merle}},
  \bibinfo {author} {\bibfnamefont {E.}~\bibnamefont {Meroni}}, \bibinfo
  {author} {\bibfnamefont {M.}~\bibnamefont {Mezzetto}}, \bibinfo {author}
  {\bibfnamefont {G.~B.}\ \bibnamefont {Mills}}, \bibinfo {author}
  {\bibfnamefont {D.}~\bibnamefont {Minic}}, \bibinfo {author} {\bibfnamefont
  {L.}~\bibnamefont {Miramonti}}, \bibinfo {author} {\bibfnamefont
  {D.}~\bibnamefont {Mohapatra}}, \bibinfo {author} {\bibfnamefont {R.~N.}\
  \bibnamefont {Mohapatra}}, \bibinfo {author} {\bibfnamefont {C.}~\bibnamefont
  {Montanari}}, \bibinfo {author} {\bibfnamefont {Y.}~\bibnamefont {Mori}},
  \bibinfo {author} {\bibfnamefont {T.~A.}\ \bibnamefont {Mueller}}, \bibinfo
  {author} {\bibfnamefont {H.~P.}\ \bibnamefont {Mumm}}, \bibinfo {author}
  {\bibfnamefont {V.}~\bibnamefont {Muratova}}, \bibinfo {author}
  {\bibfnamefont {A.~E.}\ \bibnamefont {Nelson}}, \bibinfo {author}
  {\bibfnamefont {J.~S.}\ \bibnamefont {Nico}}, \bibinfo {author}
  {\bibfnamefont {E.}~\bibnamefont {Noah}}, \bibinfo {author} {\bibfnamefont
  {J.}~\bibnamefont {Nowak}}, \bibinfo {author} {\bibfnamefont {O.~Y.}\
  \bibnamefont {Smirnov}}, \bibinfo {author} {\bibfnamefont {M.}~\bibnamefont
  {Obolensky}}, \bibinfo {author} {\bibfnamefont {S.}~\bibnamefont {Pakvasa}},
  \bibinfo {author} {\bibfnamefont {O.}~\bibnamefont {Palamara}}, \bibinfo
  {author} {\bibfnamefont {M.}~\bibnamefont {Pallavicini}}, \bibinfo {author}
  {\bibfnamefont {S.}~\bibnamefont {Pascoli}}, \bibinfo {author} {\bibfnamefont
  {L.}~\bibnamefont {Patrizii}}, \bibinfo {author} {\bibfnamefont
  {Z.}~\bibnamefont {Pavlovic}}, \bibinfo {author} {\bibfnamefont {O.~L.~G.}\
  \bibnamefont {Peres}}, \bibinfo {author} {\bibfnamefont {H.}~\bibnamefont
  {Pessard}}, \bibinfo {author} {\bibfnamefont {F.}~\bibnamefont
  {Pietropaolo}}, \bibinfo {author} {\bibfnamefont {M.~L.}\ \bibnamefont
  {Pitt}}, \bibinfo {author} {\bibfnamefont {M.}~\bibnamefont {Popovic}},
  \bibinfo {author} {\bibfnamefont {J.}~\bibnamefont {Pradler}}, \bibinfo
  {author} {\bibfnamefont {G.}~\bibnamefont {Ranucci}}, \bibinfo {author}
  {\bibfnamefont {H.}~\bibnamefont {Ray}}, \bibinfo {author} {\bibfnamefont
  {S.}~\bibnamefont {Razzaque}}, \bibinfo {author} {\bibfnamefont
  {B.}~\bibnamefont {Rebel}}, \bibinfo {author} {\bibfnamefont {R.~G.~H.}\
  \bibnamefont {Robertson}}, \bibinfo {author} {\bibfnamefont {W.}~\bibnamefont
  {Rodejohann}}, \bibinfo {author} {\bibfnamefont {S.~D.}\ \bibnamefont
  {Rountree}}, \bibinfo {author} {\bibfnamefont {C.}~\bibnamefont {Rubbia}},
  \bibinfo {author} {\bibfnamefont {O.}~\bibnamefont {Ruchayskiy}}, \bibinfo
  {author} {\bibfnamefont {P.~R.}\ \bibnamefont {Sala}}, \bibinfo {author}
  {\bibfnamefont {K.}~\bibnamefont {Scholberg}}, \bibinfo {author}
  {\bibfnamefont {T.}~\bibnamefont {Schwetz}}, \bibinfo {author} {\bibfnamefont
  {M.~H.}\ \bibnamefont {Shaevitz}}, \bibinfo {author} {\bibfnamefont
  {M.}~\bibnamefont {Shaposhnikov}}, \bibinfo {author} {\bibfnamefont
  {R.}~\bibnamefont {Shrock}}, \bibinfo {author} {\bibfnamefont
  {S.}~\bibnamefont {Simone}}, \bibinfo {author} {\bibfnamefont
  {M.}~\bibnamefont {Skorokhvatov}}, \bibinfo {author} {\bibfnamefont
  {M.}~\bibnamefont {Sorel}}, \bibinfo {author} {\bibfnamefont
  {A.}~\bibnamefont {Sousa}}, \bibinfo {author} {\bibfnamefont {D.~N.}\
  \bibnamefont {Spergel}}, \bibinfo {author} {\bibfnamefont {J.}~\bibnamefont
  {Spitz}}, \bibinfo {author} {\bibfnamefont {L.}~\bibnamefont {Stanco}},
  \bibinfo {author} {\bibfnamefont {I.}~\bibnamefont {Stancu}}, \bibinfo
  {author} {\bibfnamefont {A.}~\bibnamefont {Suzuki}}, \bibinfo {author}
  {\bibfnamefont {T.}~\bibnamefont {Takeuchi}}, \bibinfo {author}
  {\bibfnamefont {I.}~\bibnamefont {Tamborra}}, \bibinfo {author}
  {\bibfnamefont {J.}~\bibnamefont {Tang}}, \bibinfo {author} {\bibfnamefont
  {G.}~\bibnamefont {Testera}}, \bibinfo {author} {\bibfnamefont {X.~C.}\
  \bibnamefont {Tian}}, \bibinfo {author} {\bibfnamefont {A.}~\bibnamefont
  {Tonazzo}}, \bibinfo {author} {\bibfnamefont {C.~D.}\ \bibnamefont
  {Tunnell}}, \bibinfo {author} {\bibfnamefont {R.~G.}\ \bibnamefont {{Van de
  Water}}}, \bibinfo {author} {\bibfnamefont {L.}~\bibnamefont {Verde}},
  \bibinfo {author} {\bibfnamefont {E.~P.}\ \bibnamefont {Veretenkin}},
  \bibinfo {author} {\bibfnamefont {C.}~\bibnamefont {Vignoli}}, \bibinfo
  {author} {\bibfnamefont {M.}~\bibnamefont {Vivier}}, \bibinfo {author}
  {\bibfnamefont {R.~B.}\ \bibnamefont {Vogelaar}}, \bibinfo {author}
  {\bibfnamefont {M.~O.}\ \bibnamefont {Wascko}}, \bibinfo {author}
  {\bibfnamefont {J.~F.}\ \bibnamefont {Wilkerson}}, \bibinfo {author}
  {\bibfnamefont {W.}~\bibnamefont {Winter}}, \bibinfo {author} {\bibfnamefont
  {Y.~Y.~Y.}\ \bibnamefont {Wong}}, \bibinfo {author} {\bibfnamefont {T.~T.}\
  \bibnamefont {Yanagida}}, \bibinfo {author} {\bibfnamefont {O.}~\bibnamefont
  {Yasuda}}, \bibinfo {author} {\bibfnamefont {M.}~\bibnamefont {Yeh}},
  \bibinfo {author} {\bibfnamefont {F.}~\bibnamefont {Yermia}}, \bibinfo
  {author} {\bibfnamefont {Z.~W.}\ \bibnamefont {Yokley}}, \bibinfo {author}
  {\bibfnamefont {G.~P.}\ \bibnamefont {Zeller}}, \bibinfo {author}
  {\bibfnamefont {L.}~\bibnamefont {Zhan}}, \ and\ \bibinfo {author}
  {\bibfnamefont {H.}~\bibnamefont {Zhang}},\ }\href
  {http://arxiv.org/abs/1204.5379} {\  (\bibinfo {year} {2012})},\ \Eprint
  {http://arxiv.org/abs/1204.5379} {arXiv:1204.5379} \BibitemShut {NoStop}%
\bibitem [{\citenamefont {Jordan}\ \emph {et~al.}(2019)\citenamefont {Jordan},
  \citenamefont {Kahn}, \citenamefont {Krnjaic}, \citenamefont {Moschella},\
  and\ \citenamefont {Spitz}}]{Jordan2019}%
  \BibitemOpen
  \bibfield  {author} {\bibinfo {author} {\bibfnamefont {J.~R.}\ \bibnamefont
  {Jordan}}, \bibinfo {author} {\bibfnamefont {Y.}~\bibnamefont {Kahn}},
  \bibinfo {author} {\bibfnamefont {G.}~\bibnamefont {Krnjaic}}, \bibinfo
  {author} {\bibfnamefont {M.}~\bibnamefont {Moschella}}, \ and\ \bibinfo
  {author} {\bibfnamefont {J.}~\bibnamefont {Spitz}},\ }\href {\doibase
  10.1103/PhysRevLett.122.081801} {\bibfield  {journal} {\bibinfo  {journal}
  {Physical Review Letters}\ }\textbf {\bibinfo {volume} {122}},\ \bibinfo
  {pages} {081801} (\bibinfo {year} {2019})},\ \Eprint
  {http://arxiv.org/abs/1810.07185} {arXiv:1810.07185} \BibitemShut {NoStop}%
\bibitem [{\citenamefont {Hayes}\ \emph {et~al.}(2015)\citenamefont {Hayes},
  \citenamefont {Friar}, \citenamefont {Garvey}, \citenamefont {Ibeling},
  \citenamefont {Jungman}, \citenamefont {Kawano},\ and\ \citenamefont
  {Mills}}]{Hayes2015}%
  \BibitemOpen
  \bibfield  {author} {\bibinfo {author} {\bibfnamefont {A.~C.}\ \bibnamefont
  {Hayes}}, \bibinfo {author} {\bibfnamefont {J.~L.}\ \bibnamefont {Friar}},
  \bibinfo {author} {\bibfnamefont {G.~T.}\ \bibnamefont {Garvey}}, \bibinfo
  {author} {\bibfnamefont {D.}~\bibnamefont {Ibeling}}, \bibinfo {author}
  {\bibfnamefont {G.}~\bibnamefont {Jungman}}, \bibinfo {author} {\bibfnamefont
  {T.}~\bibnamefont {Kawano}}, \ and\ \bibinfo {author} {\bibfnamefont {R.~W.}\
  \bibnamefont {Mills}},\ }\href {\doibase 10.1103/PhysRevD.92.033015}
  {\bibfield  {journal} {\bibinfo  {journal} {Physical Review D}\ }\textbf
  {\bibinfo {volume} {92}},\ \bibinfo {pages} {033015} (\bibinfo {year}
  {2015})},\ \Eprint {http://arxiv.org/abs/1506.00583} {arXiv:1506.00583}
  \BibitemShut {NoStop}%
\bibitem [{\citenamefont {Huber}(2017)}]{Huber2016a}%
  \BibitemOpen
  \bibfield  {author} {\bibinfo {author} {\bibfnamefont {P.}~\bibnamefont
  {Huber}},\ }\href {\doibase 10.1103/PhysRevLett.118.042502} {\bibfield
  {journal} {\bibinfo  {journal} {Physical Review Letters}\ }\textbf {\bibinfo
  {volume} {118}},\ \bibinfo {pages} {042502} (\bibinfo {year} {2017})},\
  \Eprint {http://arxiv.org/abs/1609.03910} {arXiv:1609.03910} \BibitemShut
  {NoStop}%
\bibitem [{\citenamefont {An}\ \emph {et~al.}(2016)\citenamefont {An},
  \citenamefont {Balantekin}, \citenamefont {Band}, \citenamefont {Bishai},
  \citenamefont {Blyth}, \citenamefont {Butorov}, \citenamefont {Cao},
  \citenamefont {Cao}, \citenamefont {Cao}, \citenamefont {Cen}, \citenamefont
  {Chan}, \citenamefont {Chang}, \citenamefont {Chang}, \citenamefont {Chang},
  \citenamefont {Chen}, \citenamefont {Chen}, \citenamefont {Chen},
  \citenamefont {Chen}, \citenamefont {Chen}, \citenamefont {Cheng},
  \citenamefont {Cheng}, \citenamefont {Cheng}, \citenamefont {Cherwinka},
  \citenamefont {Chu}, \citenamefont {Cummings}, \citenamefont {de~Arcos},
  \citenamefont {Deng}, \citenamefont {Ding}, \citenamefont {Ding},
  \citenamefont {Diwan}, \citenamefont {Dove}, \citenamefont {Draeger},
  \citenamefont {Dwyer}, \citenamefont {Edwards}, \citenamefont {Ely},
  \citenamefont {Gill}, \citenamefont {Gonchar}, \citenamefont {Gong},
  \citenamefont {Gong}, \citenamefont {Grassi}, \citenamefont {Gu},
  \citenamefont {Guan}, \citenamefont {Guo}, \citenamefont {Guo}, \citenamefont
  {Hackenburg}, \citenamefont {Han}, \citenamefont {Hans}, \citenamefont {He},
  \citenamefont {Heeger}, \citenamefont {Heng}, \citenamefont {Higuera},
  \citenamefont {Hor}, \citenamefont {Hsiung}, \citenamefont {Hu},
  \citenamefont {Hu}, \citenamefont {Hu}, \citenamefont {Hu}, \citenamefont
  {Hu}, \citenamefont {Huang}, \citenamefont {Huang}, \citenamefont {Huang},
  \citenamefont {Huber}, \citenamefont {Hussain}, \citenamefont {Jaffe},
  \citenamefont {Jaffke}, \citenamefont {Jen}, \citenamefont {Jetter},
  \citenamefont {Ji}, \citenamefont {Ji}, \citenamefont {Jiao}, \citenamefont
  {Johnson}, \citenamefont {Kang}, \citenamefont {Kettell}, \citenamefont
  {Kohn}, \citenamefont {Kramer}, \citenamefont {Kwan}, \citenamefont {Kwok},
  \citenamefont {Kwok}, \citenamefont {Langford}, \citenamefont {Lau},
  \citenamefont {Lebanowski}, \citenamefont {Lee}, \citenamefont {Lei},
  \citenamefont {Leitner}, \citenamefont {Leung}, \citenamefont {Leung},
  \citenamefont {Lewis}, \citenamefont {Li}, \citenamefont {Li}, \citenamefont
  {Li}, \citenamefont {Li}, \citenamefont {Li}, \citenamefont {Li},
  \citenamefont {Li}, \citenamefont {Li}, \citenamefont {Li}, \citenamefont
  {Li}, \citenamefont {Liang}, \citenamefont {Lin}, \citenamefont {Lin},
  \citenamefont {Lin}, \citenamefont {Lin}, \citenamefont {Ling}, \citenamefont
  {Link}, \citenamefont {Littenberg}, \citenamefont {Littlejohn}, \citenamefont
  {Liu}, \citenamefont {Liu}, \citenamefont {Liu}, \citenamefont {Liu},
  \citenamefont {Liu}, \citenamefont {Lu}, \citenamefont {Lu}, \citenamefont
  {Lu}, \citenamefont {Luk}, \citenamefont {Ma}, \citenamefont {Ma},
  \citenamefont {Ma}, \citenamefont {Ma}, \citenamefont {{Martinez Caicedo}},
  \citenamefont {McDonald}, \citenamefont {McKeown}, \citenamefont {Meng},
  \citenamefont {Mitchell}, \citenamefont {{Monari Kebwaro}}, \citenamefont
  {Nakajima}, \citenamefont {Napolitano}, \citenamefont {Naumov}, \citenamefont
  {Naumova}, \citenamefont {Ngai}, \citenamefont {Ning}, \citenamefont
  {Ochoa-Ricoux}, \citenamefont {Olshevski}, \citenamefont {Pan}, \citenamefont
  {Park}, \citenamefont {Patton}, \citenamefont {Pec}, \citenamefont {Peng},
  \citenamefont {Piilonen}, \citenamefont {Pinsky}, \citenamefont {Pun},
  \citenamefont {Qi}, \citenamefont {Qi}, \citenamefont {Qian}, \citenamefont
  {Raper}, \citenamefont {Ren}, \citenamefont {Ren}, \citenamefont {Rosero},
  \citenamefont {Roskovec}, \citenamefont {Ruan}, \citenamefont {Shao},
  \citenamefont {Steiner}, \citenamefont {Sun}, \citenamefont {Sun},
  \citenamefont {Tang}, \citenamefont {Taychenachev}, \citenamefont {Tsang},
  \citenamefont {Tull}, \citenamefont {Tung}, \citenamefont {Viaux},
  \citenamefont {Viren}, \citenamefont {Vorobel}, \citenamefont {Wang},
  \citenamefont {Wang}, \citenamefont {Wang}, \citenamefont {Wang},
  \citenamefont {Wang}, \citenamefont {Wang}, \citenamefont {Wang},
  \citenamefont {Wang}, \citenamefont {Wang}, \citenamefont {Wang},
  \citenamefont {Wang}, \citenamefont {Wei}, \citenamefont {Wen}, \citenamefont
  {Whisnant}, \citenamefont {White}, \citenamefont {Whitehead}, \citenamefont
  {Wise}, \citenamefont {Wong}, \citenamefont {Wong}, \citenamefont
  {Worcester}, \citenamefont {Wu}, \citenamefont {Xia}, \citenamefont {Xia},
  \citenamefont {Xia}, \citenamefont {Xing}, \citenamefont {Xu}, \citenamefont
  {Xu}, \citenamefont {Xu}, \citenamefont {Xu}, \citenamefont {Xue},
  \citenamefont {Yan}, \citenamefont {Yang}, \citenamefont {Yang},
  \citenamefont {Yang}, \citenamefont {Yang}, \citenamefont {Ye}, \citenamefont
  {Yeh}, \citenamefont {Young}, \citenamefont {Yu}, \citenamefont {Yu},
  \citenamefont {Zang}, \citenamefont {Zhan}, \citenamefont {Zhang},
  \citenamefont {Zhang}, \citenamefont {Zhang}, \citenamefont {Zhang},
  \citenamefont {Zhang}, \citenamefont {Zhang}, \citenamefont {Zhang},
  \citenamefont {Zhang}, \citenamefont {Zhang}, \citenamefont {Zhang},
  \citenamefont {Zhao}, \citenamefont {Zhao}, \citenamefont {Zhao},
  \citenamefont {Zhao}, \citenamefont {Zheng}, \citenamefont {Zhong},
  \citenamefont {Zhou}, \citenamefont {Zhou}, \citenamefont {Zhuang},\ and\
  \citenamefont {Zou}}]{An2016}%
  \BibitemOpen
  \bibfield  {author} {\bibinfo {author} {\bibfnamefont {F.~P.}\ \bibnamefont
  {An}}, \bibinfo {author} {\bibfnamefont {A.~B.}\ \bibnamefont {Balantekin}},
  \bibinfo {author} {\bibfnamefont {H.~R.}\ \bibnamefont {Band}}, \bibinfo
  {author} {\bibfnamefont {M.}~\bibnamefont {Bishai}}, \bibinfo {author}
  {\bibfnamefont {S.}~\bibnamefont {Blyth}}, \bibinfo {author} {\bibfnamefont
  {I.}~\bibnamefont {Butorov}}, \bibinfo {author} {\bibfnamefont
  {D.}~\bibnamefont {Cao}}, \bibinfo {author} {\bibfnamefont {G.~F.}\
  \bibnamefont {Cao}}, \bibinfo {author} {\bibfnamefont {J.}~\bibnamefont
  {Cao}}, \bibinfo {author} {\bibfnamefont {W.~R.}\ \bibnamefont {Cen}},
  \bibinfo {author} {\bibfnamefont {Y.~L.}\ \bibnamefont {Chan}}, \bibinfo
  {author} {\bibfnamefont {J.~F.}\ \bibnamefont {Chang}}, \bibinfo {author}
  {\bibfnamefont {L.~C.}\ \bibnamefont {Chang}}, \bibinfo {author}
  {\bibfnamefont {Y.}~\bibnamefont {Chang}}, \bibinfo {author} {\bibfnamefont
  {H.~S.}\ \bibnamefont {Chen}}, \bibinfo {author} {\bibfnamefont {Q.~Y.}\
  \bibnamefont {Chen}}, \bibinfo {author} {\bibfnamefont {S.~M.}\ \bibnamefont
  {Chen}}, \bibinfo {author} {\bibfnamefont {Y.~X.}\ \bibnamefont {Chen}},
  \bibinfo {author} {\bibfnamefont {Y.~X.}\ \bibnamefont {Chen}}, \bibinfo
  {author} {\bibfnamefont {J.~H.}\ \bibnamefont {Cheng}}, \bibinfo {author}
  {\bibfnamefont {J.~H.}\ \bibnamefont {Cheng}}, \bibinfo {author}
  {\bibfnamefont {Y.~P.}\ \bibnamefont {Cheng}}, \bibinfo {author}
  {\bibfnamefont {J.~J.}\ \bibnamefont {Cherwinka}}, \bibinfo {author}
  {\bibfnamefont {M.~C.}\ \bibnamefont {Chu}}, \bibinfo {author} {\bibfnamefont
  {J.~P.}\ \bibnamefont {Cummings}}, \bibinfo {author} {\bibfnamefont
  {J.}~\bibnamefont {de~Arcos}}, \bibinfo {author} {\bibfnamefont {Z.~Y.}\
  \bibnamefont {Deng}}, \bibinfo {author} {\bibfnamefont {X.~F.}\ \bibnamefont
  {Ding}}, \bibinfo {author} {\bibfnamefont {Y.~Y.}\ \bibnamefont {Ding}},
  \bibinfo {author} {\bibfnamefont {M.~V.}\ \bibnamefont {Diwan}}, \bibinfo
  {author} {\bibfnamefont {J.}~\bibnamefont {Dove}}, \bibinfo {author}
  {\bibfnamefont {E.}~\bibnamefont {Draeger}}, \bibinfo {author} {\bibfnamefont
  {D.~A.}\ \bibnamefont {Dwyer}}, \bibinfo {author} {\bibfnamefont {W.~R.}\
  \bibnamefont {Edwards}}, \bibinfo {author} {\bibfnamefont {S.~R.}\
  \bibnamefont {Ely}}, \bibinfo {author} {\bibfnamefont {R.}~\bibnamefont
  {Gill}}, \bibinfo {author} {\bibfnamefont {M.}~\bibnamefont {Gonchar}},
  \bibinfo {author} {\bibfnamefont {G.~H.}\ \bibnamefont {Gong}}, \bibinfo
  {author} {\bibfnamefont {H.}~\bibnamefont {Gong}}, \bibinfo {author}
  {\bibfnamefont {M.}~\bibnamefont {Grassi}}, \bibinfo {author} {\bibfnamefont
  {W.~Q.}\ \bibnamefont {Gu}}, \bibinfo {author} {\bibfnamefont {M.~Y.}\
  \bibnamefont {Guan}}, \bibinfo {author} {\bibfnamefont {L.}~\bibnamefont
  {Guo}}, \bibinfo {author} {\bibfnamefont {X.~H.}\ \bibnamefont {Guo}},
  \bibinfo {author} {\bibfnamefont {R.~W.}\ \bibnamefont {Hackenburg}},
  \bibinfo {author} {\bibfnamefont {R.}~\bibnamefont {Han}}, \bibinfo {author}
  {\bibfnamefont {S.}~\bibnamefont {Hans}}, \bibinfo {author} {\bibfnamefont
  {M.}~\bibnamefont {He}}, \bibinfo {author} {\bibfnamefont {K.~M.}\
  \bibnamefont {Heeger}}, \bibinfo {author} {\bibfnamefont {Y.~K.}\
  \bibnamefont {Heng}}, \bibinfo {author} {\bibfnamefont {A.}~\bibnamefont
  {Higuera}}, \bibinfo {author} {\bibfnamefont {Y.~K.}\ \bibnamefont {Hor}},
  \bibinfo {author} {\bibfnamefont {Y.~B.}\ \bibnamefont {Hsiung}}, \bibinfo
  {author} {\bibfnamefont {B.~Z.}\ \bibnamefont {Hu}}, \bibinfo {author}
  {\bibfnamefont {L.~J.~M.}\ \bibnamefont {Hu}}, \bibinfo {author}
  {\bibfnamefont {L.~J.~M.}\ \bibnamefont {Hu}}, \bibinfo {author}
  {\bibfnamefont {T.}~\bibnamefont {Hu}}, \bibinfo {author} {\bibfnamefont
  {W.}~\bibnamefont {Hu}}, \bibinfo {author} {\bibfnamefont {E.~C.}\
  \bibnamefont {Huang}}, \bibinfo {author} {\bibfnamefont {H.~X.}\ \bibnamefont
  {Huang}}, \bibinfo {author} {\bibfnamefont {X.~T.}\ \bibnamefont {Huang}},
  \bibinfo {author} {\bibfnamefont {P.}~\bibnamefont {Huber}}, \bibinfo
  {author} {\bibfnamefont {G.}~\bibnamefont {Hussain}}, \bibinfo {author}
  {\bibfnamefont {D.~E.}\ \bibnamefont {Jaffe}}, \bibinfo {author}
  {\bibfnamefont {P.}~\bibnamefont {Jaffke}}, \bibinfo {author} {\bibfnamefont
  {K.~L.}\ \bibnamefont {Jen}}, \bibinfo {author} {\bibfnamefont
  {S.}~\bibnamefont {Jetter}}, \bibinfo {author} {\bibfnamefont {X.~L.~P.}\
  \bibnamefont {Ji}}, \bibinfo {author} {\bibfnamefont {X.~L.~P.}\ \bibnamefont
  {Ji}}, \bibinfo {author} {\bibfnamefont {J.~B.}\ \bibnamefont {Jiao}},
  \bibinfo {author} {\bibfnamefont {R.~A.}\ \bibnamefont {Johnson}}, \bibinfo
  {author} {\bibfnamefont {L.}~\bibnamefont {Kang}}, \bibinfo {author}
  {\bibfnamefont {S.~H.}\ \bibnamefont {Kettell}}, \bibinfo {author}
  {\bibfnamefont {S.}~\bibnamefont {Kohn}}, \bibinfo {author} {\bibfnamefont
  {M.}~\bibnamefont {Kramer}}, \bibinfo {author} {\bibfnamefont {K.~K.}\
  \bibnamefont {Kwan}}, \bibinfo {author} {\bibfnamefont {M.~W.}\ \bibnamefont
  {Kwok}}, \bibinfo {author} {\bibfnamefont {T.}~\bibnamefont {Kwok}}, \bibinfo
  {author} {\bibfnamefont {T.~J.}\ \bibnamefont {Langford}}, \bibinfo {author}
  {\bibfnamefont {K.}~\bibnamefont {Lau}}, \bibinfo {author} {\bibfnamefont
  {L.}~\bibnamefont {Lebanowski}}, \bibinfo {author} {\bibfnamefont
  {J.}~\bibnamefont {Lee}}, \bibinfo {author} {\bibfnamefont {R.~T.}\
  \bibnamefont {Lei}}, \bibinfo {author} {\bibfnamefont {R.}~\bibnamefont
  {Leitner}}, \bibinfo {author} {\bibfnamefont {K.~Y.}\ \bibnamefont {Leung}},
  \bibinfo {author} {\bibfnamefont {J.~K.~C.}\ \bibnamefont {Leung}}, \bibinfo
  {author} {\bibfnamefont {C.~A.}\ \bibnamefont {Lewis}}, \bibinfo {author}
  {\bibfnamefont {D.~J.}\ \bibnamefont {Li}}, \bibinfo {author} {\bibfnamefont
  {F.}~\bibnamefont {Li}}, \bibinfo {author} {\bibfnamefont {G.~S.}\
  \bibnamefont {Li}}, \bibinfo {author} {\bibfnamefont {Q.~J.}\ \bibnamefont
  {Li}}, \bibinfo {author} {\bibfnamefont {S.~C.}\ \bibnamefont {Li}}, \bibinfo
  {author} {\bibfnamefont {W.~D.}\ \bibnamefont {Li}}, \bibinfo {author}
  {\bibfnamefont {X.~N.~Q.}\ \bibnamefont {Li}}, \bibinfo {author}
  {\bibfnamefont {X.~N.~Q.}\ \bibnamefont {Li}}, \bibinfo {author}
  {\bibfnamefont {Y.~F.}\ \bibnamefont {Li}}, \bibinfo {author} {\bibfnamefont
  {Z.~B.}\ \bibnamefont {Li}}, \bibinfo {author} {\bibfnamefont
  {H.}~\bibnamefont {Liang}}, \bibinfo {author} {\bibfnamefont {C.~J.}\
  \bibnamefont {Lin}}, \bibinfo {author} {\bibfnamefont {G.~L.}\ \bibnamefont
  {Lin}}, \bibinfo {author} {\bibfnamefont {P.~Y.}\ \bibnamefont {Lin}},
  \bibinfo {author} {\bibfnamefont {S.~K.}\ \bibnamefont {Lin}}, \bibinfo
  {author} {\bibfnamefont {J.~J.}\ \bibnamefont {Ling}}, \bibinfo {author}
  {\bibfnamefont {J.~M.}\ \bibnamefont {Link}}, \bibinfo {author}
  {\bibfnamefont {L.}~\bibnamefont {Littenberg}}, \bibinfo {author}
  {\bibfnamefont {B.~R.}\ \bibnamefont {Littlejohn}}, \bibinfo {author}
  {\bibfnamefont {D.~W.}\ \bibnamefont {Liu}}, \bibinfo {author} {\bibfnamefont
  {H.}~\bibnamefont {Liu}}, \bibinfo {author} {\bibfnamefont {J.~C.~L.}\
  \bibnamefont {Liu}}, \bibinfo {author} {\bibfnamefont {J.~C.~L.}\
  \bibnamefont {Liu}}, \bibinfo {author} {\bibfnamefont {S.~S.}\ \bibnamefont
  {Liu}}, \bibinfo {author} {\bibfnamefont {C.}~\bibnamefont {Lu}}, \bibinfo
  {author} {\bibfnamefont {H.~Q.}\ \bibnamefont {Lu}}, \bibinfo {author}
  {\bibfnamefont {J.~S.}\ \bibnamefont {Lu}}, \bibinfo {author} {\bibfnamefont
  {K.~B.}\ \bibnamefont {Luk}}, \bibinfo {author} {\bibfnamefont {Q.~M.}\
  \bibnamefont {Ma}}, \bibinfo {author} {\bibfnamefont {X.~B.~Y.}\ \bibnamefont
  {Ma}}, \bibinfo {author} {\bibfnamefont {X.~B.~Y.}\ \bibnamefont {Ma}},
  \bibinfo {author} {\bibfnamefont {Y.~Q.}\ \bibnamefont {Ma}}, \bibinfo
  {author} {\bibfnamefont {D.~A.}\ \bibnamefont {{Martinez Caicedo}}}, \bibinfo
  {author} {\bibfnamefont {K.~T.}\ \bibnamefont {McDonald}}, \bibinfo {author}
  {\bibfnamefont {R.~D.}\ \bibnamefont {McKeown}}, \bibinfo {author}
  {\bibfnamefont {Y.}~\bibnamefont {Meng}}, \bibinfo {author} {\bibfnamefont
  {I.}~\bibnamefont {Mitchell}}, \bibinfo {author} {\bibfnamefont
  {J.}~\bibnamefont {{Monari Kebwaro}}}, \bibinfo {author} {\bibfnamefont
  {Y.}~\bibnamefont {Nakajima}}, \bibinfo {author} {\bibfnamefont
  {J.}~\bibnamefont {Napolitano}}, \bibinfo {author} {\bibfnamefont
  {D.}~\bibnamefont {Naumov}}, \bibinfo {author} {\bibfnamefont
  {E.}~\bibnamefont {Naumova}}, \bibinfo {author} {\bibfnamefont {H.~Y.}\
  \bibnamefont {Ngai}}, \bibinfo {author} {\bibfnamefont {Z.}~\bibnamefont
  {Ning}}, \bibinfo {author} {\bibfnamefont {J.~P.}\ \bibnamefont
  {Ochoa-Ricoux}}, \bibinfo {author} {\bibfnamefont {A.}~\bibnamefont
  {Olshevski}}, \bibinfo {author} {\bibfnamefont {H.-R.}\ \bibnamefont {Pan}},
  \bibinfo {author} {\bibfnamefont {J.}~\bibnamefont {Park}}, \bibinfo {author}
  {\bibfnamefont {S.}~\bibnamefont {Patton}}, \bibinfo {author} {\bibfnamefont
  {V.}~\bibnamefont {Pec}}, \bibinfo {author} {\bibfnamefont {J.~C.}\
  \bibnamefont {Peng}}, \bibinfo {author} {\bibfnamefont {L.~E.}\ \bibnamefont
  {Piilonen}}, \bibinfo {author} {\bibfnamefont {L.}~\bibnamefont {Pinsky}},
  \bibinfo {author} {\bibfnamefont {C.~S.~J.}\ \bibnamefont {Pun}}, \bibinfo
  {author} {\bibfnamefont {F.~Z.}\ \bibnamefont {Qi}}, \bibinfo {author}
  {\bibfnamefont {M.}~\bibnamefont {Qi}}, \bibinfo {author} {\bibfnamefont
  {X.}~\bibnamefont {Qian}}, \bibinfo {author} {\bibfnamefont {N.}~\bibnamefont
  {Raper}}, \bibinfo {author} {\bibfnamefont {B.}~\bibnamefont {Ren}}, \bibinfo
  {author} {\bibfnamefont {J.}~\bibnamefont {Ren}}, \bibinfo {author}
  {\bibfnamefont {R.}~\bibnamefont {Rosero}}, \bibinfo {author} {\bibfnamefont
  {B.}~\bibnamefont {Roskovec}}, \bibinfo {author} {\bibfnamefont {X.~C.}\
  \bibnamefont {Ruan}}, \bibinfo {author} {\bibfnamefont {B.~B.}\ \bibnamefont
  {Shao}}, \bibinfo {author} {\bibfnamefont {H.}~\bibnamefont {Steiner}},
  \bibinfo {author} {\bibfnamefont {G.~X.}\ \bibnamefont {Sun}}, \bibinfo
  {author} {\bibfnamefont {J.~L.}\ \bibnamefont {Sun}}, \bibinfo {author}
  {\bibfnamefont {W.}~\bibnamefont {Tang}}, \bibinfo {author} {\bibfnamefont
  {D.}~\bibnamefont {Taychenachev}}, \bibinfo {author} {\bibfnamefont {K.~V.}\
  \bibnamefont {Tsang}}, \bibinfo {author} {\bibfnamefont {C.~E.}\ \bibnamefont
  {Tull}}, \bibinfo {author} {\bibfnamefont {Y.~C.}\ \bibnamefont {Tung}},
  \bibinfo {author} {\bibfnamefont {N.}~\bibnamefont {Viaux}}, \bibinfo
  {author} {\bibfnamefont {B.}~\bibnamefont {Viren}}, \bibinfo {author}
  {\bibfnamefont {V.}~\bibnamefont {Vorobel}}, \bibinfo {author} {\bibfnamefont
  {C.~H.}\ \bibnamefont {Wang}}, \bibinfo {author} {\bibfnamefont
  {M.}~\bibnamefont {Wang}}, \bibinfo {author} {\bibfnamefont {N.~Y.}\
  \bibnamefont {Wang}}, \bibinfo {author} {\bibfnamefont {R.~G.}\ \bibnamefont
  {Wang}}, \bibinfo {author} {\bibfnamefont {W.~W.}\ \bibnamefont {Wang}},
  \bibinfo {author} {\bibfnamefont {W.~W.}\ \bibnamefont {Wang}}, \bibinfo
  {author} {\bibfnamefont {X.}~\bibnamefont {Wang}}, \bibinfo {author}
  {\bibfnamefont {Y.~F.}\ \bibnamefont {Wang}}, \bibinfo {author}
  {\bibfnamefont {Z.~M.}\ \bibnamefont {Wang}}, \bibinfo {author}
  {\bibfnamefont {Z.~M.}\ \bibnamefont {Wang}}, \bibinfo {author}
  {\bibfnamefont {Z.~M.}\ \bibnamefont {Wang}}, \bibinfo {author}
  {\bibfnamefont {H.~Y.}\ \bibnamefont {Wei}}, \bibinfo {author} {\bibfnamefont
  {L.~J.}\ \bibnamefont {Wen}}, \bibinfo {author} {\bibfnamefont
  {K.}~\bibnamefont {Whisnant}}, \bibinfo {author} {\bibfnamefont {C.~G.}\
  \bibnamefont {White}}, \bibinfo {author} {\bibfnamefont {L.}~\bibnamefont
  {Whitehead}}, \bibinfo {author} {\bibfnamefont {T.}~\bibnamefont {Wise}},
  \bibinfo {author} {\bibfnamefont {H.~L.~H.}\ \bibnamefont {Wong}}, \bibinfo
  {author} {\bibfnamefont {S.~C.~F.}\ \bibnamefont {Wong}}, \bibinfo {author}
  {\bibfnamefont {E.}~\bibnamefont {Worcester}}, \bibinfo {author}
  {\bibfnamefont {Q.}~\bibnamefont {Wu}}, \bibinfo {author} {\bibfnamefont
  {D.~M.}\ \bibnamefont {Xia}}, \bibinfo {author} {\bibfnamefont {J.~K.}\
  \bibnamefont {Xia}}, \bibinfo {author} {\bibfnamefont {X.}~\bibnamefont
  {Xia}}, \bibinfo {author} {\bibfnamefont {Z.~Z.}\ \bibnamefont {Xing}},
  \bibinfo {author} {\bibfnamefont {J.~L.~Y.}\ \bibnamefont {Xu}}, \bibinfo
  {author} {\bibfnamefont {J.~L.~Y.}\ \bibnamefont {Xu}}, \bibinfo {author}
  {\bibfnamefont {J.~L.~Y.}\ \bibnamefont {Xu}}, \bibinfo {author}
  {\bibfnamefont {Y.}~\bibnamefont {Xu}}, \bibinfo {author} {\bibfnamefont
  {T.}~\bibnamefont {Xue}}, \bibinfo {author} {\bibfnamefont {J.}~\bibnamefont
  {Yan}}, \bibinfo {author} {\bibfnamefont {C.~G.}\ \bibnamefont {Yang}},
  \bibinfo {author} {\bibfnamefont {L.}~\bibnamefont {Yang}}, \bibinfo {author}
  {\bibfnamefont {M.~T.~S.}\ \bibnamefont {Yang}}, \bibinfo {author}
  {\bibfnamefont {M.~T.~S.}\ \bibnamefont {Yang}}, \bibinfo {author}
  {\bibfnamefont {M.}~\bibnamefont {Ye}}, \bibinfo {author} {\bibfnamefont
  {M.}~\bibnamefont {Yeh}}, \bibinfo {author} {\bibfnamefont {B.~L.}\
  \bibnamefont {Young}}, \bibinfo {author} {\bibfnamefont {G.~Y.}\ \bibnamefont
  {Yu}}, \bibinfo {author} {\bibfnamefont {Z.~Y.}\ \bibnamefont {Yu}}, \bibinfo
  {author} {\bibfnamefont {S.~L.}\ \bibnamefont {Zang}}, \bibinfo {author}
  {\bibfnamefont {L.}~\bibnamefont {Zhan}}, \bibinfo {author} {\bibfnamefont
  {C.}~\bibnamefont {Zhang}}, \bibinfo {author} {\bibfnamefont {H.~H.}\
  \bibnamefont {Zhang}}, \bibinfo {author} {\bibfnamefont {J.~W.}\ \bibnamefont
  {Zhang}}, \bibinfo {author} {\bibfnamefont {Q.~M.}\ \bibnamefont {Zhang}},
  \bibinfo {author} {\bibfnamefont {Y.~M.~X.}\ \bibnamefont {Zhang}}, \bibinfo
  {author} {\bibfnamefont {Y.~M.~X.}\ \bibnamefont {Zhang}}, \bibinfo {author}
  {\bibfnamefont {Y.~M.~X.}\ \bibnamefont {Zhang}}, \bibinfo {author}
  {\bibfnamefont {Z.~Y. P.~J.}\ \bibnamefont {Zhang}}, \bibinfo {author}
  {\bibfnamefont {Z.~Y. P.~J.}\ \bibnamefont {Zhang}}, \bibinfo {author}
  {\bibfnamefont {Z.~Y. P.~J.}\ \bibnamefont {Zhang}}, \bibinfo {author}
  {\bibfnamefont {J.}~\bibnamefont {Zhao}}, \bibinfo {author} {\bibfnamefont
  {Q.~W.}\ \bibnamefont {Zhao}}, \bibinfo {author} {\bibfnamefont {Y.~B.~F.}\
  \bibnamefont {Zhao}}, \bibinfo {author} {\bibfnamefont {Y.~B.~F.}\
  \bibnamefont {Zhao}}, \bibinfo {author} {\bibfnamefont {L.}~\bibnamefont
  {Zheng}}, \bibinfo {author} {\bibfnamefont {W.~L.}\ \bibnamefont {Zhong}},
  \bibinfo {author} {\bibfnamefont {L.}~\bibnamefont {Zhou}}, \bibinfo {author}
  {\bibfnamefont {N.}~\bibnamefont {Zhou}}, \bibinfo {author} {\bibfnamefont
  {H.~L.}\ \bibnamefont {Zhuang}}, \ and\ \bibinfo {author} {\bibfnamefont
  {J.~H.}\ \bibnamefont {Zou}},\ }\href {\doibase
  10.1103/PhysRevLett.116.061801} {\bibfield  {journal} {\bibinfo  {journal}
  {Physical Review Letters}\ }\textbf {\bibinfo {volume} {116}},\ \bibinfo
  {pages} {061801} (\bibinfo {year} {2016})},\ \Eprint
  {http://arxiv.org/abs/1508.04233} {arXiv:1508.04233} \BibitemShut {NoStop}%
\bibitem [{\citenamefont {Abe}\ \emph {et~al.}(2014)\citenamefont {Abe},
  \citenamefont {dos Anjos}, \citenamefont {Barriere}, \citenamefont {Baussan},
  \citenamefont {Bekman}, \citenamefont {Bergevin}, \citenamefont {Bezerra},
  \citenamefont {Bezrukov}, \citenamefont {Blucher}, \citenamefont {Buck},
  \citenamefont {Busenitz}, \citenamefont {Cabrera}, \citenamefont {Caden},
  \citenamefont {Camilleri}, \citenamefont {Carr}, \citenamefont {Cerrada},
  \citenamefont {Chang}, \citenamefont {Chauveau}, \citenamefont {Chimenti},
  \citenamefont {Collin}, \citenamefont {Conover}, \citenamefont {Conrad},
  \citenamefont {Crespo-Anad{\'{o}}n}, \citenamefont {Crum}, \citenamefont
  {Cucoanes}, \citenamefont {Damon}, \citenamefont {Dawson}, \citenamefont
  {Dhooghe}, \citenamefont {Dietrich}, \citenamefont {Djurcic}, \citenamefont
  {Dracos}, \citenamefont {Elnimr}, \citenamefont {Etenko}, \citenamefont
  {Fallot}, \citenamefont {von Feilitzsch}, \citenamefont {Felde},
  \citenamefont {Fernandes}, \citenamefont {Fischer}, \citenamefont {Franco},
  \citenamefont {Franke}, \citenamefont {Furuta}, \citenamefont {Gil-Botella},
  \citenamefont {Giot}, \citenamefont {G{\"{o}}ger-Neff}, \citenamefont
  {Gonzalez}, \citenamefont {Goodenough}, \citenamefont {Goodman},
  \citenamefont {Grant}, \citenamefont {Haag}, \citenamefont {Hara},
  \citenamefont {Haser}, \citenamefont {Hofmann}, \citenamefont {Horton-Smith},
  \citenamefont {Hourlier}, \citenamefont {Ishitsuka}, \citenamefont {Jochum},
  \citenamefont {Jollet}, \citenamefont {Kaether}, \citenamefont {Kalousis},
  \citenamefont {Kamyshkov}, \citenamefont {Kaplan}, \citenamefont {Kawasaki},
  \citenamefont {Kemp}, \citenamefont {de~Kerret}, \citenamefont {Kryn},
  \citenamefont {Kuze}, \citenamefont {Lachenmaier}, \citenamefont {Lane},
  \citenamefont {Lasserre}, \citenamefont {Letourneau}, \citenamefont
  {Lhuillier}, \citenamefont {Lima}, \citenamefont {Lindner}, \citenamefont
  {L{\'{o}}pez-Casta{\~{n}}o}, \citenamefont {LoSecco}, \citenamefont
  {Lubsandorzhiev}, \citenamefont {Lucht}, \citenamefont {Maeda}, \citenamefont
  {Mariani}, \citenamefont {Maricic}, \citenamefont {Martino}, \citenamefont
  {Matsubara}, \citenamefont {Mention}, \citenamefont {Meregaglia},
  \citenamefont {Miletic}, \citenamefont {Milincic}, \citenamefont {Minotti},
  \citenamefont {Nagasaka}, \citenamefont {Nikitenko}, \citenamefont {Novella},
  \citenamefont {Oberauer}, \citenamefont {Obolensky}, \citenamefont {Onillon},
  \citenamefont {Osborn}, \citenamefont {Palomares}, \citenamefont {Pepe},
  \citenamefont {Perasso}, \citenamefont {Pfahler}, \citenamefont {Porta},
  \citenamefont {Pronost}, \citenamefont {Reichenbacher}, \citenamefont
  {Reinhold}, \citenamefont {R{\"{o}}hling}, \citenamefont {Roncin},
  \citenamefont {Roth}, \citenamefont {Rybolt}, \citenamefont {Sakamoto},
  \citenamefont {Santorelli}, \citenamefont {Schilithz}, \citenamefont
  {Sch{\"{o}}nert}, \citenamefont {Schoppmann}, \citenamefont {Shaevitz},
  \citenamefont {Sharankova}, \citenamefont {Shimojima}, \citenamefont
  {Shrestha}, \citenamefont {Sibille}, \citenamefont {Sinev}, \citenamefont
  {Skorokhvatov}, \citenamefont {Smith}, \citenamefont {Spitz}, \citenamefont
  {Stahl}, \citenamefont {Stancu}, \citenamefont {Stokes}, \citenamefont
  {Strait}, \citenamefont {St{\"{u}}ken}, \citenamefont {Suekane},
  \citenamefont {Sukhotin}, \citenamefont {Sumiyoshi}, \citenamefont {Sun},
  \citenamefont {Svoboda}, \citenamefont {Terao}, \citenamefont {Tonazzo},
  \citenamefont {Thi}, \citenamefont {Valdiviesso}, \citenamefont
  {Vassilopoulos}, \citenamefont {Veyssiere}, \citenamefont {Vivier},
  \citenamefont {Wagner}, \citenamefont {Walsh}, \citenamefont {Watanabe},
  \citenamefont {Wiebusch}, \citenamefont {Winslow}, \citenamefont {Wurm},
  \citenamefont {Yang}, \citenamefont {Yermia},\ and\ \citenamefont
  {Zimmer}}]{Abe2014}%
  \BibitemOpen
  \bibfield  {author} {\bibinfo {author} {\bibfnamefont {Y.}~\bibnamefont
  {Abe}}, \bibinfo {author} {\bibfnamefont {J.~C.}\ \bibnamefont {dos Anjos}},
  \bibinfo {author} {\bibfnamefont {J.~C.}\ \bibnamefont {Barriere}}, \bibinfo
  {author} {\bibfnamefont {E.}~\bibnamefont {Baussan}}, \bibinfo {author}
  {\bibfnamefont {I.}~\bibnamefont {Bekman}}, \bibinfo {author} {\bibfnamefont
  {M.}~\bibnamefont {Bergevin}}, \bibinfo {author} {\bibfnamefont {T.~J.~C.}\
  \bibnamefont {Bezerra}}, \bibinfo {author} {\bibfnamefont {L.}~\bibnamefont
  {Bezrukov}}, \bibinfo {author} {\bibfnamefont {E.}~\bibnamefont {Blucher}},
  \bibinfo {author} {\bibfnamefont {C.}~\bibnamefont {Buck}}, \bibinfo {author}
  {\bibfnamefont {J.}~\bibnamefont {Busenitz}}, \bibinfo {author}
  {\bibfnamefont {A.}~\bibnamefont {Cabrera}}, \bibinfo {author} {\bibfnamefont
  {E.}~\bibnamefont {Caden}}, \bibinfo {author} {\bibfnamefont
  {L.}~\bibnamefont {Camilleri}}, \bibinfo {author} {\bibfnamefont
  {R.}~\bibnamefont {Carr}}, \bibinfo {author} {\bibfnamefont {M.}~\bibnamefont
  {Cerrada}}, \bibinfo {author} {\bibfnamefont {P.-J.}\ \bibnamefont {Chang}},
  \bibinfo {author} {\bibfnamefont {E.}~\bibnamefont {Chauveau}}, \bibinfo
  {author} {\bibfnamefont {P.}~\bibnamefont {Chimenti}}, \bibinfo {author}
  {\bibfnamefont {A.~P.}\ \bibnamefont {Collin}}, \bibinfo {author}
  {\bibfnamefont {E.}~\bibnamefont {Conover}}, \bibinfo {author} {\bibfnamefont
  {J.~M.}\ \bibnamefont {Conrad}}, \bibinfo {author} {\bibfnamefont {J.~I.}\
  \bibnamefont {Crespo-Anad{\'{o}}n}}, \bibinfo {author} {\bibfnamefont
  {K.}~\bibnamefont {Crum}}, \bibinfo {author} {\bibfnamefont {A.~S.}\
  \bibnamefont {Cucoanes}}, \bibinfo {author} {\bibfnamefont {E.}~\bibnamefont
  {Damon}}, \bibinfo {author} {\bibfnamefont {J.~V.}\ \bibnamefont {Dawson}},
  \bibinfo {author} {\bibfnamefont {J.}~\bibnamefont {Dhooghe}}, \bibinfo
  {author} {\bibfnamefont {D.}~\bibnamefont {Dietrich}}, \bibinfo {author}
  {\bibfnamefont {Z.}~\bibnamefont {Djurcic}}, \bibinfo {author} {\bibfnamefont
  {M.}~\bibnamefont {Dracos}}, \bibinfo {author} {\bibfnamefont
  {M.}~\bibnamefont {Elnimr}}, \bibinfo {author} {\bibfnamefont
  {A.}~\bibnamefont {Etenko}}, \bibinfo {author} {\bibfnamefont
  {M.}~\bibnamefont {Fallot}}, \bibinfo {author} {\bibfnamefont
  {F.}~\bibnamefont {von Feilitzsch}}, \bibinfo {author} {\bibfnamefont
  {J.}~\bibnamefont {Felde}}, \bibinfo {author} {\bibfnamefont {S.~M.}\
  \bibnamefont {Fernandes}}, \bibinfo {author} {\bibfnamefont {V.}~\bibnamefont
  {Fischer}}, \bibinfo {author} {\bibfnamefont {D.}~\bibnamefont {Franco}},
  \bibinfo {author} {\bibfnamefont {M.}~\bibnamefont {Franke}}, \bibinfo
  {author} {\bibfnamefont {H.}~\bibnamefont {Furuta}}, \bibinfo {author}
  {\bibfnamefont {I.}~\bibnamefont {Gil-Botella}}, \bibinfo {author}
  {\bibfnamefont {L.}~\bibnamefont {Giot}}, \bibinfo {author} {\bibfnamefont
  {M.}~\bibnamefont {G{\"{o}}ger-Neff}}, \bibinfo {author} {\bibfnamefont
  {L.~F.~G.}\ \bibnamefont {Gonzalez}}, \bibinfo {author} {\bibfnamefont
  {L.}~\bibnamefont {Goodenough}}, \bibinfo {author} {\bibfnamefont {M.~C.}\
  \bibnamefont {Goodman}}, \bibinfo {author} {\bibfnamefont {C.}~\bibnamefont
  {Grant}}, \bibinfo {author} {\bibfnamefont {N.}~\bibnamefont {Haag}},
  \bibinfo {author} {\bibfnamefont {T.}~\bibnamefont {Hara}}, \bibinfo {author}
  {\bibfnamefont {J.}~\bibnamefont {Haser}}, \bibinfo {author} {\bibfnamefont
  {M.}~\bibnamefont {Hofmann}}, \bibinfo {author} {\bibfnamefont {G.~A.}\
  \bibnamefont {Horton-Smith}}, \bibinfo {author} {\bibfnamefont
  {A.}~\bibnamefont {Hourlier}}, \bibinfo {author} {\bibfnamefont
  {M.}~\bibnamefont {Ishitsuka}}, \bibinfo {author} {\bibfnamefont
  {J.}~\bibnamefont {Jochum}}, \bibinfo {author} {\bibfnamefont
  {C.}~\bibnamefont {Jollet}}, \bibinfo {author} {\bibfnamefont
  {F.}~\bibnamefont {Kaether}}, \bibinfo {author} {\bibfnamefont {L.~N.}\
  \bibnamefont {Kalousis}}, \bibinfo {author} {\bibfnamefont {Y.}~\bibnamefont
  {Kamyshkov}}, \bibinfo {author} {\bibfnamefont {D.~M.}\ \bibnamefont
  {Kaplan}}, \bibinfo {author} {\bibfnamefont {T.}~\bibnamefont {Kawasaki}},
  \bibinfo {author} {\bibfnamefont {E.}~\bibnamefont {Kemp}}, \bibinfo {author}
  {\bibfnamefont {H.}~\bibnamefont {de~Kerret}}, \bibinfo {author}
  {\bibfnamefont {D.}~\bibnamefont {Kryn}}, \bibinfo {author} {\bibfnamefont
  {M.}~\bibnamefont {Kuze}}, \bibinfo {author} {\bibfnamefont {T.}~\bibnamefont
  {Lachenmaier}}, \bibinfo {author} {\bibfnamefont {C.~E.}\ \bibnamefont
  {Lane}}, \bibinfo {author} {\bibfnamefont {T.}~\bibnamefont {Lasserre}},
  \bibinfo {author} {\bibfnamefont {A.}~\bibnamefont {Letourneau}}, \bibinfo
  {author} {\bibfnamefont {D.}~\bibnamefont {Lhuillier}}, \bibinfo {author}
  {\bibfnamefont {H.~P.}\ \bibnamefont {Lima}}, \bibinfo {author}
  {\bibfnamefont {M.}~\bibnamefont {Lindner}}, \bibinfo {author} {\bibfnamefont
  {J.~M.}\ \bibnamefont {L{\'{o}}pez-Casta{\~{n}}o}}, \bibinfo {author}
  {\bibfnamefont {J.~M.}\ \bibnamefont {LoSecco}}, \bibinfo {author}
  {\bibfnamefont {B.}~\bibnamefont {Lubsandorzhiev}}, \bibinfo {author}
  {\bibfnamefont {S.}~\bibnamefont {Lucht}}, \bibinfo {author} {\bibfnamefont
  {J.}~\bibnamefont {Maeda}}, \bibinfo {author} {\bibfnamefont
  {C.}~\bibnamefont {Mariani}}, \bibinfo {author} {\bibfnamefont
  {J.}~\bibnamefont {Maricic}}, \bibinfo {author} {\bibfnamefont
  {J.}~\bibnamefont {Martino}}, \bibinfo {author} {\bibfnamefont
  {T.}~\bibnamefont {Matsubara}}, \bibinfo {author} {\bibfnamefont
  {G.}~\bibnamefont {Mention}}, \bibinfo {author} {\bibfnamefont
  {A.}~\bibnamefont {Meregaglia}}, \bibinfo {author} {\bibfnamefont
  {T.}~\bibnamefont {Miletic}}, \bibinfo {author} {\bibfnamefont
  {R.}~\bibnamefont {Milincic}}, \bibinfo {author} {\bibfnamefont
  {A.}~\bibnamefont {Minotti}}, \bibinfo {author} {\bibfnamefont
  {Y.}~\bibnamefont {Nagasaka}}, \bibinfo {author} {\bibfnamefont
  {Y.}~\bibnamefont {Nikitenko}}, \bibinfo {author} {\bibfnamefont
  {P.}~\bibnamefont {Novella}}, \bibinfo {author} {\bibfnamefont
  {L.}~\bibnamefont {Oberauer}}, \bibinfo {author} {\bibfnamefont
  {M.}~\bibnamefont {Obolensky}}, \bibinfo {author} {\bibfnamefont
  {A.}~\bibnamefont {Onillon}}, \bibinfo {author} {\bibfnamefont
  {A.}~\bibnamefont {Osborn}}, \bibinfo {author} {\bibfnamefont
  {C.}~\bibnamefont {Palomares}}, \bibinfo {author} {\bibfnamefont {I.~M.}\
  \bibnamefont {Pepe}}, \bibinfo {author} {\bibfnamefont {S.}~\bibnamefont
  {Perasso}}, \bibinfo {author} {\bibfnamefont {P.}~\bibnamefont {Pfahler}},
  \bibinfo {author} {\bibfnamefont {A.}~\bibnamefont {Porta}}, \bibinfo
  {author} {\bibfnamefont {G.}~\bibnamefont {Pronost}}, \bibinfo {author}
  {\bibfnamefont {J.}~\bibnamefont {Reichenbacher}}, \bibinfo {author}
  {\bibfnamefont {B.}~\bibnamefont {Reinhold}}, \bibinfo {author}
  {\bibfnamefont {M.}~\bibnamefont {R{\"{o}}hling}}, \bibinfo {author}
  {\bibfnamefont {R.}~\bibnamefont {Roncin}}, \bibinfo {author} {\bibfnamefont
  {S.}~\bibnamefont {Roth}}, \bibinfo {author} {\bibfnamefont {B.}~\bibnamefont
  {Rybolt}}, \bibinfo {author} {\bibfnamefont {Y.}~\bibnamefont {Sakamoto}},
  \bibinfo {author} {\bibfnamefont {R.}~\bibnamefont {Santorelli}}, \bibinfo
  {author} {\bibfnamefont {A.~C.}\ \bibnamefont {Schilithz}}, \bibinfo {author}
  {\bibfnamefont {S.}~\bibnamefont {Sch{\"{o}}nert}}, \bibinfo {author}
  {\bibfnamefont {S.}~\bibnamefont {Schoppmann}}, \bibinfo {author}
  {\bibfnamefont {M.~H.}\ \bibnamefont {Shaevitz}}, \bibinfo {author}
  {\bibfnamefont {R.}~\bibnamefont {Sharankova}}, \bibinfo {author}
  {\bibfnamefont {S.}~\bibnamefont {Shimojima}}, \bibinfo {author}
  {\bibfnamefont {D.}~\bibnamefont {Shrestha}}, \bibinfo {author}
  {\bibfnamefont {V.}~\bibnamefont {Sibille}}, \bibinfo {author} {\bibfnamefont
  {V.}~\bibnamefont {Sinev}}, \bibinfo {author} {\bibfnamefont
  {M.}~\bibnamefont {Skorokhvatov}}, \bibinfo {author} {\bibfnamefont
  {E.}~\bibnamefont {Smith}}, \bibinfo {author} {\bibfnamefont
  {J.}~\bibnamefont {Spitz}}, \bibinfo {author} {\bibfnamefont
  {A.}~\bibnamefont {Stahl}}, \bibinfo {author} {\bibfnamefont
  {I.}~\bibnamefont {Stancu}}, \bibinfo {author} {\bibfnamefont {L.~F.~F.}\
  \bibnamefont {Stokes}}, \bibinfo {author} {\bibfnamefont {M.}~\bibnamefont
  {Strait}}, \bibinfo {author} {\bibfnamefont {A.}~\bibnamefont
  {St{\"{u}}ken}}, \bibinfo {author} {\bibfnamefont {F.}~\bibnamefont
  {Suekane}}, \bibinfo {author} {\bibfnamefont {S.}~\bibnamefont {Sukhotin}},
  \bibinfo {author} {\bibfnamefont {T.}~\bibnamefont {Sumiyoshi}}, \bibinfo
  {author} {\bibfnamefont {Y.}~\bibnamefont {Sun}}, \bibinfo {author}
  {\bibfnamefont {R.}~\bibnamefont {Svoboda}}, \bibinfo {author} {\bibfnamefont
  {K.}~\bibnamefont {Terao}}, \bibinfo {author} {\bibfnamefont
  {A.}~\bibnamefont {Tonazzo}}, \bibinfo {author} {\bibfnamefont {H.~H.~T.}\
  \bibnamefont {Thi}}, \bibinfo {author} {\bibfnamefont {G.}~\bibnamefont
  {Valdiviesso}}, \bibinfo {author} {\bibfnamefont {N.}~\bibnamefont
  {Vassilopoulos}}, \bibinfo {author} {\bibfnamefont {C.}~\bibnamefont
  {Veyssiere}}, \bibinfo {author} {\bibfnamefont {M.}~\bibnamefont {Vivier}},
  \bibinfo {author} {\bibfnamefont {S.}~\bibnamefont {Wagner}}, \bibinfo
  {author} {\bibfnamefont {N.}~\bibnamefont {Walsh}}, \bibinfo {author}
  {\bibfnamefont {H.}~\bibnamefont {Watanabe}}, \bibinfo {author}
  {\bibfnamefont {C.}~\bibnamefont {Wiebusch}}, \bibinfo {author}
  {\bibfnamefont {L.}~\bibnamefont {Winslow}}, \bibinfo {author} {\bibfnamefont
  {M.}~\bibnamefont {Wurm}}, \bibinfo {author} {\bibfnamefont {G.}~\bibnamefont
  {Yang}}, \bibinfo {author} {\bibfnamefont {F.}~\bibnamefont {Yermia}}, \ and\
  \bibinfo {author} {\bibfnamefont {V.}~\bibnamefont {Zimmer}},\ }\href
  {\doibase 10.1007/JHEP10(2014)086} {\bibfield  {journal} {\bibinfo  {journal}
  {Journal of High Energy Physics}\ }\textbf {\bibinfo {volume} {2014}},\
  \bibinfo {pages} {86} (\bibinfo {year} {2014})},\ \Eprint
  {http://arxiv.org/abs/1406.7763} {arXiv:1406.7763} \BibitemShut {NoStop}%
\bibitem [{\citenamefont {Seo}\ \emph {et~al.}(2018)\citenamefont {Seo},
  \citenamefont {Choi}, \citenamefont {Seo}, \citenamefont {Choi},
  \citenamefont {Choi}, \citenamefont {Jang}, \citenamefont {Jang},
  \citenamefont {Joo}, \citenamefont {Kim}, \citenamefont {Kim}, \citenamefont
  {Kim}, \citenamefont {Kim}, \citenamefont {Kim}, \citenamefont {Kim},
  \citenamefont {Kwon}, \citenamefont {Lee}, \citenamefont {Lee}, \citenamefont
  {Lim}, \citenamefont {Pac}, \citenamefont {Park}, \citenamefont {Park},
  \citenamefont {Park}, \citenamefont {Seon}, \citenamefont {Shin},
  \citenamefont {Yang}, \citenamefont {Yang}, \citenamefont {Yeo},\ and\
  \citenamefont {Yu}}]{Seo2018}%
  \BibitemOpen
  \bibfield  {author} {\bibinfo {author} {\bibfnamefont {S.~H.}\ \bibnamefont
  {Seo}}, \bibinfo {author} {\bibfnamefont {W.~Q.}\ \bibnamefont {Choi}},
  \bibinfo {author} {\bibfnamefont {H.}~\bibnamefont {Seo}}, \bibinfo {author}
  {\bibfnamefont {J.~H.}\ \bibnamefont {Choi}}, \bibinfo {author}
  {\bibfnamefont {Y.}~\bibnamefont {Choi}}, \bibinfo {author} {\bibfnamefont
  {H.~I.}\ \bibnamefont {Jang}}, \bibinfo {author} {\bibfnamefont {J.~S.}\
  \bibnamefont {Jang}}, \bibinfo {author} {\bibfnamefont {K.~K.}\ \bibnamefont
  {Joo}}, \bibinfo {author} {\bibfnamefont {B.~R.}\ \bibnamefont {Kim}},
  \bibinfo {author} {\bibfnamefont {H.~S.}\ \bibnamefont {Kim}}, \bibinfo
  {author} {\bibfnamefont {J.~Y.}\ \bibnamefont {Kim}}, \bibinfo {author}
  {\bibfnamefont {S.~B.}\ \bibnamefont {Kim}}, \bibinfo {author} {\bibfnamefont
  {S.~Y.}\ \bibnamefont {Kim}}, \bibinfo {author} {\bibfnamefont
  {W.}~\bibnamefont {Kim}}, \bibinfo {author} {\bibfnamefont {E.}~\bibnamefont
  {Kwon}}, \bibinfo {author} {\bibfnamefont {D.~H.}\ \bibnamefont {Lee}},
  \bibinfo {author} {\bibfnamefont {Y.~C.}\ \bibnamefont {Lee}}, \bibinfo
  {author} {\bibfnamefont {I.~T.}\ \bibnamefont {Lim}}, \bibinfo {author}
  {\bibfnamefont {M.~Y.}\ \bibnamefont {Pac}}, \bibinfo {author} {\bibfnamefont
  {I.~G.}\ \bibnamefont {Park}}, \bibinfo {author} {\bibfnamefont {J.~S.}\
  \bibnamefont {Park}}, \bibinfo {author} {\bibfnamefont {R.~G.}\ \bibnamefont
  {Park}}, \bibinfo {author} {\bibfnamefont {Y.~G.}\ \bibnamefont {Seon}},
  \bibinfo {author} {\bibfnamefont {C.~D.}\ \bibnamefont {Shin}}, \bibinfo
  {author} {\bibfnamefont {J.~H.}\ \bibnamefont {Yang}}, \bibinfo {author}
  {\bibfnamefont {J.~Y.}\ \bibnamefont {Yang}}, \bibinfo {author}
  {\bibfnamefont {I.~S.}\ \bibnamefont {Yeo}}, \ and\ \bibinfo {author}
  {\bibfnamefont {I.}~\bibnamefont {Yu}},\ }\href {\doibase
  10.1103/PhysRevD.98.012002} {\bibfield  {journal} {\bibinfo  {journal}
  {Physical Review D}\ }\textbf {\bibinfo {volume} {98}},\ \bibinfo {pages}
  {012002} (\bibinfo {year} {2018})},\ \Eprint
  {http://arxiv.org/abs/1610.04326} {arXiv:1610.04326} \BibitemShut {NoStop}%
\bibitem [{\citenamefont {Hayes}\ and\ \citenamefont
  {Vogel}(2016)}]{Hayes2016}%
  \BibitemOpen
  \bibfield  {author} {\bibinfo {author} {\bibfnamefont {A.~C.}\ \bibnamefont
  {Hayes}}\ and\ \bibinfo {author} {\bibfnamefont {P.}~\bibnamefont {Vogel}},\
  }\href {\doibase 10.1146/annurev-nucl-102115-044826} {\bibfield  {journal}
  {\bibinfo  {journal} {Annual Review of Nuclear and Particle Science}\
  }\textbf {\bibinfo {volume} {66}},\ \bibinfo {pages} {219} (\bibinfo {year}
  {2016})},\ \Eprint {http://arxiv.org/abs/1605.02047} {arXiv:1605.02047}
  \BibitemShut {NoStop}%
\bibitem [{\citenamefont {Huber}(2011)}]{Huber2011}%
  \BibitemOpen
  \bibfield  {author} {\bibinfo {author} {\bibfnamefont {P.}~\bibnamefont
  {Huber}},\ }\href {\doibase 10.1103/PhysRevC.84.024617} {\bibfield  {journal}
  {\bibinfo  {journal} {Physical Review C}\ }\textbf {\bibinfo {volume} {84}},\
  \bibinfo {pages} {24617} (\bibinfo {year} {2011})}\BibitemShut {NoStop}%
\bibitem [{\citenamefont {Huber}(2012)}]{Huber2012}%
  \BibitemOpen
  \bibfield  {author} {\bibinfo {author} {\bibfnamefont {P.}~\bibnamefont
  {Huber}},\ }\href {\doibase 10.1103/PhysRevC.85.029901} {\bibfield  {journal}
  {\bibinfo  {journal} {Physical Review C}\ }\textbf {\bibinfo {volume} {85}},\
  \bibinfo {pages} {029901} (\bibinfo {year} {2012})}\BibitemShut {NoStop}%
\bibitem [{\citenamefont {Mueller}\ \emph {et~al.}(2011)\citenamefont
  {Mueller}, \citenamefont {Lhuillier}, \citenamefont {Fallot}, \citenamefont
  {Letourneau}, \citenamefont {Cormon}, \citenamefont {Fechner}, \citenamefont
  {Giot}, \citenamefont {Lasserre}, \citenamefont {Martino}, \citenamefont
  {Mention}, \citenamefont {Porta},\ and\ \citenamefont
  {Yermia}}]{Mueller2011}%
  \BibitemOpen
  \bibfield  {author} {\bibinfo {author} {\bibfnamefont {T.~A.}\ \bibnamefont
  {Mueller}}, \bibinfo {author} {\bibfnamefont {D.}~\bibnamefont {Lhuillier}},
  \bibinfo {author} {\bibfnamefont {M.}~\bibnamefont {Fallot}}, \bibinfo
  {author} {\bibfnamefont {A.}~\bibnamefont {Letourneau}}, \bibinfo {author}
  {\bibfnamefont {S.}~\bibnamefont {Cormon}}, \bibinfo {author} {\bibfnamefont
  {M.}~\bibnamefont {Fechner}}, \bibinfo {author} {\bibfnamefont
  {L.}~\bibnamefont {Giot}}, \bibinfo {author} {\bibfnamefont {T.}~\bibnamefont
  {Lasserre}}, \bibinfo {author} {\bibfnamefont {J.}~\bibnamefont {Martino}},
  \bibinfo {author} {\bibfnamefont {G.}~\bibnamefont {Mention}}, \bibinfo
  {author} {\bibfnamefont {A.}~\bibnamefont {Porta}}, \ and\ \bibinfo {author}
  {\bibfnamefont {F.}~\bibnamefont {Yermia}},\ }\href {\doibase
  10.1103/PhysRevC.83.054615} {\bibfield  {journal} {\bibinfo  {journal}
  {Physical Review C - Nuclear Physics}\ }\textbf {\bibinfo {volume} {83}},\
  \bibinfo {pages} {054615} (\bibinfo {year} {2011})},\ \Eprint
  {http://arxiv.org/abs/1101.2663} {arXiv:1101.2663} \BibitemShut {NoStop}%
\bibitem [{\citenamefont {Hayes}\ \emph {et~al.}(2014)\citenamefont {Hayes},
  \citenamefont {Friar}, \citenamefont {Garvey}, \citenamefont {Jungman},\ and\
  \citenamefont {Jonkmans}}]{Hayes2014}%
  \BibitemOpen
  \bibfield  {author} {\bibinfo {author} {\bibfnamefont {A.~C.}\ \bibnamefont
  {Hayes}}, \bibinfo {author} {\bibfnamefont {J.~L.}\ \bibnamefont {Friar}},
  \bibinfo {author} {\bibfnamefont {G.~T.}\ \bibnamefont {Garvey}}, \bibinfo
  {author} {\bibfnamefont {G.}~\bibnamefont {Jungman}}, \ and\ \bibinfo
  {author} {\bibfnamefont {G.}~\bibnamefont {Jonkmans}},\ }\href {\doibase
  10.1103/PhysRevLett.112.202501} {\bibfield  {journal} {\bibinfo  {journal}
  {Physical Review Letters}\ }\textbf {\bibinfo {volume} {112}},\ \bibinfo
  {pages} {202501} (\bibinfo {year} {2014})},\ \Eprint
  {http://arxiv.org/abs/1309.4146} {arXiv:1309.4146} \BibitemShut {NoStop}%
\bibitem [{\citenamefont {Fang}\ and\ \citenamefont {Brown}(2015)}]{Fang2015}%
  \BibitemOpen
  \bibfield  {author} {\bibinfo {author} {\bibfnamefont {D.-L.}\ \bibnamefont
  {Fang}}\ and\ \bibinfo {author} {\bibfnamefont {B.~A.}\ \bibnamefont
  {Brown}},\ }\href {\doibase 10.1103/PhysRevC.91.025503} {\bibfield  {journal}
  {\bibinfo  {journal} {Physical Review C}\ }\textbf {\bibinfo {volume} {91}},\
  \bibinfo {pages} {025503} (\bibinfo {year} {2015})},\ \Eprint
  {http://arxiv.org/abs/1502.02246} {arXiv:1502.02246} \BibitemShut {NoStop}%
\bibitem [{\citenamefont {Hayen}\ \emph {et~al.}(2019)\citenamefont {Hayen},
  \citenamefont {Kostensalo}, \citenamefont {Severijns},\ and\ \citenamefont
  {Suhonen}}]{Hayen2019}%
  \BibitemOpen
  \bibfield  {author} {\bibinfo {author} {\bibfnamefont {L.}~\bibnamefont
  {Hayen}}, \bibinfo {author} {\bibfnamefont {J.}~\bibnamefont {Kostensalo}},
  \bibinfo {author} {\bibfnamefont {N.}~\bibnamefont {Severijns}}, \ and\
  \bibinfo {author} {\bibfnamefont {J.}~\bibnamefont {Suhonen}},\ }\href
  {\doibase 10.1103/PhysRevC.99.031301} {\bibfield  {journal} {\bibinfo
  {journal} {Physical Review C}\ }\textbf {\bibinfo {volume} {99}},\ \bibinfo
  {pages} {031301(R)} (\bibinfo {year} {2019})}\BibitemShut {NoStop}%
\bibitem [{\citenamefont {Stech}\ and\ \citenamefont
  {Sch{\"{u}}lke}(1964)}]{Stech1964}%
  \BibitemOpen
  \bibfield  {author} {\bibinfo {author} {\bibfnamefont {B.}~\bibnamefont
  {Stech}}\ and\ \bibinfo {author} {\bibfnamefont {L.}~\bibnamefont
  {Sch{\"{u}}lke}},\ }\href {\doibase 10.1007/BF01381649} {\bibfield  {journal}
  {\bibinfo  {journal} {Zeitschrift f{\"{u}}r Physik}\ }\textbf {\bibinfo
  {volume} {179}},\ \bibinfo {pages} {314} (\bibinfo {year}
  {1964})}\BibitemShut {NoStop}%
\bibitem [{\citenamefont {Holstein}(1979)}]{Holstein1979b}%
  \BibitemOpen
  \bibfield  {author} {\bibinfo {author} {\bibfnamefont {B.~R.}\ \bibnamefont
  {Holstein}},\ }\href {\doibase 10.1103/PhysRevC.19.1544} {\bibfield
  {journal} {\bibinfo  {journal} {Physical Review C}\ }\textbf {\bibinfo
  {volume} {19}},\ \bibinfo {pages} {1544} (\bibinfo {year}
  {1979})}\BibitemShut {NoStop}%
\bibitem [{\citenamefont {Hayen}\ \emph {et~al.}(2018)\citenamefont {Hayen},
  \citenamefont {Severijns}, \citenamefont {Bodek}, \citenamefont {Rozpedzik},\
  and\ \citenamefont {Mougeot}}]{Hayen2018}%
  \BibitemOpen
  \bibfield  {author} {\bibinfo {author} {\bibfnamefont {L.}~\bibnamefont
  {Hayen}}, \bibinfo {author} {\bibfnamefont {N.}~\bibnamefont {Severijns}},
  \bibinfo {author} {\bibfnamefont {K.}~\bibnamefont {Bodek}}, \bibinfo
  {author} {\bibfnamefont {D.}~\bibnamefont {Rozpedzik}}, \ and\ \bibinfo
  {author} {\bibfnamefont {X.}~\bibnamefont {Mougeot}},\ }\href {\doibase
  10.1103/RevModPhys.90.015008} {\bibfield  {journal} {\bibinfo  {journal}
  {Reviews of Modern Physics}\ }\textbf {\bibinfo {volume} {90}},\ \bibinfo
  {pages} {015008} (\bibinfo {year} {2018})},\ \Eprint
  {http://arxiv.org/abs/1709.07530} {arXiv:1709.07530} \BibitemShut {NoStop}%
\bibitem [{\citenamefont {Behrens}\ and\ \citenamefont
  {B{\"{u}}hring}(1971)}]{Behrens1971}%
  \BibitemOpen
  \bibfield  {author} {\bibinfo {author} {\bibfnamefont {H.}~\bibnamefont
  {Behrens}}\ and\ \bibinfo {author} {\bibfnamefont {W.}~\bibnamefont
  {B{\"{u}}hring}},\ }\href {\doibase 10.1016/0375-9474(71)90489-1} {\bibfield
  {journal} {\bibinfo  {journal} {Nuclear Physics A}\ }\textbf {\bibinfo
  {volume} {162}},\ \bibinfo {pages} {111} (\bibinfo {year}
  {1971})}\BibitemShut {NoStop}%
\bibitem [{\citenamefont {Behrens}\ and\ \citenamefont
  {B{\"{u}}hring}(1982)}]{Behrens1982}%
  \BibitemOpen
  \bibfield  {author} {\bibinfo {author} {\bibfnamefont {H.}~\bibnamefont
  {Behrens}}\ and\ \bibinfo {author} {\bibfnamefont {W.}~\bibnamefont
  {B{\"{u}}hring}},\ }\href@noop {} {\emph {\bibinfo {title} {{Electron radial
  wave functions and nuclear beta-decay}}}}\ (\bibinfo  {publisher} {Clarendon
  Press, Oxford},\ \bibinfo {year} {1982})\BibitemShut {NoStop}%
\bibitem [{\citenamefont {Warburton}\ and\ \citenamefont
  {Towner}(1994)}]{Warburton1994}%
  \BibitemOpen
  \bibfield  {author} {\bibinfo {author} {\bibfnamefont {E.~K.}\ \bibnamefont
  {Warburton}}\ and\ \bibinfo {author} {\bibfnamefont {I.~S.}\ \bibnamefont
  {Towner}},\ }\href {\doibase 10.1016/0370-1573(94)90144-9} {\bibfield
  {journal} {\bibinfo  {journal} {Physics Reports}\ }\textbf {\bibinfo {volume}
  {242}},\ \bibinfo {pages} {103} (\bibinfo {year} {1994})}\BibitemShut
  {NoStop}%
\bibitem [{\citenamefont {Baumann}\ \emph {et~al.}(1998)\citenamefont
  {Baumann}, \citenamefont {Bounajma}, \citenamefont {Didierjean},
  \citenamefont {Huck}, \citenamefont {Knipper}, \citenamefont {Ramdhane},
  \citenamefont {Walter}, \citenamefont {Marguier}, \citenamefont
  {Richard-Serre},\ and\ \citenamefont {Brown}}]{Baumann1998}%
  \BibitemOpen
  \bibfield  {author} {\bibinfo {author} {\bibfnamefont {P.}~\bibnamefont
  {Baumann}}, \bibinfo {author} {\bibfnamefont {M.}~\bibnamefont {Bounajma}},
  \bibinfo {author} {\bibfnamefont {F.}~\bibnamefont {Didierjean}}, \bibinfo
  {author} {\bibfnamefont {a.}~\bibnamefont {Huck}}, \bibinfo {author}
  {\bibfnamefont {a.}~\bibnamefont {Knipper}}, \bibinfo {author} {\bibfnamefont
  {M.}~\bibnamefont {Ramdhane}}, \bibinfo {author} {\bibfnamefont
  {G.}~\bibnamefont {Walter}}, \bibinfo {author} {\bibfnamefont
  {G.}~\bibnamefont {Marguier}}, \bibinfo {author} {\bibfnamefont
  {C.}~\bibnamefont {Richard-Serre}}, \ and\ \bibinfo {author} {\bibfnamefont
  {B.}~\bibnamefont {Brown}},\ }\href {\doibase 10.1103/PhysRevC.58.1970}
  {\bibfield  {journal} {\bibinfo  {journal} {Physical Review C}\ }\textbf
  {\bibinfo {volume} {58}},\ \bibinfo {pages} {1970} (\bibinfo {year}
  {1998})}\BibitemShut {NoStop}%
\bibitem [{\citenamefont {Behrens}\ and\ \citenamefont
  {J{\"{a}}necke}(1969)}]{Behrens1969}%
  \BibitemOpen
  \bibfield  {author} {\bibinfo {author} {\bibfnamefont {H.}~\bibnamefont
  {Behrens}}\ and\ \bibinfo {author} {\bibfnamefont {J.}~\bibnamefont
  {J{\"{a}}necke}},\ }\href@noop {} {\emph {\bibinfo {title}
  {{Landolt-B{\"{o}}rnstein Tables, Gruppe I, Band 4}}}}\ (\bibinfo
  {publisher} {Springer},\ \bibinfo {year} {1969})\BibitemShut {NoStop}%
\bibitem [{\citenamefont {Morita}\ and\ \citenamefont
  {Morita}(1958)}]{Morita1958}%
  \BibitemOpen
  \bibfield  {author} {\bibinfo {author} {\bibfnamefont {M.}~\bibnamefont
  {Morita}}\ and\ \bibinfo {author} {\bibfnamefont {R.~S.}\ \bibnamefont
  {Morita}},\ }\href {\doibase 10.1103/PhysRev.109.2048} {\bibfield  {journal}
  {\bibinfo  {journal} {Physical Review}\ }\textbf {\bibinfo {volume} {109}},\
  \bibinfo {pages} {2048} (\bibinfo {year} {1958})}\BibitemShut {NoStop}%
\bibitem [{\citenamefont {Schopper}(1966)}]{Schopper1966}%
  \BibitemOpen
  \bibfield  {author} {\bibinfo {author} {\bibfnamefont {H.~F.}\ \bibnamefont
  {Schopper}},\ }\href@noop {} {\emph {\bibinfo {title} {{Weak Interactions and
  Nuclear Beta Decay}}}}\ (\bibinfo  {publisher} {North-Holland Publishing
  Company},\ \bibinfo {year} {1966})\BibitemShut {NoStop}%
\bibitem [{\citenamefont {Kotani}(1959)}]{Kotani1959}%
  \BibitemOpen
  \bibfield  {author} {\bibinfo {author} {\bibfnamefont {T.}~\bibnamefont
  {Kotani}},\ }\href {\doibase 10.1103/PhysRev.114.795} {\bibfield  {journal}
  {\bibinfo  {journal} {Physical Review}\ }\textbf {\bibinfo {volume} {114}},\
  \bibinfo {pages} {795} (\bibinfo {year} {1959})}\BibitemShut {NoStop}%
\bibitem [{\citenamefont {B{\"{u}}hring}(1984)}]{Buhring1984}%
  \BibitemOpen
  \bibfield  {author} {\bibinfo {author} {\bibfnamefont {W.}~\bibnamefont
  {B{\"{u}}hring}},\ }\href {\doibase 10.1016/0375-9474(84)90190-8} {\bibfield
  {journal} {\bibinfo  {journal} {Nuclear Physics A}\ }\textbf {\bibinfo
  {volume} {430}},\ \bibinfo {pages} {1} (\bibinfo {year} {1984})}\BibitemShut
  {NoStop}%
\bibitem [{\citenamefont {Holstein}(1974)}]{Holstein1974}%
  \BibitemOpen
  \bibfield  {author} {\bibinfo {author} {\bibfnamefont {B.}~\bibnamefont
  {Holstein}},\ }\href
  {http://journals.aps.org/rmp/abstract/10.1103/RevModPhys.46.789} {\bibfield
  {journal} {\bibinfo  {journal} {Reviews of Modern Physics}\ }\textbf
  {\bibinfo {volume} {46}},\ \bibinfo {pages} {789} (\bibinfo {year}
  {1974})}\BibitemShut {NoStop}%
\bibitem [{\citenamefont {Wang}\ and\ \citenamefont {Hayes}(2017)}]{Wang2017}%
  \BibitemOpen
  \bibfield  {author} {\bibinfo {author} {\bibfnamefont {X.~B.}\ \bibnamefont
  {Wang}}\ and\ \bibinfo {author} {\bibfnamefont {A.~C.}\ \bibnamefont
  {Hayes}},\ }\href {\doibase 10.1103/PhysRevC.95.064313} {\bibfield  {journal}
  {\bibinfo  {journal} {Physical Review C}\ }\textbf {\bibinfo {volume} {95}},\
  \bibinfo {pages} {064313} (\bibinfo {year} {2017})},\ \Eprint
  {http://arxiv.org/abs/1702.07520} {arXiv:1702.07520} \BibitemShut {NoStop}%
\bibitem [{\citenamefont {Fallot}\ \emph {et~al.}(2012)\citenamefont {Fallot},
  \citenamefont {Cormon}, \citenamefont {Estienne}, \citenamefont {Algora},
  \citenamefont {Bui}, \citenamefont {Cucoanes}, \citenamefont {Elnimr},
  \citenamefont {Giot}, \citenamefont {Jordan}, \citenamefont {Martino},
  \citenamefont {Onillon}, \citenamefont {Porta}, \citenamefont {Pronost},
  \citenamefont {Remoto}, \citenamefont {Ta{\'{i}}n}, \citenamefont {Yermia},\
  and\ \citenamefont {Zakari-Issoufou}}]{Fallot2012}%
  \BibitemOpen
  \bibfield  {author} {\bibinfo {author} {\bibfnamefont {M.}~\bibnamefont
  {Fallot}}, \bibinfo {author} {\bibfnamefont {S.}~\bibnamefont {Cormon}},
  \bibinfo {author} {\bibfnamefont {M.}~\bibnamefont {Estienne}}, \bibinfo
  {author} {\bibfnamefont {A.}~\bibnamefont {Algora}}, \bibinfo {author}
  {\bibfnamefont {V.~M.}\ \bibnamefont {Bui}}, \bibinfo {author} {\bibfnamefont
  {A.}~\bibnamefont {Cucoanes}}, \bibinfo {author} {\bibfnamefont
  {M.}~\bibnamefont {Elnimr}}, \bibinfo {author} {\bibfnamefont
  {L.}~\bibnamefont {Giot}}, \bibinfo {author} {\bibfnamefont {D.}~\bibnamefont
  {Jordan}}, \bibinfo {author} {\bibfnamefont {J.}~\bibnamefont {Martino}},
  \bibinfo {author} {\bibfnamefont {A.}~\bibnamefont {Onillon}}, \bibinfo
  {author} {\bibfnamefont {A.}~\bibnamefont {Porta}}, \bibinfo {author}
  {\bibfnamefont {G.}~\bibnamefont {Pronost}}, \bibinfo {author} {\bibfnamefont
  {A.}~\bibnamefont {Remoto}}, \bibinfo {author} {\bibfnamefont {J.~L.}\
  \bibnamefont {Ta{\'{i}}n}}, \bibinfo {author} {\bibfnamefont
  {F.}~\bibnamefont {Yermia}}, \ and\ \bibinfo {author} {\bibfnamefont {A.-A.}\
  \bibnamefont {Zakari-Issoufou}},\ }\href {\doibase
  10.1103/PhysRevLett.109.202504} {\bibfield  {journal} {\bibinfo  {journal}
  {Physical Review Letters}\ }\textbf {\bibinfo {volume} {109}},\ \bibinfo
  {pages} {202504} (\bibinfo {year} {2012})},\ \Eprint
  {http://arxiv.org/abs/1208.3877} {arXiv:1208.3877} \BibitemShut {NoStop}%
\bibitem [{\citenamefont {Algora}\ \emph {et~al.}(2010)\citenamefont {Algora},
  \citenamefont {Jordan}, \citenamefont {Ta{\'{i}}n}, \citenamefont {Rubio},
  \citenamefont {Agramunt}, \citenamefont {Perez-Cerdan}, \citenamefont
  {Molina}, \citenamefont {Caballero}, \citenamefont {N{\'{a}}cher},
  \citenamefont {Krasznahorkay}, \citenamefont {Hunyadi}, \citenamefont
  {Guly{\'{a}}s}, \citenamefont {Vit{\'{e}}z}, \citenamefont {Csatl{\'{o}}s},
  \citenamefont {Csige}, \citenamefont {{\"{A}}ysto}, \citenamefont
  {Penttil{\"{a}}}, \citenamefont {Moore}, \citenamefont {Eronen},
  \citenamefont {Jokinen}, \citenamefont {Nieminen}, \citenamefont {Hakala},
  \citenamefont {Karvonen}, \citenamefont {Kankainen}, \citenamefont
  {Saastamoinen}, \citenamefont {Rissanen}, \citenamefont {Kessler},
  \citenamefont {Weber}, \citenamefont {Ronkainen}, \citenamefont {Rahaman},
  \citenamefont {Elomaa}, \citenamefont {Rinta-Antila}, \citenamefont {Hager},
  \citenamefont {Sonoda}, \citenamefont {Burkard}, \citenamefont
  {H{\"{u}}ller}, \citenamefont {Batist}, \citenamefont {Gelletly},
  \citenamefont {Nichols}, \citenamefont {Yoshida}, \citenamefont {Sonzogni},\
  and\ \citenamefont {Per{\"{a}}j{\"{a}}rvi}}]{Algora2010}%
  \BibitemOpen
  \bibfield  {author} {\bibinfo {author} {\bibfnamefont {A.}~\bibnamefont
  {Algora}}, \bibinfo {author} {\bibfnamefont {D.}~\bibnamefont {Jordan}},
  \bibinfo {author} {\bibfnamefont {J.~L.}\ \bibnamefont {Ta{\'{i}}n}},
  \bibinfo {author} {\bibfnamefont {B.}~\bibnamefont {Rubio}}, \bibinfo
  {author} {\bibfnamefont {J.}~\bibnamefont {Agramunt}}, \bibinfo {author}
  {\bibfnamefont {A.~B.}\ \bibnamefont {Perez-Cerdan}}, \bibinfo {author}
  {\bibfnamefont {F.}~\bibnamefont {Molina}}, \bibinfo {author} {\bibfnamefont
  {L.}~\bibnamefont {Caballero}}, \bibinfo {author} {\bibfnamefont
  {E.}~\bibnamefont {N{\'{a}}cher}}, \bibinfo {author} {\bibfnamefont
  {A.}~\bibnamefont {Krasznahorkay}}, \bibinfo {author} {\bibfnamefont {M.~D.}\
  \bibnamefont {Hunyadi}}, \bibinfo {author} {\bibfnamefont {J.}~\bibnamefont
  {Guly{\'{a}}s}}, \bibinfo {author} {\bibfnamefont {A.}~\bibnamefont
  {Vit{\'{e}}z}}, \bibinfo {author} {\bibfnamefont {M.}~\bibnamefont
  {Csatl{\'{o}}s}}, \bibinfo {author} {\bibfnamefont {L.}~\bibnamefont
  {Csige}}, \bibinfo {author} {\bibfnamefont {J.}~\bibnamefont {{\"{A}}ysto}},
  \bibinfo {author} {\bibfnamefont {H.}~\bibnamefont {Penttil{\"{a}}}},
  \bibinfo {author} {\bibfnamefont {I.~D.}\ \bibnamefont {Moore}}, \bibinfo
  {author} {\bibfnamefont {T.}~\bibnamefont {Eronen}}, \bibinfo {author}
  {\bibfnamefont {A.}~\bibnamefont {Jokinen}}, \bibinfo {author} {\bibfnamefont
  {A.}~\bibnamefont {Nieminen}}, \bibinfo {author} {\bibfnamefont
  {J.}~\bibnamefont {Hakala}}, \bibinfo {author} {\bibfnamefont
  {P.}~\bibnamefont {Karvonen}}, \bibinfo {author} {\bibfnamefont
  {A.}~\bibnamefont {Kankainen}}, \bibinfo {author} {\bibfnamefont
  {A.}~\bibnamefont {Saastamoinen}}, \bibinfo {author} {\bibfnamefont
  {J.}~\bibnamefont {Rissanen}}, \bibinfo {author} {\bibfnamefont
  {T.}~\bibnamefont {Kessler}}, \bibinfo {author} {\bibfnamefont
  {C.}~\bibnamefont {Weber}}, \bibinfo {author} {\bibfnamefont
  {J.}~\bibnamefont {Ronkainen}}, \bibinfo {author} {\bibfnamefont
  {S.}~\bibnamefont {Rahaman}}, \bibinfo {author} {\bibfnamefont
  {V.}~\bibnamefont {Elomaa}}, \bibinfo {author} {\bibfnamefont
  {S.}~\bibnamefont {Rinta-Antila}}, \bibinfo {author} {\bibfnamefont
  {U.}~\bibnamefont {Hager}}, \bibinfo {author} {\bibfnamefont
  {T.}~\bibnamefont {Sonoda}}, \bibinfo {author} {\bibfnamefont
  {K.}~\bibnamefont {Burkard}}, \bibinfo {author} {\bibfnamefont
  {W.}~\bibnamefont {H{\"{u}}ller}}, \bibinfo {author} {\bibfnamefont
  {L.}~\bibnamefont {Batist}}, \bibinfo {author} {\bibfnamefont
  {W.}~\bibnamefont {Gelletly}}, \bibinfo {author} {\bibfnamefont {A.~L.}\
  \bibnamefont {Nichols}}, \bibinfo {author} {\bibfnamefont {T.}~\bibnamefont
  {Yoshida}}, \bibinfo {author} {\bibfnamefont {A.~A.}\ \bibnamefont
  {Sonzogni}}, \ and\ \bibinfo {author} {\bibfnamefont {K.}~\bibnamefont
  {Per{\"{a}}j{\"{a}}rvi}},\ }\href {\doibase 10.1103/PhysRevLett.105.202501}
  {\bibfield  {journal} {\bibinfo  {journal} {Physical Review Letters}\
  }\textbf {\bibinfo {volume} {105}},\ \bibinfo {pages} {202501} (\bibinfo
  {year} {2010})}\BibitemShut {NoStop}%
\bibitem [{\citenamefont {Zakari-Issoufou}\ \emph {et~al.}(2015)\citenamefont
  {Zakari-Issoufou}, \citenamefont {Fallot}, \citenamefont {Porta},
  \citenamefont {Algora}, \citenamefont {Tain}, \citenamefont {Valencia},
  \citenamefont {Rice}, \citenamefont {Bui}, \citenamefont {Cormon},
  \citenamefont {Estienne}, \citenamefont {Agramunt}, \citenamefont
  {{\"{A}}yst{\"{o}}}, \citenamefont {Bowry}, \citenamefont {Briz},
  \citenamefont {Caballero-Folch}, \citenamefont {Cano-Ott}, \citenamefont
  {Cucoanes}, \citenamefont {Elomaa}, \citenamefont {Eronen}, \citenamefont
  {Est{\'{e}}vez}, \citenamefont {Farrelly}, \citenamefont {Garcia},
  \citenamefont {Gelletly}, \citenamefont {Gomez-Hornillos}, \citenamefont
  {Gorlychev}, \citenamefont {Hakala}, \citenamefont {Jokinen}, \citenamefont
  {Jordan}, \citenamefont {Kankainen}, \citenamefont {Karvonen}, \citenamefont
  {Kolhinen}, \citenamefont {Kondev}, \citenamefont {Martinez}, \citenamefont
  {Mendoza}, \citenamefont {Molina}, \citenamefont {Moore}, \citenamefont
  {Perez-Cerd{\'{a}}n}, \citenamefont {Podoly{\'{a}}k}, \citenamefont
  {Penttil{\"{a}}}, \citenamefont {Regan}, \citenamefont {Reponen},
  \citenamefont {Rissanen}, \citenamefont {Rubio}, \citenamefont {Shiba},
  \citenamefont {Sonzogni},\ and\ \citenamefont {Weber}}]{Zakari-Issoufou2015}%
  \BibitemOpen
  \bibfield  {author} {\bibinfo {author} {\bibfnamefont {A.-A.}\ \bibnamefont
  {Zakari-Issoufou}}, \bibinfo {author} {\bibfnamefont {M.}~\bibnamefont
  {Fallot}}, \bibinfo {author} {\bibfnamefont {A.}~\bibnamefont {Porta}},
  \bibinfo {author} {\bibfnamefont {A.}~\bibnamefont {Algora}}, \bibinfo
  {author} {\bibfnamefont {J.~L.}\ \bibnamefont {Tain}}, \bibinfo {author}
  {\bibfnamefont {E.}~\bibnamefont {Valencia}}, \bibinfo {author}
  {\bibfnamefont {S.}~\bibnamefont {Rice}}, \bibinfo {author} {\bibfnamefont
  {V.~M.}\ \bibnamefont {Bui}}, \bibinfo {author} {\bibfnamefont
  {S.}~\bibnamefont {Cormon}}, \bibinfo {author} {\bibfnamefont
  {M.}~\bibnamefont {Estienne}}, \bibinfo {author} {\bibfnamefont
  {J.}~\bibnamefont {Agramunt}}, \bibinfo {author} {\bibfnamefont
  {J.}~\bibnamefont {{\"{A}}yst{\"{o}}}}, \bibinfo {author} {\bibfnamefont
  {M.}~\bibnamefont {Bowry}}, \bibinfo {author} {\bibfnamefont {J.~A.}\
  \bibnamefont {Briz}}, \bibinfo {author} {\bibfnamefont {R.}~\bibnamefont
  {Caballero-Folch}}, \bibinfo {author} {\bibfnamefont {D.}~\bibnamefont
  {Cano-Ott}}, \bibinfo {author} {\bibfnamefont {A.}~\bibnamefont {Cucoanes}},
  \bibinfo {author} {\bibfnamefont {V.-V.}\ \bibnamefont {Elomaa}}, \bibinfo
  {author} {\bibfnamefont {T.}~\bibnamefont {Eronen}}, \bibinfo {author}
  {\bibfnamefont {E.}~\bibnamefont {Est{\'{e}}vez}}, \bibinfo {author}
  {\bibfnamefont {G.~F.}\ \bibnamefont {Farrelly}}, \bibinfo {author}
  {\bibfnamefont {A.~R.}\ \bibnamefont {Garcia}}, \bibinfo {author}
  {\bibfnamefont {W.}~\bibnamefont {Gelletly}}, \bibinfo {author}
  {\bibfnamefont {M.~B.}\ \bibnamefont {Gomez-Hornillos}}, \bibinfo {author}
  {\bibfnamefont {V.}~\bibnamefont {Gorlychev}}, \bibinfo {author}
  {\bibfnamefont {J.}~\bibnamefont {Hakala}}, \bibinfo {author} {\bibfnamefont
  {A.}~\bibnamefont {Jokinen}}, \bibinfo {author} {\bibfnamefont {M.~D.}\
  \bibnamefont {Jordan}}, \bibinfo {author} {\bibfnamefont {A.}~\bibnamefont
  {Kankainen}}, \bibinfo {author} {\bibfnamefont {P.}~\bibnamefont {Karvonen}},
  \bibinfo {author} {\bibfnamefont {V.~S.}\ \bibnamefont {Kolhinen}}, \bibinfo
  {author} {\bibfnamefont {F.~G.}\ \bibnamefont {Kondev}}, \bibinfo {author}
  {\bibfnamefont {T.}~\bibnamefont {Martinez}}, \bibinfo {author}
  {\bibfnamefont {E.}~\bibnamefont {Mendoza}}, \bibinfo {author} {\bibfnamefont
  {F.}~\bibnamefont {Molina}}, \bibinfo {author} {\bibfnamefont
  {I.}~\bibnamefont {Moore}}, \bibinfo {author} {\bibfnamefont {A.~B.}\
  \bibnamefont {Perez-Cerd{\'{a}}n}}, \bibinfo {author} {\bibfnamefont
  {Z.}~\bibnamefont {Podoly{\'{a}}k}}, \bibinfo {author} {\bibfnamefont
  {H.}~\bibnamefont {Penttil{\"{a}}}}, \bibinfo {author} {\bibfnamefont
  {P.~H.}\ \bibnamefont {Regan}}, \bibinfo {author} {\bibfnamefont
  {M.}~\bibnamefont {Reponen}}, \bibinfo {author} {\bibfnamefont
  {J.}~\bibnamefont {Rissanen}}, \bibinfo {author} {\bibfnamefont
  {B.}~\bibnamefont {Rubio}}, \bibinfo {author} {\bibfnamefont
  {T.}~\bibnamefont {Shiba}}, \bibinfo {author} {\bibfnamefont {A.~A.}\
  \bibnamefont {Sonzogni}}, \ and\ \bibinfo {author} {\bibfnamefont
  {C.}~\bibnamefont {Weber}},\ }\href {\doibase 10.1103/PhysRevLett.115.102503}
  {\bibfield  {journal} {\bibinfo  {journal} {Physical Review Letters}\
  }\textbf {\bibinfo {volume} {115}},\ \bibinfo {pages} {102503} (\bibinfo
  {year} {2015})},\ \Eprint {http://arxiv.org/abs/1504.05812}
  {arXiv:1504.05812} \BibitemShut {NoStop}%
\bibitem [{\citenamefont {Rice}\ \emph {et~al.}(2017)\citenamefont {Rice},
  \citenamefont {Algora}, \citenamefont {Tain}, \citenamefont {Valencia},
  \citenamefont {Agramunt}, \citenamefont {Rubio}, \citenamefont {Gelletly},
  \citenamefont {Regan}, \citenamefont {Zakari-Issoufou}, \citenamefont
  {Fallot}, \citenamefont {Porta}, \citenamefont {Rissanen}, \citenamefont
  {Eronen}, \citenamefont {{\"{A}}yst{\"{o}}}, \citenamefont {Batist},
  \citenamefont {Bowry}, \citenamefont {Bui}, \citenamefont {Caballero-Folch},
  \citenamefont {Cano-Ott}, \citenamefont {Elomaa}, \citenamefont {Estevez},
  \citenamefont {Farrelly}, \citenamefont {Garcia}, \citenamefont
  {Gomez-Hornillos}, \citenamefont {Gorlychev}, \citenamefont {Hakala},
  \citenamefont {Jordan}, \citenamefont {Jokinen}, \citenamefont {Kolhinen},
  \citenamefont {Kondev}, \citenamefont {Mart{\'{i}}nez}, \citenamefont
  {Mason}, \citenamefont {Mendoza}, \citenamefont {Moore}, \citenamefont
  {Penttil{\"{a}}}, \citenamefont {Podoly{\'{a}}k}, \citenamefont {Reponen},
  \citenamefont {Sonnenschein}, \citenamefont {Sonzogni},\ and\ \citenamefont
  {Sarriguren}}]{Rice2017}%
  \BibitemOpen
  \bibfield  {author} {\bibinfo {author} {\bibfnamefont {S.}~\bibnamefont
  {Rice}}, \bibinfo {author} {\bibfnamefont {A.}~\bibnamefont {Algora}},
  \bibinfo {author} {\bibfnamefont {J.~L.}\ \bibnamefont {Tain}}, \bibinfo
  {author} {\bibfnamefont {E.}~\bibnamefont {Valencia}}, \bibinfo {author}
  {\bibfnamefont {J.}~\bibnamefont {Agramunt}}, \bibinfo {author}
  {\bibfnamefont {B.}~\bibnamefont {Rubio}}, \bibinfo {author} {\bibfnamefont
  {W.}~\bibnamefont {Gelletly}}, \bibinfo {author} {\bibfnamefont {P.~H.}\
  \bibnamefont {Regan}}, \bibinfo {author} {\bibfnamefont {A.-A.}\ \bibnamefont
  {Zakari-Issoufou}}, \bibinfo {author} {\bibfnamefont {M.}~\bibnamefont
  {Fallot}}, \bibinfo {author} {\bibfnamefont {A.}~\bibnamefont {Porta}},
  \bibinfo {author} {\bibfnamefont {J.}~\bibnamefont {Rissanen}}, \bibinfo
  {author} {\bibfnamefont {T.}~\bibnamefont {Eronen}}, \bibinfo {author}
  {\bibfnamefont {J.}~\bibnamefont {{\"{A}}yst{\"{o}}}}, \bibinfo {author}
  {\bibfnamefont {L.}~\bibnamefont {Batist}}, \bibinfo {author} {\bibfnamefont
  {M.}~\bibnamefont {Bowry}}, \bibinfo {author} {\bibfnamefont {V.~M.}\
  \bibnamefont {Bui}}, \bibinfo {author} {\bibfnamefont {R.}~\bibnamefont
  {Caballero-Folch}}, \bibinfo {author} {\bibfnamefont {D.}~\bibnamefont
  {Cano-Ott}}, \bibinfo {author} {\bibfnamefont {V.-V.}\ \bibnamefont
  {Elomaa}}, \bibinfo {author} {\bibfnamefont {E.}~\bibnamefont {Estevez}},
  \bibinfo {author} {\bibfnamefont {G.~F.}\ \bibnamefont {Farrelly}}, \bibinfo
  {author} {\bibfnamefont {A.~R.}\ \bibnamefont {Garcia}}, \bibinfo {author}
  {\bibfnamefont {B.}~\bibnamefont {Gomez-Hornillos}}, \bibinfo {author}
  {\bibfnamefont {V.}~\bibnamefont {Gorlychev}}, \bibinfo {author}
  {\bibfnamefont {J.}~\bibnamefont {Hakala}}, \bibinfo {author} {\bibfnamefont
  {M.~D.}\ \bibnamefont {Jordan}}, \bibinfo {author} {\bibfnamefont
  {A.}~\bibnamefont {Jokinen}}, \bibinfo {author} {\bibfnamefont {V.~S.}\
  \bibnamefont {Kolhinen}}, \bibinfo {author} {\bibfnamefont {F.~G.}\
  \bibnamefont {Kondev}}, \bibinfo {author} {\bibfnamefont {T.}~\bibnamefont
  {Mart{\'{i}}nez}}, \bibinfo {author} {\bibfnamefont {P.}~\bibnamefont
  {Mason}}, \bibinfo {author} {\bibfnamefont {E.}~\bibnamefont {Mendoza}},
  \bibinfo {author} {\bibfnamefont {I.}~\bibnamefont {Moore}}, \bibinfo
  {author} {\bibfnamefont {H.}~\bibnamefont {Penttil{\"{a}}}}, \bibinfo
  {author} {\bibfnamefont {Z.}~\bibnamefont {Podoly{\'{a}}k}}, \bibinfo
  {author} {\bibfnamefont {M.}~\bibnamefont {Reponen}}, \bibinfo {author}
  {\bibfnamefont {V.}~\bibnamefont {Sonnenschein}}, \bibinfo {author}
  {\bibfnamefont {A.~A.}\ \bibnamefont {Sonzogni}}, \ and\ \bibinfo {author}
  {\bibfnamefont {P.}~\bibnamefont {Sarriguren}},\ }\href {\doibase
  10.1103/PhysRevC.96.014320} {\bibfield  {journal} {\bibinfo  {journal}
  {Physical Review C}\ }\textbf {\bibinfo {volume} {96}},\ \bibinfo {pages}
  {014320} (\bibinfo {year} {2017})}\BibitemShut {NoStop}%
\bibitem [{\citenamefont {Hardy}\ \emph {et~al.}(1977)\citenamefont {Hardy},
  \citenamefont {Carraz}, \citenamefont {Jonson},\ and\ \citenamefont
  {Hansen}}]{Hardy1977}%
  \BibitemOpen
  \bibfield  {author} {\bibinfo {author} {\bibfnamefont {J.~C.}\ \bibnamefont
  {Hardy}}, \bibinfo {author} {\bibfnamefont {L.~C.}\ \bibnamefont {Carraz}},
  \bibinfo {author} {\bibfnamefont {B.}~\bibnamefont {Jonson}}, \ and\ \bibinfo
  {author} {\bibfnamefont {P.~G.}\ \bibnamefont {Hansen}},\ }\href {\doibase
  10.1016/0370-2693(77)90223-4} {\bibfield  {journal} {\bibinfo  {journal}
  {Physics Letters B}\ }\textbf {\bibinfo {volume} {71}},\ \bibinfo {pages}
  {307} (\bibinfo {year} {1977})}\BibitemShut {NoStop}%
\bibitem [{\citenamefont {Navratil}\ \emph {et~al.}(2018)\citenamefont
  {Navratil}, \citenamefont {Dunn}, \citenamefont {Sin}, \citenamefont
  {Sirakov}, \citenamefont {Nobre}, \citenamefont {Wright}, \citenamefont
  {Sleaford}, \citenamefont {Bates}, \citenamefont {Conlin}, \citenamefont
  {Koning}, \citenamefont {Pronyaev}, \citenamefont {Brown}, \citenamefont
  {Lestone}, \citenamefont {Neudecker}, \citenamefont {Herman}, \citenamefont
  {Kopecky}, \citenamefont {Gaines}, \citenamefont {Romano}, \citenamefont
  {Soukhovitskii}, \citenamefont {Sonzogni}, \citenamefont {Zerkle},
  \citenamefont {Welser-Sherrill}, \citenamefont {Becker}, \citenamefont
  {Pigni}, \citenamefont {{\v{Z}}erovnik}, \citenamefont {Noguere},
  \citenamefont {Arcilla}, \citenamefont {Rochman}, \citenamefont {Guber},
  \citenamefont {Sobes}, \citenamefont {Hawari}, \citenamefont {Kiedrowski},
  \citenamefont {Talou}, \citenamefont {Simakov}, \citenamefont {Roubtsov},
  \citenamefont {Mattoon}, \citenamefont {Thompson}, \citenamefont {Plompen},
  \citenamefont {Lubitz}, \citenamefont {Hale}, \citenamefont {Arbanas},
  \citenamefont {Brown}, \citenamefont {van~der Marck}, \citenamefont
  {Chadwick}, \citenamefont {McCutchan}, \citenamefont {Pritychenko},
  \citenamefont {Carlson}, \citenamefont {Smith}, \citenamefont
  {Schillebeeckx}, \citenamefont {White}, \citenamefont {Johnson},
  \citenamefont {Beck}, \citenamefont {Zhu}, \citenamefont {Descalle},
  \citenamefont {Wiarda}, \citenamefont {Kawano}, \citenamefont {Casperson},
  \citenamefont {Trkov}, \citenamefont {Firestone}, \citenamefont {Danon},
  \citenamefont {Capote}, \citenamefont {Mughabghab}, \citenamefont {Holmes},
  \citenamefont {{M{\'{a}}rquez Dami{\'{a}}n}}, \citenamefont {Cullen},
  \citenamefont {Wormald}, \citenamefont {Paris}, \citenamefont {Stetcu},
  \citenamefont {Kahler},\ and\ \citenamefont {Leal}}]{Navratil2018}%
  \BibitemOpen
  \bibfield  {author} {\bibinfo {author} {\bibfnamefont {P.}~\bibnamefont
  {Navratil}}, \bibinfo {author} {\bibfnamefont {M.}~\bibnamefont {Dunn}},
  \bibinfo {author} {\bibfnamefont {M.}~\bibnamefont {Sin}}, \bibinfo {author}
  {\bibfnamefont {I.}~\bibnamefont {Sirakov}}, \bibinfo {author} {\bibfnamefont
  {G.}~\bibnamefont {Nobre}}, \bibinfo {author} {\bibfnamefont
  {R.}~\bibnamefont {Wright}}, \bibinfo {author} {\bibfnamefont
  {B.}~\bibnamefont {Sleaford}}, \bibinfo {author} {\bibfnamefont
  {C.}~\bibnamefont {Bates}}, \bibinfo {author} {\bibfnamefont
  {J.}~\bibnamefont {Conlin}}, \bibinfo {author} {\bibfnamefont
  {A.}~\bibnamefont {Koning}}, \bibinfo {author} {\bibfnamefont
  {V.}~\bibnamefont {Pronyaev}}, \bibinfo {author} {\bibfnamefont
  {F.}~\bibnamefont {Brown}}, \bibinfo {author} {\bibfnamefont
  {J.}~\bibnamefont {Lestone}}, \bibinfo {author} {\bibfnamefont
  {D.}~\bibnamefont {Neudecker}}, \bibinfo {author} {\bibfnamefont
  {M.}~\bibnamefont {Herman}}, \bibinfo {author} {\bibfnamefont
  {S.}~\bibnamefont {Kopecky}}, \bibinfo {author} {\bibfnamefont
  {T.}~\bibnamefont {Gaines}}, \bibinfo {author} {\bibfnamefont
  {P.}~\bibnamefont {Romano}}, \bibinfo {author} {\bibfnamefont
  {E.}~\bibnamefont {Soukhovitskii}}, \bibinfo {author} {\bibfnamefont
  {A.}~\bibnamefont {Sonzogni}}, \bibinfo {author} {\bibfnamefont
  {M.}~\bibnamefont {Zerkle}}, \bibinfo {author} {\bibfnamefont
  {L.}~\bibnamefont {Welser-Sherrill}}, \bibinfo {author} {\bibfnamefont
  {B.}~\bibnamefont {Becker}}, \bibinfo {author} {\bibfnamefont
  {M.}~\bibnamefont {Pigni}}, \bibinfo {author} {\bibfnamefont
  {G.}~\bibnamefont {{\v{Z}}erovnik}}, \bibinfo {author} {\bibfnamefont
  {G.}~\bibnamefont {Noguere}}, \bibinfo {author} {\bibfnamefont
  {R.}~\bibnamefont {Arcilla}}, \bibinfo {author} {\bibfnamefont
  {D.}~\bibnamefont {Rochman}}, \bibinfo {author} {\bibfnamefont
  {K.}~\bibnamefont {Guber}}, \bibinfo {author} {\bibfnamefont
  {V.}~\bibnamefont {Sobes}}, \bibinfo {author} {\bibfnamefont
  {A.}~\bibnamefont {Hawari}}, \bibinfo {author} {\bibfnamefont
  {B.}~\bibnamefont {Kiedrowski}}, \bibinfo {author} {\bibfnamefont
  {P.}~\bibnamefont {Talou}}, \bibinfo {author} {\bibfnamefont
  {S.}~\bibnamefont {Simakov}}, \bibinfo {author} {\bibfnamefont
  {D.}~\bibnamefont {Roubtsov}}, \bibinfo {author} {\bibfnamefont
  {C.}~\bibnamefont {Mattoon}}, \bibinfo {author} {\bibfnamefont
  {I.}~\bibnamefont {Thompson}}, \bibinfo {author} {\bibfnamefont
  {A.}~\bibnamefont {Plompen}}, \bibinfo {author} {\bibfnamefont
  {C.}~\bibnamefont {Lubitz}}, \bibinfo {author} {\bibfnamefont
  {G.}~\bibnamefont {Hale}}, \bibinfo {author} {\bibfnamefont {G.}~\bibnamefont
  {Arbanas}}, \bibinfo {author} {\bibfnamefont {D.}~\bibnamefont {Brown}},
  \bibinfo {author} {\bibfnamefont {S.}~\bibnamefont {van~der Marck}}, \bibinfo
  {author} {\bibfnamefont {M.}~\bibnamefont {Chadwick}}, \bibinfo {author}
  {\bibfnamefont {E.}~\bibnamefont {McCutchan}}, \bibinfo {author}
  {\bibfnamefont {B.}~\bibnamefont {Pritychenko}}, \bibinfo {author}
  {\bibfnamefont {A.}~\bibnamefont {Carlson}}, \bibinfo {author} {\bibfnamefont
  {D.}~\bibnamefont {Smith}}, \bibinfo {author} {\bibfnamefont
  {P.}~\bibnamefont {Schillebeeckx}}, \bibinfo {author} {\bibfnamefont
  {M.}~\bibnamefont {White}}, \bibinfo {author} {\bibfnamefont
  {T.}~\bibnamefont {Johnson}}, \bibinfo {author} {\bibfnamefont
  {B.}~\bibnamefont {Beck}}, \bibinfo {author} {\bibfnamefont {Y.}~\bibnamefont
  {Zhu}}, \bibinfo {author} {\bibfnamefont {M.-A.}\ \bibnamefont {Descalle}},
  \bibinfo {author} {\bibfnamefont {D.}~\bibnamefont {Wiarda}}, \bibinfo
  {author} {\bibfnamefont {T.}~\bibnamefont {Kawano}}, \bibinfo {author}
  {\bibfnamefont {R.}~\bibnamefont {Casperson}}, \bibinfo {author}
  {\bibfnamefont {A.}~\bibnamefont {Trkov}}, \bibinfo {author} {\bibfnamefont
  {R.}~\bibnamefont {Firestone}}, \bibinfo {author} {\bibfnamefont
  {Y.}~\bibnamefont {Danon}}, \bibinfo {author} {\bibfnamefont
  {R.}~\bibnamefont {Capote}}, \bibinfo {author} {\bibfnamefont
  {S.}~\bibnamefont {Mughabghab}}, \bibinfo {author} {\bibfnamefont
  {J.}~\bibnamefont {Holmes}}, \bibinfo {author} {\bibfnamefont
  {J.}~\bibnamefont {{M{\'{a}}rquez Dami{\'{a}}n}}}, \bibinfo {author}
  {\bibfnamefont {D.}~\bibnamefont {Cullen}}, \bibinfo {author} {\bibfnamefont
  {J.}~\bibnamefont {Wormald}}, \bibinfo {author} {\bibfnamefont
  {M.}~\bibnamefont {Paris}}, \bibinfo {author} {\bibfnamefont
  {I.}~\bibnamefont {Stetcu}}, \bibinfo {author} {\bibfnamefont
  {A.}~\bibnamefont {Kahler}}, \ and\ \bibinfo {author} {\bibfnamefont
  {L.}~\bibnamefont {Leal}},\ }\href {\doibase 10.1016/j.nds.2018.02.001}
  {\bibfield  {journal} {\bibinfo  {journal} {Nuclear Data Sheets}\ }\textbf
  {\bibinfo {volume} {148}},\ \bibinfo {pages} {1} (\bibinfo {year}
  {2018})}\BibitemShut {NoStop}%
\bibitem [{\citenamefont {{Van der Marck}}(2012)}]{VanderMarck2012}%
  \BibitemOpen
  \bibfield  {author} {\bibinfo {author} {\bibfnamefont {S.~C.}\ \bibnamefont
  {{Van der Marck}}},\ }\href {\doibase 10.1016/j.nds.2012.11.003} {\bibfield
  {journal} {\bibinfo  {journal} {Nuclear Data Sheets}\ }\textbf {\bibinfo
  {volume} {113}},\ \bibinfo {pages} {2935} (\bibinfo {year}
  {2012})}\BibitemShut {NoStop}%
\bibitem [{\citenamefont {Kellett}\ \emph {et~al.}(2009)\citenamefont
  {Kellett}, \citenamefont {Bersillon},\ and\ \citenamefont
  {Mills}}]{Kellett2009}%
  \BibitemOpen
  \bibfield  {author} {\bibinfo {author} {\bibfnamefont {M.}~\bibnamefont
  {Kellett}}, \bibinfo {author} {\bibfnamefont {O.}~\bibnamefont {Bersillon}},
  \ and\ \bibinfo {author} {\bibfnamefont {R.}~\bibnamefont {Mills}},\ }\href
  {https://www.oecd-nea.org/dbdata/nds{\_}jefreports/jefreport-20/nea6287-jeff-20.pdf}
  {\emph {\bibinfo {title} {{The JEFF-3.1/-3.1.1 RadioActive Decay Data and
  Fission Yields Sub-Libraries}}}},\ \bibinfo {type} {Tech. Rep.}\ \bibinfo
  {number} {JEFF Report 20}\ (\bibinfo {year} {2009})\BibitemShut {NoStop}%
\bibitem [{\citenamefont {Fallot}\ and\ \citenamefont {Sonzogni}()}]{FallotPC}%
  \BibitemOpen
  \bibfield  {author} {\bibinfo {author} {\bibfnamefont {M.}~\bibnamefont
  {Fallot}}\ and\ \bibinfo {author} {\bibfnamefont {A.}~\bibnamefont
  {Sonzogni}},\ }\href@noop {} {\enquote {\bibinfo {title} {{Private
  Communication}},}\ }\BibitemShut {NoStop}%
\bibitem [{\citenamefont {Herman}\ and\ \citenamefont
  {Trkov}(2010)}]{Herman2010}%
  \BibitemOpen
  \bibfield  {author} {\bibinfo {author} {\bibfnamefont {M.}~\bibnamefont
  {Herman}}\ and\ \bibinfo {author} {\bibfnamefont {A.}~\bibnamefont {Trkov}},\
  }\href
  {http://scholar.google.com/scholar?hl=en{\&}btnG=Search{\&}q=intitle:ENDF-6+Formats+Manual.+Data+Formats+and+Procedures+for+the+Evaluated+Nuclear+Data+Files+ENDF/B-VI+and+ENDF/B-VII{\#}0}
  {\emph {\bibinfo {title} {Brookhaven National Laboratory}}},\ \bibinfo {type}
  {Tech. Rep.}\ (\bibinfo {year} {2010})\BibitemShut {NoStop}%
\bibitem [{\citenamefont {Chadwick}\ \emph {et~al.}(2011)\citenamefont
  {Chadwick}, \citenamefont {Herman}, \citenamefont {Oblo{\v{z}}insk{\'{y}}},
  \citenamefont {Dunn}, \citenamefont {Danon}, \citenamefont {Kahler},
  \citenamefont {Smith}, \citenamefont {Pritychenko}, \citenamefont {Arbanas},
  \citenamefont {Arcilla}, \citenamefont {Brewer}, \citenamefont {Brown},
  \citenamefont {Capote}, \citenamefont {Carlson}, \citenamefont {Cho},
  \citenamefont {Derrien}, \citenamefont {Guber}, \citenamefont {Hale},
  \citenamefont {Hoblit}, \citenamefont {Holloway}, \citenamefont {Johnson},
  \citenamefont {Kawano}, \citenamefont {Kiedrowski}, \citenamefont {Kim},
  \citenamefont {Kunieda}, \citenamefont {Larson}, \citenamefont {Leal},
  \citenamefont {Lestone}, \citenamefont {Little}, \citenamefont {McCutchan},
  \citenamefont {MacFarlane}, \citenamefont {MacInnes}, \citenamefont
  {Mattoon}, \citenamefont {McKnight}, \citenamefont {Mughabghab},
  \citenamefont {Nobre}, \citenamefont {Palmiotti}, \citenamefont {Palumbo},
  \citenamefont {Pigni}, \citenamefont {Pronyaev}, \citenamefont {Sayer},
  \citenamefont {Sonzogni}, \citenamefont {Summers}, \citenamefont {Talou},
  \citenamefont {Thompson}, \citenamefont {Trkov}, \citenamefont {Vogt},
  \citenamefont {van~der Marck}, \citenamefont {Wallner}, \citenamefont
  {White}, \citenamefont {Wiarda},\ and\ \citenamefont {Young}}]{Chadwick2011}%
  \BibitemOpen
  \bibfield  {author} {\bibinfo {author} {\bibfnamefont {M.}~\bibnamefont
  {Chadwick}}, \bibinfo {author} {\bibfnamefont {M.}~\bibnamefont {Herman}},
  \bibinfo {author} {\bibfnamefont {P.}~\bibnamefont {Oblo{\v{z}}insk{\'{y}}}},
  \bibinfo {author} {\bibfnamefont {M.}~\bibnamefont {Dunn}}, \bibinfo {author}
  {\bibfnamefont {Y.}~\bibnamefont {Danon}}, \bibinfo {author} {\bibfnamefont
  {A.}~\bibnamefont {Kahler}}, \bibinfo {author} {\bibfnamefont
  {D.}~\bibnamefont {Smith}}, \bibinfo {author} {\bibfnamefont
  {B.}~\bibnamefont {Pritychenko}}, \bibinfo {author} {\bibfnamefont
  {G.}~\bibnamefont {Arbanas}}, \bibinfo {author} {\bibfnamefont
  {R.}~\bibnamefont {Arcilla}}, \bibinfo {author} {\bibfnamefont
  {R.}~\bibnamefont {Brewer}}, \bibinfo {author} {\bibfnamefont
  {D.}~\bibnamefont {Brown}}, \bibinfo {author} {\bibfnamefont
  {R.}~\bibnamefont {Capote}}, \bibinfo {author} {\bibfnamefont
  {A.}~\bibnamefont {Carlson}}, \bibinfo {author} {\bibfnamefont
  {Y.}~\bibnamefont {Cho}}, \bibinfo {author} {\bibfnamefont {H.}~\bibnamefont
  {Derrien}}, \bibinfo {author} {\bibfnamefont {K.}~\bibnamefont {Guber}},
  \bibinfo {author} {\bibfnamefont {G.}~\bibnamefont {Hale}}, \bibinfo {author}
  {\bibfnamefont {S.}~\bibnamefont {Hoblit}}, \bibinfo {author} {\bibfnamefont
  {S.}~\bibnamefont {Holloway}}, \bibinfo {author} {\bibfnamefont
  {T.}~\bibnamefont {Johnson}}, \bibinfo {author} {\bibfnamefont
  {T.}~\bibnamefont {Kawano}}, \bibinfo {author} {\bibfnamefont
  {B.}~\bibnamefont {Kiedrowski}}, \bibinfo {author} {\bibfnamefont
  {H.}~\bibnamefont {Kim}}, \bibinfo {author} {\bibfnamefont {S.}~\bibnamefont
  {Kunieda}}, \bibinfo {author} {\bibfnamefont {N.}~\bibnamefont {Larson}},
  \bibinfo {author} {\bibfnamefont {L.}~\bibnamefont {Leal}}, \bibinfo {author}
  {\bibfnamefont {J.}~\bibnamefont {Lestone}}, \bibinfo {author} {\bibfnamefont
  {R.}~\bibnamefont {Little}}, \bibinfo {author} {\bibfnamefont
  {E.}~\bibnamefont {McCutchan}}, \bibinfo {author} {\bibfnamefont
  {R.}~\bibnamefont {MacFarlane}}, \bibinfo {author} {\bibfnamefont
  {M.}~\bibnamefont {MacInnes}}, \bibinfo {author} {\bibfnamefont
  {C.}~\bibnamefont {Mattoon}}, \bibinfo {author} {\bibfnamefont
  {R.}~\bibnamefont {McKnight}}, \bibinfo {author} {\bibfnamefont
  {S.}~\bibnamefont {Mughabghab}}, \bibinfo {author} {\bibfnamefont
  {G.}~\bibnamefont {Nobre}}, \bibinfo {author} {\bibfnamefont
  {G.}~\bibnamefont {Palmiotti}}, \bibinfo {author} {\bibfnamefont
  {A.}~\bibnamefont {Palumbo}}, \bibinfo {author} {\bibfnamefont
  {M.}~\bibnamefont {Pigni}}, \bibinfo {author} {\bibfnamefont
  {V.}~\bibnamefont {Pronyaev}}, \bibinfo {author} {\bibfnamefont
  {R.}~\bibnamefont {Sayer}}, \bibinfo {author} {\bibfnamefont
  {A.}~\bibnamefont {Sonzogni}}, \bibinfo {author} {\bibfnamefont
  {N.}~\bibnamefont {Summers}}, \bibinfo {author} {\bibfnamefont
  {P.}~\bibnamefont {Talou}}, \bibinfo {author} {\bibfnamefont
  {I.}~\bibnamefont {Thompson}}, \bibinfo {author} {\bibfnamefont
  {A.}~\bibnamefont {Trkov}}, \bibinfo {author} {\bibfnamefont
  {R.}~\bibnamefont {Vogt}}, \bibinfo {author} {\bibfnamefont {S.}~\bibnamefont
  {van~der Marck}}, \bibinfo {author} {\bibfnamefont {A.}~\bibnamefont
  {Wallner}}, \bibinfo {author} {\bibfnamefont {M.}~\bibnamefont {White}},
  \bibinfo {author} {\bibfnamefont {D.}~\bibnamefont {Wiarda}}, \ and\ \bibinfo
  {author} {\bibfnamefont {P.}~\bibnamefont {Young}},\ }\href {\doibase
  10.1016/j.nds.2011.11.002} {\bibfield  {journal} {\bibinfo  {journal}
  {Nuclear Data Sheets}\ }\textbf {\bibinfo {volume} {112}},\ \bibinfo {pages}
  {2887} (\bibinfo {year} {2011})}\BibitemShut {NoStop}%
\bibitem [{\citenamefont {Kawano}\ \emph {et~al.}(2008)\citenamefont {Kawano},
  \citenamefont {M{\"{o}}ller},\ and\ \citenamefont {Wilson}}]{Kawano2008}%
  \BibitemOpen
  \bibfield  {author} {\bibinfo {author} {\bibfnamefont {T.}~\bibnamefont
  {Kawano}}, \bibinfo {author} {\bibfnamefont {P.}~\bibnamefont
  {M{\"{o}}ller}}, \ and\ \bibinfo {author} {\bibfnamefont {W.~B.}\
  \bibnamefont {Wilson}},\ }\href {\doibase 10.1103/PhysRevC.78.054601}
  {\bibfield  {journal} {\bibinfo  {journal} {Physical Review C}\ }\textbf
  {\bibinfo {volume} {78}},\ \bibinfo {pages} {054601} (\bibinfo {year}
  {2008})}\BibitemShut {NoStop}%
\bibitem [{\citenamefont {Sonzogni}\ \emph {et~al.}(2015)\citenamefont
  {Sonzogni}, \citenamefont {Johnson},\ and\ \citenamefont
  {McCutchan}}]{Sonzogni2015}%
  \BibitemOpen
  \bibfield  {author} {\bibinfo {author} {\bibfnamefont {A.~A.}\ \bibnamefont
  {Sonzogni}}, \bibinfo {author} {\bibfnamefont {T.~D.}\ \bibnamefont
  {Johnson}}, \ and\ \bibinfo {author} {\bibfnamefont {E.~A.}\ \bibnamefont
  {McCutchan}},\ }\href {\doibase 10.1103/PhysRevC.91.011301} {\bibfield
  {journal} {\bibinfo  {journal} {Physical Review C}\ }\textbf {\bibinfo
  {volume} {91}},\ \bibinfo {pages} {011301} (\bibinfo {year}
  {2015})}\BibitemShut {NoStop}%
\bibitem [{\citenamefont {Hayen}\ and\ \citenamefont
  {Severijns}(2019)}]{Hayen2019a}%
  \BibitemOpen
  \bibfield  {author} {\bibinfo {author} {\bibfnamefont {L.}~\bibnamefont
  {Hayen}}\ and\ \bibinfo {author} {\bibfnamefont {N.}~\bibnamefont
  {Severijns}},\ }\href {\doibase 10.1016/j.cpc.2019.02.012} {\bibfield
  {journal} {\bibinfo  {journal} {Computer Physics Communications}\ }\textbf
  {\bibinfo {volume} {240}},\ \bibinfo {pages} {152} (\bibinfo {year}
  {2019})},\ \Eprint {http://arxiv.org/abs/1803.00525} {arXiv:1803.00525}
  \BibitemShut {NoStop}%
\bibitem [{\citenamefont {Haag}\ \emph {et~al.}(2014)\citenamefont {Haag},
  \citenamefont {Gelletly}, \citenamefont {von Feilitzsch}, \citenamefont
  {Oberauer}, \citenamefont {Potzel}, \citenamefont {Schreckenbach},\ and\
  \citenamefont {Sonzogni}}]{Haag2014}%
  \BibitemOpen
  \bibfield  {author} {\bibinfo {author} {\bibfnamefont {N.}~\bibnamefont
  {Haag}}, \bibinfo {author} {\bibfnamefont {W.}~\bibnamefont {Gelletly}},
  \bibinfo {author} {\bibfnamefont {F.}~\bibnamefont {von Feilitzsch}},
  \bibinfo {author} {\bibfnamefont {L.}~\bibnamefont {Oberauer}}, \bibinfo
  {author} {\bibfnamefont {W.}~\bibnamefont {Potzel}}, \bibinfo {author}
  {\bibfnamefont {K.}~\bibnamefont {Schreckenbach}}, \ and\ \bibinfo {author}
  {\bibfnamefont {A.~A.}\ \bibnamefont {Sonzogni}},\ }\href
  {http://arxiv.org/abs/1405.3501} {\  (\bibinfo {year} {2014})},\ \Eprint
  {http://arxiv.org/abs/1405.3501} {arXiv:1405.3501} \BibitemShut {NoStop}%
\bibitem [{\citenamefont {Brown}\ and\ \citenamefont {Rae}(2014)}]{Brown2014}%
  \BibitemOpen
  \bibfield  {author} {\bibinfo {author} {\bibfnamefont {B.~A.}\ \bibnamefont
  {Brown}}\ and\ \bibinfo {author} {\bibfnamefont {W.~D.~M.}\ \bibnamefont
  {Rae}},\ }\href {\doibase 10.1016/j.nds.2014.07.022} {\bibfield  {journal}
  {\bibinfo  {journal} {Nuclear Data Sheets}\ }\textbf {\bibinfo {volume}
  {120}},\ \bibinfo {pages} {115} (\bibinfo {year} {2014})}\BibitemShut
  {NoStop}%
\bibitem [{\citenamefont {Mach}\ \emph {et~al.}(1990)\citenamefont {Mach},
  \citenamefont {Warburton}, \citenamefont {Gill}, \citenamefont {Casten},
  \citenamefont {Becker}, \citenamefont {Brown},\ and\ \citenamefont
  {Winger}}]{Mach1990}%
  \BibitemOpen
  \bibfield  {author} {\bibinfo {author} {\bibfnamefont {H.}~\bibnamefont
  {Mach}}, \bibinfo {author} {\bibfnamefont {E.~K.}\ \bibnamefont {Warburton}},
  \bibinfo {author} {\bibfnamefont {R.~L.}\ \bibnamefont {Gill}}, \bibinfo
  {author} {\bibfnamefont {R.~F.}\ \bibnamefont {Casten}}, \bibinfo {author}
  {\bibfnamefont {J.~A.}\ \bibnamefont {Becker}}, \bibinfo {author}
  {\bibfnamefont {B.~A.}\ \bibnamefont {Brown}}, \ and\ \bibinfo {author}
  {\bibfnamefont {J.~A.}\ \bibnamefont {Winger}},\ }\href {\doibase
  10.1103/PhysRevC.41.226} {\bibfield  {journal} {\bibinfo  {journal} {Physical
  Review C}\ }\textbf {\bibinfo {volume} {41}},\ \bibinfo {pages} {226}
  (\bibinfo {year} {1990})}\BibitemShut {NoStop}%
\bibitem [{\citenamefont {Machleidt}(2001)}]{Machleidt2001}%
  \BibitemOpen
  \bibfield  {author} {\bibinfo {author} {\bibfnamefont {R.}~\bibnamefont
  {Machleidt}},\ }\href {\doibase 10.1103/PhysRevC.63.024001} {\bibfield
  {journal} {\bibinfo  {journal} {Physical Review C}\ }\textbf {\bibinfo
  {volume} {63}},\ \bibinfo {pages} {024001} (\bibinfo {year} {2001})},\
  \Eprint {http://arxiv.org/abs/0006014} {arXiv:0006014 [nucl-th]} \BibitemShut
  {NoStop}%
\bibitem [{\citenamefont {Lalkovski}\ \emph {et~al.}(2013)\citenamefont
  {Lalkovski}, \citenamefont {Bruce}, \citenamefont {Jungclaus}, \citenamefont
  {G{\'{o}}rska}, \citenamefont {Pf{\"{u}}tzner}, \citenamefont
  {C{\'{a}}ceres}, \citenamefont {Naqvi}, \citenamefont {Pietri}, \citenamefont
  {Podoly{\'{a}}k}, \citenamefont {Simpson}, \citenamefont {Andgren},
  \citenamefont {Bednarczyk}, \citenamefont {Beck}, \citenamefont {Benlliure},
  \citenamefont {Benzoni}, \citenamefont {Casarejos}, \citenamefont
  {Cederwall}, \citenamefont {Crespi}, \citenamefont {Cuenca-Garc{\'{i}}a},
  \citenamefont {Cullen}, \citenamefont {{Denis Bacelar}}, \citenamefont
  {Detistov}, \citenamefont {Doornenbal}, \citenamefont {Farrelly},
  \citenamefont {Garnsworthy}, \citenamefont {Geissel}, \citenamefont
  {Gelletly}, \citenamefont {Gerl}, \citenamefont {Grebosz}, \citenamefont
  {Hadinia}, \citenamefont {Hellstr{\"{o}}m}, \citenamefont {Hinke},
  \citenamefont {Hoischen}, \citenamefont {Ilie}, \citenamefont {Jaworski},
  \citenamefont {Jolie}, \citenamefont {Khaplanov}, \citenamefont {Kisyov},
  \citenamefont {Kmiecik}, \citenamefont {Kojouharov}, \citenamefont {Kumar},
  \citenamefont {Kurz}, \citenamefont {Maj}, \citenamefont {Mandal},
  \citenamefont {Modamio}, \citenamefont {Montes}, \citenamefont {Myalski},
  \citenamefont {Palacz}, \citenamefont {Prokopowicz}, \citenamefont {Reiter},
  \citenamefont {Regan}, \citenamefont {Rudolph}, \citenamefont {Schaffner},
  \citenamefont {Sohler}, \citenamefont {Steer}, \citenamefont {Tashenov},
  \citenamefont {Walker}, \citenamefont {Walker}, \citenamefont {Weick},
  \citenamefont {Werner-Malento}, \citenamefont {Wieland}, \citenamefont
  {Wollersheim},\ and\ \citenamefont {Zhekova}}]{Lalkovski2013}%
  \BibitemOpen
  \bibfield  {author} {\bibinfo {author} {\bibfnamefont {S.}~\bibnamefont
  {Lalkovski}}, \bibinfo {author} {\bibfnamefont {A.~M.}\ \bibnamefont
  {Bruce}}, \bibinfo {author} {\bibfnamefont {A.}~\bibnamefont {Jungclaus}},
  \bibinfo {author} {\bibfnamefont {M.}~\bibnamefont {G{\'{o}}rska}}, \bibinfo
  {author} {\bibfnamefont {M.}~\bibnamefont {Pf{\"{u}}tzner}}, \bibinfo
  {author} {\bibfnamefont {L.}~\bibnamefont {C{\'{a}}ceres}}, \bibinfo {author}
  {\bibfnamefont {F.}~\bibnamefont {Naqvi}}, \bibinfo {author} {\bibfnamefont
  {S.}~\bibnamefont {Pietri}}, \bibinfo {author} {\bibfnamefont
  {Z.}~\bibnamefont {Podoly{\'{a}}k}}, \bibinfo {author} {\bibfnamefont
  {G.~S.}\ \bibnamefont {Simpson}}, \bibinfo {author} {\bibfnamefont
  {K.}~\bibnamefont {Andgren}}, \bibinfo {author} {\bibfnamefont
  {P.}~\bibnamefont {Bednarczyk}}, \bibinfo {author} {\bibfnamefont
  {T.}~\bibnamefont {Beck}}, \bibinfo {author} {\bibfnamefont {J.}~\bibnamefont
  {Benlliure}}, \bibinfo {author} {\bibfnamefont {G.}~\bibnamefont {Benzoni}},
  \bibinfo {author} {\bibfnamefont {E.}~\bibnamefont {Casarejos}}, \bibinfo
  {author} {\bibfnamefont {B.}~\bibnamefont {Cederwall}}, \bibinfo {author}
  {\bibfnamefont {F.~C.~L.}\ \bibnamefont {Crespi}}, \bibinfo {author}
  {\bibfnamefont {J.~J.}\ \bibnamefont {Cuenca-Garc{\'{i}}a}}, \bibinfo
  {author} {\bibfnamefont {I.~J.}\ \bibnamefont {Cullen}}, \bibinfo {author}
  {\bibfnamefont {A.~M.}\ \bibnamefont {{Denis Bacelar}}}, \bibinfo {author}
  {\bibfnamefont {P.}~\bibnamefont {Detistov}}, \bibinfo {author}
  {\bibfnamefont {P.}~\bibnamefont {Doornenbal}}, \bibinfo {author}
  {\bibfnamefont {G.~F.}\ \bibnamefont {Farrelly}}, \bibinfo {author}
  {\bibfnamefont {A.~B.}\ \bibnamefont {Garnsworthy}}, \bibinfo {author}
  {\bibfnamefont {H.}~\bibnamefont {Geissel}}, \bibinfo {author} {\bibfnamefont
  {W.}~\bibnamefont {Gelletly}}, \bibinfo {author} {\bibfnamefont
  {J.}~\bibnamefont {Gerl}}, \bibinfo {author} {\bibfnamefont {J.}~\bibnamefont
  {Grebosz}}, \bibinfo {author} {\bibfnamefont {B.}~\bibnamefont {Hadinia}},
  \bibinfo {author} {\bibfnamefont {M.}~\bibnamefont {Hellstr{\"{o}}m}},
  \bibinfo {author} {\bibfnamefont {C.}~\bibnamefont {Hinke}}, \bibinfo
  {author} {\bibfnamefont {R.}~\bibnamefont {Hoischen}}, \bibinfo {author}
  {\bibfnamefont {G.}~\bibnamefont {Ilie}}, \bibinfo {author} {\bibfnamefont
  {G.}~\bibnamefont {Jaworski}}, \bibinfo {author} {\bibfnamefont
  {J.}~\bibnamefont {Jolie}}, \bibinfo {author} {\bibfnamefont
  {A.}~\bibnamefont {Khaplanov}}, \bibinfo {author} {\bibfnamefont
  {S.}~\bibnamefont {Kisyov}}, \bibinfo {author} {\bibfnamefont
  {M.}~\bibnamefont {Kmiecik}}, \bibinfo {author} {\bibfnamefont
  {I.}~\bibnamefont {Kojouharov}}, \bibinfo {author} {\bibfnamefont
  {R.}~\bibnamefont {Kumar}}, \bibinfo {author} {\bibfnamefont
  {N.}~\bibnamefont {Kurz}}, \bibinfo {author} {\bibfnamefont {A.}~\bibnamefont
  {Maj}}, \bibinfo {author} {\bibfnamefont {S.}~\bibnamefont {Mandal}},
  \bibinfo {author} {\bibfnamefont {V.}~\bibnamefont {Modamio}}, \bibinfo
  {author} {\bibfnamefont {F.}~\bibnamefont {Montes}}, \bibinfo {author}
  {\bibfnamefont {S.}~\bibnamefont {Myalski}}, \bibinfo {author} {\bibfnamefont
  {M.}~\bibnamefont {Palacz}}, \bibinfo {author} {\bibfnamefont
  {W.}~\bibnamefont {Prokopowicz}}, \bibinfo {author} {\bibfnamefont
  {P.}~\bibnamefont {Reiter}}, \bibinfo {author} {\bibfnamefont {P.~H.}\
  \bibnamefont {Regan}}, \bibinfo {author} {\bibfnamefont {D.}~\bibnamefont
  {Rudolph}}, \bibinfo {author} {\bibfnamefont {H.}~\bibnamefont {Schaffner}},
  \bibinfo {author} {\bibfnamefont {D.}~\bibnamefont {Sohler}}, \bibinfo
  {author} {\bibfnamefont {S.~J.}\ \bibnamefont {Steer}}, \bibinfo {author}
  {\bibfnamefont {S.}~\bibnamefont {Tashenov}}, \bibinfo {author}
  {\bibfnamefont {J.}~\bibnamefont {Walker}}, \bibinfo {author} {\bibfnamefont
  {P.~M.}\ \bibnamefont {Walker}}, \bibinfo {author} {\bibfnamefont
  {H.}~\bibnamefont {Weick}}, \bibinfo {author} {\bibfnamefont
  {E.}~\bibnamefont {Werner-Malento}}, \bibinfo {author} {\bibfnamefont
  {O.}~\bibnamefont {Wieland}}, \bibinfo {author} {\bibfnamefont {H.~J.}\
  \bibnamefont {Wollersheim}}, \ and\ \bibinfo {author} {\bibfnamefont
  {M.}~\bibnamefont {Zhekova}},\ }\href {\doibase 10.1103/PhysRevC.87.034308}
  {\bibfield  {journal} {\bibinfo  {journal} {Physical Review C}\ }\textbf
  {\bibinfo {volume} {87}},\ \bibinfo {pages} {034308} (\bibinfo {year}
  {2013})},\ \Eprint {http://arxiv.org/abs/1212.4961} {arXiv:1212.4961}
  \BibitemShut {NoStop}%
\bibitem [{\citenamefont {Brown}(2012)}]{Brown2012}%
  \BibitemOpen
  \bibfield  {author} {\bibinfo {author} {\bibfnamefont {B.~A.}\ \bibnamefont
  {Brown}},\ }\href@noop {} {\bibfield  {journal} {\bibinfo  {journal}
  {Unpublished}\ } (\bibinfo {year} {2012})}\BibitemShut {NoStop}%
\bibitem [{\citenamefont {Suhonen}(2017)}]{Suhonen2017a}%
  \BibitemOpen
  \bibfield  {author} {\bibinfo {author} {\bibfnamefont {J.}~\bibnamefont
  {Suhonen}},\ }\href {\doibase 10.3389/fphy.2017.00055} {\bibfield  {journal}
  {\bibinfo  {journal} {Frontiers in Physics}\ }\textbf {\bibinfo {volume}
  {5}},\ \bibinfo {pages} {55} (\bibinfo {year} {2017})},\ \Eprint
  {http://arxiv.org/abs/1712.01565} {arXiv:1712.01565} \BibitemShut {NoStop}%
\bibitem [{\citenamefont {Kostensalo}\ and\ \citenamefont
  {Suhonen}(2018)}]{Kostensalo2018}%
  \BibitemOpen
  \bibfield  {author} {\bibinfo {author} {\bibfnamefont {J.}~\bibnamefont
  {Kostensalo}}\ and\ \bibinfo {author} {\bibfnamefont {J.}~\bibnamefont
  {Suhonen}},\ }\href {\doibase 10.1016/j.physletb.2018.02.053} {\bibfield
  {journal} {\bibinfo  {journal} {Physics Letters, Section B: Nuclear,
  Elementary Particle and High-Energy Physics}\ }\textbf {\bibinfo {volume}
  {781}},\ \bibinfo {pages} {480} (\bibinfo {year} {2018})}\BibitemShut
  {NoStop}%
\bibitem [{\citenamefont {Kostensalo}\ \emph {et~al.}(2017)\citenamefont
  {Kostensalo}, \citenamefont {Haaranen},\ and\ \citenamefont
  {Suhonen}}]{Kostensalo2017}%
  \BibitemOpen
  \bibfield  {author} {\bibinfo {author} {\bibfnamefont {J.}~\bibnamefont
  {Kostensalo}}, \bibinfo {author} {\bibfnamefont {M.}~\bibnamefont
  {Haaranen}}, \ and\ \bibinfo {author} {\bibfnamefont {J.}~\bibnamefont
  {Suhonen}},\ }\href {\doibase 10.1103/PhysRevC.95.044313} {\bibfield
  {journal} {\bibinfo  {journal} {Physical Review C}\ }\textbf {\bibinfo
  {volume} {95}},\ \bibinfo {pages} {044313} (\bibinfo {year}
  {2017})}\BibitemShut {NoStop}%
\bibitem [{\citenamefont {Haaranen}\ \emph {et~al.}(2017)\citenamefont
  {Haaranen}, \citenamefont {Kotila},\ and\ \citenamefont
  {Suhonen}}]{Haaranen2017}%
  \BibitemOpen
  \bibfield  {author} {\bibinfo {author} {\bibfnamefont {M.}~\bibnamefont
  {Haaranen}}, \bibinfo {author} {\bibfnamefont {J.}~\bibnamefont {Kotila}}, \
  and\ \bibinfo {author} {\bibfnamefont {J.}~\bibnamefont {Suhonen}},\ }\href
  {\doibase 10.1103/PhysRevC.95.024327} {\bibfield  {journal} {\bibinfo
  {journal} {Physical Review C}\ }\textbf {\bibinfo {volume} {95}},\ \bibinfo
  {pages} {024327} (\bibinfo {year} {2017})}\BibitemShut {NoStop}%
\bibitem [{\citenamefont {Bodenstein-Dresler}\ \emph
  {et~al.}(2018)\citenamefont {Bodenstein-Dresler}, \citenamefont {Chu},
  \citenamefont {Gehre}, \citenamefont {G{\"{o}}{\ss}ling}, \citenamefont
  {Heimbold}, \citenamefont {Herrmann}, \citenamefont {Hodak}, \citenamefont
  {Kostensalo}, \citenamefont {Kr{\"{o}}ninger}, \citenamefont {K{\"{u}}ttler},
  \citenamefont {Nitsch}, \citenamefont {Quante}, \citenamefont {Rukhadze},
  \citenamefont {Stekl}, \citenamefont {Suhonen}, \citenamefont
  {Tebr{\"{u}}gge}, \citenamefont {Temminghoff}, \citenamefont {Volkmer},
  \citenamefont {Zatschler},\ and\ \citenamefont
  {Zuber}}]{Bodenstein-Dresler2018}%
  \BibitemOpen
  \bibfield  {author} {\bibinfo {author} {\bibfnamefont {L.}~\bibnamefont
  {Bodenstein-Dresler}}, \bibinfo {author} {\bibfnamefont {Y.}~\bibnamefont
  {Chu}}, \bibinfo {author} {\bibfnamefont {D.}~\bibnamefont {Gehre}}, \bibinfo
  {author} {\bibfnamefont {C.}~\bibnamefont {G{\"{o}}{\ss}ling}}, \bibinfo
  {author} {\bibfnamefont {A.}~\bibnamefont {Heimbold}}, \bibinfo {author}
  {\bibfnamefont {C.}~\bibnamefont {Herrmann}}, \bibinfo {author}
  {\bibfnamefont {R.}~\bibnamefont {Hodak}}, \bibinfo {author} {\bibfnamefont
  {J.}~\bibnamefont {Kostensalo}}, \bibinfo {author} {\bibfnamefont
  {K.}~\bibnamefont {Kr{\"{o}}ninger}}, \bibinfo {author} {\bibfnamefont
  {J.}~\bibnamefont {K{\"{u}}ttler}}, \bibinfo {author} {\bibfnamefont
  {C.}~\bibnamefont {Nitsch}}, \bibinfo {author} {\bibfnamefont
  {T.}~\bibnamefont {Quante}}, \bibinfo {author} {\bibfnamefont
  {E.}~\bibnamefont {Rukhadze}}, \bibinfo {author} {\bibfnamefont
  {I.}~\bibnamefont {Stekl}}, \bibinfo {author} {\bibfnamefont
  {J.}~\bibnamefont {Suhonen}}, \bibinfo {author} {\bibfnamefont
  {J.}~\bibnamefont {Tebr{\"{u}}gge}}, \bibinfo {author} {\bibfnamefont
  {R.}~\bibnamefont {Temminghoff}}, \bibinfo {author} {\bibfnamefont
  {J.}~\bibnamefont {Volkmer}}, \bibinfo {author} {\bibfnamefont
  {S.}~\bibnamefont {Zatschler}}, \ and\ \bibinfo {author} {\bibfnamefont
  {K.}~\bibnamefont {Zuber}},\ }\href {http://arxiv.org/abs/1806.02254} {\
  (\bibinfo {year} {2018})},\ \Eprint {http://arxiv.org/abs/1806.02254}
  {arXiv:1806.02254} \BibitemShut {NoStop}%
\bibitem [{\citenamefont {Davidson}(1968)}]{Davidson1968}%
  \BibitemOpen
  \bibfield  {author} {\bibinfo {author} {\bibfnamefont {J.~P.}\ \bibnamefont
  {Davidson}},\ }\href@noop {} {\emph {\bibinfo {title} {{Collective Models of
  the Nucleus}}}}\ (\bibinfo  {publisher} {Academic Press},\ \bibinfo {address}
  {New York and London},\ \bibinfo {year} {1968})\BibitemShut {NoStop}%
\bibitem [{\citenamefont {Suhonen}\ and\ \citenamefont
  {Kostensalo}(2019)}]{Suhonen2019}%
  \BibitemOpen
  \bibfield  {author} {\bibinfo {author} {\bibfnamefont {J.}~\bibnamefont
  {Suhonen}}\ and\ \bibinfo {author} {\bibfnamefont {J.}~\bibnamefont
  {Kostensalo}},\ }\href {\doibase 10.3389/fphy.2019.00029} {\bibfield
  {journal} {\bibinfo  {journal} {Frontiers in Physics}\ }\textbf {\bibinfo
  {volume} {7}},\ \bibinfo {pages} {29} (\bibinfo {year} {2019})}\BibitemShut
  {NoStop}%
\bibitem [{\citenamefont {Sonzogni}\ \emph {et~al.}(2017)\citenamefont
  {Sonzogni}, \citenamefont {McCutchan},\ and\ \citenamefont
  {Hayes}}]{Sonzogni2017}%
  \BibitemOpen
  \bibfield  {author} {\bibinfo {author} {\bibfnamefont {A.~A.}\ \bibnamefont
  {Sonzogni}}, \bibinfo {author} {\bibfnamefont {E.~A.}\ \bibnamefont
  {McCutchan}}, \ and\ \bibinfo {author} {\bibfnamefont {A.~C.}\ \bibnamefont
  {Hayes}},\ }\href {\doibase 10.1103/PhysRevLett.119.112501} {\bibfield
  {journal} {\bibinfo  {journal} {Physical Review Letters}\ }\textbf {\bibinfo
  {volume} {119}},\ \bibinfo {pages} {112501} (\bibinfo {year}
  {2017})}\BibitemShut {NoStop}%
\bibitem [{\citenamefont {Fang}()}]{FangPC}%
  \BibitemOpen
  \bibfield  {author} {\bibinfo {author} {\bibfnamefont {D.-L.}\ \bibnamefont
  {Fang}},\ }\href@noop {} {\bibinfo  {journal} {Private Communication}\
  }\BibitemShut {NoStop}%
\bibitem [{\citenamefont {Yoshida}\ \emph {et~al.}(2018)\citenamefont
  {Yoshida}, \citenamefont {Tachibana}, \citenamefont {Okumura},\ and\
  \citenamefont {Chiba}}]{Yoshida2018}%
  \BibitemOpen
\bibfield  {journal} {  }\bibfield  {author} {\bibinfo {author} {\bibfnamefont
  {T.}~\bibnamefont {Yoshida}}, \bibinfo {author} {\bibfnamefont
  {T.}~\bibnamefont {Tachibana}}, \bibinfo {author} {\bibfnamefont
  {S.}~\bibnamefont {Okumura}}, \ and\ \bibinfo {author} {\bibfnamefont
  {S.}~\bibnamefont {Chiba}},\ }\href {\doibase 10.1103/PhysRevC.98.041303}
  {\bibfield  {journal} {\bibinfo  {journal} {Physical Review C}\ }\textbf
  {\bibinfo {volume} {98}},\ \bibinfo {pages} {41303} (\bibinfo {year}
  {2018})}\BibitemShut {NoStop}%
\bibitem [{\citenamefont {Scott}(1992)}]{Scott1992}%
  \BibitemOpen
  \bibfield  {author} {\bibinfo {author} {\bibfnamefont {D.~W.}\ \bibnamefont
  {Scott}},\ }\href@noop {} {\emph {\bibinfo {title} {{Multivariate Density
  Estimation: Theory, Practice, and Visualization}}}}\ (\bibinfo  {publisher}
  {John Wiley $\backslash${\&} Sons},\ \bibinfo {year} {1992})\BibitemShut
  {NoStop}%
\bibitem [{\citenamefont {Foreman-Mackey}(2016)}]{Foreman-Mackey2016}%
  \BibitemOpen
  \bibfield  {author} {\bibinfo {author} {\bibfnamefont {D.}~\bibnamefont
  {Foreman-Mackey}},\ }\href {\doibase 10.21105/joss.00024} {\bibfield
  {journal} {\bibinfo  {journal} {The Journal of Open Source Software}\
  }\textbf {\bibinfo {volume} {1}},\ \bibinfo {pages} {24} (\bibinfo {year}
  {2016})}\BibitemShut {NoStop}%
\bibitem [{\citenamefont {Dwyer}\ and\ \citenamefont
  {Langford}(2015)}]{Dwyer2015}%
  \BibitemOpen
  \bibfield  {author} {\bibinfo {author} {\bibfnamefont {D.~A.}\ \bibnamefont
  {Dwyer}}\ and\ \bibinfo {author} {\bibfnamefont {T.~J.}\ \bibnamefont
  {Langford}},\ }\href {\doibase 10.1103/PhysRevLett.114.012502} {\bibfield
  {journal} {\bibinfo  {journal} {Physical Review Letters}\ }\textbf {\bibinfo
  {volume} {114}},\ \bibinfo {pages} {012502} (\bibinfo {year} {2015})},\
  \Eprint {http://arxiv.org/abs/1407.1281} {arXiv:1407.1281} \BibitemShut
  {NoStop}%
\bibitem [{\citenamefont {Estienne}\ \emph {et~al.}(2019)\citenamefont
  {Estienne}, \citenamefont {Fallot}, \citenamefont {Algora}, \citenamefont
  {Briz-Monago}, \citenamefont {Bui}, \citenamefont {Cormon}, \citenamefont
  {Gelletly}, \citenamefont {Giot}, \citenamefont {Guadilla}, \citenamefont
  {Jordan}, \citenamefont {{Le Meur}}, \citenamefont {Porta}, \citenamefont
  {Rice}, \citenamefont {Rubio}, \citenamefont {Ta{\'{i}}n}, \citenamefont
  {Valencia},\ and\ \citenamefont {Zakari-Issoufou}}]{Estienne2019}%
  \BibitemOpen
  \bibfield  {author} {\bibinfo {author} {\bibfnamefont {M.}~\bibnamefont
  {Estienne}}, \bibinfo {author} {\bibfnamefont {M.}~\bibnamefont {Fallot}},
  \bibinfo {author} {\bibfnamefont {A.}~\bibnamefont {Algora}}, \bibinfo
  {author} {\bibfnamefont {J.}~\bibnamefont {Briz-Monago}}, \bibinfo {author}
  {\bibfnamefont {V.~M.}\ \bibnamefont {Bui}}, \bibinfo {author} {\bibfnamefont
  {S.}~\bibnamefont {Cormon}}, \bibinfo {author} {\bibfnamefont
  {W.}~\bibnamefont {Gelletly}}, \bibinfo {author} {\bibfnamefont
  {L.}~\bibnamefont {Giot}}, \bibinfo {author} {\bibfnamefont {V.}~\bibnamefont
  {Guadilla}}, \bibinfo {author} {\bibfnamefont {D.}~\bibnamefont {Jordan}},
  \bibinfo {author} {\bibfnamefont {L.}~\bibnamefont {{Le Meur}}}, \bibinfo
  {author} {\bibfnamefont {A.}~\bibnamefont {Porta}}, \bibinfo {author}
  {\bibfnamefont {S.}~\bibnamefont {Rice}}, \bibinfo {author} {\bibfnamefont
  {B.}~\bibnamefont {Rubio}}, \bibinfo {author} {\bibfnamefont {J.~L.}\
  \bibnamefont {Ta{\'{i}}n}}, \bibinfo {author} {\bibfnamefont
  {E.}~\bibnamefont {Valencia}}, \ and\ \bibinfo {author} {\bibfnamefont
  {A.-A.}\ \bibnamefont {Zakari-Issoufou}},\ }\href {\doibase
  10.1103/PhysRevLett.123.022502} {\bibfield  {journal} {\bibinfo  {journal}
  {Physical Review Letters}\ }\textbf {\bibinfo {volume} {123}},\ \bibinfo
  {pages} {022502} (\bibinfo {year} {2019})},\ \Eprint
  {http://arxiv.org/abs/1904.09358} {arXiv:1904.09358} \BibitemShut {NoStop}%
\bibitem [{\citenamefont {Vogel}\ and\ \citenamefont
  {Beacom}(1999)}]{Vogel1999}%
  \BibitemOpen
  \bibfield  {author} {\bibinfo {author} {\bibfnamefont {P.}~\bibnamefont
  {Vogel}}\ and\ \bibinfo {author} {\bibfnamefont {J.~F.}\ \bibnamefont
  {Beacom}},\ }\href {\doibase 10.1103/PhysRevD.60.053003} {\bibfield
  {journal} {\bibinfo  {journal} {Physical Review D}\ }\textbf {\bibinfo
  {volume} {60}},\ \bibinfo {pages} {10} (\bibinfo {year} {1999})},\ \Eprint
  {http://arxiv.org/abs/9903554} {arXiv:9903554 [hep-ph]} \BibitemShut
  {NoStop}%
\bibitem [{\citenamefont {Strumia}\ and\ \citenamefont
  {Vissani}(2003)}]{Strumia2003}%
  \BibitemOpen
  \bibfield  {author} {\bibinfo {author} {\bibfnamefont {A.}~\bibnamefont
  {Strumia}}\ and\ \bibinfo {author} {\bibfnamefont {F.}~\bibnamefont
  {Vissani}},\ }\href {\doibase 10.1016/S0370-2693(03)00616-6} {\bibfield
  {journal} {\bibinfo  {journal} {Physics Letters, Section B: Nuclear,
  Elementary Particle and High-Energy Physics}\ }\textbf {\bibinfo {volume}
  {564}},\ \bibinfo {pages} {42} (\bibinfo {year} {2003})},\ \Eprint
  {http://arxiv.org/abs/0302055} {arXiv:0302055 [astro-ph]} \BibitemShut
  {NoStop}%
\bibitem [{\citenamefont {Kurylov}\ \emph {et~al.}(2003)\citenamefont
  {Kurylov}, \citenamefont {Ramsey-Musolf},\ and\ \citenamefont
  {Vogel}}]{Kurylov2003}%
  \BibitemOpen
  \bibfield  {author} {\bibinfo {author} {\bibfnamefont {A.}~\bibnamefont
  {Kurylov}}, \bibinfo {author} {\bibfnamefont {M.~J.}\ \bibnamefont
  {Ramsey-Musolf}}, \ and\ \bibinfo {author} {\bibfnamefont {P.}~\bibnamefont
  {Vogel}},\ }\href {\doibase 10.1103/PhysRevC.67.035502} {\bibfield  {journal}
  {\bibinfo  {journal} {Physical Review C}\ }\textbf {\bibinfo {volume} {67}},\
  \bibinfo {pages} {035502} (\bibinfo {year} {2003})},\ \Eprint
  {http://arxiv.org/abs/0211306v1} {arXiv:0211306v1 [arXiv:hep-ph]}
  \BibitemShut {NoStop}%
\bibitem [{\citenamefont {An}\ \emph {et~al.}(2012)\citenamefont {An},
  \citenamefont {Bai}, \citenamefont {Balantekin}, \citenamefont {Band},
  \citenamefont {Beavis}, \citenamefont {Beriguete}, \citenamefont {Bishai},
  \citenamefont {Blyth}, \citenamefont {Boddy}, \citenamefont {Brown},
  \citenamefont {Cai}, \citenamefont {Cao}, \citenamefont {Cao}, \citenamefont
  {Carr}, \citenamefont {Chan}, \citenamefont {Chang}, \citenamefont {Chang},
  \citenamefont {Chasman}, \citenamefont {Chen}, \citenamefont {Chen},
  \citenamefont {Chen}, \citenamefont {Chen}, \citenamefont {Chen},
  \citenamefont {Chen}, \citenamefont {Chen}, \citenamefont {Chen},
  \citenamefont {Chen}, \citenamefont {Cherwinka}, \citenamefont {Chu},
  \citenamefont {Cummings}, \citenamefont {Deng}, \citenamefont {Ding},
  \citenamefont {Diwan}, \citenamefont {Dong}, \citenamefont {Draeger},
  \citenamefont {Du}, \citenamefont {Dwyer}, \citenamefont {Edwards},
  \citenamefont {Ely}, \citenamefont {Fang}, \citenamefont {Fu}, \citenamefont
  {Fu}, \citenamefont {Ge}, \citenamefont {Ghazikhanian}, \citenamefont {Gill},
  \citenamefont {Goett}, \citenamefont {Gonchar}, \citenamefont {Gong},
  \citenamefont {Gong}, \citenamefont {Gornushkin}, \citenamefont {Greenler},
  \citenamefont {Gu}, \citenamefont {Guan}, \citenamefont {Guo}, \citenamefont
  {Hackenburg}, \citenamefont {Hahn}, \citenamefont {Hans}, \citenamefont {He},
  \citenamefont {He}, \citenamefont {He}, \citenamefont {Heeger}, \citenamefont
  {Heng}, \citenamefont {Hinrichs}, \citenamefont {Ho}, \citenamefont {Hor},
  \citenamefont {Hsiung}, \citenamefont {Hu}, \citenamefont {Hu}, \citenamefont
  {Hu}, \citenamefont {Huang}, \citenamefont {Huang}, \citenamefont {Huang},
  \citenamefont {Huang}, \citenamefont {Huang}, \citenamefont {Huber},
  \citenamefont {Isvan}, \citenamefont {Jaffe}, \citenamefont {Jetter},
  \citenamefont {Ji}, \citenamefont {Ji}, \citenamefont {Jiang}, \citenamefont
  {Jiang}, \citenamefont {Jiao}, \citenamefont {Johnson}, \citenamefont {Kang},
  \citenamefont {Kettell}, \citenamefont {Kramer}, \citenamefont {Kwan},
  \citenamefont {Kwok}, \citenamefont {Kwok}, \citenamefont {Lai},
  \citenamefont {Lai}, \citenamefont {Lai}, \citenamefont {Lau}, \citenamefont
  {Lebanowski}, \citenamefont {Lee}, \citenamefont {Lee}, \citenamefont
  {Leitner}, \citenamefont {Leung}, \citenamefont {Leung}, \citenamefont
  {Lewis}, \citenamefont {Li}, \citenamefont {Li}, \citenamefont {Li},
  \citenamefont {Li}, \citenamefont {Li}, \citenamefont {Li}, \citenamefont
  {Li}, \citenamefont {Li}, \citenamefont {Li}, \citenamefont {Li},
  \citenamefont {Li}, \citenamefont {Li}, \citenamefont {Liang}, \citenamefont
  {Liang}, \citenamefont {Lin}, \citenamefont {Lin}, \citenamefont {Lin},
  \citenamefont {Lin}, \citenamefont {Lin}, \citenamefont {Ling}, \citenamefont
  {Link}, \citenamefont {Littenberg}, \citenamefont {Littlejohn}, \citenamefont
  {Liu}, \citenamefont {Liu}, \citenamefont {Liu}, \citenamefont {Liu},
  \citenamefont {Liu}, \citenamefont {Liu}, \citenamefont {Liu}, \citenamefont
  {Liu}, \citenamefont {Liu}, \citenamefont {Lu}, \citenamefont {Lu},
  \citenamefont {Luk}, \citenamefont {Luk}, \citenamefont {Luo}, \citenamefont
  {Luo}, \citenamefont {Ma}, \citenamefont {Ma}, \citenamefont {Ma},
  \citenamefont {Ma}, \citenamefont {Ma}, \citenamefont {Mayes}, \citenamefont
  {McDonald}, \citenamefont {McFarlane}, \citenamefont {McKeown}, \citenamefont
  {Meng}, \citenamefont {Mohapatra}, \citenamefont {Morgan}, \citenamefont
  {Nakajima}, \citenamefont {Napolitano}, \citenamefont {Naumov}, \citenamefont
  {Nemchenok}, \citenamefont {Newsom}, \citenamefont {Ngai}, \citenamefont
  {Ngai}, \citenamefont {Nie}, \citenamefont {Ning}, \citenamefont
  {Ochoa-Ricoux}, \citenamefont {Oh}, \citenamefont {Olshevski}, \citenamefont
  {Pagac}, \citenamefont {Patton}, \citenamefont {Pearson}, \citenamefont
  {Pec}, \citenamefont {Peng}, \citenamefont {Piilonen}, \citenamefont
  {Pinsky}, \citenamefont {Pun}, \citenamefont {Qi}, \citenamefont {Qi},
  \citenamefont {Qian}, \citenamefont {Raper}, \citenamefont {Rosero},
  \citenamefont {Roskovec}, \citenamefont {Ruan}, \citenamefont {Seilhan},
  \citenamefont {Shao}, \citenamefont {Shih}, \citenamefont {Steiner},
  \citenamefont {Stoler}, \citenamefont {Sun}, \citenamefont {Sun},
  \citenamefont {Tam}, \citenamefont {Tanaka}, \citenamefont {Tang},
  \citenamefont {Themann}, \citenamefont {Torun}, \citenamefont {Trentalange},
  \citenamefont {Tsai}, \citenamefont {Tsang}, \citenamefont {Tsang},
  \citenamefont {Tull}, \citenamefont {Viren}, \citenamefont {Virostek},
  \citenamefont {Vorobel}, \citenamefont {Wang}, \citenamefont {Wang},
  \citenamefont {Wang}, \citenamefont {Wang}, \citenamefont {Wang},
  \citenamefont {Wang}, \citenamefont {Wang}, \citenamefont {Wang},
  \citenamefont {Wang}, \citenamefont {Wang}, \citenamefont {Wang},
  \citenamefont {Wang}, \citenamefont {Wang}, \citenamefont {Wang},
  \citenamefont {Wang}, \citenamefont {Webber}, \citenamefont {Wei},
  \citenamefont {Wen}, \citenamefont {Wenman}, \citenamefont {Whisnant},
  \citenamefont {White}, \citenamefont {Whitehead}, \citenamefont {Whitten},
  \citenamefont {Wilhelmi}, \citenamefont {Wise}, \citenamefont {Wong},
  \citenamefont {Wong}, \citenamefont {Wong}, \citenamefont {Worcester},
  \citenamefont {Wu}, \citenamefont {Wu}, \citenamefont {Xia}, \citenamefont
  {Xiang}, \citenamefont {Xiao}, \citenamefont {Xing}, \citenamefont {Xu},
  \citenamefont {Xu}, \citenamefont {Xu}, \citenamefont {Xu}, \citenamefont
  {Xu}, \citenamefont {Xu}, \citenamefont {Xue}, \citenamefont {Yang},
  \citenamefont {Yang}, \citenamefont {Ye}, \citenamefont {Yeh}, \citenamefont
  {Yeh}, \citenamefont {Yip}, \citenamefont {Young}, \citenamefont {Yu},
  \citenamefont {Zhan}, \citenamefont {Zhang}, \citenamefont {Zhang},
  \citenamefont {Zhang}, \citenamefont {Zhang}, \citenamefont {Zhang},
  \citenamefont {Zhang}, \citenamefont {Zhang}, \citenamefont {Zhang},
  \citenamefont {Zhang}, \citenamefont {Zhang}, \citenamefont {Zhang},
  \citenamefont {Zhang}, \citenamefont {Zhang}, \citenamefont {Zhao},
  \citenamefont {Zhao}, \citenamefont {Zhao}, \citenamefont {Zheng},
  \citenamefont {Zhong}, \citenamefont {Zhou}, \citenamefont {Zhou},
  \citenamefont {Zhuang},\ and\ \citenamefont {Zou}}]{An2012}%
  \BibitemOpen
  \bibfield  {author} {\bibinfo {author} {\bibfnamefont {F.~P.}\ \bibnamefont
  {An}}, \bibinfo {author} {\bibfnamefont {J.~Z.}\ \bibnamefont {Bai}},
  \bibinfo {author} {\bibfnamefont {A.~B.}\ \bibnamefont {Balantekin}},
  \bibinfo {author} {\bibfnamefont {H.~R.}\ \bibnamefont {Band}}, \bibinfo
  {author} {\bibfnamefont {D.}~\bibnamefont {Beavis}}, \bibinfo {author}
  {\bibfnamefont {W.}~\bibnamefont {Beriguete}}, \bibinfo {author}
  {\bibfnamefont {M.}~\bibnamefont {Bishai}}, \bibinfo {author} {\bibfnamefont
  {S.}~\bibnamefont {Blyth}}, \bibinfo {author} {\bibfnamefont
  {K.}~\bibnamefont {Boddy}}, \bibinfo {author} {\bibfnamefont {R.~L.}\
  \bibnamefont {Brown}}, \bibinfo {author} {\bibfnamefont {B.}~\bibnamefont
  {Cai}}, \bibinfo {author} {\bibfnamefont {G.~F.}\ \bibnamefont {Cao}},
  \bibinfo {author} {\bibfnamefont {J.}~\bibnamefont {Cao}}, \bibinfo {author}
  {\bibfnamefont {R.}~\bibnamefont {Carr}}, \bibinfo {author} {\bibfnamefont
  {W.~T.}\ \bibnamefont {Chan}}, \bibinfo {author} {\bibfnamefont {J.~F.}\
  \bibnamefont {Chang}}, \bibinfo {author} {\bibfnamefont {Y.}~\bibnamefont
  {Chang}}, \bibinfo {author} {\bibfnamefont {C.}~\bibnamefont {Chasman}},
  \bibinfo {author} {\bibfnamefont {H.~S.}\ \bibnamefont {Chen}}, \bibinfo
  {author} {\bibfnamefont {H.~Y.}\ \bibnamefont {Chen}}, \bibinfo {author}
  {\bibfnamefont {S.~J.}\ \bibnamefont {Chen}}, \bibinfo {author}
  {\bibfnamefont {S.~M.}\ \bibnamefont {Chen}}, \bibinfo {author}
  {\bibfnamefont {X.~C.}\ \bibnamefont {Chen}}, \bibinfo {author}
  {\bibfnamefont {X.~H.}\ \bibnamefont {Chen}}, \bibinfo {author}
  {\bibfnamefont {X.~S.}\ \bibnamefont {Chen}}, \bibinfo {author}
  {\bibfnamefont {Y.}~\bibnamefont {Chen}}, \bibinfo {author} {\bibfnamefont
  {Y.~X.}\ \bibnamefont {Chen}}, \bibinfo {author} {\bibfnamefont {J.~J.}\
  \bibnamefont {Cherwinka}}, \bibinfo {author} {\bibfnamefont {M.~C.}\
  \bibnamefont {Chu}}, \bibinfo {author} {\bibfnamefont {J.~P.}\ \bibnamefont
  {Cummings}}, \bibinfo {author} {\bibfnamefont {Z.~Y.}\ \bibnamefont {Deng}},
  \bibinfo {author} {\bibfnamefont {Y.~Y.}\ \bibnamefont {Ding}}, \bibinfo
  {author} {\bibfnamefont {M.~V.}\ \bibnamefont {Diwan}}, \bibinfo {author}
  {\bibfnamefont {L.}~\bibnamefont {Dong}}, \bibinfo {author} {\bibfnamefont
  {E.}~\bibnamefont {Draeger}}, \bibinfo {author} {\bibfnamefont {X.~F.}\
  \bibnamefont {Du}}, \bibinfo {author} {\bibfnamefont {D.~A.}\ \bibnamefont
  {Dwyer}}, \bibinfo {author} {\bibfnamefont {W.~R.}\ \bibnamefont {Edwards}},
  \bibinfo {author} {\bibfnamefont {S.~R.}\ \bibnamefont {Ely}}, \bibinfo
  {author} {\bibfnamefont {S.~D.}\ \bibnamefont {Fang}}, \bibinfo {author}
  {\bibfnamefont {J.~Y.}\ \bibnamefont {Fu}}, \bibinfo {author} {\bibfnamefont
  {Z.~W.}\ \bibnamefont {Fu}}, \bibinfo {author} {\bibfnamefont {L.~Q.}\
  \bibnamefont {Ge}}, \bibinfo {author} {\bibfnamefont {V.}~\bibnamefont
  {Ghazikhanian}}, \bibinfo {author} {\bibfnamefont {R.~L.}\ \bibnamefont
  {Gill}}, \bibinfo {author} {\bibfnamefont {J.}~\bibnamefont {Goett}},
  \bibinfo {author} {\bibfnamefont {M.}~\bibnamefont {Gonchar}}, \bibinfo
  {author} {\bibfnamefont {G.~H.}\ \bibnamefont {Gong}}, \bibinfo {author}
  {\bibfnamefont {H.}~\bibnamefont {Gong}}, \bibinfo {author} {\bibfnamefont
  {Y.~A.}\ \bibnamefont {Gornushkin}}, \bibinfo {author} {\bibfnamefont
  {L.~S.}\ \bibnamefont {Greenler}}, \bibinfo {author} {\bibfnamefont {W.~Q.}\
  \bibnamefont {Gu}}, \bibinfo {author} {\bibfnamefont {M.~Y.}\ \bibnamefont
  {Guan}}, \bibinfo {author} {\bibfnamefont {X.~H.}\ \bibnamefont {Guo}},
  \bibinfo {author} {\bibfnamefont {R.~W.}\ \bibnamefont {Hackenburg}},
  \bibinfo {author} {\bibfnamefont {R.~L.}\ \bibnamefont {Hahn}}, \bibinfo
  {author} {\bibfnamefont {S.}~\bibnamefont {Hans}}, \bibinfo {author}
  {\bibfnamefont {M.}~\bibnamefont {He}}, \bibinfo {author} {\bibfnamefont
  {Q.}~\bibnamefont {He}}, \bibinfo {author} {\bibfnamefont {W.~S.}\
  \bibnamefont {He}}, \bibinfo {author} {\bibfnamefont {K.~M.}\ \bibnamefont
  {Heeger}}, \bibinfo {author} {\bibfnamefont {Y.~K.}\ \bibnamefont {Heng}},
  \bibinfo {author} {\bibfnamefont {P.}~\bibnamefont {Hinrichs}}, \bibinfo
  {author} {\bibfnamefont {T.~H.}\ \bibnamefont {Ho}}, \bibinfo {author}
  {\bibfnamefont {Y.~K.}\ \bibnamefont {Hor}}, \bibinfo {author} {\bibfnamefont
  {Y.~B.}\ \bibnamefont {Hsiung}}, \bibinfo {author} {\bibfnamefont {B.~Z.}\
  \bibnamefont {Hu}}, \bibinfo {author} {\bibfnamefont {T.}~\bibnamefont {Hu}},
  \bibinfo {author} {\bibfnamefont {T.}~\bibnamefont {Hu}}, \bibinfo {author}
  {\bibfnamefont {H.~X.}\ \bibnamefont {Huang}}, \bibinfo {author}
  {\bibfnamefont {H.~Z.}\ \bibnamefont {Huang}}, \bibinfo {author}
  {\bibfnamefont {P.~W.}\ \bibnamefont {Huang}}, \bibinfo {author}
  {\bibfnamefont {X.}~\bibnamefont {Huang}}, \bibinfo {author} {\bibfnamefont
  {X.~T.}\ \bibnamefont {Huang}}, \bibinfo {author} {\bibfnamefont
  {P.}~\bibnamefont {Huber}}, \bibinfo {author} {\bibfnamefont
  {Z.}~\bibnamefont {Isvan}}, \bibinfo {author} {\bibfnamefont {D.~E.}\
  \bibnamefont {Jaffe}}, \bibinfo {author} {\bibfnamefont {S.}~\bibnamefont
  {Jetter}}, \bibinfo {author} {\bibfnamefont {X.~L.}\ \bibnamefont {Ji}},
  \bibinfo {author} {\bibfnamefont {X.~P.}\ \bibnamefont {Ji}}, \bibinfo
  {author} {\bibfnamefont {H.~J.}\ \bibnamefont {Jiang}}, \bibinfo {author}
  {\bibfnamefont {W.~Q.}\ \bibnamefont {Jiang}}, \bibinfo {author}
  {\bibfnamefont {J.~B.}\ \bibnamefont {Jiao}}, \bibinfo {author}
  {\bibfnamefont {R.~A.}\ \bibnamefont {Johnson}}, \bibinfo {author}
  {\bibfnamefont {L.}~\bibnamefont {Kang}}, \bibinfo {author} {\bibfnamefont
  {S.~H.}\ \bibnamefont {Kettell}}, \bibinfo {author} {\bibfnamefont
  {M.}~\bibnamefont {Kramer}}, \bibinfo {author} {\bibfnamefont {K.~K.}\
  \bibnamefont {Kwan}}, \bibinfo {author} {\bibfnamefont {M.~W.}\ \bibnamefont
  {Kwok}}, \bibinfo {author} {\bibfnamefont {T.}~\bibnamefont {Kwok}}, \bibinfo
  {author} {\bibfnamefont {C.~Y.}\ \bibnamefont {Lai}}, \bibinfo {author}
  {\bibfnamefont {W.~C.}\ \bibnamefont {Lai}}, \bibinfo {author} {\bibfnamefont
  {W.~H.}\ \bibnamefont {Lai}}, \bibinfo {author} {\bibfnamefont
  {K.}~\bibnamefont {Lau}}, \bibinfo {author} {\bibfnamefont {L.}~\bibnamefont
  {Lebanowski}}, \bibinfo {author} {\bibfnamefont {J.}~\bibnamefont {Lee}},
  \bibinfo {author} {\bibfnamefont {M.~K.~P.}\ \bibnamefont {Lee}}, \bibinfo
  {author} {\bibfnamefont {R.}~\bibnamefont {Leitner}}, \bibinfo {author}
  {\bibfnamefont {J.~K.~C.}\ \bibnamefont {Leung}}, \bibinfo {author}
  {\bibfnamefont {K.~Y.}\ \bibnamefont {Leung}}, \bibinfo {author}
  {\bibfnamefont {C.~A.}\ \bibnamefont {Lewis}}, \bibinfo {author}
  {\bibfnamefont {B.}~\bibnamefont {Li}}, \bibinfo {author} {\bibfnamefont
  {F.}~\bibnamefont {Li}}, \bibinfo {author} {\bibfnamefont {G.~S.}\
  \bibnamefont {Li}}, \bibinfo {author} {\bibfnamefont {J.}~\bibnamefont {Li}},
  \bibinfo {author} {\bibfnamefont {Q.~J.}\ \bibnamefont {Li}}, \bibinfo
  {author} {\bibfnamefont {S.~F.}\ \bibnamefont {Li}}, \bibinfo {author}
  {\bibfnamefont {W.~D.}\ \bibnamefont {Li}}, \bibinfo {author} {\bibfnamefont
  {X.~B.}\ \bibnamefont {Li}}, \bibinfo {author} {\bibfnamefont {X.~N.}\
  \bibnamefont {Li}}, \bibinfo {author} {\bibfnamefont {X.~Q.}\ \bibnamefont
  {Li}}, \bibinfo {author} {\bibfnamefont {Y.}~\bibnamefont {Li}}, \bibinfo
  {author} {\bibfnamefont {Z.~B.}\ \bibnamefont {Li}}, \bibinfo {author}
  {\bibfnamefont {H.}~\bibnamefont {Liang}}, \bibinfo {author} {\bibfnamefont
  {J.}~\bibnamefont {Liang}}, \bibinfo {author} {\bibfnamefont {C.~J.}\
  \bibnamefont {Lin}}, \bibinfo {author} {\bibfnamefont {G.~L.}\ \bibnamefont
  {Lin}}, \bibinfo {author} {\bibfnamefont {S.~K.}\ \bibnamefont {Lin}},
  \bibinfo {author} {\bibfnamefont {S.~X.}\ \bibnamefont {Lin}}, \bibinfo
  {author} {\bibfnamefont {Y.~C.}\ \bibnamefont {Lin}}, \bibinfo {author}
  {\bibfnamefont {J.~J.}\ \bibnamefont {Ling}}, \bibinfo {author}
  {\bibfnamefont {J.~M.}\ \bibnamefont {Link}}, \bibinfo {author}
  {\bibfnamefont {L.}~\bibnamefont {Littenberg}}, \bibinfo {author}
  {\bibfnamefont {B.~R.}\ \bibnamefont {Littlejohn}}, \bibinfo {author}
  {\bibfnamefont {B.~J.}\ \bibnamefont {Liu}}, \bibinfo {author} {\bibfnamefont
  {C.}~\bibnamefont {Liu}}, \bibinfo {author} {\bibfnamefont {D.~W.}\
  \bibnamefont {Liu}}, \bibinfo {author} {\bibfnamefont {H.}~\bibnamefont
  {Liu}}, \bibinfo {author} {\bibfnamefont {J.~C.}\ \bibnamefont {Liu}},
  \bibinfo {author} {\bibfnamefont {J.~L.}\ \bibnamefont {Liu}}, \bibinfo
  {author} {\bibfnamefont {S.}~\bibnamefont {Liu}}, \bibinfo {author}
  {\bibfnamefont {X.}~\bibnamefont {Liu}}, \bibinfo {author} {\bibfnamefont
  {Y.~B.}\ \bibnamefont {Liu}}, \bibinfo {author} {\bibfnamefont
  {C.}~\bibnamefont {Lu}}, \bibinfo {author} {\bibfnamefont {H.~Q.}\
  \bibnamefont {Lu}}, \bibinfo {author} {\bibfnamefont {A.}~\bibnamefont
  {Luk}}, \bibinfo {author} {\bibfnamefont {K.~B.}\ \bibnamefont {Luk}},
  \bibinfo {author} {\bibfnamefont {T.}~\bibnamefont {Luo}}, \bibinfo {author}
  {\bibfnamefont {X.~L.}\ \bibnamefont {Luo}}, \bibinfo {author} {\bibfnamefont
  {L.~H.}\ \bibnamefont {Ma}}, \bibinfo {author} {\bibfnamefont {Q.~M.}\
  \bibnamefont {Ma}}, \bibinfo {author} {\bibfnamefont {X.~B.}\ \bibnamefont
  {Ma}}, \bibinfo {author} {\bibfnamefont {X.~Y.}\ \bibnamefont {Ma}}, \bibinfo
  {author} {\bibfnamefont {Y.~Q.}\ \bibnamefont {Ma}}, \bibinfo {author}
  {\bibfnamefont {B.}~\bibnamefont {Mayes}}, \bibinfo {author} {\bibfnamefont
  {K.~T.}\ \bibnamefont {McDonald}}, \bibinfo {author} {\bibfnamefont {M.~C.}\
  \bibnamefont {McFarlane}}, \bibinfo {author} {\bibfnamefont {R.~D.}\
  \bibnamefont {McKeown}}, \bibinfo {author} {\bibfnamefont {Y.}~\bibnamefont
  {Meng}}, \bibinfo {author} {\bibfnamefont {D.}~\bibnamefont {Mohapatra}},
  \bibinfo {author} {\bibfnamefont {J.~E.}\ \bibnamefont {Morgan}}, \bibinfo
  {author} {\bibfnamefont {Y.}~\bibnamefont {Nakajima}}, \bibinfo {author}
  {\bibfnamefont {J.}~\bibnamefont {Napolitano}}, \bibinfo {author}
  {\bibfnamefont {D.}~\bibnamefont {Naumov}}, \bibinfo {author} {\bibfnamefont
  {I.}~\bibnamefont {Nemchenok}}, \bibinfo {author} {\bibfnamefont
  {C.}~\bibnamefont {Newsom}}, \bibinfo {author} {\bibfnamefont {H.~Y.}\
  \bibnamefont {Ngai}}, \bibinfo {author} {\bibfnamefont {W.~K.}\ \bibnamefont
  {Ngai}}, \bibinfo {author} {\bibfnamefont {Y.~B.}\ \bibnamefont {Nie}},
  \bibinfo {author} {\bibfnamefont {Z.}~\bibnamefont {Ning}}, \bibinfo {author}
  {\bibfnamefont {J.~P.}\ \bibnamefont {Ochoa-Ricoux}}, \bibinfo {author}
  {\bibfnamefont {D.}~\bibnamefont {Oh}}, \bibinfo {author} {\bibfnamefont
  {A.}~\bibnamefont {Olshevski}}, \bibinfo {author} {\bibfnamefont
  {A.}~\bibnamefont {Pagac}}, \bibinfo {author} {\bibfnamefont
  {S.}~\bibnamefont {Patton}}, \bibinfo {author} {\bibfnamefont
  {C.}~\bibnamefont {Pearson}}, \bibinfo {author} {\bibfnamefont
  {V.}~\bibnamefont {Pec}}, \bibinfo {author} {\bibfnamefont {J.~C.}\
  \bibnamefont {Peng}}, \bibinfo {author} {\bibfnamefont {L.~E.}\ \bibnamefont
  {Piilonen}}, \bibinfo {author} {\bibfnamefont {L.}~\bibnamefont {Pinsky}},
  \bibinfo {author} {\bibfnamefont {C.~S.~J.}\ \bibnamefont {Pun}}, \bibinfo
  {author} {\bibfnamefont {F.~Z.}\ \bibnamefont {Qi}}, \bibinfo {author}
  {\bibfnamefont {M.}~\bibnamefont {Qi}}, \bibinfo {author} {\bibfnamefont
  {X.}~\bibnamefont {Qian}}, \bibinfo {author} {\bibfnamefont {N.}~\bibnamefont
  {Raper}}, \bibinfo {author} {\bibfnamefont {R.}~\bibnamefont {Rosero}},
  \bibinfo {author} {\bibfnamefont {B.}~\bibnamefont {Roskovec}}, \bibinfo
  {author} {\bibfnamefont {X.~C.}\ \bibnamefont {Ruan}}, \bibinfo {author}
  {\bibfnamefont {B.}~\bibnamefont {Seilhan}}, \bibinfo {author} {\bibfnamefont
  {B.~B.}\ \bibnamefont {Shao}}, \bibinfo {author} {\bibfnamefont
  {K.}~\bibnamefont {Shih}}, \bibinfo {author} {\bibfnamefont {H.}~\bibnamefont
  {Steiner}}, \bibinfo {author} {\bibfnamefont {P.}~\bibnamefont {Stoler}},
  \bibinfo {author} {\bibfnamefont {G.~X.}\ \bibnamefont {Sun}}, \bibinfo
  {author} {\bibfnamefont {J.~L.}\ \bibnamefont {Sun}}, \bibinfo {author}
  {\bibfnamefont {Y.~H.}\ \bibnamefont {Tam}}, \bibinfo {author} {\bibfnamefont
  {H.~K.}\ \bibnamefont {Tanaka}}, \bibinfo {author} {\bibfnamefont
  {X.}~\bibnamefont {Tang}}, \bibinfo {author} {\bibfnamefont {H.}~\bibnamefont
  {Themann}}, \bibinfo {author} {\bibfnamefont {Y.}~\bibnamefont {Torun}},
  \bibinfo {author} {\bibfnamefont {S.}~\bibnamefont {Trentalange}}, \bibinfo
  {author} {\bibfnamefont {O.}~\bibnamefont {Tsai}}, \bibinfo {author}
  {\bibfnamefont {K.~V.}\ \bibnamefont {Tsang}}, \bibinfo {author}
  {\bibfnamefont {R.~H.~M.}\ \bibnamefont {Tsang}}, \bibinfo {author}
  {\bibfnamefont {C.}~\bibnamefont {Tull}}, \bibinfo {author} {\bibfnamefont
  {B.}~\bibnamefont {Viren}}, \bibinfo {author} {\bibfnamefont
  {S.}~\bibnamefont {Virostek}}, \bibinfo {author} {\bibfnamefont
  {V.}~\bibnamefont {Vorobel}}, \bibinfo {author} {\bibfnamefont {C.~H.}\
  \bibnamefont {Wang}}, \bibinfo {author} {\bibfnamefont {L.~S.}\ \bibnamefont
  {Wang}}, \bibinfo {author} {\bibfnamefont {L.~Y.}\ \bibnamefont {Wang}},
  \bibinfo {author} {\bibfnamefont {L.~Z.}\ \bibnamefont {Wang}}, \bibinfo
  {author} {\bibfnamefont {M.}~\bibnamefont {Wang}}, \bibinfo {author}
  {\bibfnamefont {N.~Y.}\ \bibnamefont {Wang}}, \bibinfo {author}
  {\bibfnamefont {R.~G.}\ \bibnamefont {Wang}}, \bibinfo {author}
  {\bibfnamefont {T.}~\bibnamefont {Wang}}, \bibinfo {author} {\bibfnamefont
  {W.}~\bibnamefont {Wang}}, \bibinfo {author} {\bibfnamefont {X.}~\bibnamefont
  {Wang}}, \bibinfo {author} {\bibfnamefont {X.}~\bibnamefont {Wang}}, \bibinfo
  {author} {\bibfnamefont {Y.~F.}\ \bibnamefont {Wang}}, \bibinfo {author}
  {\bibfnamefont {Z.}~\bibnamefont {Wang}}, \bibinfo {author} {\bibfnamefont
  {Z.}~\bibnamefont {Wang}}, \bibinfo {author} {\bibfnamefont {Z.~M.}\
  \bibnamefont {Wang}}, \bibinfo {author} {\bibfnamefont {D.~M.}\ \bibnamefont
  {Webber}}, \bibinfo {author} {\bibfnamefont {Y.~D.}\ \bibnamefont {Wei}},
  \bibinfo {author} {\bibfnamefont {L.~J.}\ \bibnamefont {Wen}}, \bibinfo
  {author} {\bibfnamefont {D.~L.}\ \bibnamefont {Wenman}}, \bibinfo {author}
  {\bibfnamefont {K.}~\bibnamefont {Whisnant}}, \bibinfo {author}
  {\bibfnamefont {C.~G.}\ \bibnamefont {White}}, \bibinfo {author}
  {\bibfnamefont {L.}~\bibnamefont {Whitehead}}, \bibinfo {author}
  {\bibfnamefont {C.~A.}\ \bibnamefont {Whitten}}, \bibinfo {author}
  {\bibfnamefont {J.}~\bibnamefont {Wilhelmi}}, \bibinfo {author}
  {\bibfnamefont {T.}~\bibnamefont {Wise}}, \bibinfo {author} {\bibfnamefont
  {H.~C.}\ \bibnamefont {Wong}}, \bibinfo {author} {\bibfnamefont {H.~L.~H.}\
  \bibnamefont {Wong}}, \bibinfo {author} {\bibfnamefont {J.}~\bibnamefont
  {Wong}}, \bibinfo {author} {\bibfnamefont {E.~T.}\ \bibnamefont {Worcester}},
  \bibinfo {author} {\bibfnamefont {F.~F.}\ \bibnamefont {Wu}}, \bibinfo
  {author} {\bibfnamefont {Q.}~\bibnamefont {Wu}}, \bibinfo {author}
  {\bibfnamefont {D.~M.}\ \bibnamefont {Xia}}, \bibinfo {author} {\bibfnamefont
  {S.~T.}\ \bibnamefont {Xiang}}, \bibinfo {author} {\bibfnamefont
  {Q.}~\bibnamefont {Xiao}}, \bibinfo {author} {\bibfnamefont {Z.~Z.}\
  \bibnamefont {Xing}}, \bibinfo {author} {\bibfnamefont {G.}~\bibnamefont
  {Xu}}, \bibinfo {author} {\bibfnamefont {J.}~\bibnamefont {Xu}}, \bibinfo
  {author} {\bibfnamefont {J.}~\bibnamefont {Xu}}, \bibinfo {author}
  {\bibfnamefont {J.~L.}\ \bibnamefont {Xu}}, \bibinfo {author} {\bibfnamefont
  {W.}~\bibnamefont {Xu}}, \bibinfo {author} {\bibfnamefont {Y.}~\bibnamefont
  {Xu}}, \bibinfo {author} {\bibfnamefont {T.}~\bibnamefont {Xue}}, \bibinfo
  {author} {\bibfnamefont {C.~G.}\ \bibnamefont {Yang}}, \bibinfo {author}
  {\bibfnamefont {L.}~\bibnamefont {Yang}}, \bibinfo {author} {\bibfnamefont
  {M.}~\bibnamefont {Ye}}, \bibinfo {author} {\bibfnamefont {M.}~\bibnamefont
  {Yeh}}, \bibinfo {author} {\bibfnamefont {Y.~S.}\ \bibnamefont {Yeh}},
  \bibinfo {author} {\bibfnamefont {K.}~\bibnamefont {Yip}}, \bibinfo {author}
  {\bibfnamefont {B.~L.}\ \bibnamefont {Young}}, \bibinfo {author}
  {\bibfnamefont {Z.~Y.}\ \bibnamefont {Yu}}, \bibinfo {author} {\bibfnamefont
  {L.}~\bibnamefont {Zhan}}, \bibinfo {author} {\bibfnamefont {C.}~\bibnamefont
  {Zhang}}, \bibinfo {author} {\bibfnamefont {F.~H.}\ \bibnamefont {Zhang}},
  \bibinfo {author} {\bibfnamefont {J.~W.}\ \bibnamefont {Zhang}}, \bibinfo
  {author} {\bibfnamefont {Q.~M.}\ \bibnamefont {Zhang}}, \bibinfo {author}
  {\bibfnamefont {K.}~\bibnamefont {Zhang}}, \bibinfo {author} {\bibfnamefont
  {Q.~X.}\ \bibnamefont {Zhang}}, \bibinfo {author} {\bibfnamefont {S.~H.}\
  \bibnamefont {Zhang}}, \bibinfo {author} {\bibfnamefont {Y.~C.}\ \bibnamefont
  {Zhang}}, \bibinfo {author} {\bibfnamefont {Y.~H.}\ \bibnamefont {Zhang}},
  \bibinfo {author} {\bibfnamefont {Y.~X.}\ \bibnamefont {Zhang}}, \bibinfo
  {author} {\bibfnamefont {Z.~J.}\ \bibnamefont {Zhang}}, \bibinfo {author}
  {\bibfnamefont {Z.~P.}\ \bibnamefont {Zhang}}, \bibinfo {author}
  {\bibfnamefont {Z.~Y.}\ \bibnamefont {Zhang}}, \bibinfo {author}
  {\bibfnamefont {J.}~\bibnamefont {Zhao}}, \bibinfo {author} {\bibfnamefont
  {Q.~W.}\ \bibnamefont {Zhao}}, \bibinfo {author} {\bibfnamefont {Y.~B.}\
  \bibnamefont {Zhao}}, \bibinfo {author} {\bibfnamefont {L.}~\bibnamefont
  {Zheng}}, \bibinfo {author} {\bibfnamefont {W.~L.}\ \bibnamefont {Zhong}},
  \bibinfo {author} {\bibfnamefont {L.}~\bibnamefont {Zhou}}, \bibinfo {author}
  {\bibfnamefont {Z.~Y.}\ \bibnamefont {Zhou}}, \bibinfo {author}
  {\bibfnamefont {H.~L.}\ \bibnamefont {Zhuang}}, \ and\ \bibinfo {author}
  {\bibfnamefont {J.~H.}\ \bibnamefont {Zou}},\ }\href {\doibase
  10.1103/PhysRevLett.108.171803} {\bibfield  {journal} {\bibinfo  {journal}
  {Physical Review Letters}\ }\textbf {\bibinfo {volume} {108}},\ \bibinfo
  {pages} {171803} (\bibinfo {year} {2012})},\ \Eprint
  {http://arxiv.org/abs/1203.1669} {arXiv:1203.1669} \BibitemShut {NoStop}%
\bibitem [{\citenamefont {Ahn}\ \emph {et~al.}(2012)\citenamefont {Ahn},
  \citenamefont {Chebotaryov}, \citenamefont {Choi}, \citenamefont {Choi},
  \citenamefont {Choi}, \citenamefont {Choi}, \citenamefont {Jang},
  \citenamefont {Jang}, \citenamefont {Jeon}, \citenamefont {Jeong},
  \citenamefont {Joo}, \citenamefont {Kim}, \citenamefont {Kim}, \citenamefont
  {Kim}, \citenamefont {Kim}, \citenamefont {Kim}, \citenamefont {Kim},
  \citenamefont {Kim}, \citenamefont {Kim}, \citenamefont {Kim}, \citenamefont
  {Lee}, \citenamefont {Lee}, \citenamefont {Lim}, \citenamefont {Ma},
  \citenamefont {Pac}, \citenamefont {Park}, \citenamefont {Park},
  \citenamefont {Park}, \citenamefont {Shin}, \citenamefont {Siyeon},
  \citenamefont {Yang}, \citenamefont {Yeo}, \citenamefont {Yi},\ and\
  \citenamefont {Yu}}]{Ahn2012}%
  \BibitemOpen
  \bibfield  {author} {\bibinfo {author} {\bibfnamefont {J.~K.}\ \bibnamefont
  {Ahn}}, \bibinfo {author} {\bibfnamefont {S.}~\bibnamefont {Chebotaryov}},
  \bibinfo {author} {\bibfnamefont {J.~H.}\ \bibnamefont {Choi}}, \bibinfo
  {author} {\bibfnamefont {S.}~\bibnamefont {Choi}}, \bibinfo {author}
  {\bibfnamefont {W.}~\bibnamefont {Choi}}, \bibinfo {author} {\bibfnamefont
  {Y.}~\bibnamefont {Choi}}, \bibinfo {author} {\bibfnamefont {H.~I.}\
  \bibnamefont {Jang}}, \bibinfo {author} {\bibfnamefont {J.~S.}\ \bibnamefont
  {Jang}}, \bibinfo {author} {\bibfnamefont {E.~J.}\ \bibnamefont {Jeon}},
  \bibinfo {author} {\bibfnamefont {I.~S.}\ \bibnamefont {Jeong}}, \bibinfo
  {author} {\bibfnamefont {K.~K.}\ \bibnamefont {Joo}}, \bibinfo {author}
  {\bibfnamefont {B.~C.~R.}\ \bibnamefont {Kim}}, \bibinfo {author}
  {\bibfnamefont {B.~C.~R.}\ \bibnamefont {Kim}}, \bibinfo {author}
  {\bibfnamefont {H.~S.}\ \bibnamefont {Kim}}, \bibinfo {author} {\bibfnamefont
  {J.~Y.}\ \bibnamefont {Kim}}, \bibinfo {author} {\bibfnamefont {S.~H. B.~Y.}\
  \bibnamefont {Kim}}, \bibinfo {author} {\bibfnamefont {S.~H. B.~Y.}\
  \bibnamefont {Kim}}, \bibinfo {author} {\bibfnamefont {S.~H. B.~Y.}\
  \bibnamefont {Kim}}, \bibinfo {author} {\bibfnamefont {W.}~\bibnamefont
  {Kim}}, \bibinfo {author} {\bibfnamefont {Y.~D.}\ \bibnamefont {Kim}},
  \bibinfo {author} {\bibfnamefont {J.~K.}\ \bibnamefont {Lee}}, \bibinfo
  {author} {\bibfnamefont {J.~K.}\ \bibnamefont {Lee}}, \bibinfo {author}
  {\bibfnamefont {I.~T.}\ \bibnamefont {Lim}}, \bibinfo {author} {\bibfnamefont
  {K.~J.}\ \bibnamefont {Ma}}, \bibinfo {author} {\bibfnamefont {M.~Y.}\
  \bibnamefont {Pac}}, \bibinfo {author} {\bibfnamefont {I.~G.}\ \bibnamefont
  {Park}}, \bibinfo {author} {\bibfnamefont {J.~S.}\ \bibnamefont {Park}},
  \bibinfo {author} {\bibfnamefont {K.~S.}\ \bibnamefont {Park}}, \bibinfo
  {author} {\bibfnamefont {J.~W.}\ \bibnamefont {Shin}}, \bibinfo {author}
  {\bibfnamefont {K.}~\bibnamefont {Siyeon}}, \bibinfo {author} {\bibfnamefont
  {B.~S.}\ \bibnamefont {Yang}}, \bibinfo {author} {\bibfnamefont {I.~S.}\
  \bibnamefont {Yeo}}, \bibinfo {author} {\bibfnamefont {S.~H.}\ \bibnamefont
  {Yi}}, \ and\ \bibinfo {author} {\bibfnamefont {I.}~\bibnamefont {Yu}},\
  }\href {\doibase 10.1103/PhysRevLett.108.191802} {\bibfield  {journal}
  {\bibinfo  {journal} {Physical Review Letters}\ }\textbf {\bibinfo {volume}
  {108}},\ \bibinfo {pages} {191802} (\bibinfo {year} {2012})},\ \Eprint
  {http://arxiv.org/abs/1204.0626} {arXiv:1204.0626} \BibitemShut {NoStop}%
\end{thebibliography}%

\end{document}